\documentclass[%
reprint,
superscriptaddress,
amsmath,amssymb,
aps,
prb,
]{revtex4-1}

\usepackage{graphicx}
\usepackage{dcolumn}
\usepackage{bm}
\usepackage{mathtools}
\usepackage{bbold}

\begin{document}

\preprint{}

\title{Revisiting Quasiparticle Scattering Interference in High-Temperature Superconductors: The Problem of Narrow Peaks}

\author{Miguel Antonio Sulangi}
\affiliation{%
	Instituut-Lorentz for Theoretical Physics, Leiden University, Leiden, Netherlands 2333 CA
}
\author{Milan P. Allan}
 \affiliation{
 	Leiden Institute of Physics, Leiden University, Leiden, Netherlands 2333 CA}
\author{Jan Zaanen}%
\affiliation{%
	Instituut-Lorentz for Theoretical Physics, Leiden University, Leiden, Netherlands 2333 CA
}

\date{\today}

\begin{abstract}

We revisit the interpretation of quasiparticle scattering interference in cuprate high-$T_c$ superconductors. This phenomenon has been very successful in reconstructing the dispersions of $d$-wave Bogoliubov excitations, but the successful identification and interpretation of QPI in scanning tunneling spectroscopy (STS) experiments rely on theoretical results obtained for the case of isolated impurities. We introduce a highly flexible technique to simulate STS measurements by computing the local density of states using real-space Green's functions defined on two-dimensional lattices with as many as 100,000 sites.  We focus on the following question: to what extent can the experimental results be reproduced when various forms of distributed disorder are present?  We consider randomly distributed point-like impurities, smooth ``Coulombic'' disorder, and disorder arising from random on-site energies and superconducting gaps. We find an apparent paradox: the QPI peaks in the Fourier-transformed local density of states appear to be sharper and better defined in experiment than those seen in our simulations. We arrive at a no-go result for smooth-potential disorder since this does not reproduce the QPI peaks associated with large-momentum scattering.  An ensemble of point-like impurities gets closest to experiment, but this goes hand in hand with impurity cores that are not seen in experiment. We also study the effects of possible measurement artifacts (the ``fork mechanism''), which turn out to be of relatively minor consequence. It appears that a more microscopic model of the tunneling process needs to be incorporated in order to account for the sharpness of the experimentally-obtained QPI peaks. 
\end{abstract}

\maketitle

\section{\label{sec:level1}Introduction}

Scanning tunneling spectroscopy (STS) has matured into one of the most powerful techniques for studying complex electron systems. It has been most successful in the study of high-$T_c$ superconductors, where it has revealed a spectacular array of new phenomena to be present in the cuprates.\cite{schmidt2011electronic} Prominent examples of such phenomena include ordering in the pseudogap\cite{vershinin2004local,hanaguri2004checkerboard,kohsaka2007intrinsic,lawler2010intra}, inhomogeneities in the superconducting gap and pseudogap\cite{lang2002imaging,fang2004periodic,mcelroy2005atomic}, and quasiparticle interference (QPI)\cite{hoffman2002imaging,mcelroy2003relating}. 

\begin{figure}[ht]
	\centering
	\includegraphics[height=.3\textwidth]{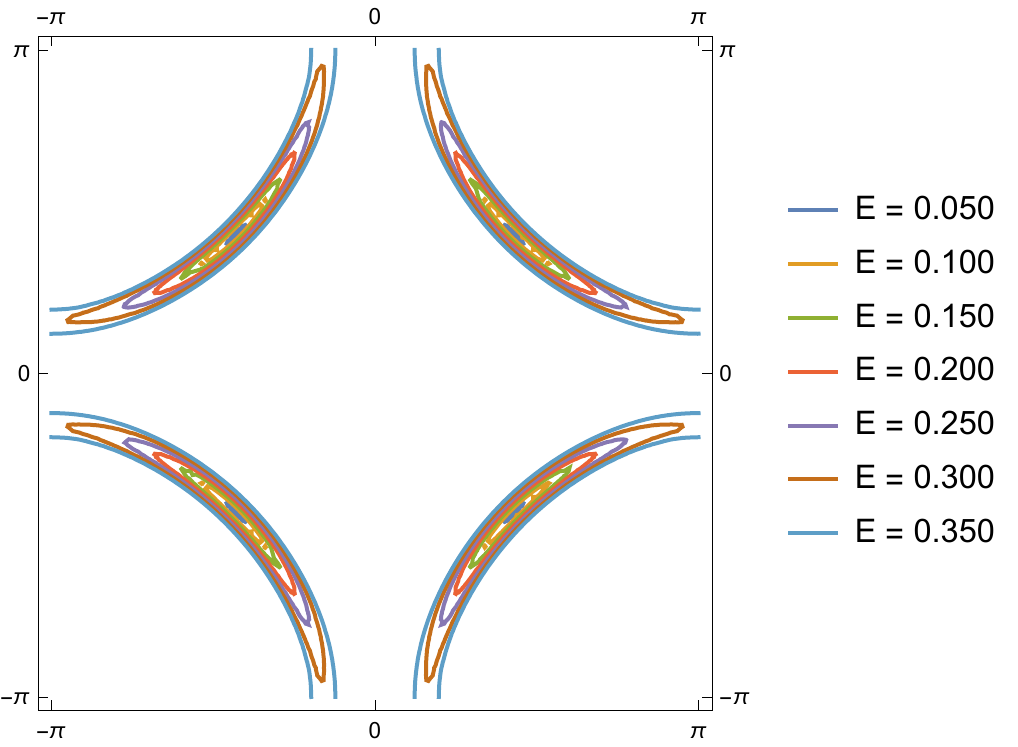}\hfill
	\caption{Contours of constant energy for a $d$-wave superconductor for different energies $E$, in units where $t = 1$. Observe that energies from $E = 0.050$ to $E = 0.300$ feature closed, banana-shaped CCEs, while for higher energies such as $E = 0.350$ the CCE changes topology and becomes open.}
	\label{fig:CCE}
\end{figure}

\begin{figure}[ht]
	\centering
	\includegraphics[height=.3\textwidth]{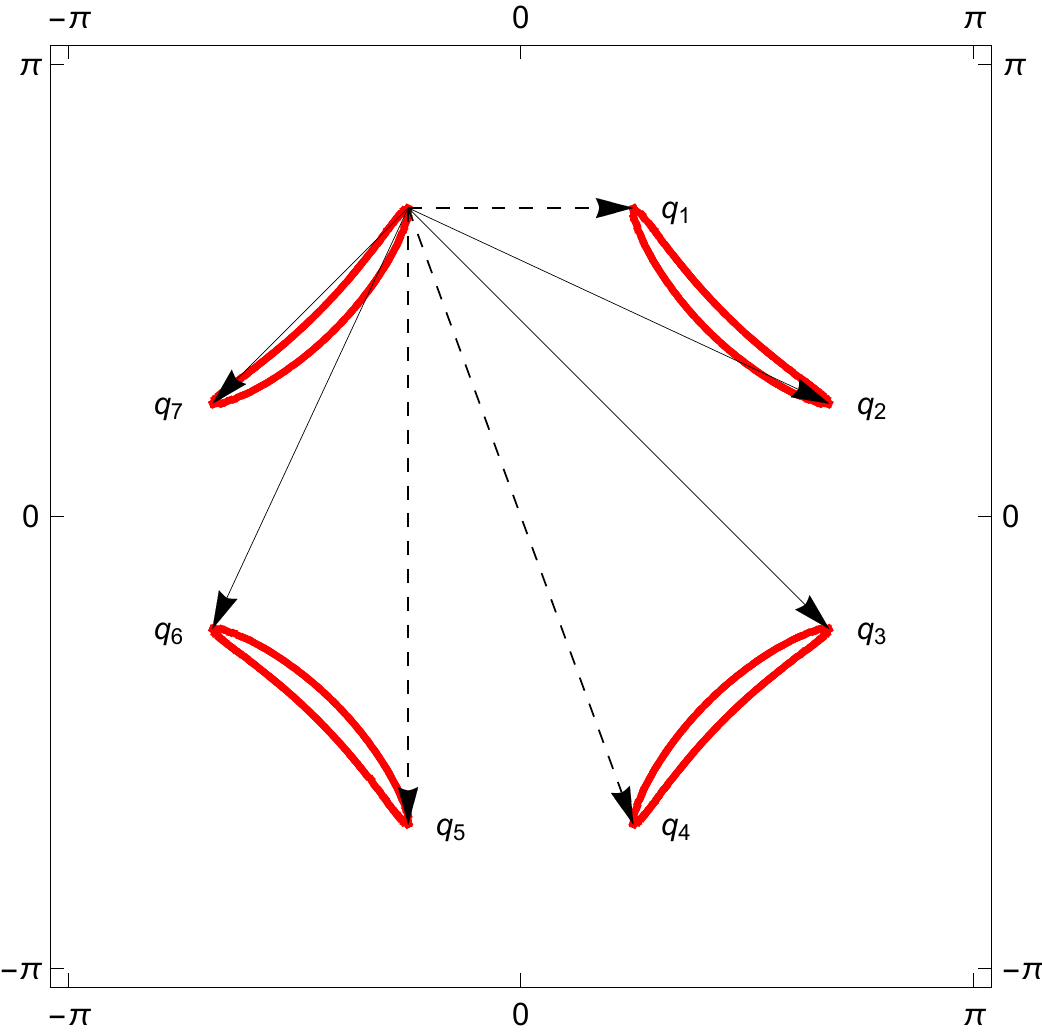}\hfill
	\caption{The octet model in $\mathbf{k}$-space. Shown are the seven wavevectors connecting one tip of a ``banana'' to another when $E = 0.200$. Dashed arrows denote wavevectors connecting states where the superconducting gap has the same sign, while undashed ones connect states where the gap changes sign. }
	\label{fig:octet}
\end{figure}

\begin{figure}[ht]
	\centering
	\includegraphics[height=.3\textwidth]{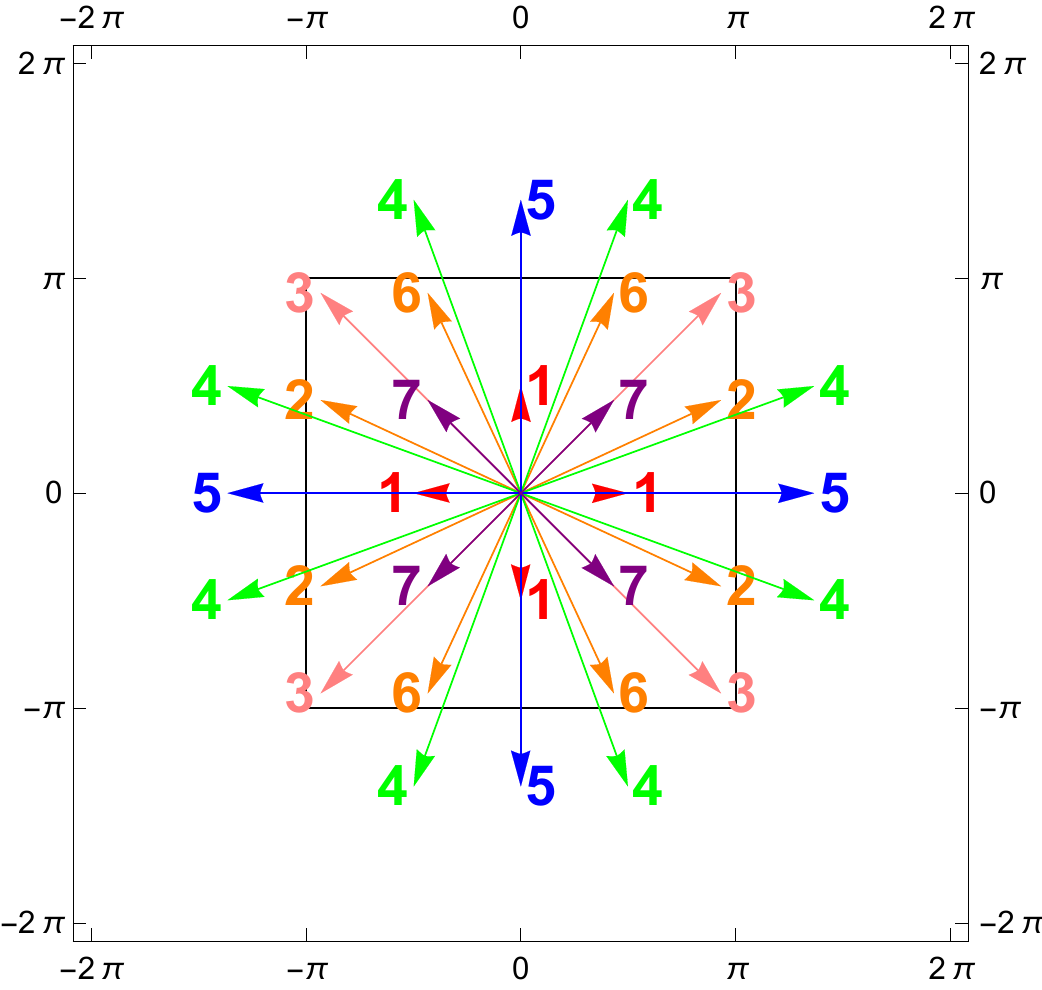}\hfill
	\caption{Locations of the special $\mathbf{q_i}$ wavevectors in extended $\mathbf{q}$-space. The energy is $E  = 0.200$, same as in Fig.~\ref{fig:octet}. The octet model predicts that peaks in the Fourier-transformed LDOS will be present at these locations. A square demarcating the boundary of the first Brillouin zone (\emph{i.e.}, $-\pi \leq q_x, q_y \leq \pi$) is shown. Note that certain wavevectors (in this particular case, $\mathbf{q_4}$ and $\mathbf{q_5}$) may extend beyond the first Brillouin zone. In our lattice simulations these peaks will be folded back into the first Brillouin zone.}
	\label{fig:qspace}
\end{figure}

Here we wish to revisit the interpretation of the QPI phenomenon. This was first observed in the cuprates when STS measurements done on superconducting Bi$_2$Sr$_2$CaCu$_2$O$_{8+\delta}$ found that spatial modulations in the local density of states (LDOS) were present in the real-space maps. A particular category of these modulations is found to be \emph{incommensurate} and, more importantly, \emph{dispersive}---that is, the wavevector peaks in the Fourier power spectrum corresponding to these modulations are found to be energy-dependent.\cite{hoffman2002imaging,mcelroy2003relating,kohsaka2008cooper} In the underdoped regime, these coexist with peaks which are non-dispersing and are attributed to the presence of ``stripy'' charge-density-wave order\cite{howald2003coexistence, howald2003periodic} or an electronic glass\cite{kohsaka2007intrinsic}. In a remarkable advance, these were explained in a series of papers laying out the theory as understood for a single point-like scatterer.\cite{hoffman2002imaging, wang2003quasiparticle, capriotti2003wave} In essence, the effect can be understood in terms of interference fringes associated with the coherent Bogoliubov quasiparticles of the $d$-wave superconductor, which behave like quantum-mechanical waves that diffract in the presence of quenched disorder.\cite{zaanen2003superconductivity} Given their quasi-relativistic dispersion, this scattering is strongly enhanced at wavevectors associated with the extrema of the dispersions at a given energy. This is illustrated in Figs.~\ref{fig:CCE} and~\ref{fig:octet}. With increasing energy, the contours of constant energy (CCEs) of the Bogoliubov excitations in momentum space change shape (Fig.~\ref{fig:CCE}). The scattering is strongly enhanced at the tips of the banana-shaped contours (Fig.~\ref{fig:octet}), defining an octet of characteristic momenta.  Upon Fourier-transforming the real-space STS maps, one finds peaks at these momenta, which disperse as function of energy (Figs.~\ref{fig:qspace} and \ref{fig:dispersion}). This forms a set of data that allows one to reconstruct the dispersion relations of the Bogoliubov quasiparticles. These are strikingly consistent with results from ARPES, where these single-particle dispersions are measured directly in momentum space. It is beyond doubt that this ``octet model'' interpretation is correct for the cuprates, especially as additional evidence for QPI has also been obtained from Ca$_{2 - x}$Na$_{x}$CuO$_{2}$Cl$_{2}$.\cite{hanaguri2007quasiparticle} The effect has also been observed in iron-based superconductors\cite{allan2012anisotropic,allan2013anisotropic,allan2015identifying} and heavy-fermion materials\cite{lee2009heavy,schmidt2010imaging,allan2013imaging,van2014direct}. The success of the octet model has spurred a considerable amount of theoretical work on the signatures of QPI in related states of matter such as the pseudogap phase of the cuprates,\cite{pereg2003theory,pereg2005quasiparticle,misra2004failure,bena2004quasiparticle} as well as in systems without a gap, such as graphite\cite{bena2005quasiparticle} and the surface states of three-dimensional topological insulators\cite{fu2009hexagonal,roushan2009topological,guo2010theory}. The ubiquity of QPI in gapless systems is not surprising, as its signatures were in fact first imaged in conventional metals.\cite{crommie1993imaging,sprunger1997giant,hofmann1997anisotropic,petersen1998direct}


The octet model is simply a kinematical picture describing the scattering of quasiparticles in the presence of disorder. It is another matter to explain how well-defined patterns of QPIs can arise under realistic conditions. This was intensely studied theoretically, at first starting from models describing $d$-wave fermions scattering from a single isolated impurity potential.\cite{wang2003quasiparticle, capriotti2003wave,zhu2004power,nunner2006fourier,vishik2009momentum,kreisel2015interpretation}  In Section \ref{point}, we will reproduce a typical result involving a single point scatterer. One infers from the results that there is an overall similarity between these theoretical results and the experimental data. However, even on a qualitative level it is not completely satisfactory. In our numerically obtained Fourier-space maps, the ``peaks'' are actually associated with intensity enhancements of intersecting diffuse streaks and blurry regions. In contrast, the experimental QPI signals are remarkably well-defined \emph{peaks}. 

A caveat is that microscopic details do matter when taking into account the actual measurement process involved in STS experiments. This was anticipated early on by the observation that the mismatch between the $s$-wave orbital emanating from the tunneling tip and the microscopic $d_{x^2-y^2}$ copper-centred orbitals in the perovskite planes implies that the tunneling current enters the \emph{nearest neighbors} of the copper site over which the tip is positioned.\cite{martin2002impurity} This ``fork mechanism''  was recently confirmed by an impressive first-principles model of the tunneling process.\cite{kreisel2015interpretation} We will study the effects of this ``fork'' on the QPIs in Section \ref{point}. We will find that this is actually only a minor concern for the overall interpretation. Kreisel \emph{et al.} also find that modifications coming from a realistic description of the tunneling process have the potential to resolve the apparent paradox that we will demonstrate. We will come back to this issue at the end of this paper. 

The serious problem with the point-like scatterer model lies in its inconsistency with the actual chemistry of the cuprates. Point-like impurities are naturally explained in terms of substitional defects in the cuprate planes. However the CuO$_2$ planes are well-established to be very clean with regard to their stoichiometry. In fact, zinc and nickel can be substituted for copper in the CuO$_2$ planes. Since such chemical defects correspond to strong potentials, this gives rise to a major modification of the electronic structure at the impurity core. This is indeed seen in STS, as the zinc impurities show up very prominently in the LDOS maps of zinc-doped BSCCO.\cite{pan2000imaging,balatsky2006impurity} The details of these core states were in fact instrumental in identifying the ``fork'' mechanism.\cite{martin2002impurity, kreisel2015interpretation} Nickel impurities were found to be similarly visible in the case of nickel-doped BSCCO, the difference in this case being that nickel impurities are magnetic scatterers.\cite{hudson2001interplay}  On the other hand, the STS spectra of pristine cuprates do not show any of these localized impurity states. 

Instead, it appears that disorder in the cuprates should be of a more distributed and smooth kind. Doping occurs away from the CuO$_2$ planes. These are charged impurities, and given the poor screening along the $c$-axis, one then expects smooth, Coulombic disorder, similar to what is realized in modulation doping of semiconductors.\cite{pan2001microscopic} Such off-plane dopants have indeed been imaged in STS experiments on BSCCO.\cite{mcelroy2005atomic} Similarly, dopants might modulate the tilting patterns in the CuO$_2$ planes, resulting in a similar form of distributed disorder.\cite{eisaki2004effect} This involves \emph{inherently many-impurity} effects that are not easy to study using the standard single-impurity $T$-matrix method. We note that multiple point-like impurities have indeed been considered before in the literature.\cite{zhu2004power, capriotti2003wave, atkinson2000details} However, the most general many-impurity problem is technically very demanding, especially when one tries to consider forms of disorder other than point impurities, or when one tries to scale up the system size. 

Given these difficulties, we take advantage of an alternative numerical method to directly compute the LDOS, inspired by methods heavily in use in the quantum transport community. This is outlined in Section \ref{model}. Our point of departure is a tight-binding Hamiltonian on a square lattice describing a $d$-wave superconductor. Instead of diagonalizing this real-space Hamiltonian, we compute the Green's function directly by \emph{inverting} the Hamiltonian, which can be done efficiently, and from the Green's function we obtain the LDOS. Superconducting gap functions and even full self-energies can be straightfowardly incorporated. Any form of spatial inhomogeneities can be modeled efficiently using this method, and our system sizes can be made very large---for instance, LDOS maps of systems with size $1000 \times 120$, which we use, can be obtained in a matter of minutes---the better to approach the same large field of view as current experiments have. We originally aimed to use this to study more complex phenomena such as the gap inhomogeneities (``quantum mayonnaise'') found in the pseudogap regime, as well as the effects of the electronic self-energies on STS results.\cite{dalla2016friedel} However, we found out that issues arise already on the most fundamental level of the theory of QPI deep in the superconducting state of the cuprates, which is the subject of this paper. 

Using this method, we can address any conceivable form of spatial disorder and study its effects on the QPI spectra. We set the stage in Section \ref{point}, focusing on the case of a single weak point-like impurity. We then insert a large number of such weak point-like impurities at random positions and examine QPI with and without the filter effect. We then examine in detail the related case where many \emph{unitary} scatterers are present.  We next turn our attention to a single Coulombic impurity and subsequently to a densely distributed random ensemble of such smooth scatterers. Although the real-space patterns appear to be suggestively similar to the stripe-like textures seen in experiment, this runs into a very serious problem: the peaks in the power spectra involving large momenta disappear very rapidly, and this holds even if the range of the potential is shortened. We consider then the case of a random on-site potential, similar to Anderson's model of disorder.  Although the effects of quasiparticle scattering interference can indeed be seen in the real-space and Fourier-transformed maps, this form of disorder results in power spectra which show considerable fuzziness, in contrast to the well-defined peaks seen in experiment. We end by considering a simple model of superconducting gap disorder. Although this works quite well for the simplified case we consider, the problem is that, for more realistic smooth gap inhomogenieties, large-momenta peaks will be suppressed. 

By eyeballing the numerous plots present in this paper, the reader may already have convinced himself or herself that there is a serious problem with the standard explanation of QPIs. By making the model of disorder more and more realistic, the correspondence with experiment deteriorates. As we will discuss in the final section, it is an interesting open challenge to explain the sharpness of the QPI peaks as seen in STS measurements.

\section{Model and Methods} \label{model}

Two important requirements in theoretically reproducing results from STS experiments are large system sizes and the ability to model general forms of inhomogeneities. Modern STS experiments feature a very large field of view, which allows large-scale inhomogeneities present in materials to be visualized. Replicating this large field of view numerically is a challenge because simulations with large system sizes require sizable amounts of computational effort. Most numerical work on disordered high-temperature superconductors has centered around two methods: the $T$-matrix method and exact diagonalization. The $T$-matrix approach has the advantage of being exact for the case of point-like impurities and requires minimal numerical effort, even for large system sizes, but is restricted in its applicability---smooth potential scatterers, for instance, are not accessible in this formalism.  On the other hand, exact diagonalization allows any form of disorder to be modeled, but at the expense of being restricted to relatively small system sizes.

In this paper we utilize a method---a novel one as far as its application to both disordered $d$-wave superconductors and the modeling of STS experiments is concerned---that is formally \emph{exact},  allows any form of disorder to be modeled, gives access to very large system sizes, and is computationally efficient. In addition, since it is based on Green's functions, it is straightforward to include the effects of self-energies; this will be the subject of an upcoming paper.\cite{sulangi2017upcoming} Before introducing the method, we will first discuss the lattice model of the cuprates that we will use in this paper.  Our starting point is the following tight-binding Hamiltonian for a $d$-wave superconductor on a square lattice:
\begin{equation}
H = \sum_{\langle i, j \rangle} \Big[ -\sum_{\sigma} t_{ij}c_{i\sigma}^{\dagger}c_{j\sigma} + \Delta_{ij}c_{i \uparrow}^{\dagger}c_{j \downarrow}^{\dagger} + \Delta_{ij}^{\ast}c_{i \uparrow}c_{j \downarrow} \Big].
\end{equation}
We include nearest-neighbor and next-nearest-neighbor hopping (specified by the amplitudes $t$ and $t'$, respectively) and a chemical potential $\mu$. $d$-wave pairing is incorporated by ensuring that the gap function has the form $\Delta_{ij} = \pm\Delta_0$, where $(i,j)$ are two nearest-neighbor sites and the positive and negative values of $\Delta_{ij}$ are chosen for pairs of sites along the $x$- and $y$-directions, respectively. This is a mean-field Hamiltonian for the $d$-wave superconducting state of the cuprates. We set the lattice spacing $a = 1$ and the nearest-neighbor hopping $t = 1$---\emph{i.e.}, we will thus measure all energies in units of $t$. 

In the clean limit, the Hamiltonian can be diagonalized by going to momentum space. The quasiparticle energies are given by 
\begin{equation}
E(\mathbf{k}) = \sqrt{\epsilon^2_k + \Delta_k^2},
\label{eq:CCE}
\end{equation}
where 
\begin{equation}
\epsilon_k = -2t(\cos k_x + \cos k_y ) - 4t'\cos k_x \cos k_y  - \mu
\end{equation}
and
\begin{equation}
\Delta_k = 2\Delta_0(\cos k_x - \cos k_y).
\end{equation} 
Eq.~\ref{eq:CCE} describes the dispersion of the Bogoliubov quasiparticles of a $d$-wave superconductor. At $E = 0$ there are four points in momentum space at which zero-energy excitations exist. For the purposes of our calculations we take the band-structure and pairing parameters relative to $t = 1$ as $t' = -0.3$, $\mu = -0.8$, and $\Delta_0 = 0.08$ throughout this paper. We note that while these band-structure parameters cover hoppings only up to the next-nearest-neighbor level, we selected them to be close to the phenomenological values obtained by Norman \emph{et al}. for optimally-doped BSCCO.\cite{norman1995phenomenological} Our results will turn out not to depend sensitively on band-structure details.

\subsection{Green's Functions and the Local Density of States}
The central quantity of interest in our study is the local density of states (LDOS) of a superconductor in the presence of disorder. The LDOS at position $\mathbf{r}$ and energy $E$ can be expressed as 
\begin{equation}
\rho(\mathbf{r},E) = -\frac{1}{\pi}\operatorname{Im}G_{11}(\mathbf{r}, \mathbf{r}, E + i0^{+}),
\label{eq:LDOS}
\end{equation}
where $G$ is simply the full Green's function corresponding to $H$ in Nambu space, given by
\begin{equation}
G = (\omega \mathbb{1} - H)^{-1},
\end{equation}
and $G_{11}$ is the particle Green's function. One can observe from Eq.~\ref{eq:LDOS} that to obtain the LDOS  we do not need all the elements of $G$---the bare LDOS can be obtained from just the diagonal elements of $G$. (Note however that when we will come to include nontrivial tunneling processes, more elements of $G$ will be needed; this will be described in detail in the next subsection.)

We proceed by noting that $H$, in a real-space basis, can be written as a block tridiagonal matrix---\emph{without any approximations}---when periodic boundary conditions are imposed along the $y$-direction and open boundary conditions are placed along the $x$-direction. $H$ exhibits the following structure:
\begin{equation}
\mathbf{H} = 
 \left( \begin{array}{lllllllll}
\mathbf{a}_1 & \mathbf{b}_{1} & \mathbf{0}  & \mathbf{0} & \hdots& \mathbf{0} & \mathbf{0} \\
\mathbf{b}^{\dagger}_{1} & \mathbf{a}_2 & \mathbf{b}_{2}  & \mathbf{0} & \hdots  & \mathbf{0} & \mathbf{0}\\
\mathbf{0} & \mathbf{b}^{\dagger}_{2} & \mathbf{a}_3  & \mathbf{b}_{3} & \hdots & \mathbf{0} & \mathbf{0} \\
\mathbf{0} & \mathbf{0} & \mathbf{b}^{\dagger}_{3} & \mathbf{a}_4 & \hdots& \mathbf{0} & \mathbf{0} \\
\vdots & \vdots & \vdots & \vdots & \ddots & \mathbf{b}_{N_x - 2} & \mathbf{0} \\
\mathbf{0} & \mathbf{0} & \mathbf{0} & \mathbf{0} & \mathbf{b}^{\dagger}_{N_x - 2} &\mathbf{a}_{N_x - 1} & \mathbf{b}_{N_x  -1} \\
\mathbf{0} & \mathbf{0} & \mathbf{0} & \mathbf{0} & \mathbf{0} & \mathbf{b}^{\dagger}_{N_x - 1}& \mathbf{a}_{N_x}
\end{array} \right).
\end{equation}
$N_x$ and $N_y$ denote the number of sites in the $x$- and $y$-directions, respectively. $\mathbf{a}_i$ is a $2N_y \times 2N_y$ block containing all hoppings,  pairings, and on-site energies along the $y$-direction at the $i$th column. $\mathbf{b}_i$ meanwhile is a $2N_y \times 2N_y$ block that contains hopping and pairing terms along the $x$-direction between the $i$th and $(i+1)$th columns. 

By construction the inverse Green's function $G^{-1} = \omega \mathbb{1} - H$ is block tridiagonal as well. A well-known result states that one can obtain the diagonal blocks of $G$, and hence the LDOS, using the following block-by-block algorithm:\cite{godfrin1991method,reuter2012efficient,hod2006first}
\begin{equation}
\mathbf{G}_{i i} = [\omega\mathbf{1} - \mathbf{a}_i - \mathbf{C}_i - \mathbf{D}_i]^{-1}.
\end{equation}
$\mathbf{C}_i$ and $\mathbf{D}_i$ are calculated from the following expressions:
\begin{eqnarray}
\mathbf{C}_i = 
\begin{cases}
\mathbf{0} & \text{if}\ i = 1 \\
\mathbf{b}^{\dagger}_{i-1}[\omega\mathbf{1} - \mathbf{a}_{i-1} - \mathbf{C}_{i-1}]^{-1}\mathbf{b}_{i-1} & \text{if}\ 1 < i \leq N_x
\end{cases}
\end{eqnarray}
and
\begin{eqnarray}
\mathbf{D}_i = 
\begin{cases}
\mathbf{0} & \text{if}\ i = N_x \\
\mathbf{b}_{i}[\omega\mathbf{1} - \mathbf{a}_{i+1} - \mathbf{D}_{i+1}]^{-1}\mathbf{b}^{\dagger}_{i} & \text{if}\ 1 \leq i < N_x.
\end{cases}
\end{eqnarray}

This algorithm is very fast compared to full exact diagonalization. Taking into account the block matrix inversions needed, the computational complexity of this algorithm is $O(N_x N^3_y)$. This allows us to make $N_x$ very large without significantly impacting performance, and this results in reducing finite-size effects in that direction considerably. In contrast, because the complexity scales as the cube of the length along the $y$-direction, $N_y$ is taken to be considerably smaller than $N_x$. However, even in that case the scaling of the complexity with $N_y$ is still very favorable compared to other methods. $N_y$ in turn can be made much larger than the typical length of the system in exact diagonalization studies. We again reiterate that this procedure is exact---no approximations or truncations have been performed at any stage of the computation. Recursive techniques such as this, which make use of the sparsity of the Hamiltonian matrix, are very widely used in the quantum transport community to compute Green's functions.\cite{drouvelis2006parallel,petersen2008block,wimmer2009optimal,li2012extension,li2013fast,kuzmin2013fast}

We then obtain the LDOS of the full system from the diagonal blocks $\mathbf{G}_{ii}$ using Eq.~\ref{eq:LDOS}. For our computations we took $N_x = 1000$ and $N_y = 120$. The LDOS maps were then extracted from the middle $100 \times 100$ subsection of the system. We note that this $100 \times 100$ field of view is similar to what present-day STS measurements are capable of. While minor artifacts from the open boundary condition along the $x$-direction remain, the very large value of $N_x$ and taking the LDOS maps from the middlemost segment of the system combine to ensure that these effects are minimized. In obtaining the LDOS we used a small finite inverse quasiparticle lifetime given by $\eta = 0.01$, expressed in units of $t$.

The power spectrum can then be straightfowardly computed by performing a fast Fourier transform on the real-space maps. The quantity we are interested in is the amplitude of the Fourier-transformed maps, $|\rho(\mathbf{q}, E)|$. 

\subsection{Modeling the Measurement Process}

Our discussion beforehand neglected the specifics of the tunneling process between the tip and the CuO$_2$ plane. Here we will discuss how to incorporate the ``fork mechanism,'' an effective description of the tunneling process, in our computations. This mechanism was first proposed as an attempt to account for some inconsistencies between experimentally- and theoretically-obtained maps for zinc-doped BSCCO.\cite{pan2000imaging} The motivation was the observation that, for zinc-doped BSCCO, the LDOS maps show no suppression \emph{at} the impurity site, whereas theory predicts that maximal suppression should occur precisely there. One possibility is that some kind of filtering mechanism occurs when an electron tunnels from the STM tip to the copper-oxide plane. Martin \emph{et al.} argued that the tunneling matrix element is actually of a $d$-wave nature.\cite{martin2002impurity} Because the electron would have to tunnel through an insulating BiO layer before reaching the CuO$_2$ layer, the most dominant tunneling process involves \emph{nearest-neighbor} $3d_{x^2 - y^2}$ orbitals. The filtered LDOS at a site thus consists of a sum of the LDOS at the four nearest-neighbor sites \emph{and} multiple pairwise interference factors. Such a filtering mechanism has been put on rigorous footing in recent first-principles work.\cite{kreisel2015interpretation} 

Here we adopt the simplest form of the fork mechanism and recast this into the Green's function formalism we use in our computations. We introduce a filter function $f(\mathbf{r}, \mathbf{r'})$ which incorporates the tunneling matrix elements between the STM tip and the the CuO$_2$ plane. The filtered LDOS, $\rho_{f}(\mathbf{r})$, can therefore be expressed as a generalized convolution between the two-point Green's function $G$ and $f$:
\begin{eqnarray}
\rho_{f}(\mathbf{r}, E) = -\frac{1}{\pi}\operatorname{Im}\sum_{\mathbf{r_1, r_2}} f(\mathbf{r - r_1}, \mathbf{r - r_2}) \\ \times G_{11}(\mathbf{r_1}, \mathbf{r_2}, E + i0^{+}).
\label{eq:LDOSfiltered}
\end{eqnarray}

The filtering mechanism can be incorporated by a suitable choice of $f$. For instance, to have $s$-wave filtering (\emph{i.e.}, direct tunneling, which should result in the \emph{bare} LDOS), the filter function is simply given by
\begin{equation}
f(\mathbf{r}, \mathbf{r'}) = \delta_{\mathbf{r,0}}\delta_{\mathbf{r', 0}},
\end{equation} 
which would simply result in Eq.~\ref{eq:LDOS}. To have the desired $d$-wave fork effect, the following choice of $f$ is needed:
\begin{eqnarray}
f(\mathbf{r}, \mathbf{r'}) = (\delta_{\mathbf{r, \hat{x}}} + \delta_{\mathbf{r, -\hat{x}}} - \delta_{\mathbf{r, \hat{y}}} - \delta_{\mathbf{r, -\hat{y}}}) \nonumber \\
\times  (\delta_{\mathbf{r', \hat{x}}} + \delta_{\mathbf{r', -\hat{x}}} - \delta_{\mathbf{r', \hat{y}}} - \delta_{\mathbf{r', -\hat{y}}}).
\end{eqnarray}
Here $\mathbf{\hat{x}}$ and $\mathbf{\hat{y}}$ are unit vectors in the $x$- and $y$-directions, respectively.

Now we discuss how this is implemented in our computations. Observe that Eq.~\ref{eq:LDOSfiltered} with a $d$-wave filter has sixteen terms. This presents a complication in our block-by-block algorithm, because now we will have to obtain the first and second block diagonals above and below the main block diagonal. To be more precise, in addition to $\mathbf{G}_{ii}$, we will need the following eight other blocks to calculate $\rho_f(\mathbf{r}, E)$: $\mathbf{G}_{i-1,i-1}$, $\mathbf{G}_{i-1,i}$, $\mathbf{G}_{i-1,i+1}$, $\mathbf{G}_{i,i-1}$, $\mathbf{G}_{i,i+1}$, $\mathbf{G}_{i+1,i-1}$, $\mathbf{G}_{i+1,i}$, and $\mathbf{G}_{i+1,i+1}$. Fortunately all off-diagonal blocks are calculable recursively using the following expressions:\cite{godfrin1991method,reuter2012efficient}
\begin{eqnarray}
\mathbf{G}_{i j} = 
\begin{cases}
-[\omega\mathbf{1} - \mathbf{a}_i - \mathbf{D}_i]^{-1}\mathbf{b}^{\dagger}_{i-1}\mathbf{G}_{i-1,j} & \text{if}\ i > j, \\
-[\omega\mathbf{1} - \mathbf{a}_i - \mathbf{C}_i]^{-1}\mathbf{b}_{i}\mathbf{G}_{i+1,j} & \text{if}\ i < j.
\end{cases}
\end{eqnarray}
Here, $\mathbf{a}_i$, $\mathbf{b}_i$, $\mathbf{C}_i$, and $\mathbf{D}_i$ are defined in the same way as before.

\section{Point-like scatterers} \label{point}
\begin{figure}[t]
	\centering
	\includegraphics[width=.5\textwidth]{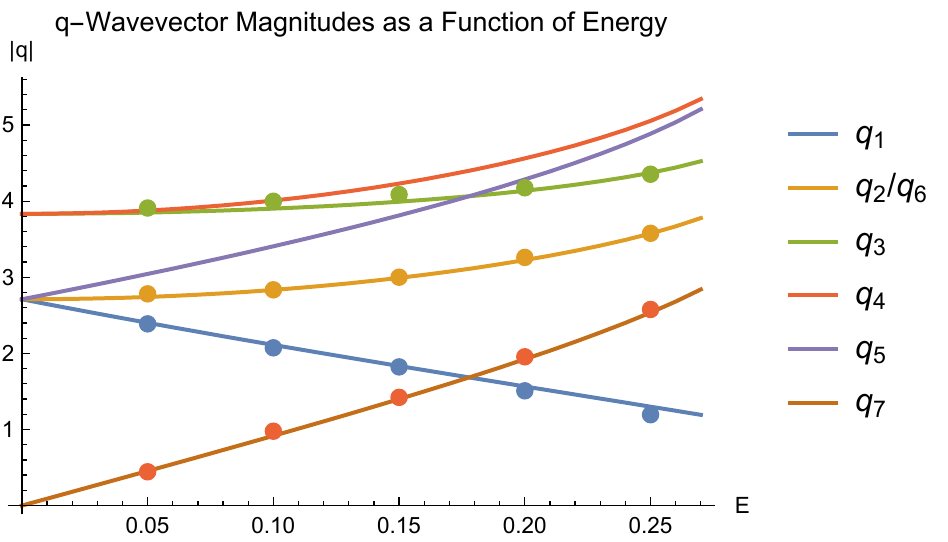}\hfill
	\caption{Plots of the magnitudes of the various $\mathbf{q}_i$ wavevectors as a function of energy $E$. Lines denote the expected dispersions of the $\mathbf{q}_i$ wavevectors as predicted by the octet model. Points show observed peaks for the case of a single weak point-like impurity with $V = 0.5$ at selected energies. Note that the dispersions for the large-wavevector peaks are shown without backfolding. We do not show peaks associated with $\mathbf{q}_4$ and $\mathbf{q}_5$, as these cannot be discerned clearly from the numerically-obtained power spectrum for a weak impurity. These dispersions are consistent with the behavior of peaks as observed in experiment.}
	\label{fig:dispersion}
\end{figure}

We first consider QPI arising from point-like impurities. This is by far the most comprehensively studied form of disorder in the cuprates. QPI was first understood theoretically by considering the effect of a single isolated impurity on the LDOS of the cuprates.\cite{wang2003quasiparticle,capriotti2003wave}  We revisit this single-impurity case first in order to lay down a reference template in the form of this well-known case to facilitate comparisons with new results. We will then turn to the case of many point-like impurities distributed randomly on the plane. 

The phenomenological octet model is an empirical success---in experiment one can clearly identify a set of seven dispersing peaks in the Fourier transform of the LDOS maps. Given the knowledge of the dispersion of the $d$-wave Bogoliubov quasiparticles, one can construct, for a given bias voltage,  contours of constant energy (CCEs) in the first Brillouin zone, which are given by solutions to Eq.~\ref{eq:CCE} for a given energy $E$. These CCEs are closed banana-shaped contours until $E$ is such that their tips reach the Brillouin zone boundary. Each of these four ``bananas'' is centered around a \emph{node}---\emph{i.e.}, one of four points along the normal-state Fermi surface where $\Delta_k$ vanishes. Plots of these CCEs with the parameters we set are shown in Fig.~\ref{fig:CCE}. Within the octet model, scattering processes from one tip of a banana to another become dominant, owing to the large joint density of states between any two such points. These dominant scattering processes manifest themselves in a set of visible peaks at seven characteristic momenta $\mathbf{q}_i$, with $i = 1, 2, \ldots 7$ in the power spectrum. These momenta are shown in Fig.~\ref{fig:octet}. 

Because the banana-shaped contours change their shape as $E$ changes, these $\mathbf{q}_i$'s should disperse; $|\mathbf{q}_7|$, for instance, should increase with increasing $|E|$. In Fig.~\ref{fig:dispersion} we reproduce the dispersions of the various $\mathbf{q}_i$ wavevectors as predicted by the octet model and compare them with peaks obtained from exact numerical calculations involving a single weak point-like scatterer. The expected dispersions are easily calculated from Eq.~\ref{eq:CCE}, making use of the fact that the density of states at energy $E$ is strongly enhanced by contributions at points in momentum space where $|\nabla_{\mathbf{k}}E|$ is a minimum, which are precisely at the tips of the ``bananas.''\cite{mcelroy2003relating} Here it can be seen that most of the peaks from our numerics match quite well with the predictions of the octet model.  The behavior of the peaks as one varies the energy matches very closely with what is seen in experiment.

\begin{figure*}[ht]
	\centering
	\includegraphics[width=.25\textwidth]{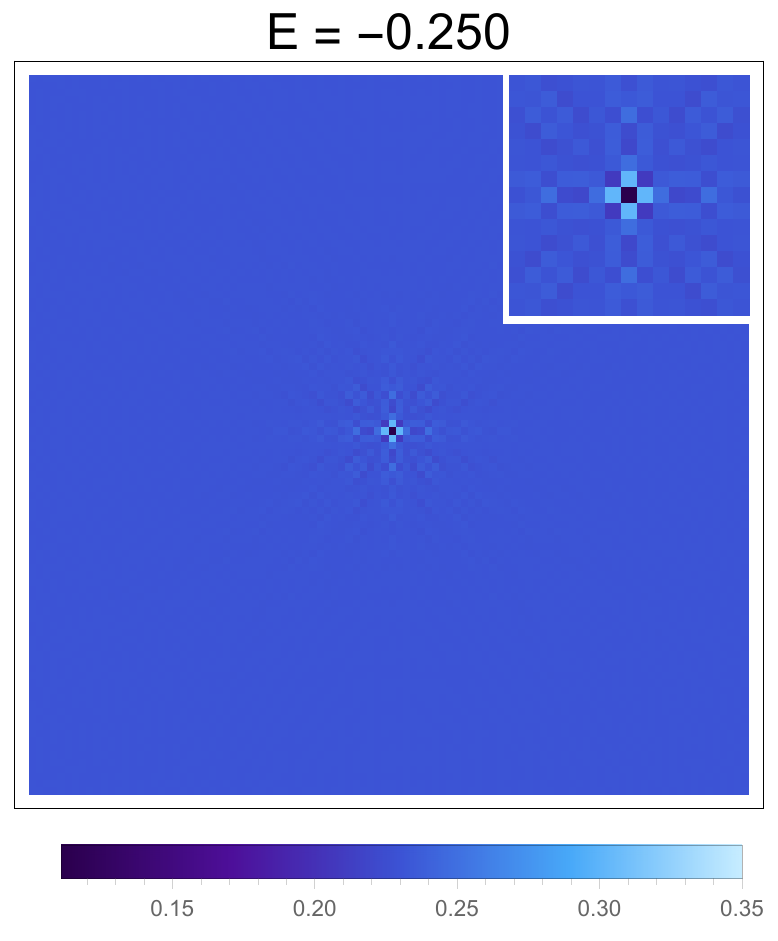}\hfill
	\includegraphics[width=.25\textwidth]{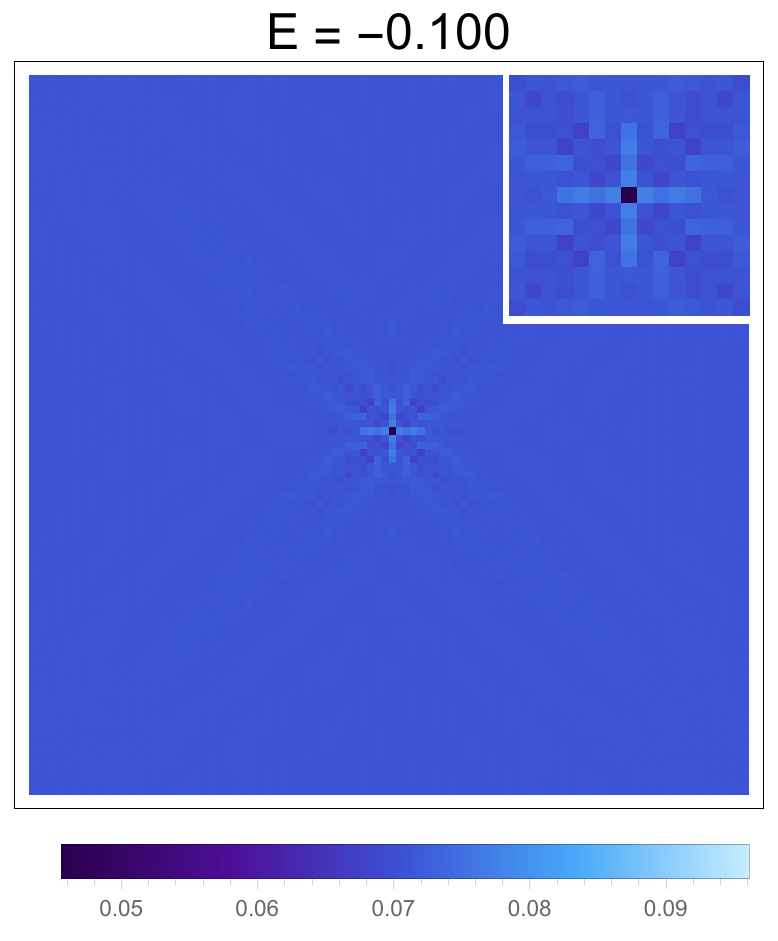}\hfill
	\includegraphics[width=.25\textwidth]{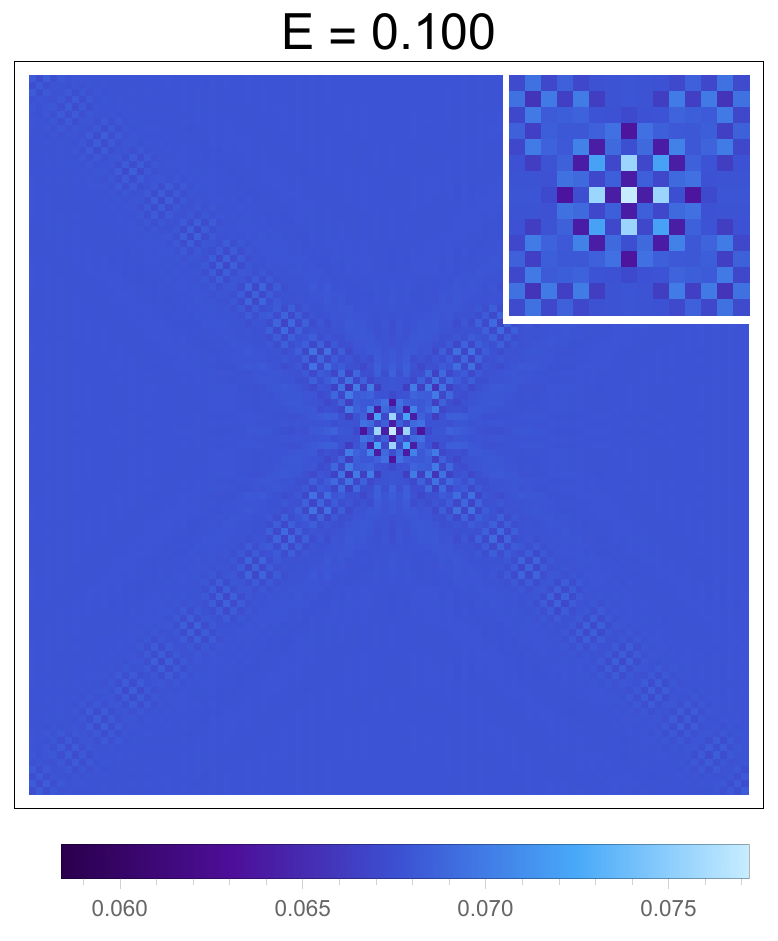}\hfill
	\includegraphics[width=.25\textwidth]{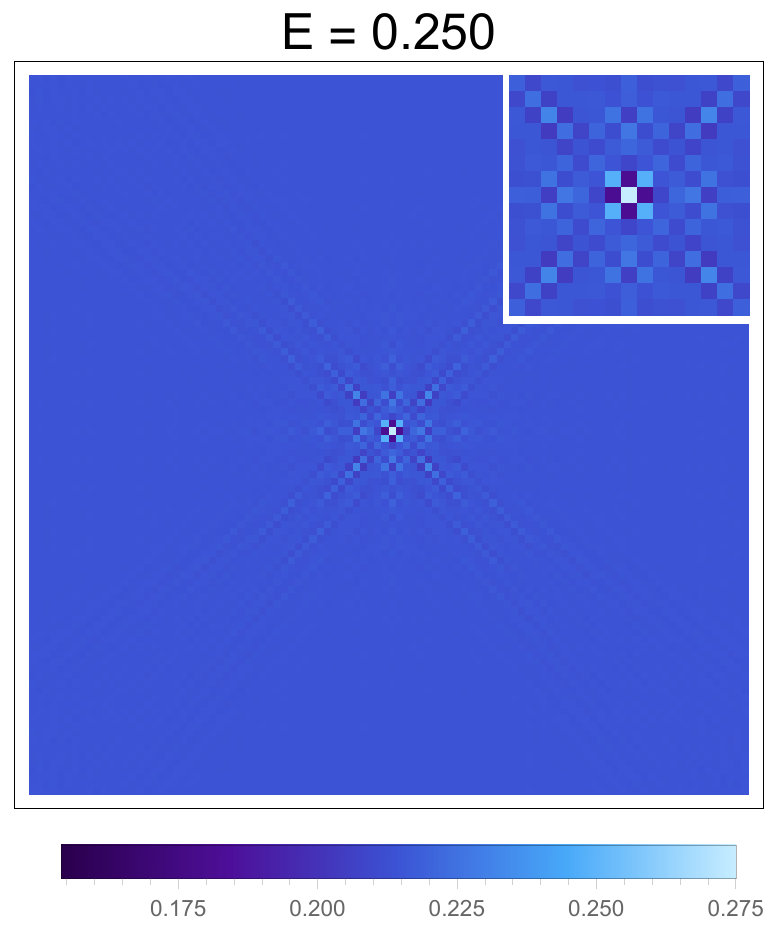}\hfill
	\caption{Real-space LDOS maps for the single weak point-like scatterer case. Here an isolated point-like impurity ($V = 0.5$) is placed in the middle of the sample. The field of view is $100 \times 100$. Shown are maps corresponding to energies $E = \pm0.100$ and $E = \pm0.250$. Inset: a close-up view of the impurity.}
	\label{fig:pwreal}
\end{figure*}

\begin{figure*}
	\centering
	\includegraphics[width=.2\textwidth]{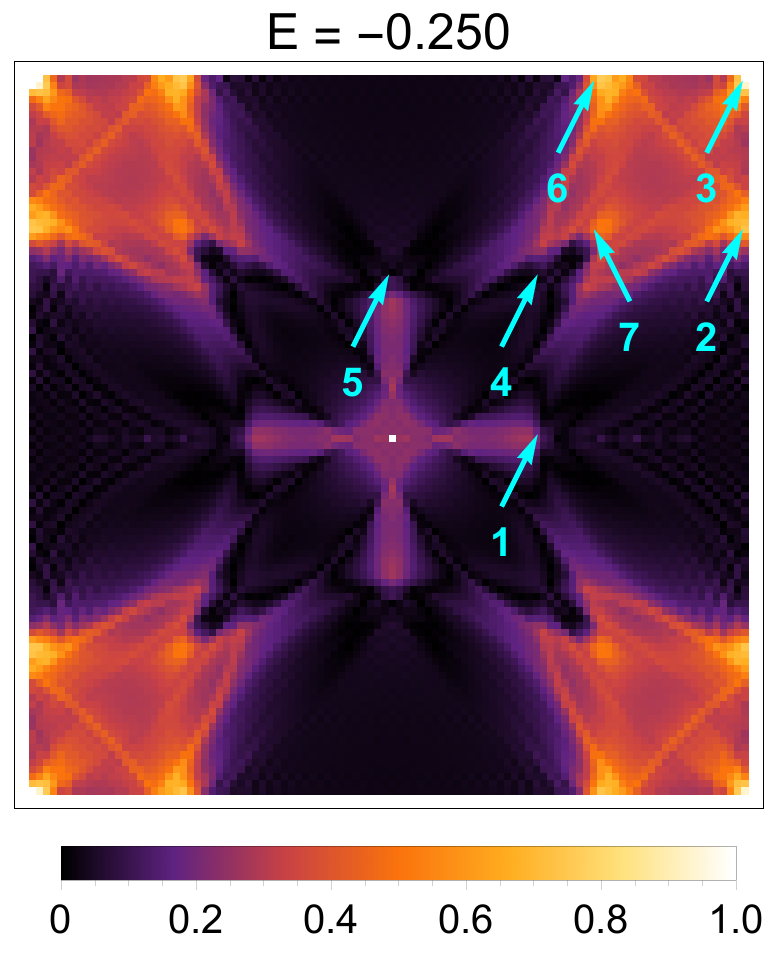}\hfill
	\includegraphics[width=.2\textwidth]{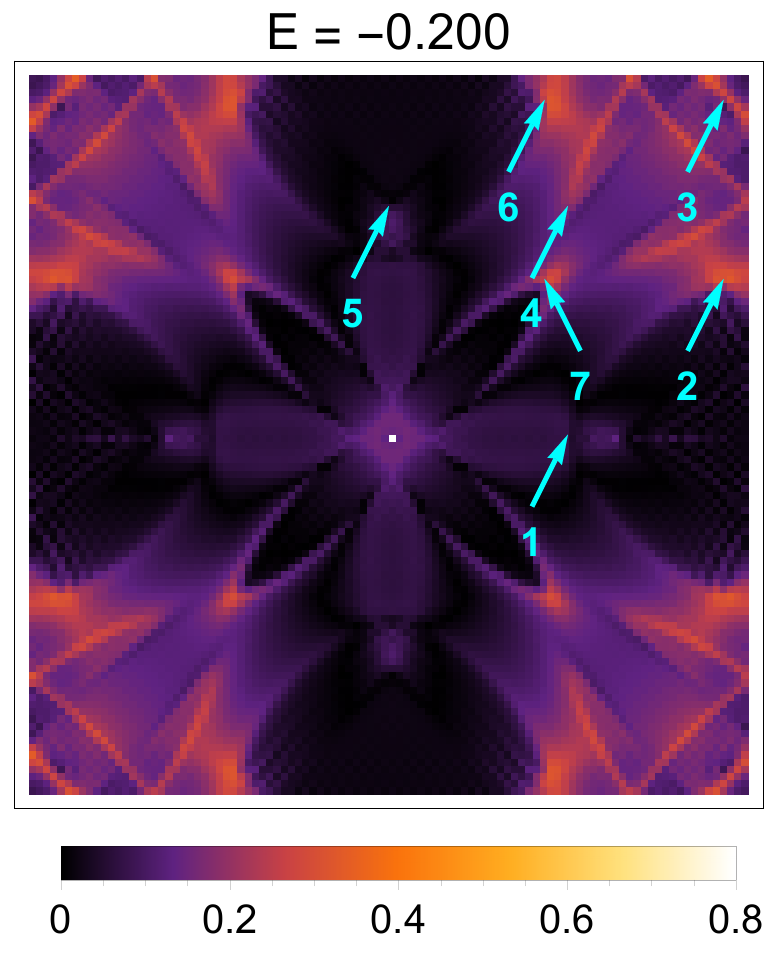}\hfill
	\includegraphics[width=.2\textwidth]{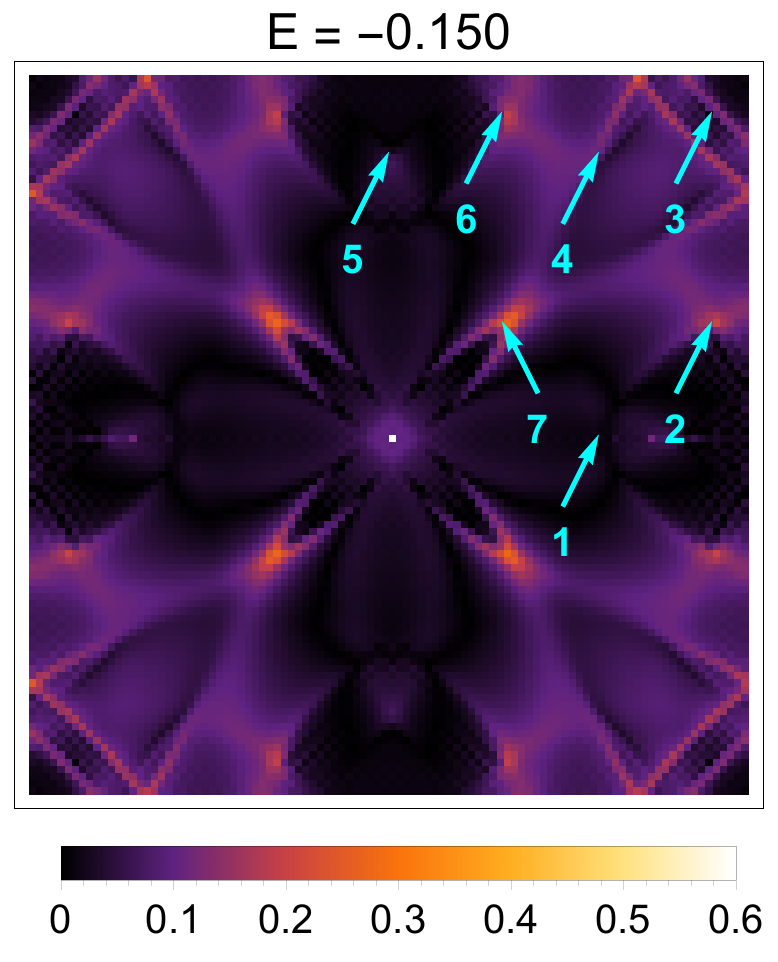}\hfill
	\includegraphics[width=.2\textwidth]{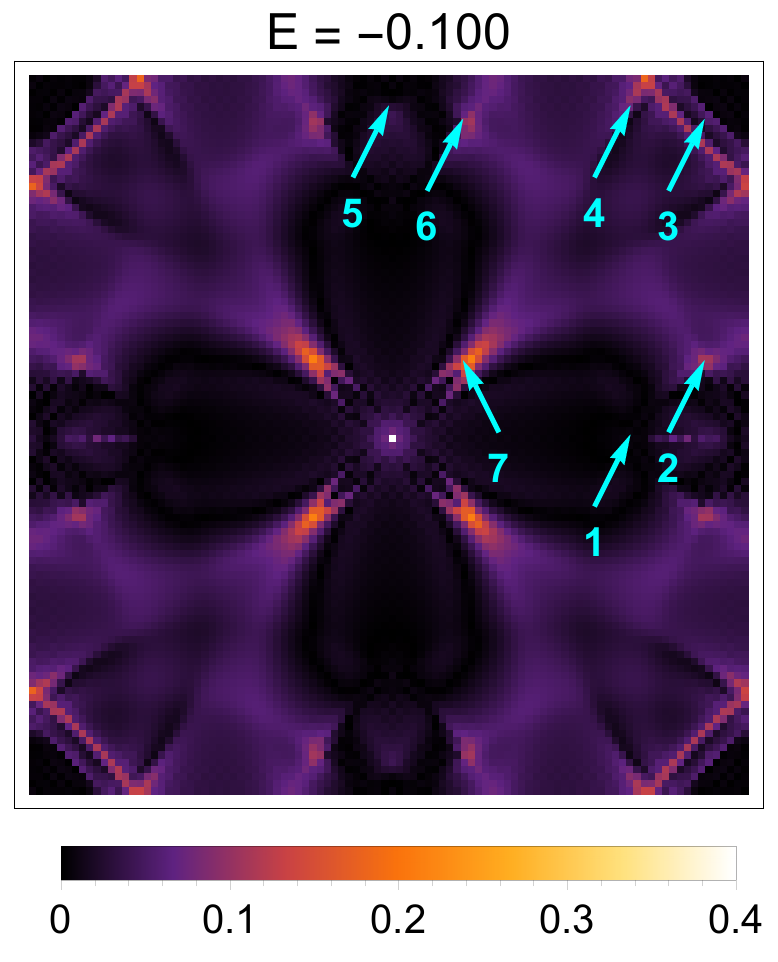}\hfill
	\includegraphics[width=.2\textwidth]{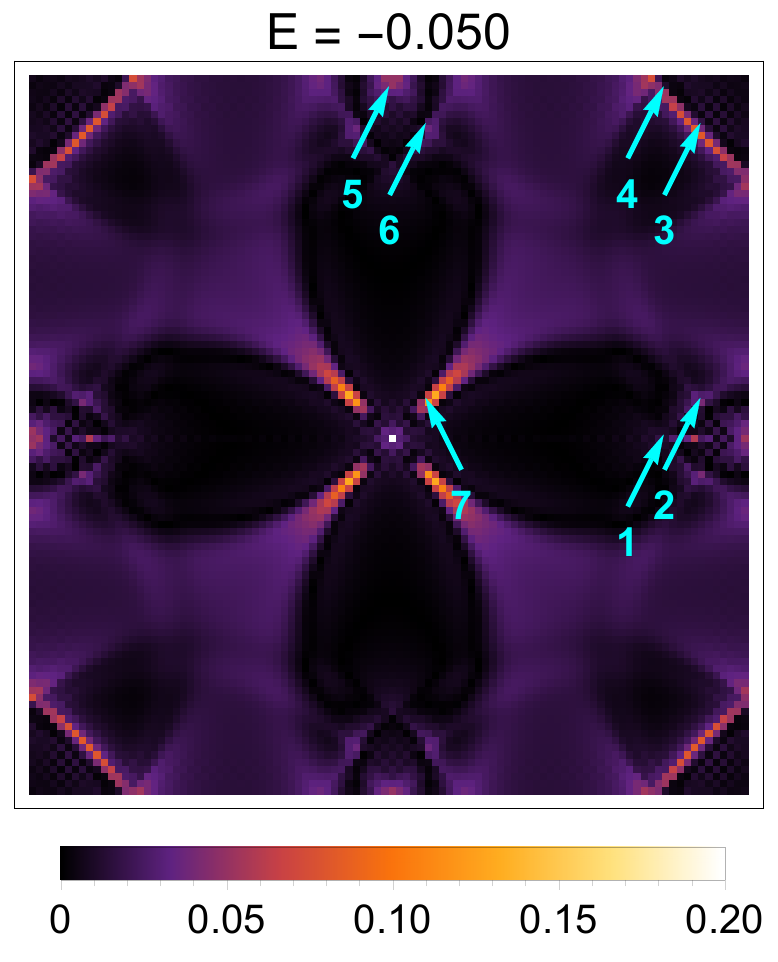}\\
	\includegraphics[width=.2\textwidth]{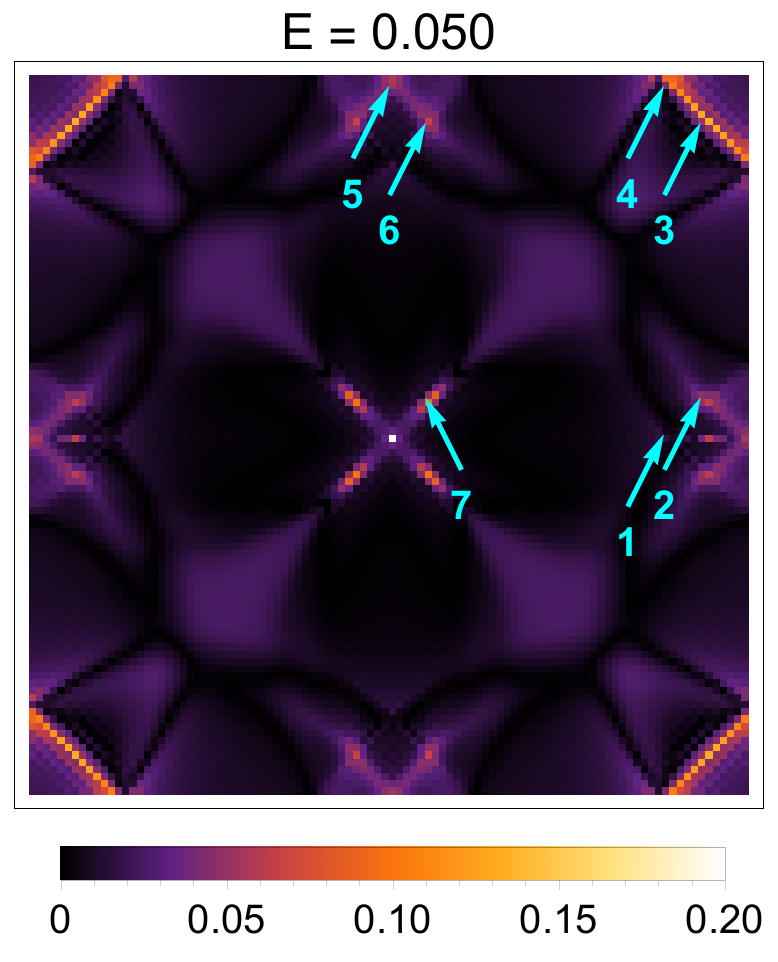}\hfill
	\includegraphics[width=.2\textwidth]{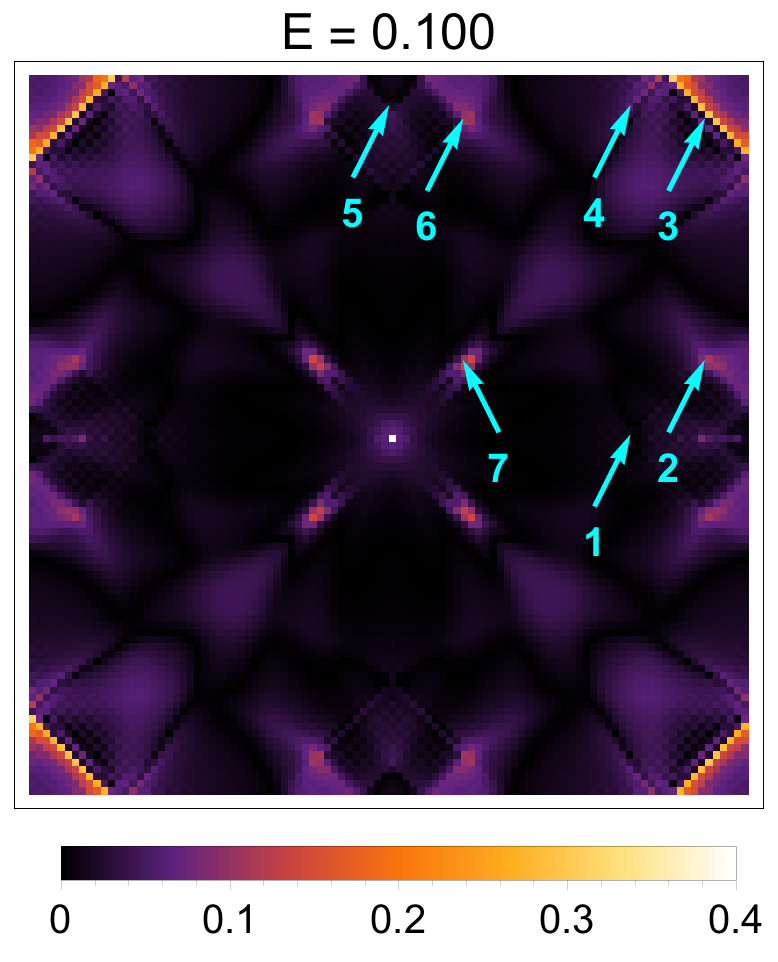}\hfill
	\includegraphics[width=.2\textwidth]{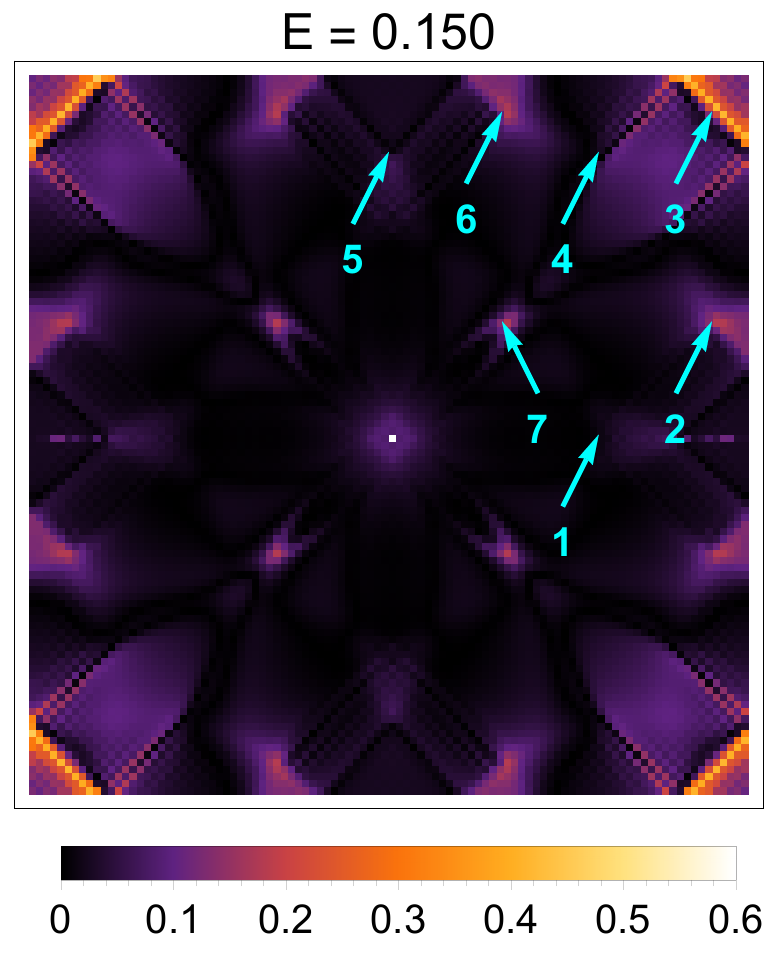}\hfill
	\includegraphics[width=.2\textwidth]{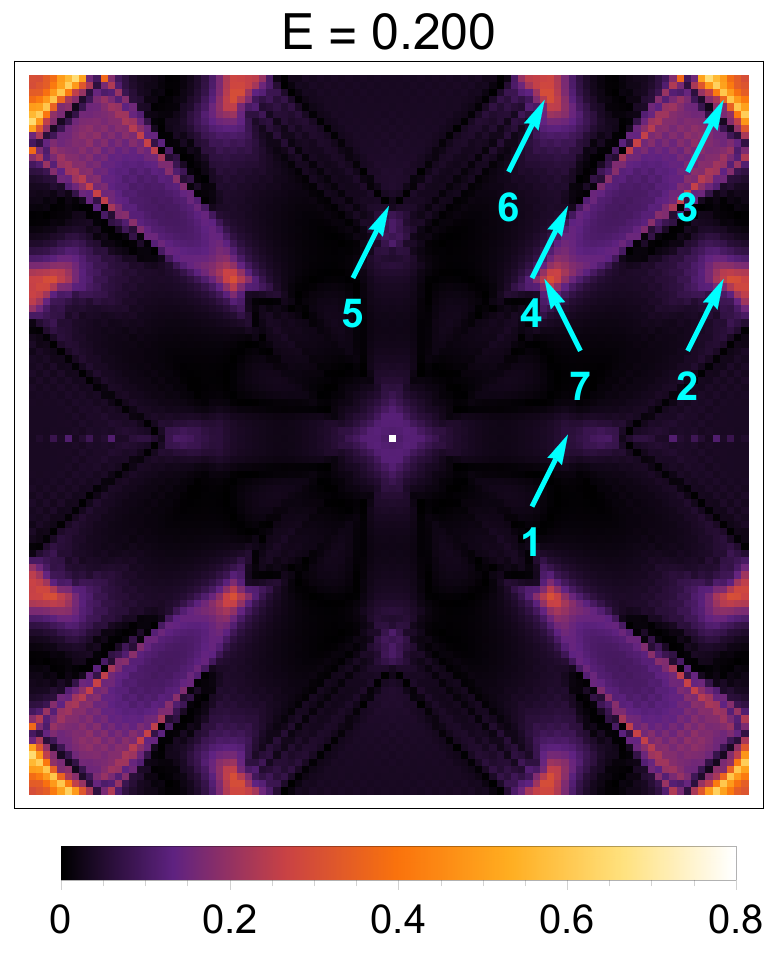}\hfill
	\includegraphics[width=.2\textwidth]{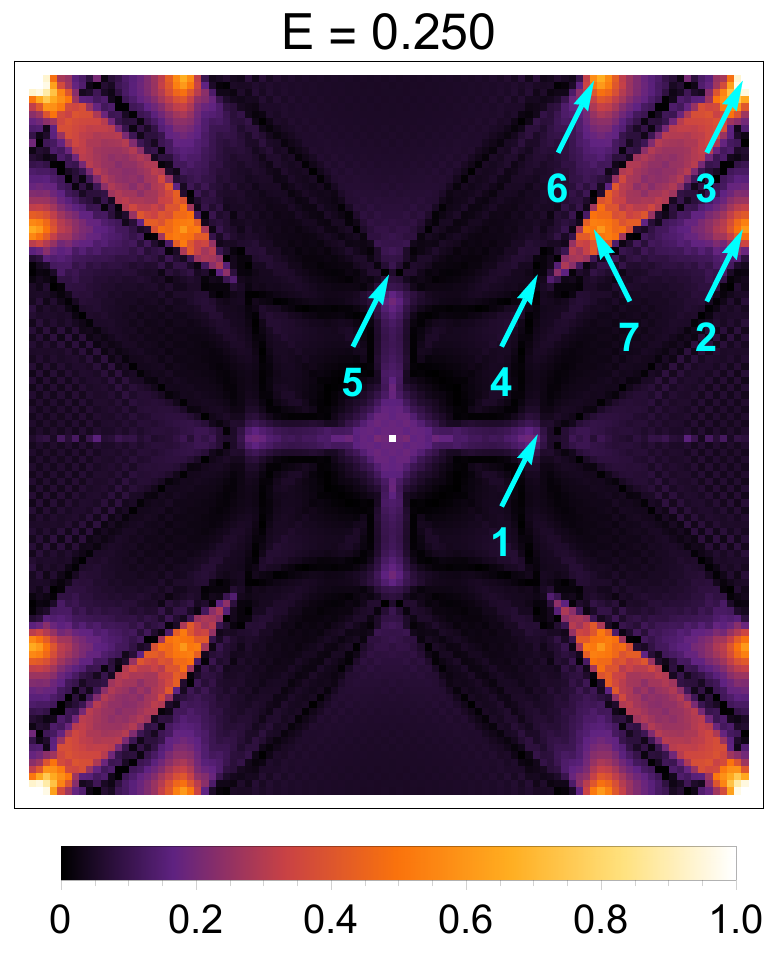}\hfill
	\caption{Fourier-transformed maps for the single weak point-like scatterer case, with $V = 0.5$. Power spectra for both positive and negative bias voltages are shown for energies ranging from $E = \pm0.050 $ to $E = \pm0.250$. Arrows indicate where the peaks corresponding to the characteristic momenta of the octet model show up in the upper-right quadrant. The color scaling varies linearly with energy.}
	\label{fig:pwfourier}
\end{figure*}

\subsection{Single Weak Point-Like Impurity}

We first start with the best-case scenario as far as reproducing the phenomenology of the octet model is concerned: the case of a single point-like scatterer. To examine this more clearly, we add an on-site energy of $V=0.5$ to a single site in the middle of the field of view. This is a weak, \emph{non-unitary} potential, so this would not induce resonances at zero energy. The LDOS maps results are shown in  Fig.~\ref{fig:pwreal}. In the real-space images, one can see clear, energy-dependent oscillations in the LDOS which emanate from the impurity core. Despite the weakness of the potential, these oscillations dominate the signal at all energies, and the isolated impurity itself can be easily seen. It should be noted that at the impurity site the LDOS is not suppressed, but instead has a finite value for the energies we considered.

In contrast to the rather limited information conveyed by the real-space maps, the Fourier-transformed maps, shown in Fig.~\ref{fig:pwfourier}, display considerably more information. These are identical to the Fourier maps computed using the standard single-impurity $T$-matrix method---as it should, since that is a different method of solving the same problem. These show peaks with positions that are indeed consistent with the octet model. However, one also sees that these peaks are little more than enhanced regions in a more diffuse background. Even when the potential is weak, the spectra are dominated by momenta that connect different segments of the bananas, giving rise to patterns consisting of diffuse streaks, blurry regions, and propeller-shaped sections. The special momenta of the octet model merely correspond to points at which the spectral weight is enhanced relative to the background. That is, these points coexist alongside these background patterns that arise from other scattering processes. A noteworthy feature of the power spectra of the case of a weak point potential is that $\mathbf{q}_4$ and $\mathbf{q}_5$ are not discernable at all. The most dominant peaks are $\mathbf{q}_2$, $\mathbf{q}_3$, $\mathbf{q}_6$, and $\mathbf{q}_7$, which become even more pronounced at higher energies. It is quite telling that, even at the idealized single point-impurity level, the correspondence between the full numerics and the expectations from the octet model is not fully realized---we remind the reader yet again that experimental Fourier maps show \emph{all} seven peaks.

As we have emphasized before, impurity cores are not seen in the data, which excludes the possibility that QPI is caused by strong local impurity potentials. However our real-space results suggest that even a weak impurity gives rise to telltale patterns in the LDOS that point to its existence, and that these weak impurities can be easily identified in real space. The Fourier-transformed maps featuring a single weak impurity also show rather imperfect correspondence with experiment---power spectra from STS show far sharper peaks than our theoretically-obtained maps display.  As we will subsequently argue, the addition of any realistic details to this idealized case will have the effect of further blurring the sharp features in the Fourier spectra. The presence of these complicating factors compounds the difficulty of explaining the sharpness of the octet model QPI peaks as seen in experiments. 

\subsection{Multiple Weak Point-Like Impurities}

\begin{figure*}[ht]
	\centering
	\includegraphics[width=.25\textwidth]{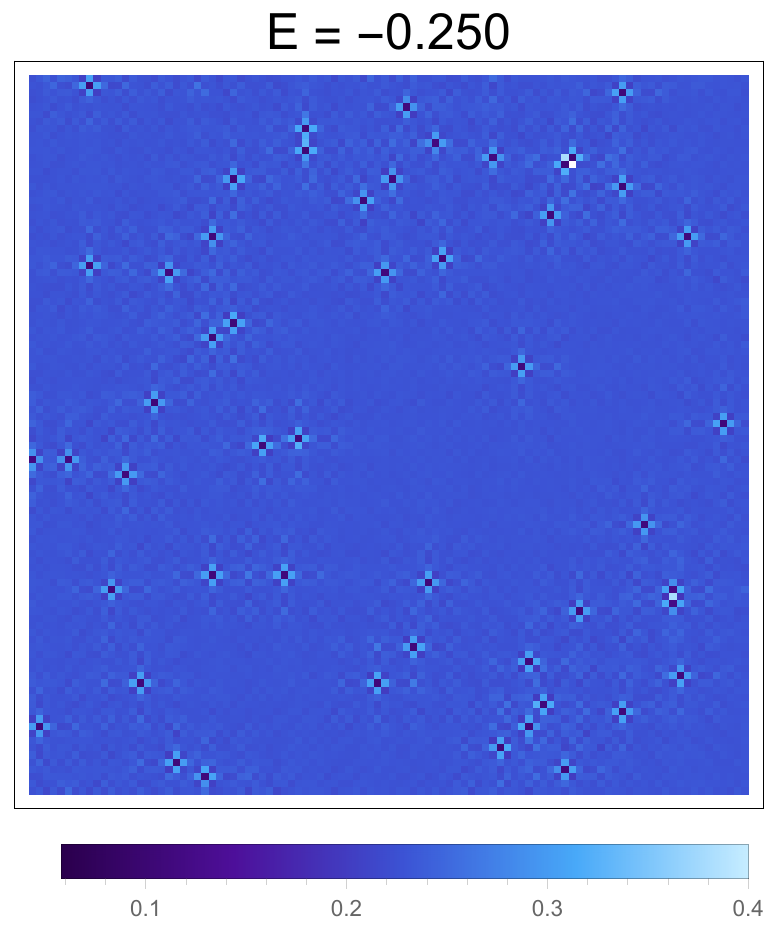}\hfill
	\includegraphics[width=.25\textwidth]{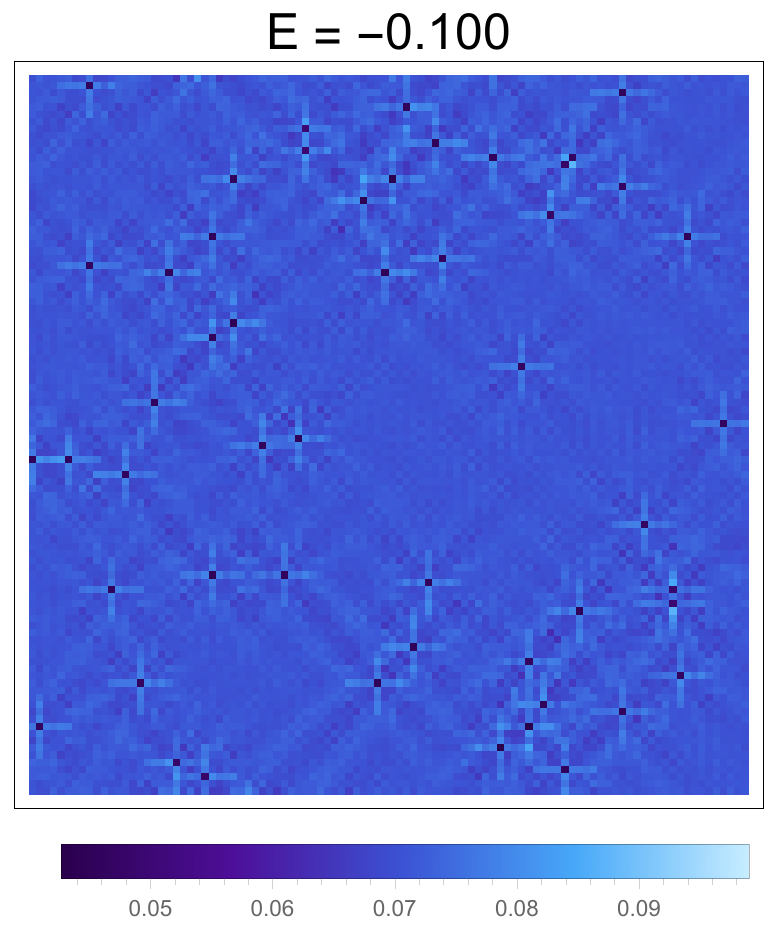}\hfill
	\includegraphics[width=.25\textwidth]{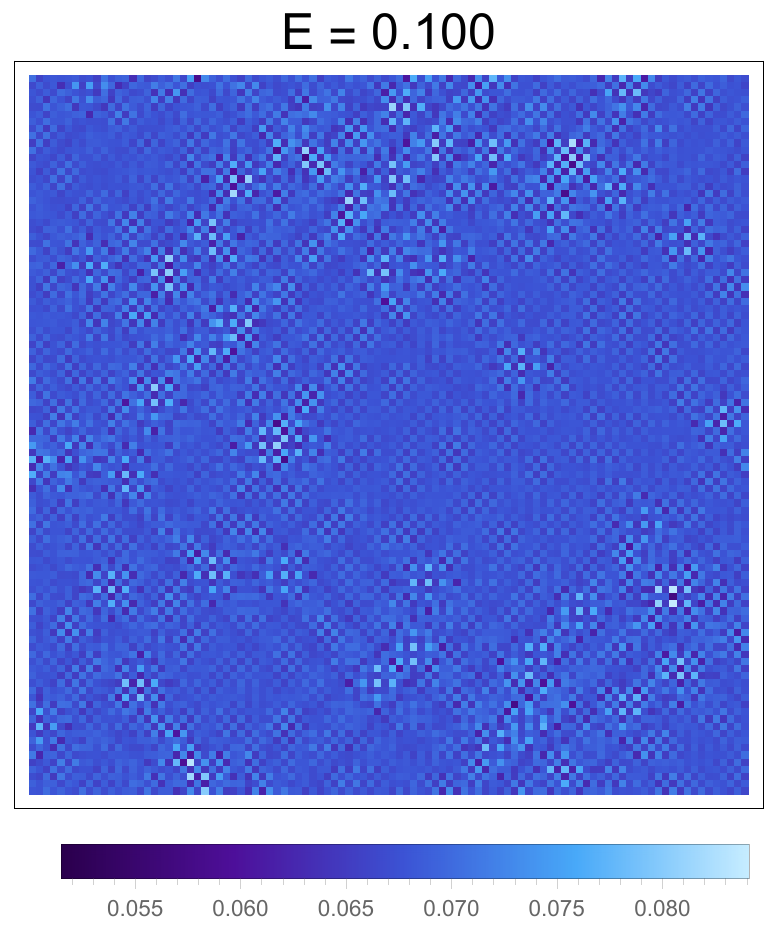}\hfill
	\includegraphics[width=.25\textwidth]{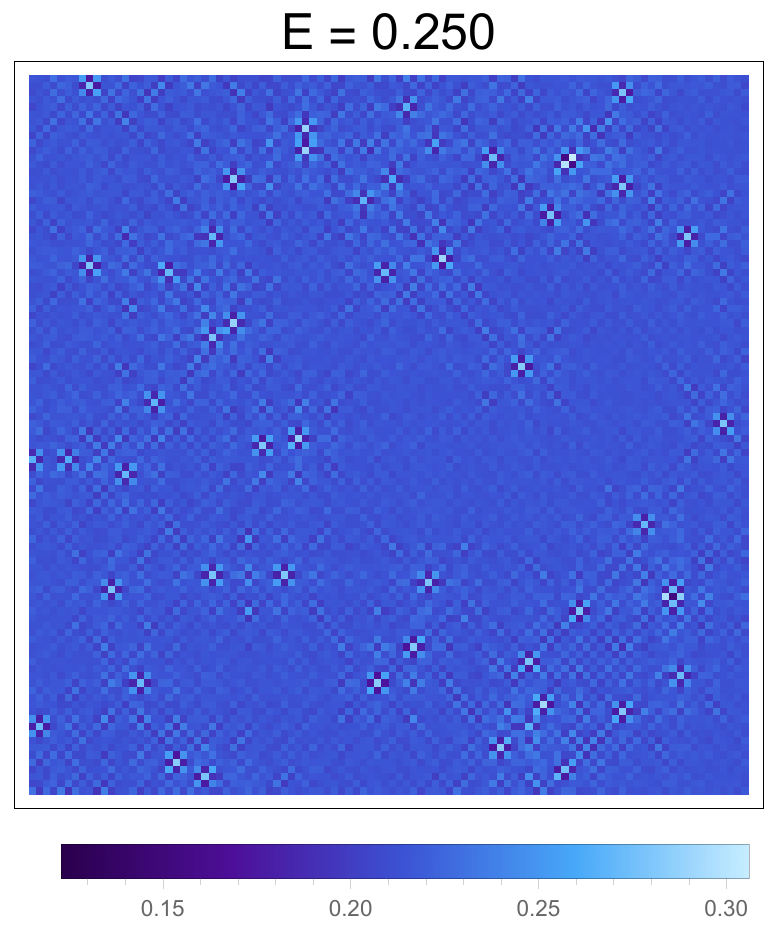}\hfill
	\caption{Real-space LDOS maps for a $d$-wave superconductor with a 0.5\% concentration of weak point-like scatterers ($V = 0.5$) distributed randomly across the CuO$_2$ plane. The field of view is $100 \times 100$, and the energies shown are $E = \pm0.100$ and $E = \pm0.250$.}
	\label{fig:spwreal}
\end{figure*}

\begin{figure*}
	\centering
	\includegraphics[width=.2\textwidth]{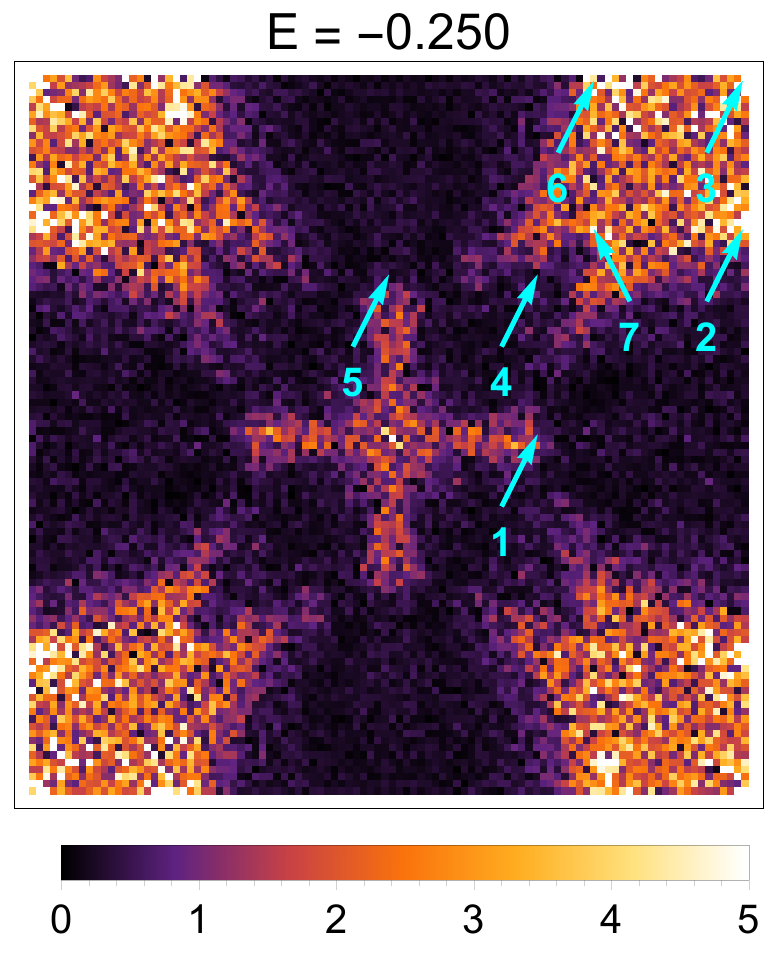}\hfill
	\includegraphics[width=.2\textwidth]{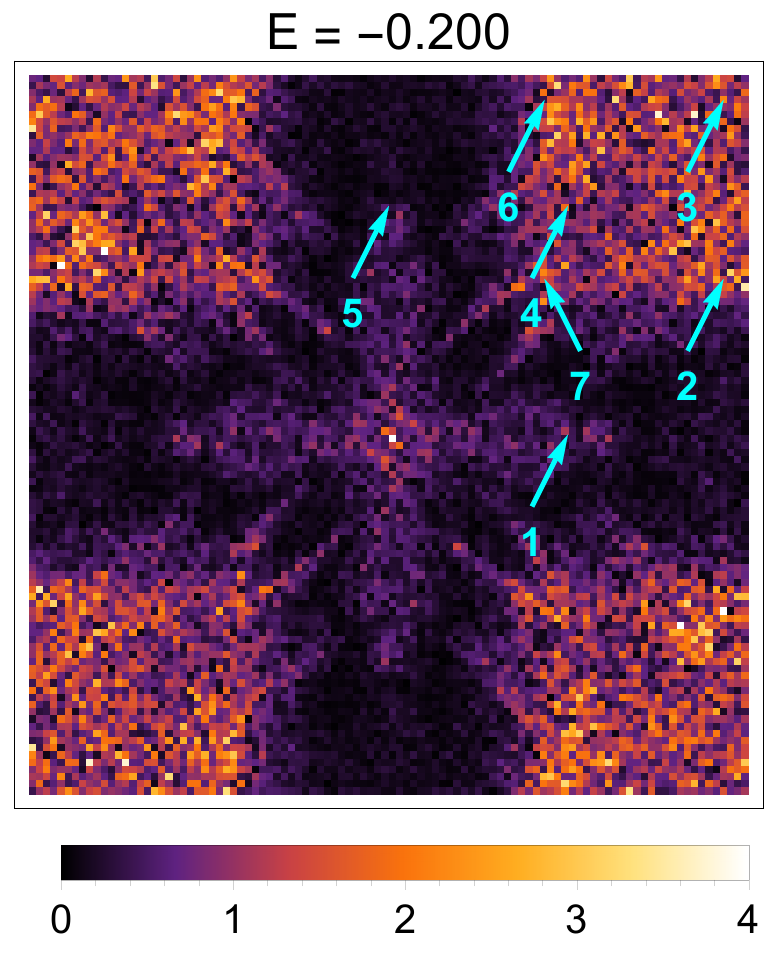}\hfill
	\includegraphics[width=.2\textwidth]{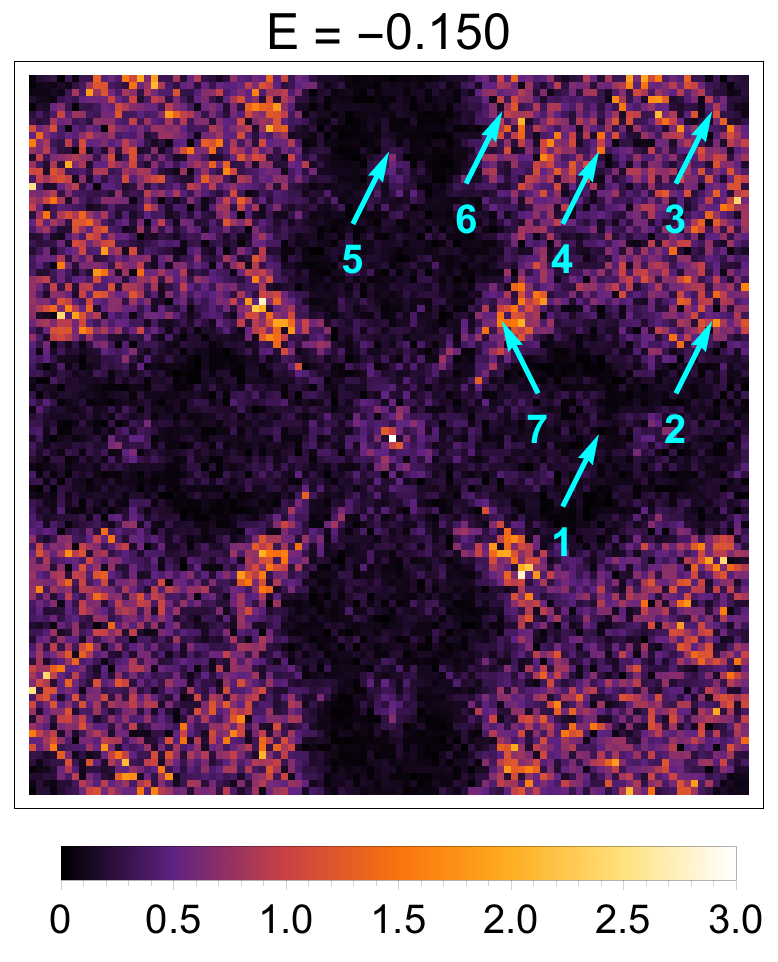}\hfill
	\includegraphics[width=.2\textwidth]{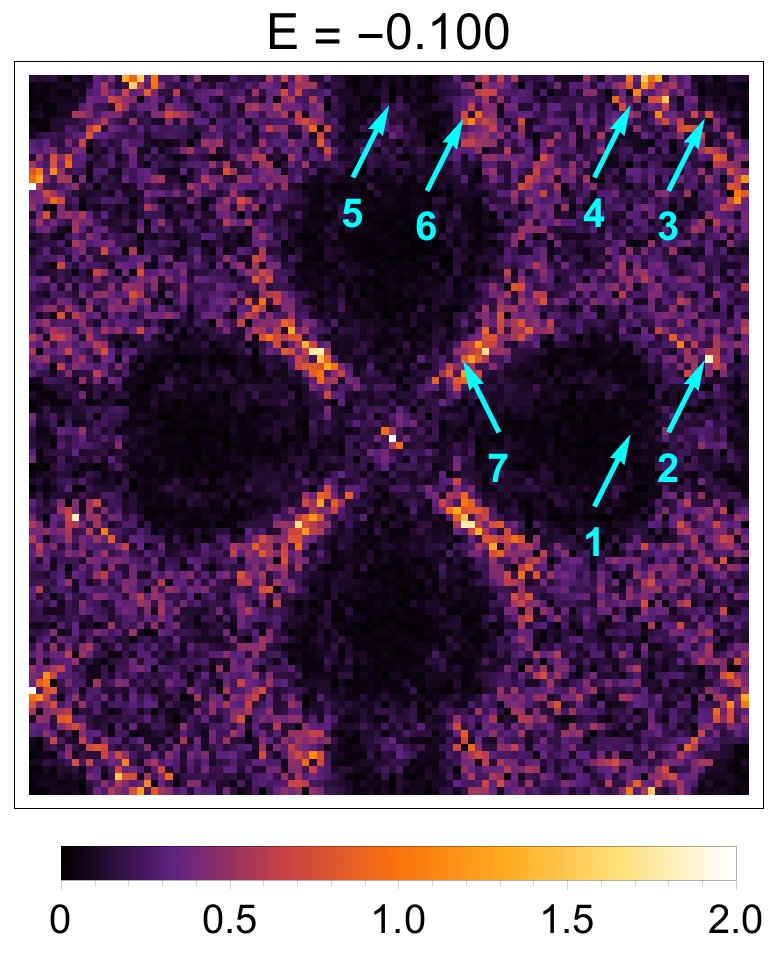}\hfill
	\includegraphics[width=.2\textwidth]{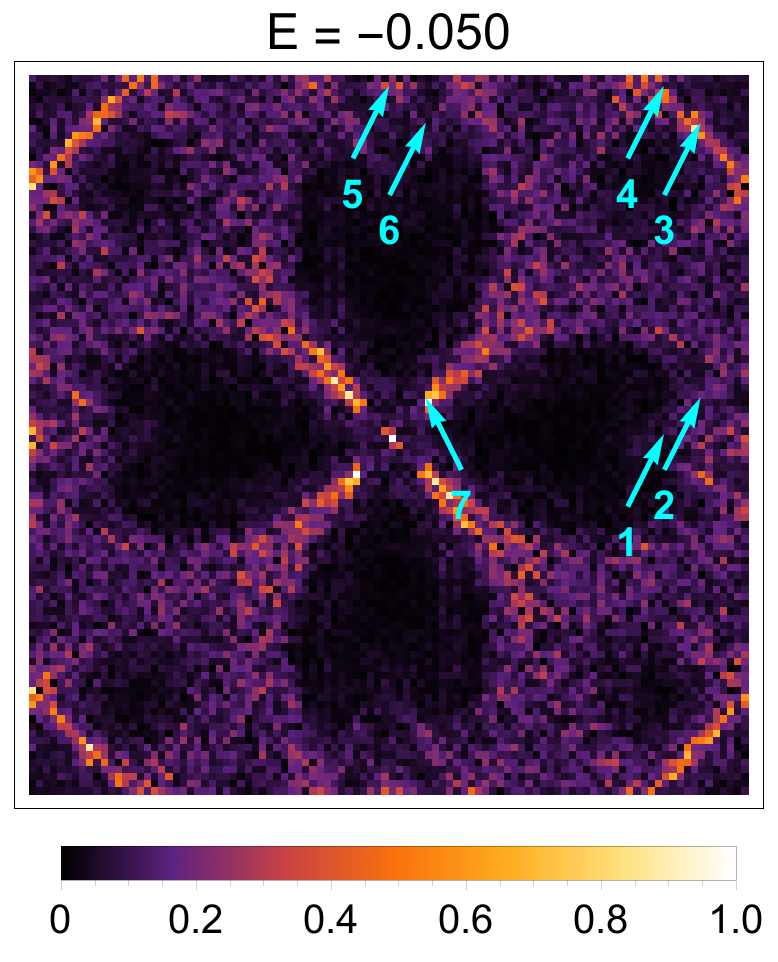}\\
	\includegraphics[width=.2\textwidth]{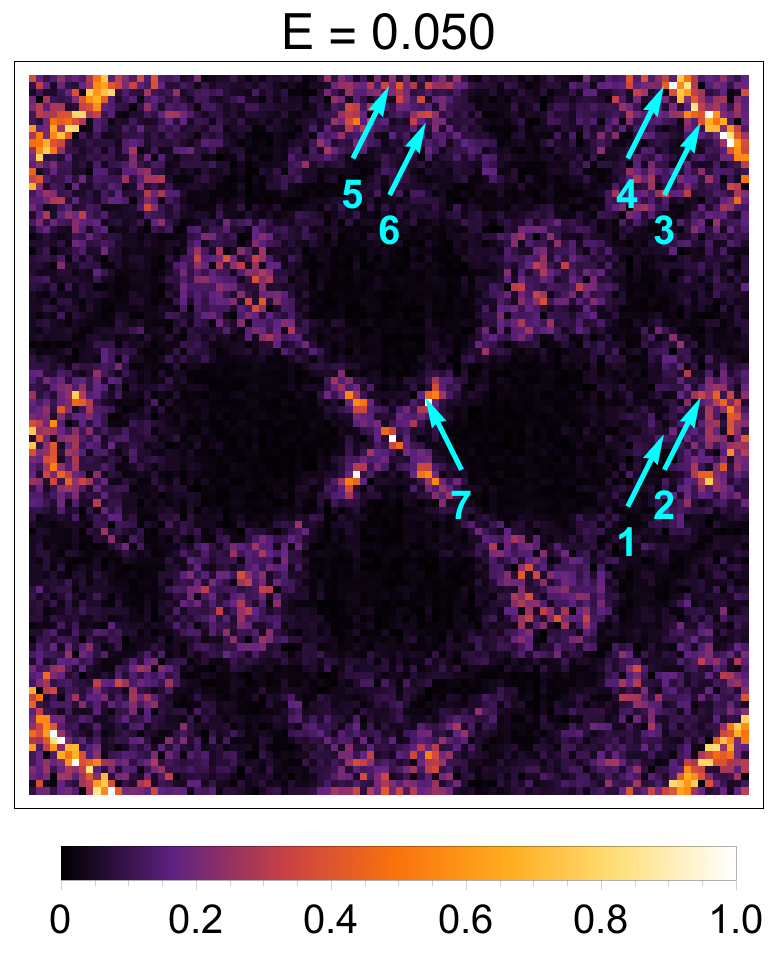}\hfill
	\includegraphics[width=.2\textwidth]{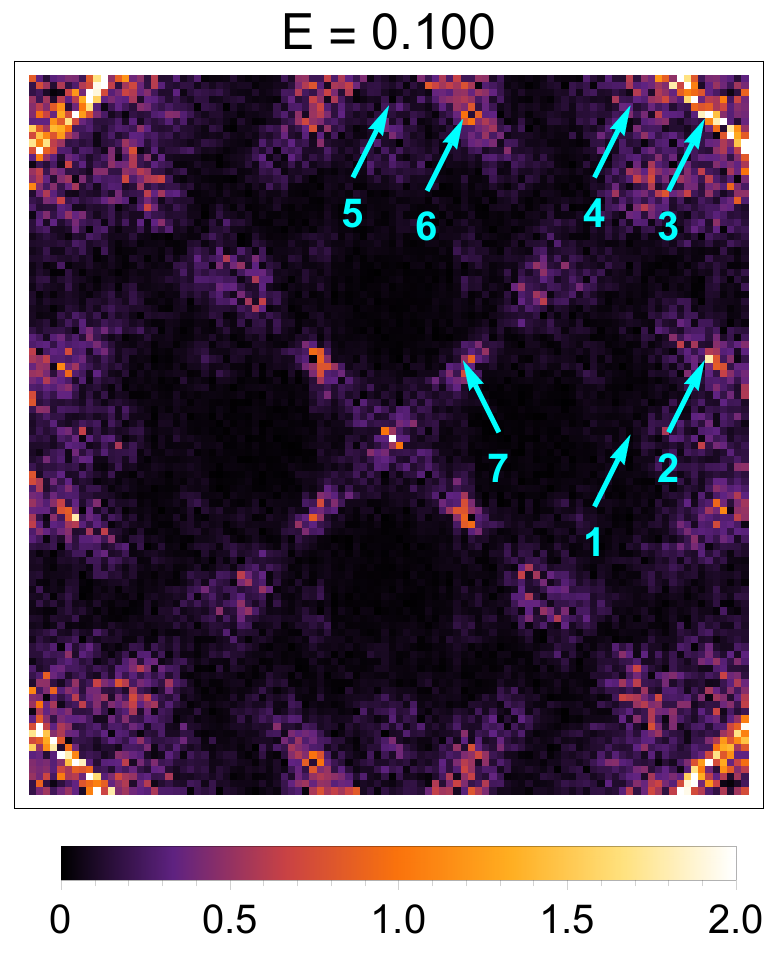}\hfill
	\includegraphics[width=.2\textwidth]{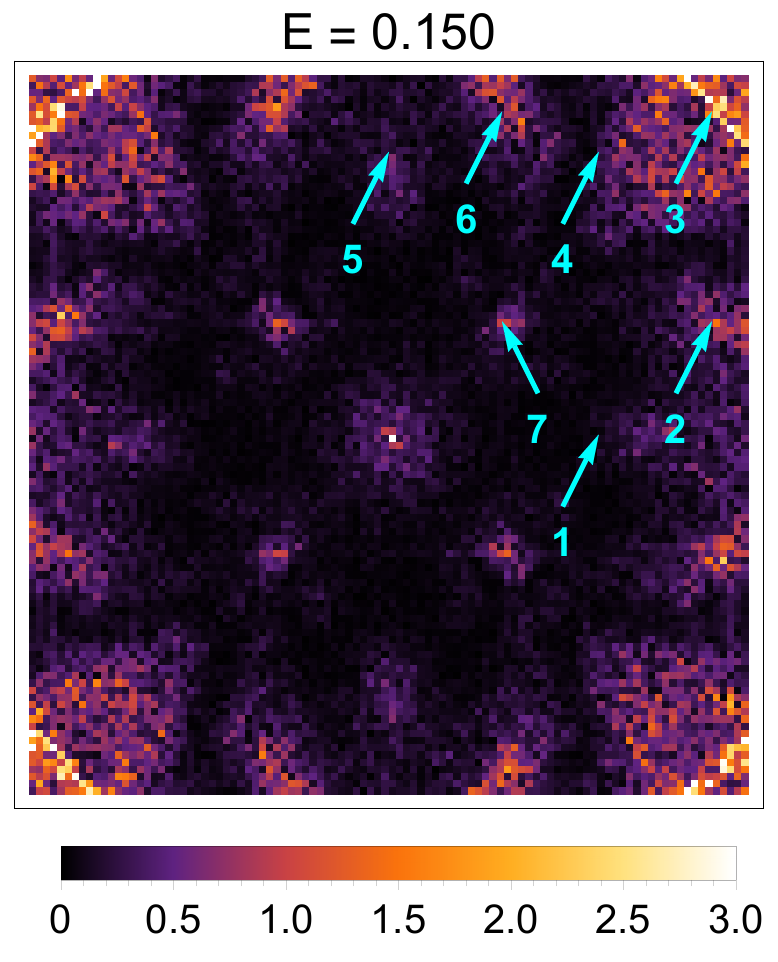}\hfill
	\includegraphics[width=.2\textwidth]{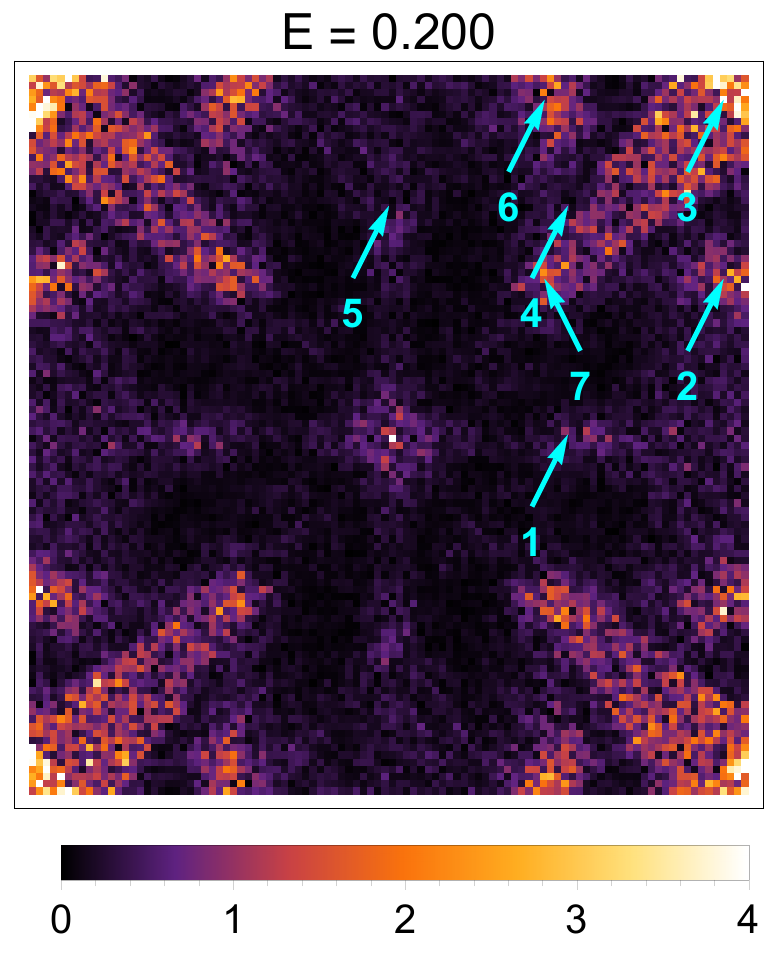}\hfill
	\includegraphics[width=.2\textwidth]{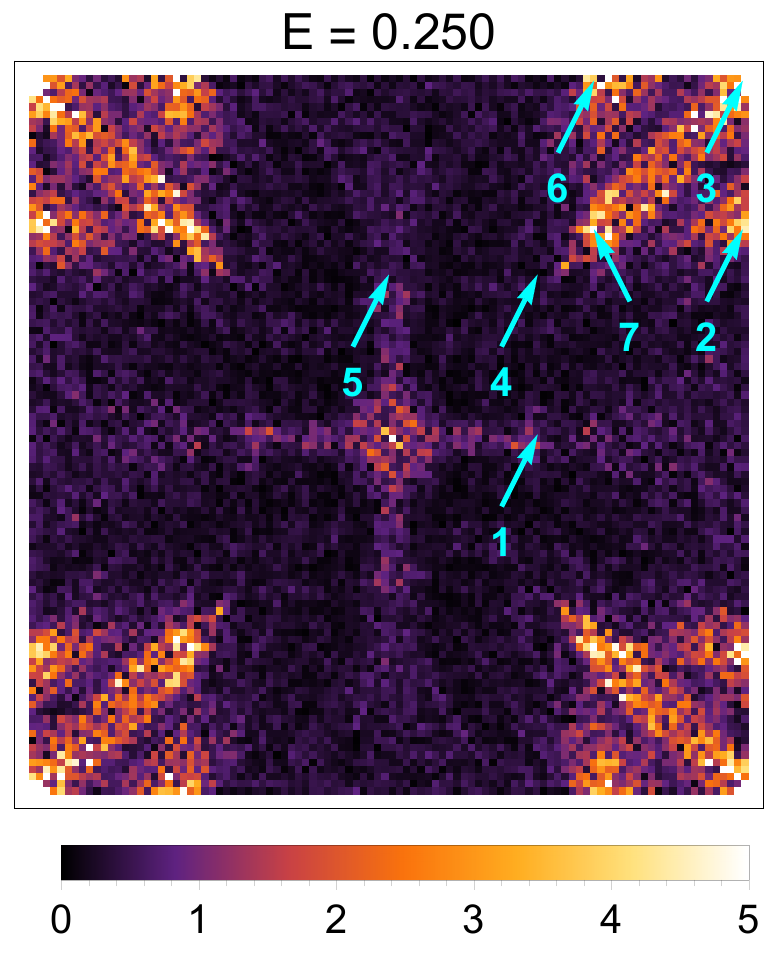}\hfill
	\caption{Fourier-transformed maps for a system with a 0.5\% concentration of weak point-like scatterers ($V =0.5$). Shown are energies ranging from  $E = \pm0.050$ to  $E = \pm0.250$, along with arrows showing where the octet wavevectors are expected to be found. The color scaling varies linearly with energy.}
	\label{fig:spwfourier}
\end{figure*}

The many-impurity case is the next case we will consider. This has in fact been considered before using either a multiple-scattering $T$-matrix approach\cite{zhu2004power} or exact diagonalization of the Bogoliubov-de Gennes Hamiltonian for small system sizes.\cite{atkinson2000details} Here we take advantage of the flexibility of the numerical method we use and obtain \emph{exact} results for large system sizes. We randomly distribute many weak point-like scatterers in our system, and to optimize the correspondence with experimental results, we take the concentration of such weak scatterers to be low, with only 0.5\% of lattice sites possessing such an impurity. As in the isolated-impurity case, we take the strength of each impurity to be $V = 0.5$.

 As in the single-impurity case, the impurities are easily visible in the real-space images, but in addition we also see stripe-like patterns covering the entire field of view, which are seen to depend on the energy (Fig.~\ref{fig:spwreal}). At first glance these look strikingly similar to the real-space patterns due to QPI seen in the raw experimental data. It is worth noting that the original real-space QPI results were initially misidentified as stripy charge-density waves. On closer inspection, novel multiple-scattering effects are seen when impurities get close together, as already discussed in the literature.\cite{atkinson2003quantum,zhu2003two,zhu2004power} For instance, when two impurities line up such that their diagonal streaks overlap each other neatly, the streaks constructively interfere and have the effect that they become more intense. 

The Fourier-transformed maps are themselves quite illuminating. The consequence of the randomness of the impurity positions is that the Fourier maps show \emph{speckle patterns}, as demonstrated in Fig. \ref{fig:spwfourier}. This is just in line with our expectations: the familiar speckle patterns produced by laser light scattering against a random medium have precisely the same origin. Not surprisingly, one sees very similar speckle in the \emph{experimental} Fourier maps. At these low impurity concentrations, the outcome is a speckled version of the single-impurity results. This looks much more like the real data, and the special momenta of the octet model are by and large still discernible in this case. The peaks that are prominent in the single-impurity case are similarly visible, with the difference that there is much more fuzziness present in these regions. However, because this is simply a many-impurity version of the single weak-scatterer case, this inherits the fact that no large spectral weight is associated with the $\mathbf{q}_4$ and $\mathbf{q}_5$ wavevectors.

\begin{figure*}[ht]
	\centering
	\includegraphics[width=.25\textwidth]{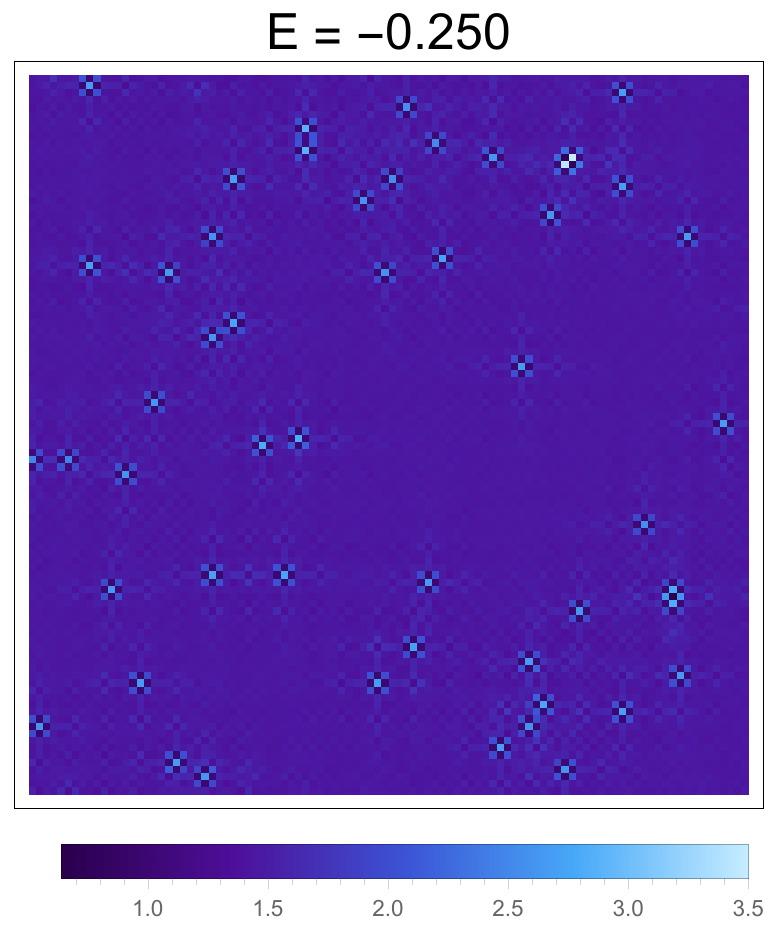}\hfill
	\includegraphics[width=.25\textwidth]{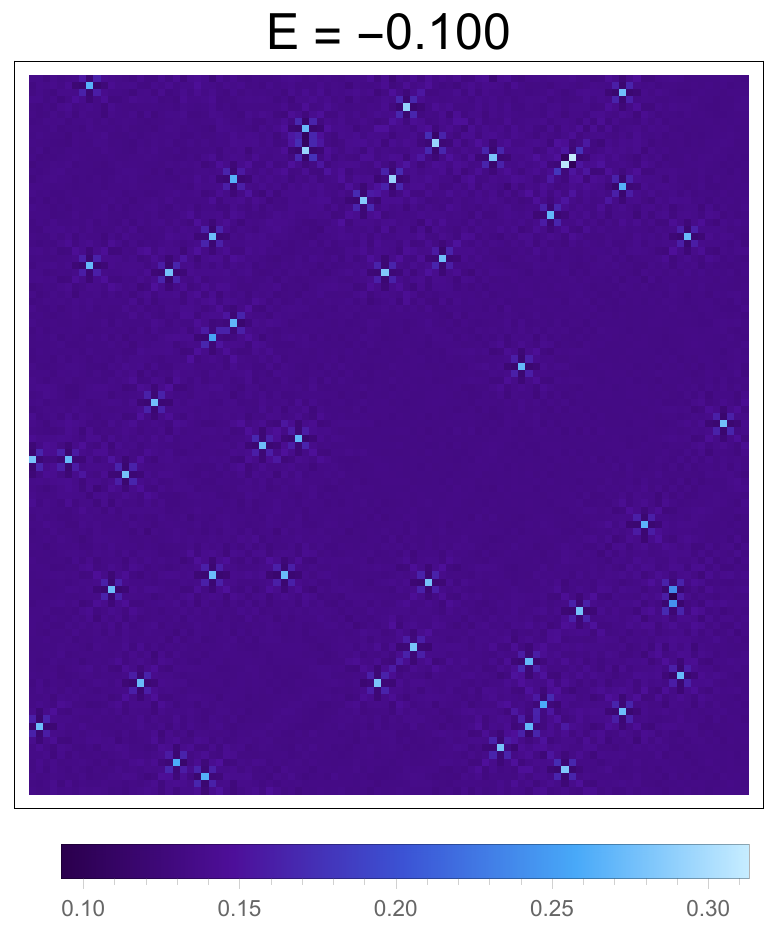}\hfill
	\includegraphics[width=.25\textwidth]{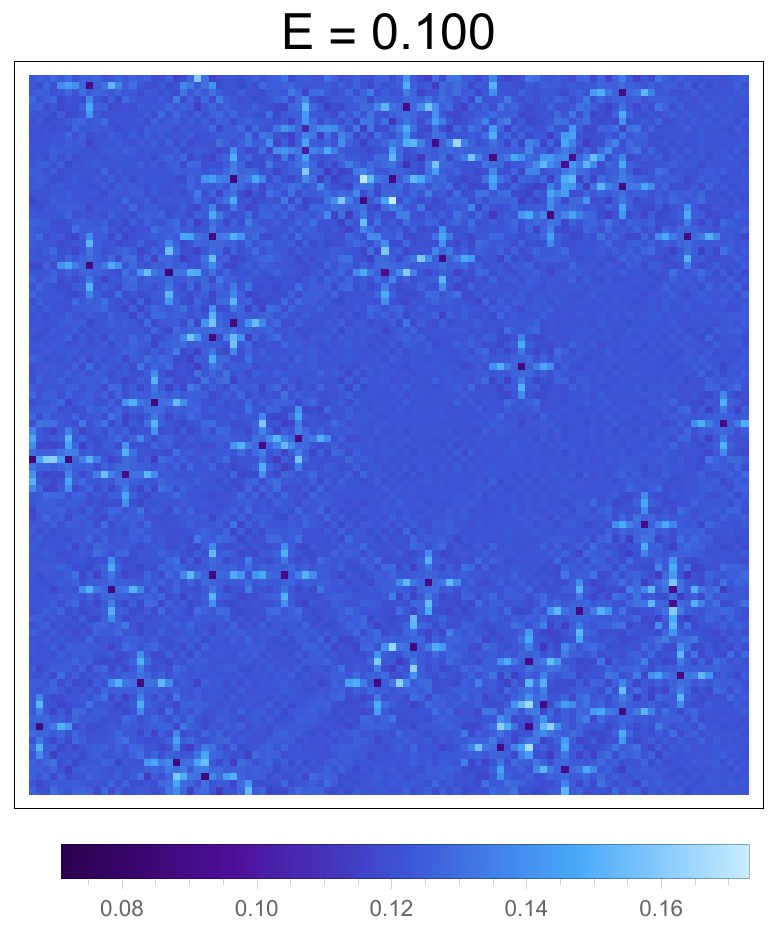}\hfill
	\includegraphics[width=.25\textwidth]{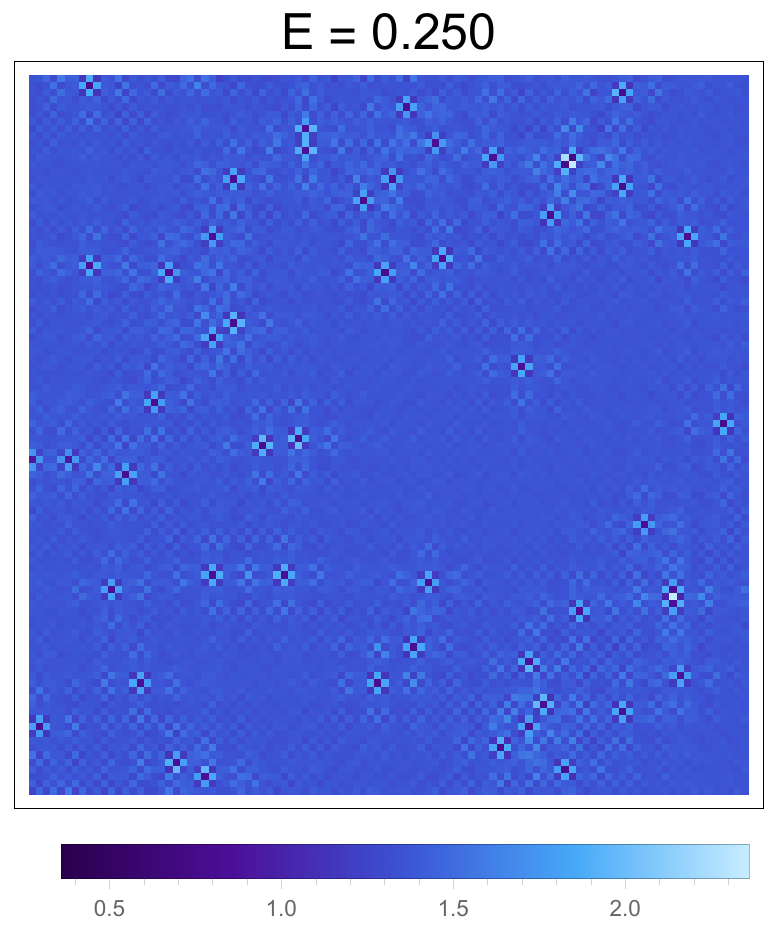}\hfill
	\caption{Filtered real-space LDOS maps for a $d$-wave superconductor with a 0.5\% concentration of weak point-like scatterers ($V = 0.5$) distributed randomly across the CuO$_2$ plane. The field of view is $100 \times 100$, and the energies shown are $E = \pm0.100$ and $E = \pm0.250$.}
	\label{fig:fspwreal}
\end{figure*}

\begin{figure*}
	\centering
	\includegraphics[width=.2\textwidth]{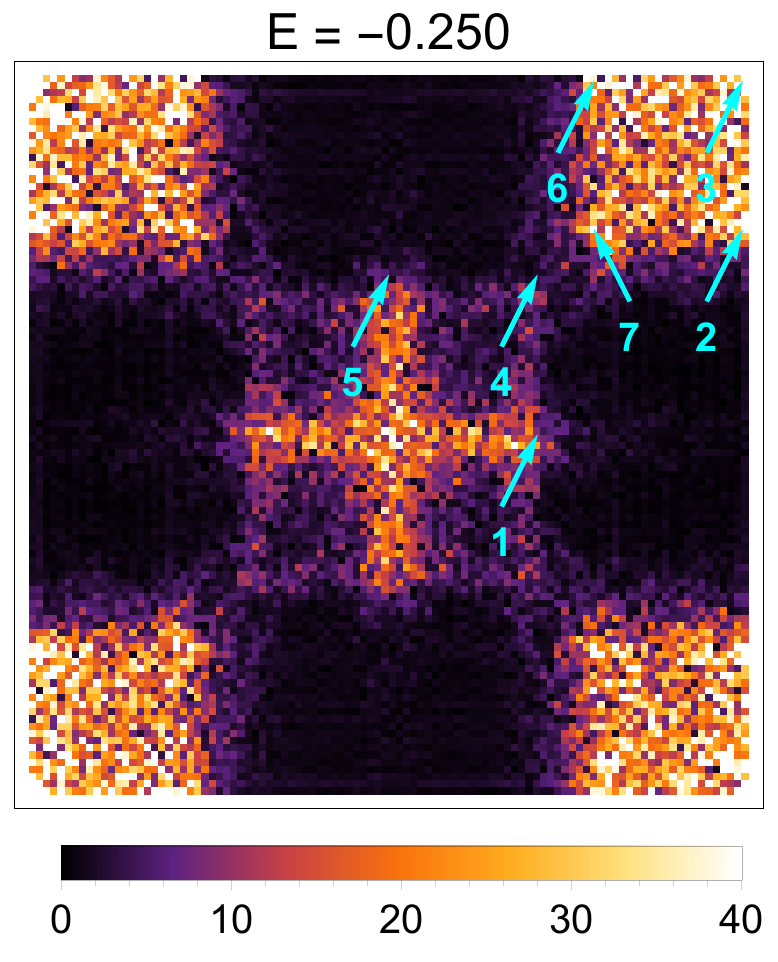}\hfill
	\includegraphics[width=.2\textwidth]{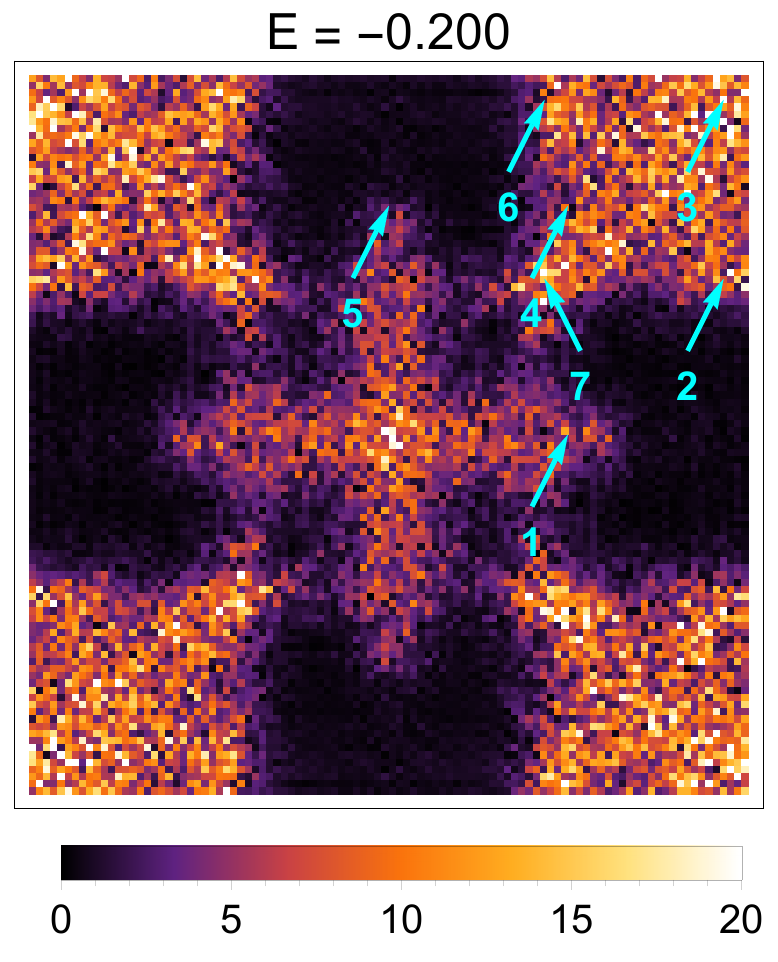}\hfill
	\includegraphics[width=.2\textwidth]{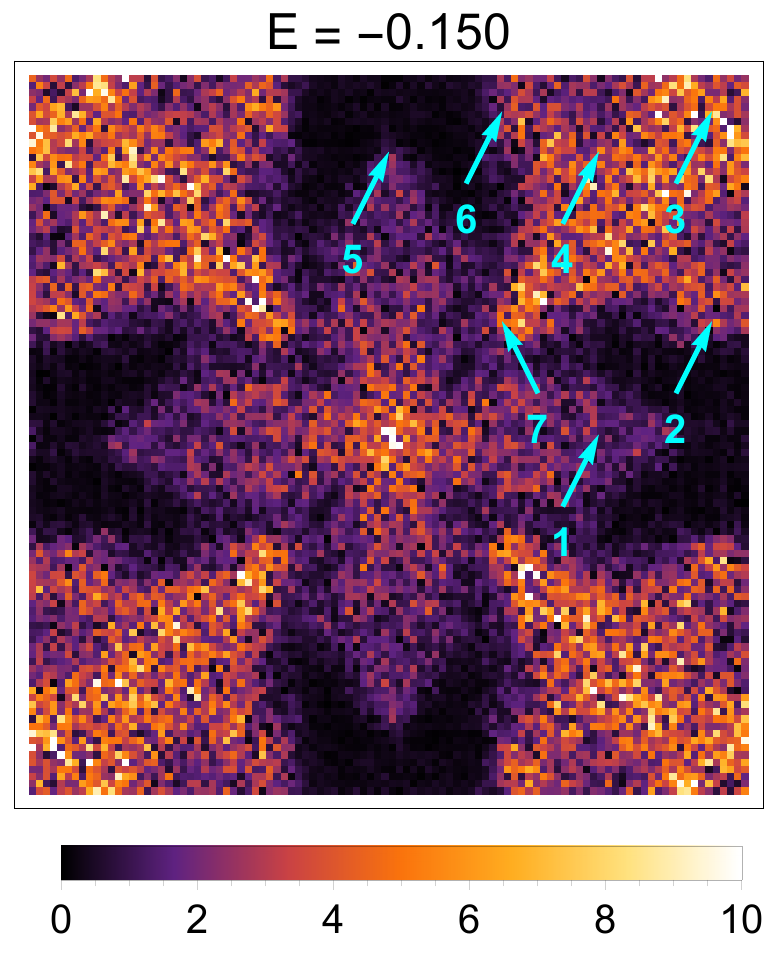}\hfill
	\includegraphics[width=.2\textwidth]{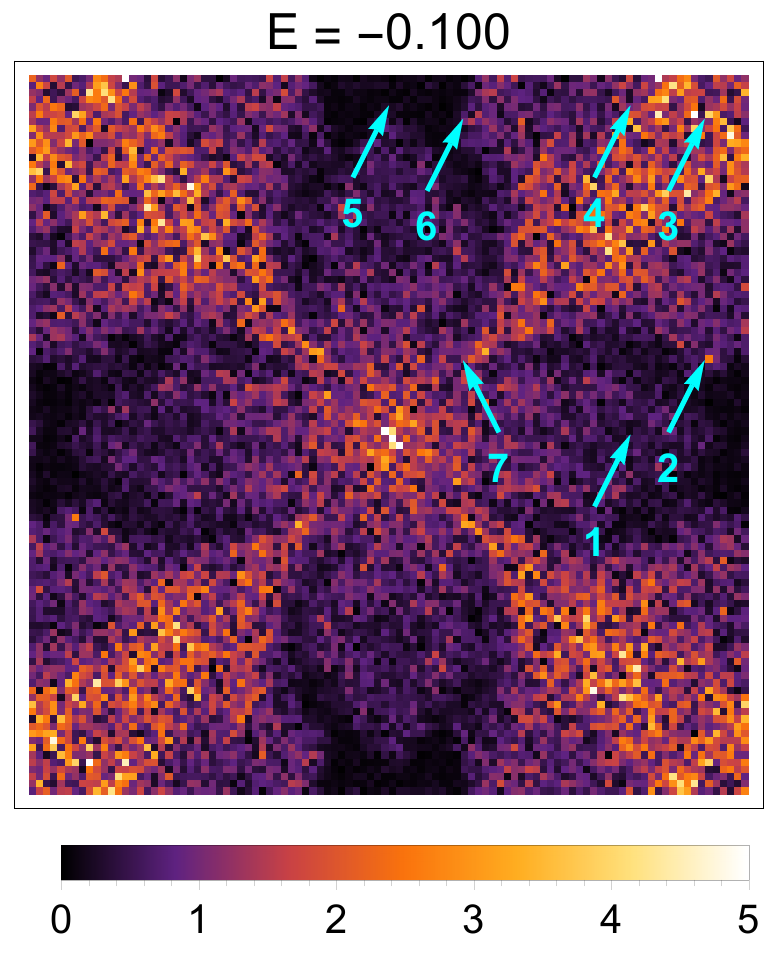}\hfill
	\includegraphics[width=.2\textwidth]{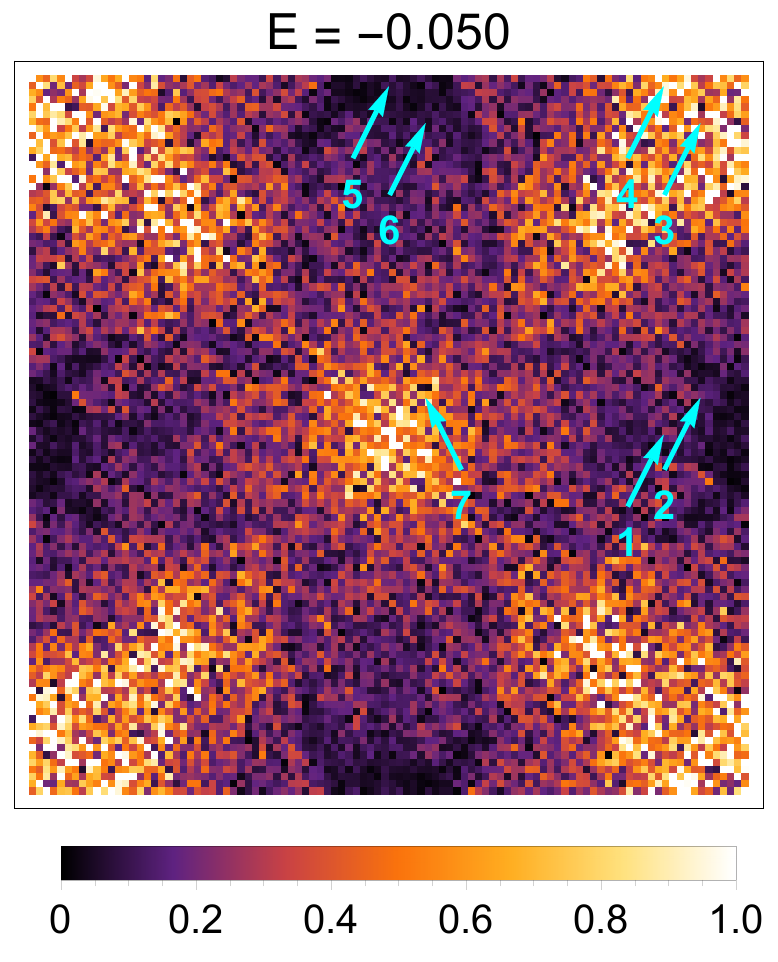}\\
	\includegraphics[width=.2\textwidth]{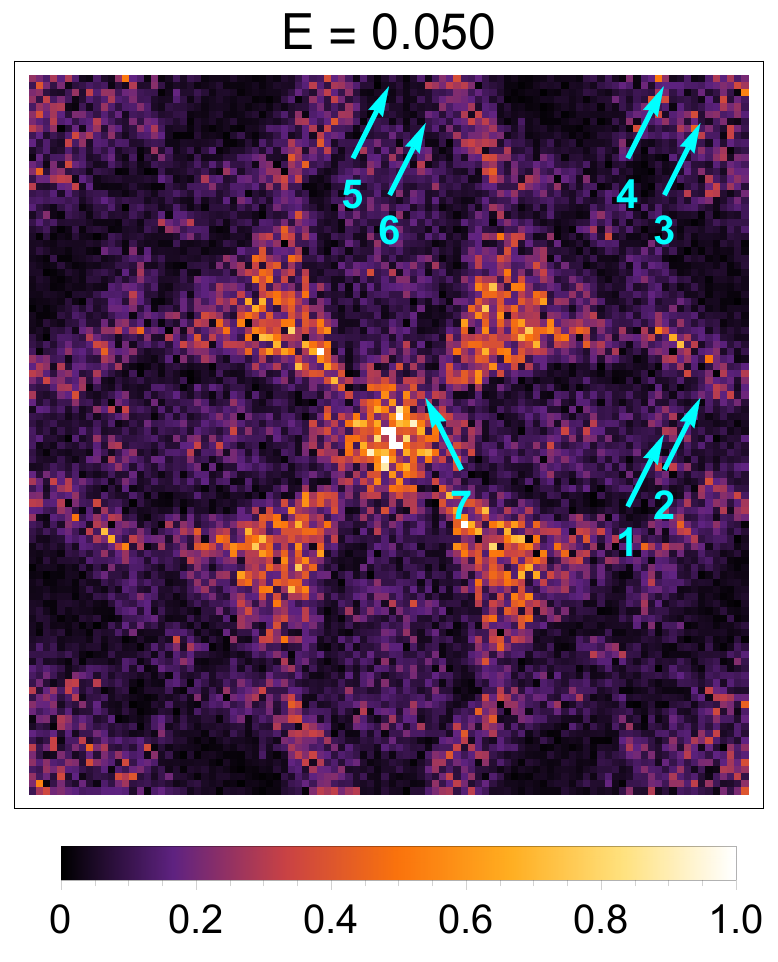}\hfill
	\includegraphics[width=.2\textwidth]{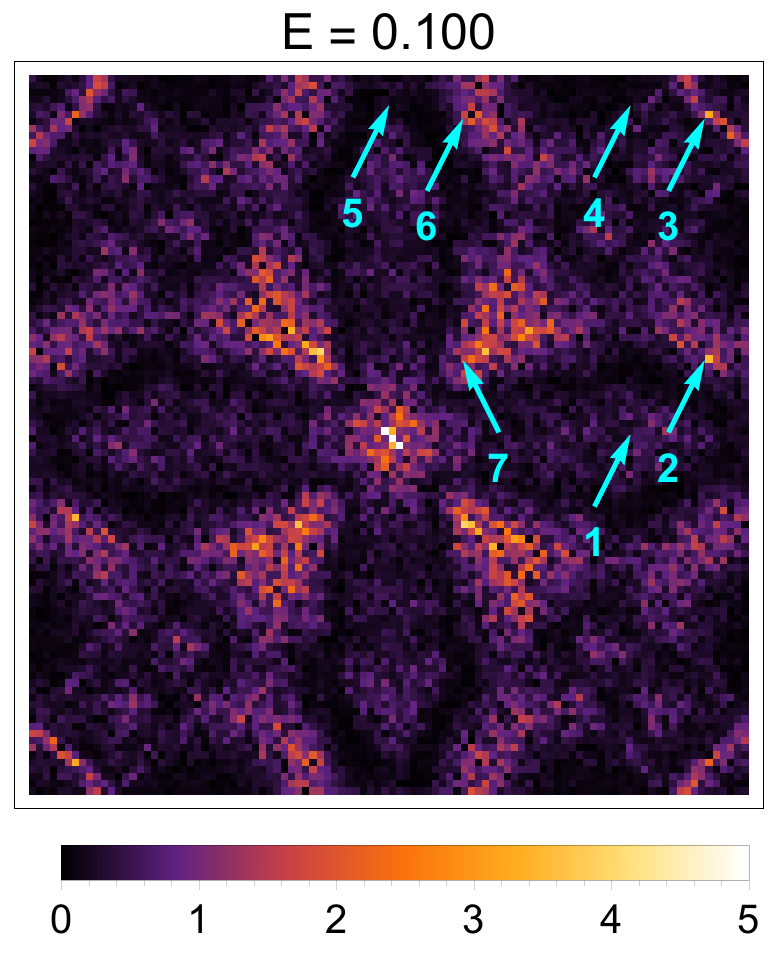}\hfill
	\includegraphics[width=.2\textwidth]{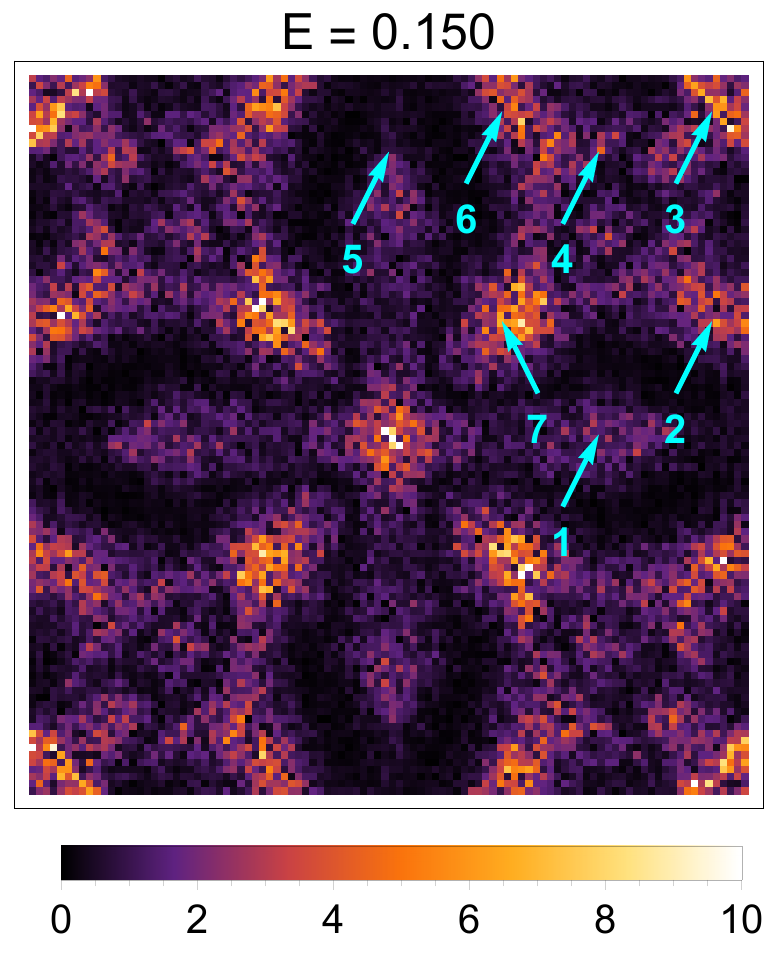}\hfill
	\includegraphics[width=.2\textwidth]{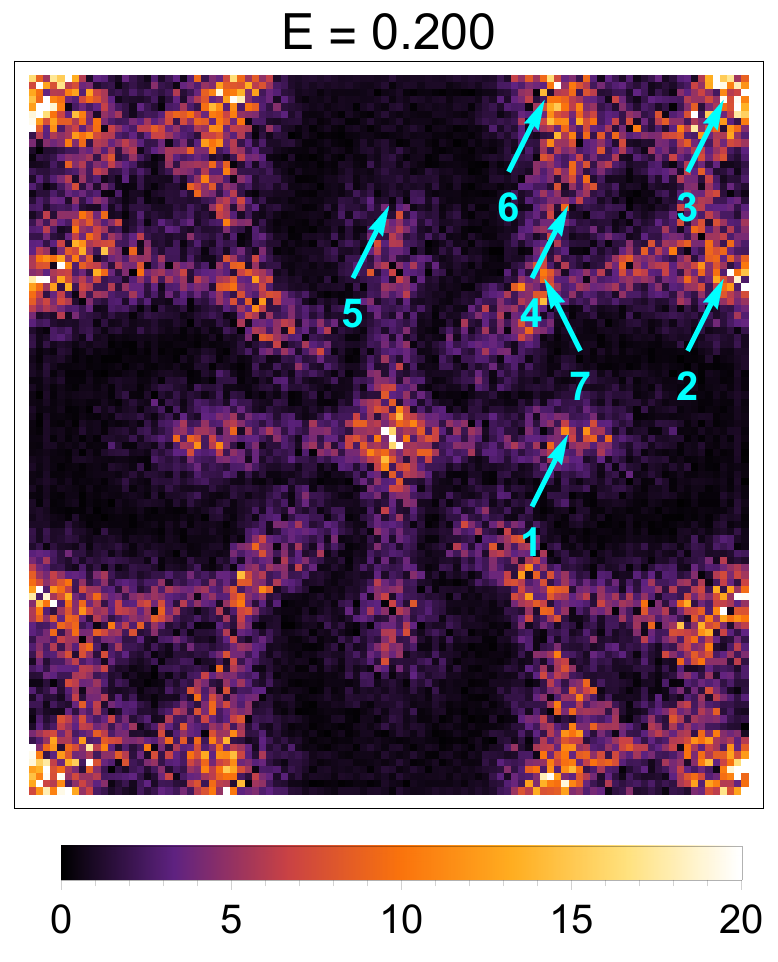}\hfill
	\includegraphics[width=.2\textwidth]{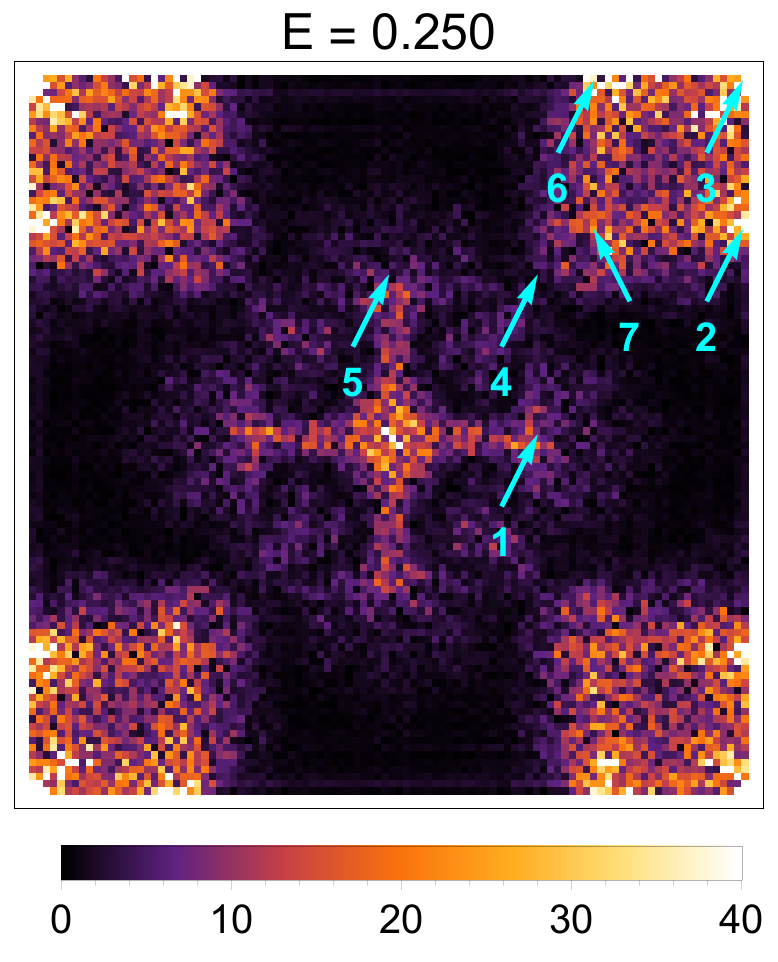}\hfill
	\caption{Fourier-transformed filtered maps for a system with a 0.5\% concentration of weak point-like scatterers ($V =0.5$). Shown are energies ranging from  $E = \pm0.050$ to  $E = \pm0.250$, along with arrows showing where the octet wavevectors are expected to be found. The color scaling varies with energy.}
	\label{fig:fspwfourier}
\end{figure*}

To complete the discussion of the multiple weak-impurity case, we will include the fork effect, discussed earlier in Section \ref{model}, and see whether this leads to dramatic differences in the observed real-space and Fourier-space maps.  In Figs.~\ref{fig:fspwreal} and~\ref{fig:fspwfourier} we show plots with the filtered LDOS for the weak-impurity case. It can be seen that the impurities are considerably more visible in the filtered real-space maps than in the unfiltered ones. The patterns in the filtered real-space maps resemble those found in the bare cases. One takeaway from this case is that for weak impurities the individual impurities remain visible whether the fork effect is present or not.  

The Fourier transforms of the filtered maps have a number of interesting features. Most of the momenta predicted by the octet model do show up in the power spectrum, and, notably, the locations of the peaks are \emph{not} altered relative to the unfiltered case. This is not surprising, as the fork effect does not alter the dispersion of the Bogoliubov quasiparticles, so the basic physics of the octet model remains in place. The main qualitative effect of the fork mechanism is the shifting of spectral weight from one part of momentum space to another, resulting in some differences from the unfiltered case---but nothing that results in the complete suppression of peaks expected from the octet model. The fork effect preserves the special momenta of the octet model. The shifting of the spectral weight however results in fuzzier peaks than in the unfiltered case.

The overall effect of the fork mechanism, at least in our simple treatment, is to amplify or suppress portions of the power spectrum without altering the presence of peaks that the octet model predicts will be present. In this sense the fork mechanism, while indeed a crucial phenomenon that one must ultimately incorporate in any description of the tunneling process, plays only a minor role in the overall description of quasiparticle interference in BSCCO. The issues associated with the point-like impurity case sans the fork effect---that the impurities are visible in real space and that the peaks seen in experiment are sharper than seen in numerical simulations---remain even when the fork effect is taken into account. In this sense the issues we discussed require a resolution beyond simply accounting for filter effects, and require examining whether the form of disorder we had used---namely, weak point-like scatterers---is indeed correct.

\subsection{Multiple Unitary Point-Like Impurities}

\begin{figure*}[ht]
	\centering
	\includegraphics[width=.25\textwidth]{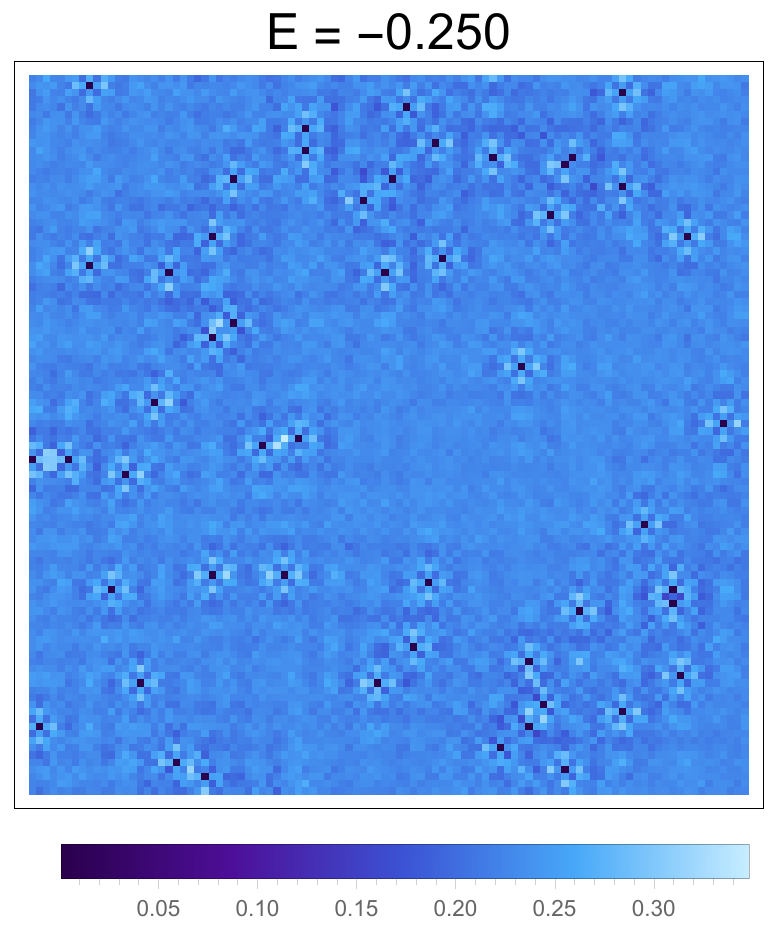}\hfill
	\includegraphics[width=.25\textwidth]{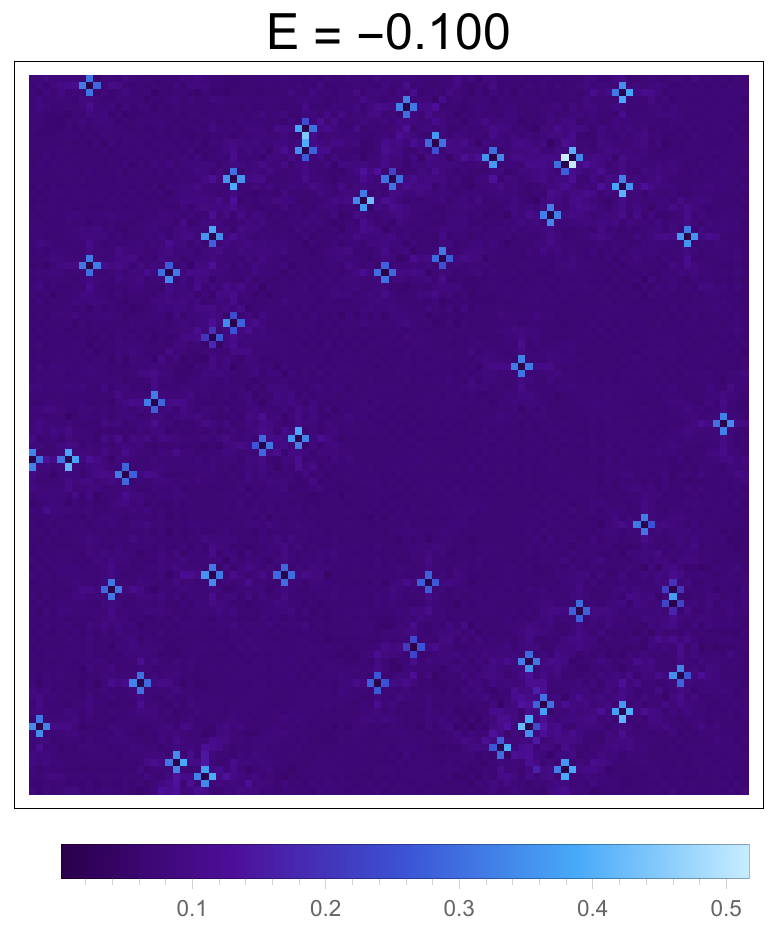}\hfill
	\includegraphics[width=.25\textwidth]{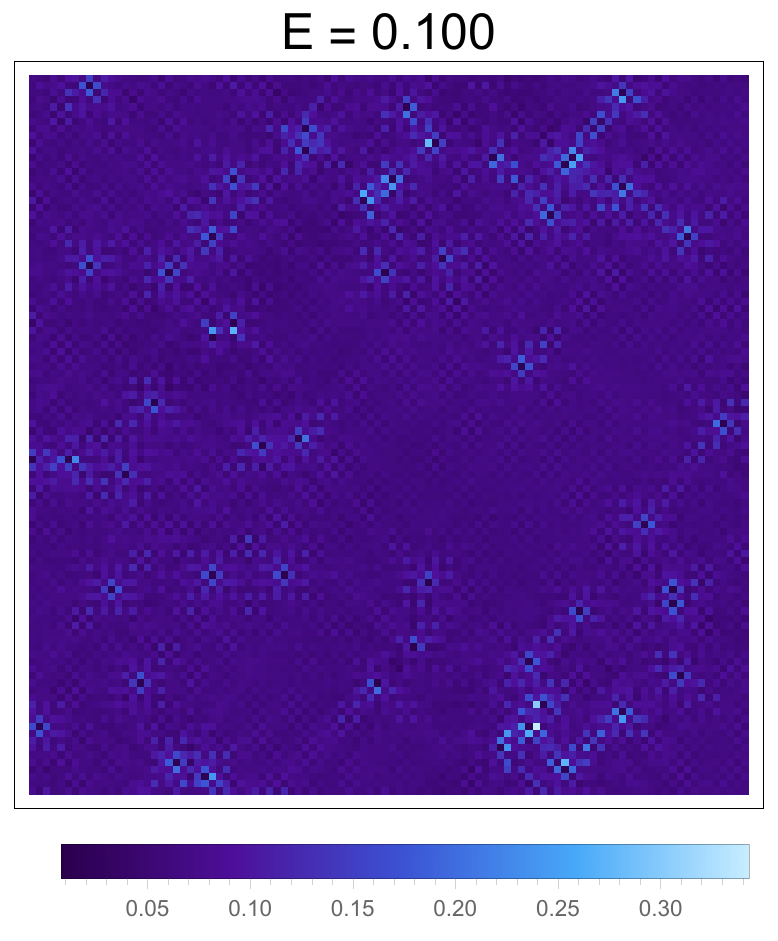}\hfill
	\includegraphics[width=.25\textwidth]{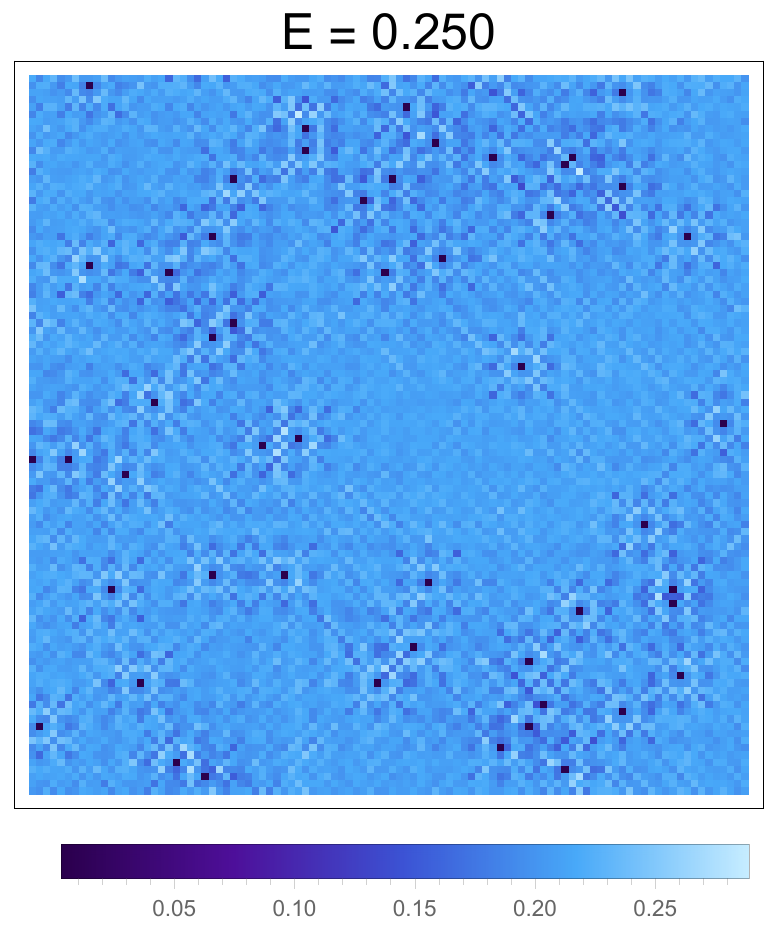}\hfill
	\caption{Real-space LDOS maps for a $d$-wave superconductor with a 0.5\% concentration of unitary point-like scatterers ($V = 10$) distributed randomly across the CuO$_2$ plane. The field of view is $100 \times 100$, and the energies shown are $E = \pm0.100$ and $E = \pm0.250$.}
	\label{fig:mpreal}
\end{figure*}

\begin{figure*}
	\centering
	\includegraphics[width=.2\textwidth]{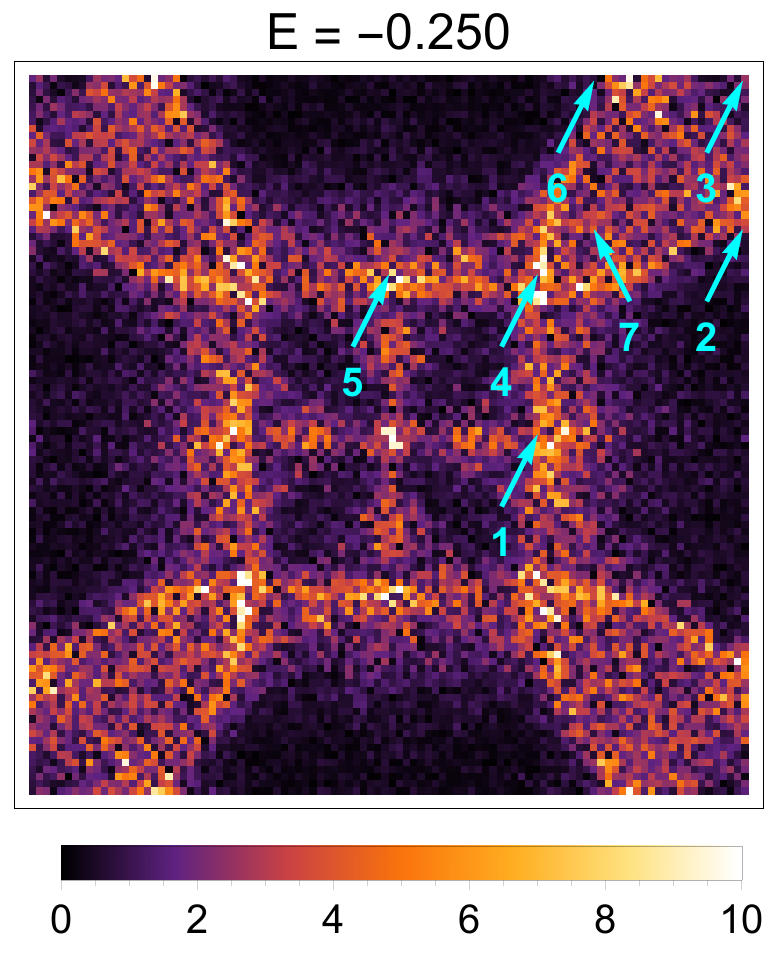}\hfill
	\includegraphics[width=.2\textwidth]{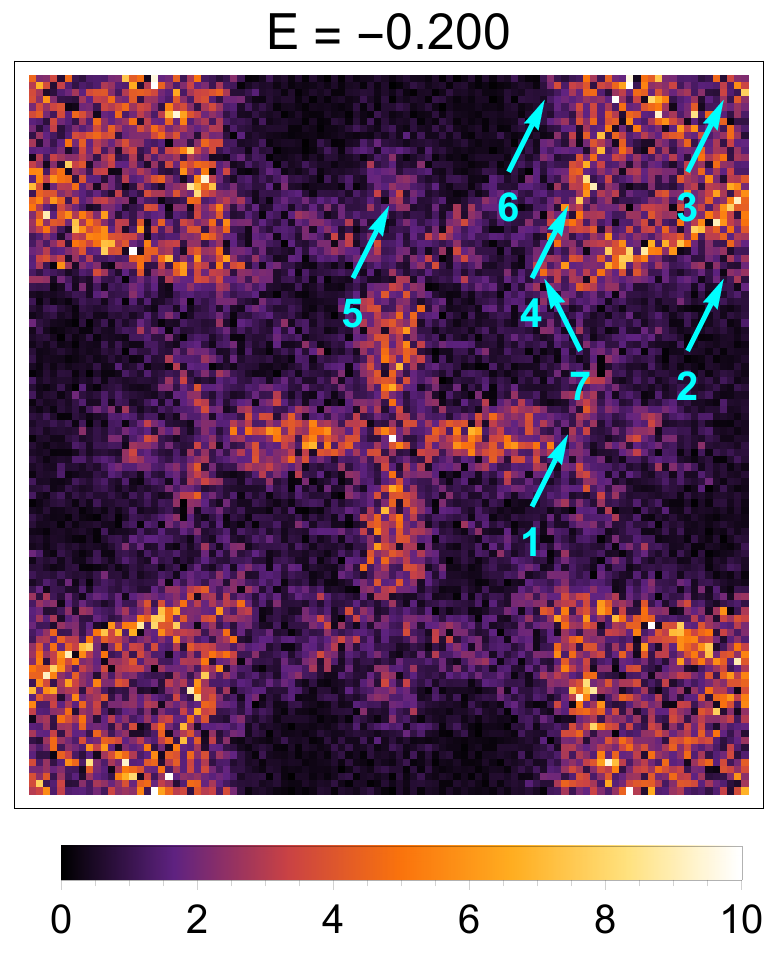}\hfill
	\includegraphics[width=.2\textwidth]{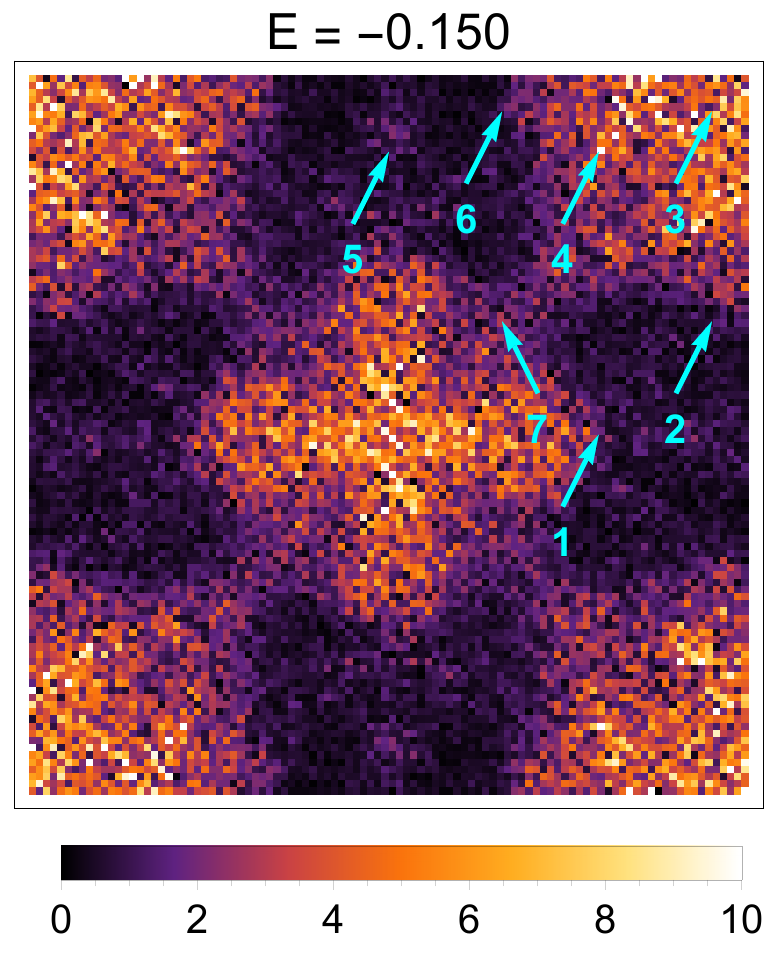}\hfill
	\includegraphics[width=.2\textwidth]{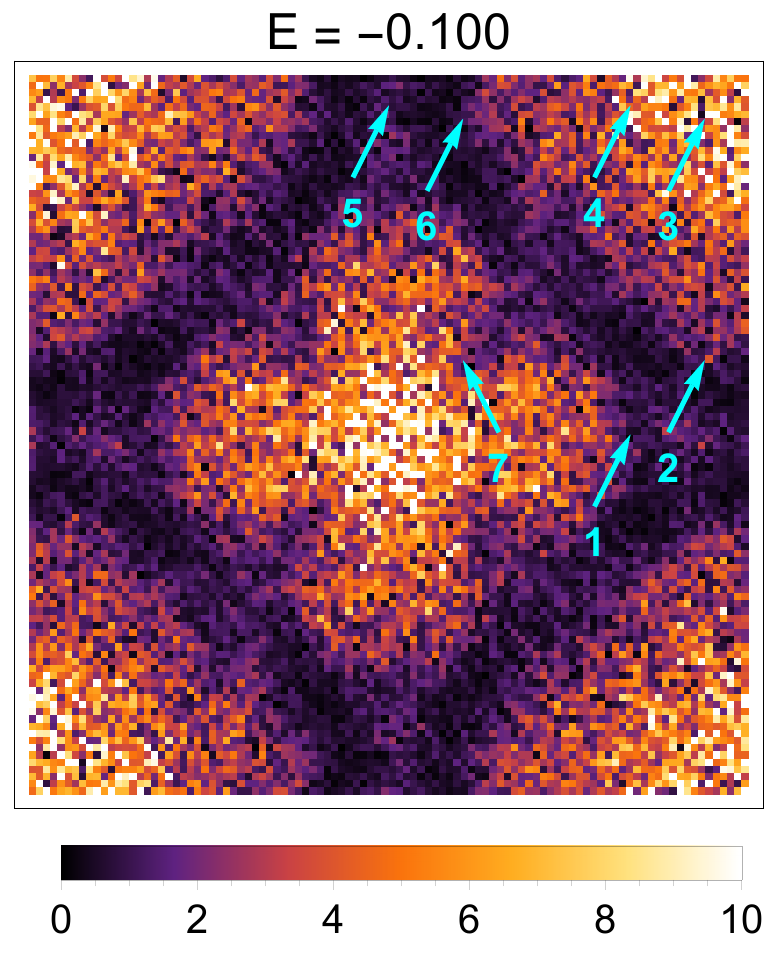}\hfill
	\includegraphics[width=.2\textwidth]{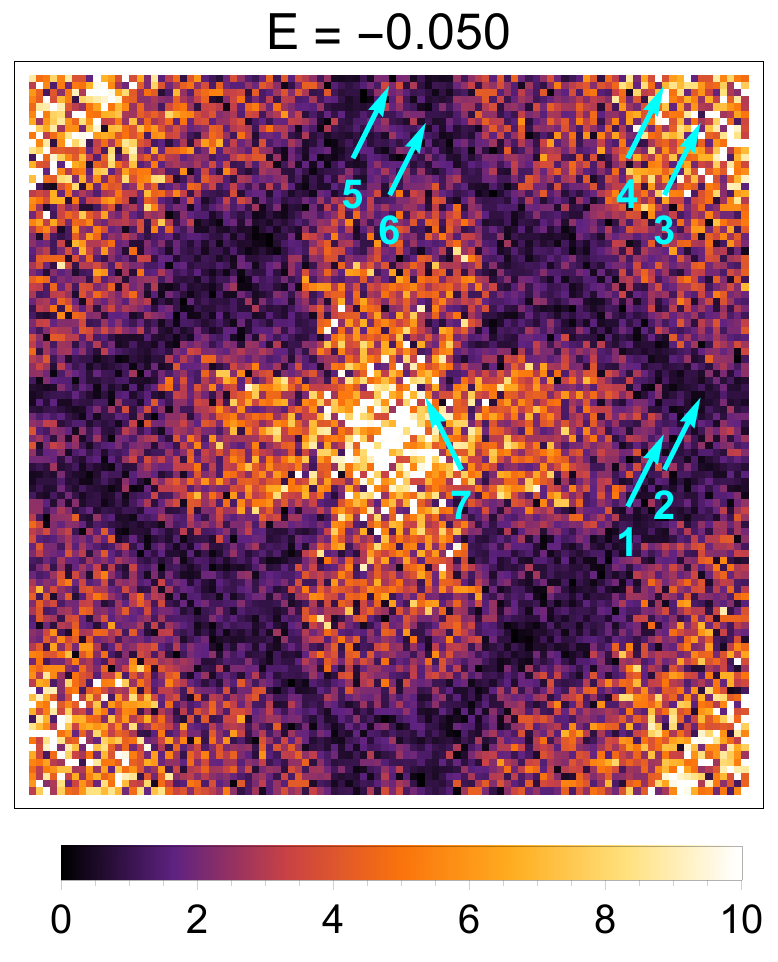}\\
	\includegraphics[width=.2\textwidth]{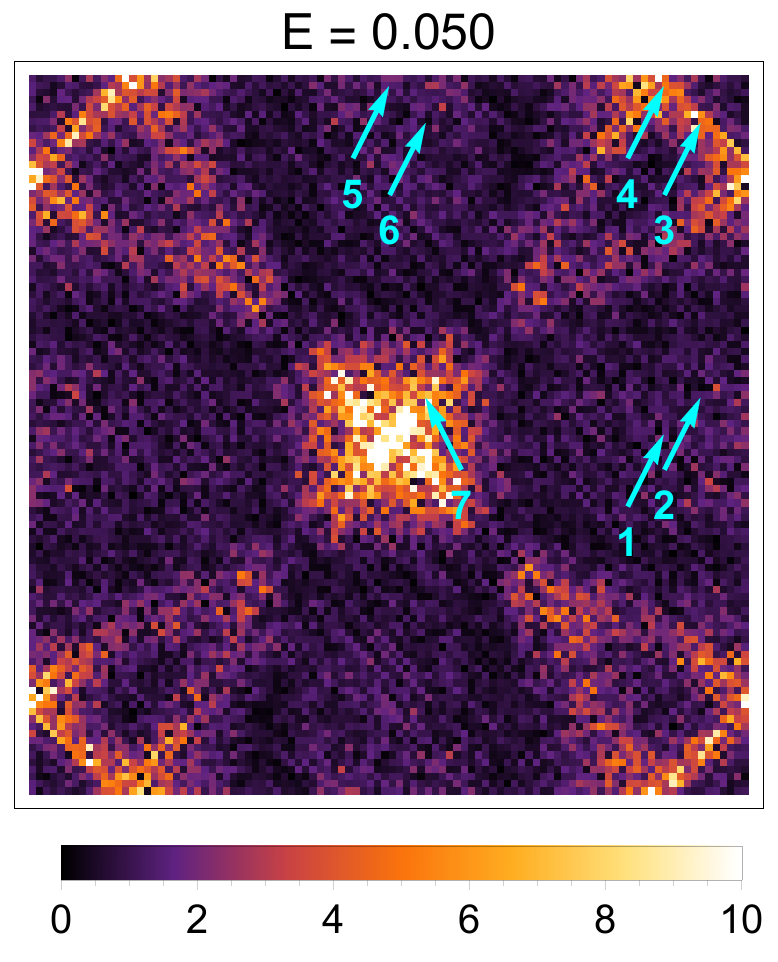}\hfill
	\includegraphics[width=.2\textwidth]{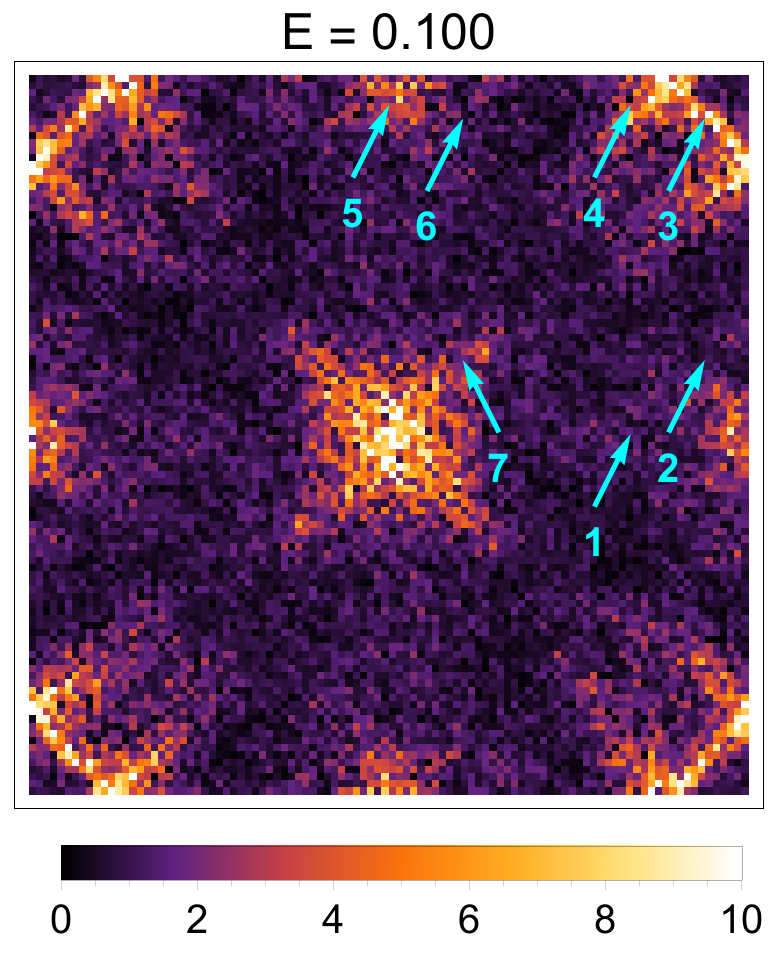}\hfill
	\includegraphics[width=.2\textwidth]{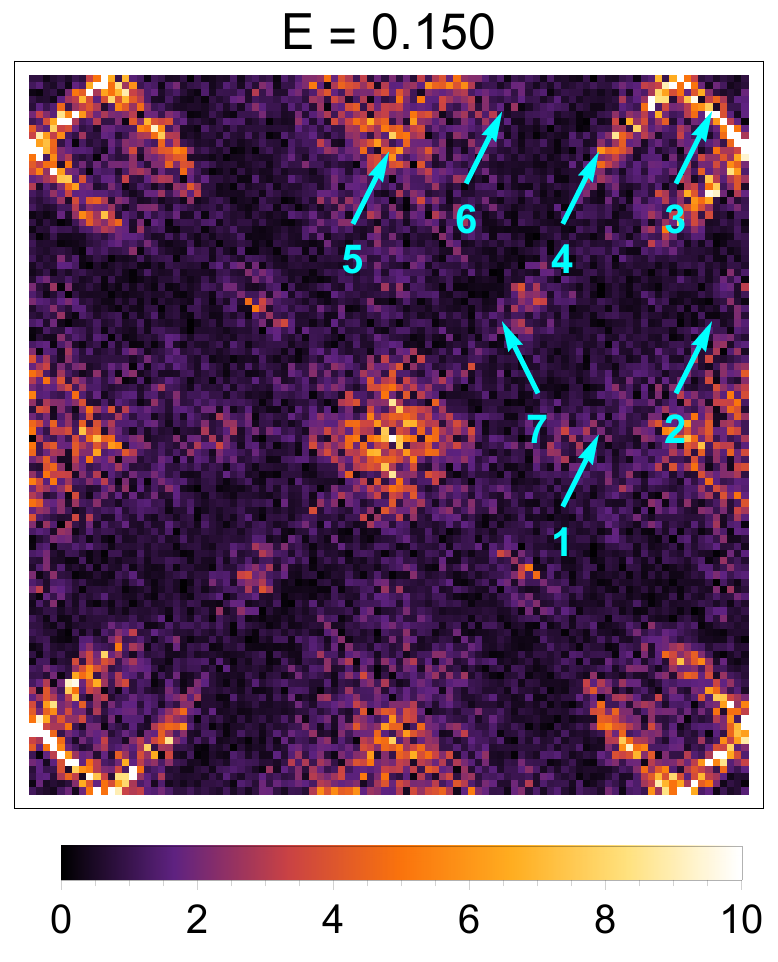}\hfill
	\includegraphics[width=.2\textwidth]{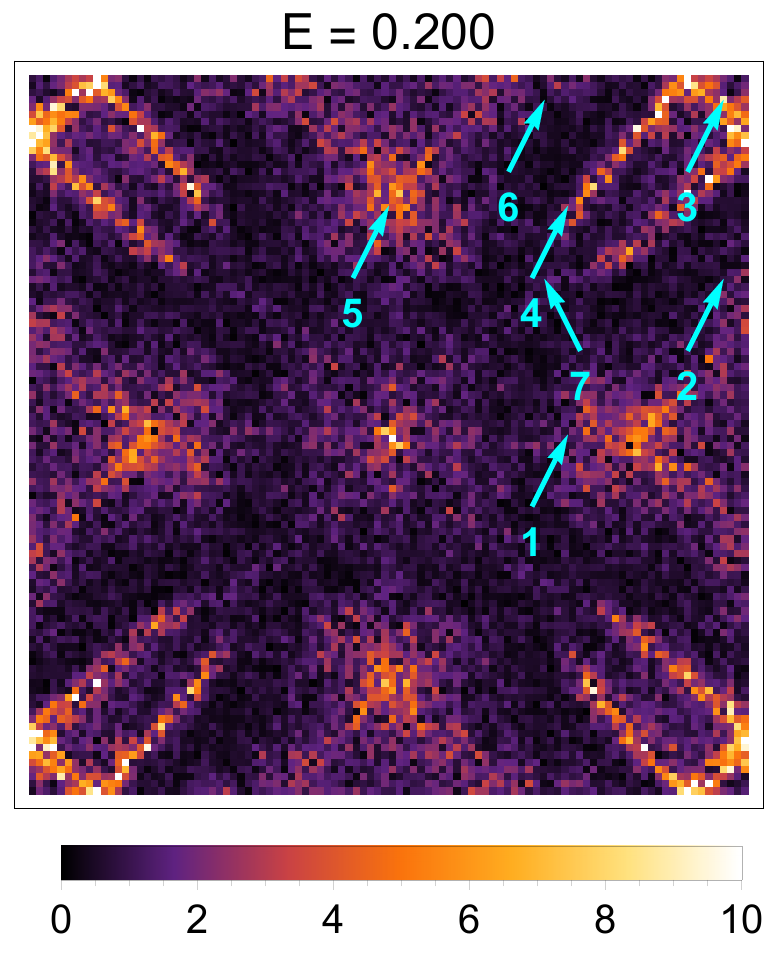}\hfill
	\includegraphics[width=.2\textwidth]{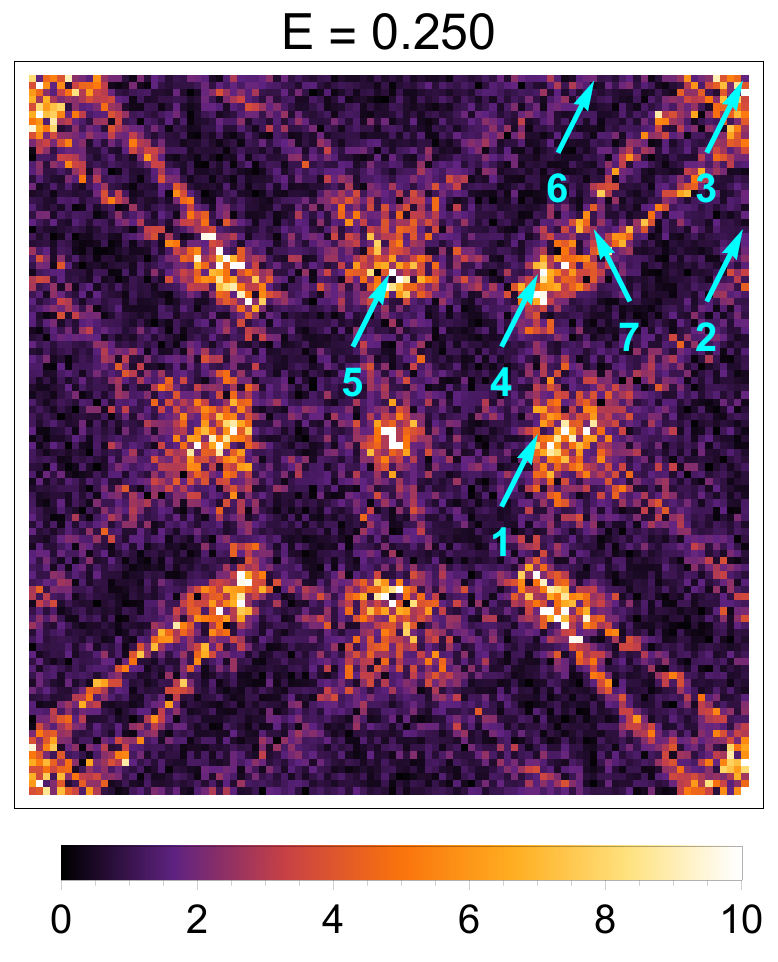}\hfill
	\caption{Fourier-transformed maps for a system with a 0.5\% concentration of unitary point-like scatterers ($V =10$). Shown are energies ranging from  $E = \pm0.050$ to  $E = \pm0.250$, along with arrows showing where the octet wavevectors are expected to be found. The color scale is the same for all energies}
	\label{fig:mpfourier}
\end{figure*}

For completeness we discuss the case where many \emph{unitary} point-like scatterers are present, especially in relation to the weak-potential case we previously tackled. Plots are shown in Figs.~\ref{fig:mpreal} and~\ref{fig:mpfourier}. In these plots we took the many-impurity disorder configuration to be the same as in the weak case, and we set $V = 10$. This form of disorder provides a realistic description of zinc-doped BSCCO, as zinc impurities are known to behave as unitary scatterers.\cite{pan2000imaging}

It is worth noting the similarities and differences between the weak-impurity and unitary-impurity cases. The real-space pictures for both cases are similar in that the individual impurities themselves can be easily detected. There is a difference, however: in the unitary case, the LDOS is heavily suppressed at the impurity site, whereas in the weak-impurity case it is generally not so. Real-space maps from both weak- and unitary-scatterer cases feature long-ranged diagonal streaks, but the modulations for the unitary-scatterer case are much more pronounced than in the weak case. The power spectra of the unitary-impurity case also display considerable differences from those of the weak case. While peaks at the same locations and with similar dispersive behavior can be observed in both cases, the weights of those peaks are different. In particular, $\mathbf{q}_1$, $\mathbf{q}_4$, and $\mathbf{q}_5$ are much stronger than in the weak case, and in fact become the most prominent wavevectors in the power spectrum as energies increase. That said, the Fourier maps are far noisier than in the weak case, and as a consequence of strong scattering due to the large size of $V$, the main feature of the power spectrum is a series of diffuse streaks originating from scattering between points on CCEs. In a manner similar to that of the weak-impurity case, the peaks corresponding to the octet momenta emerge as the special points along these streaks with the highest spectral weight. These streaks in the unitary case are a considerably more prominent feature of the power spectrum than in the weak-impurity case.

\subsection{Dependence of the Power Spectrum on the Impurity Strength}
	
\begin{figure}[ht]
	\centering
	\includegraphics[width=.5\textwidth]{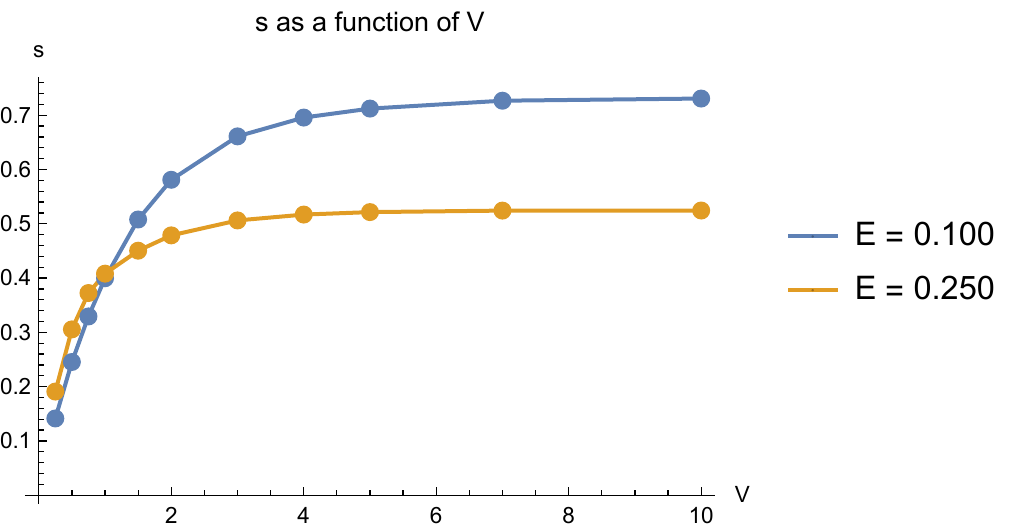}\hfill
	\caption{Plot of the impurity weight $s$ (defined in Eq.~\ref{eq:sdef}) versus the the impurity strength $V$ for $E = 0.100$ and $E = 0.250$. Here we consider a single point-like impurity located in the center of the sample.}
	\label{fig:sv}
\end{figure}	

While we have restricted ourselves to the case of point-like impurities, our results for weak and unitary impurities suggest that even within the point-like model of disorder, qualitatively different behavior can be observed by varying the impurity strength. One could then ask if it is possible to identify whether the QPI observed in experiment is due primarily to unitary or weak scatterers. We will attempt to provide a measure that quantifies the impact of the impurity strength $V$ on the power spectrum.

Our main measurable of interest will be a quantity $s$, which we dub the \emph{impurity weight} and define in the following way:
\begin{equation}
s(V, E) = \frac{\sum_{\mathbf{q \in \text{BZ}}} |\rho(\mathbf{q}, V, E)| - |\rho(\mathbf{q = 0}, V, E)|}{\sum_{\mathbf{q \in \text{BZ}}} |\rho(\mathbf{q}, V, E)|}.
\label{eq:sdef}
\end{equation}
Here $\rho(\mathbf{q}, V, E)$ is the Fourier transform of the LDOS map at energy $E$ of a $d$-wave superconductor with a single point-like impurity with strength $V$ positioned in the middle of the field of view. As Eq.~\ref{eq:sdef} shows, the impurity weight is simply the ratio of the integrated power spectrum \emph{without} the $\mathbf{q = 0}$ contribution to the \emph{total} integrated power spectrum (\emph{i.e.}, \emph{with} the $\mathbf{q = 0}$ contribution). $\rho(\mathbf{q = 0}, V, E)$ is removed from the numerator because that contribution is what one obtains when Fourier-transforming an LDOS map of a spatially homogeneous $d$-wave superconductor. The numerator of Eq.~\ref{eq:sdef} thus describes only the contributions of the inhomogeneities to the power spectrum. One then expects that in the limit where the impurity is very weak, the power spectrum is dominated by the $\mathbf{q = 0}$ contribution and hence the impurity weight $s$ is very small. We note that because of the underlying lattice the power spectrum is backfolded into the first Brillouin zone. We consider only unfiltered LDOS maps and their Fourier transforms.

We plot $s$ as a function of $V$ for two representative energies in Fig.~\ref{fig:sv}. We let $V$ vary from $V = 0.25$ to $V = 10$, covering the unitary- and weak-scatterer cases discussed in depth earlier, and consider $E = 0.100$ and $E = 0.250$. It can be seen that when the impurity is weak, $s$ is a small quantity that depends approximately linearly on $V$. There is a broad crossover region around $V \approx 2$ where $s$ begins to increase more slowly with $V$. For larger values of $V$ corresponding to unitary scatterers, $s$ does not show any dependence on $V$ and saturates to a fixed value.

As a tool for potentially identifying the nature of point-like scatterers in experiment, the impurity weight is admittedly limited, unless one already knows this for cuprates that are already firmly identified as hosting unitary scatterers, such as zinc-doped BSCCO. The main takeaway from these results is that for weak scatterers the impurity weight is less than for unitary ones. In this light it would be interesting to revisit data from BSCCO with and without zinc impurities and calculate the impurity weight for various bias voltages. One identifying signal that QPI in BSCCO is caused by weak impurities is an $s$-value that is less than that obtained from zinc-doped BSCCO.

\section{Smooth Disorder} \label{smooth}
\begin{figure*}[ht]
	\centering
	\includegraphics[width=.25\textwidth]{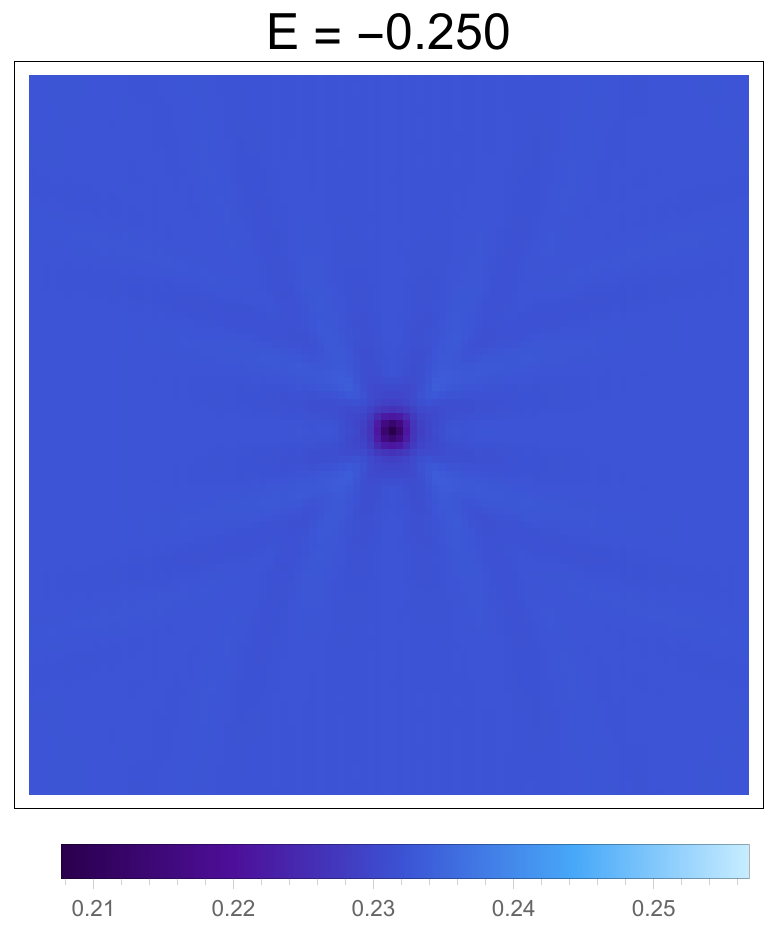}\hfill
	\includegraphics[width=.25\textwidth]{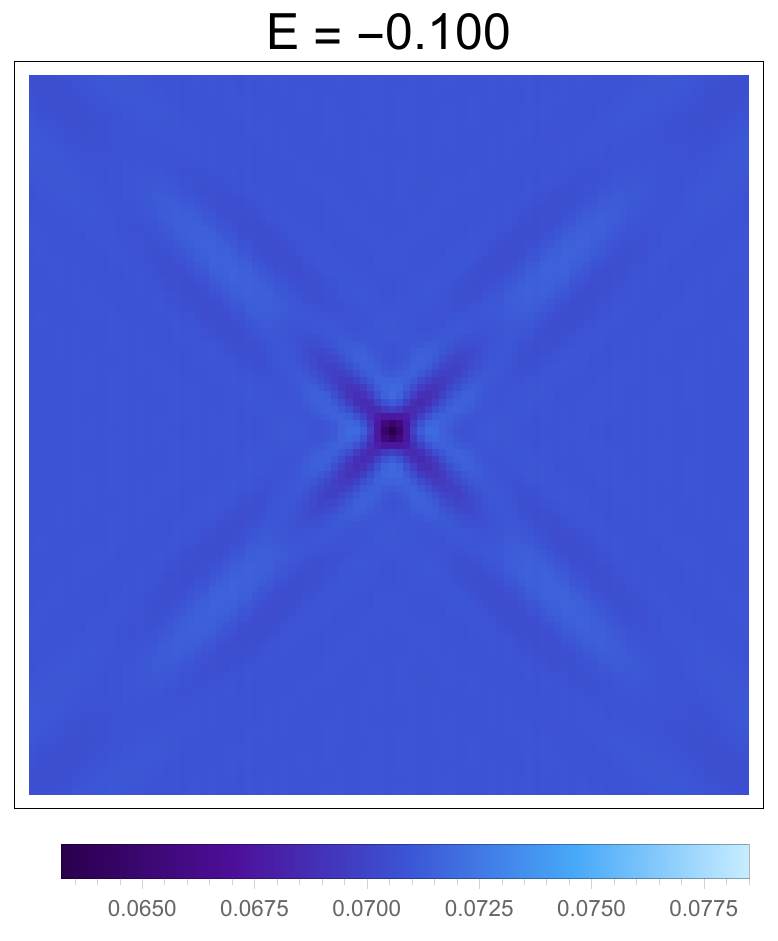}\hfill
	\includegraphics[width=.25\textwidth]{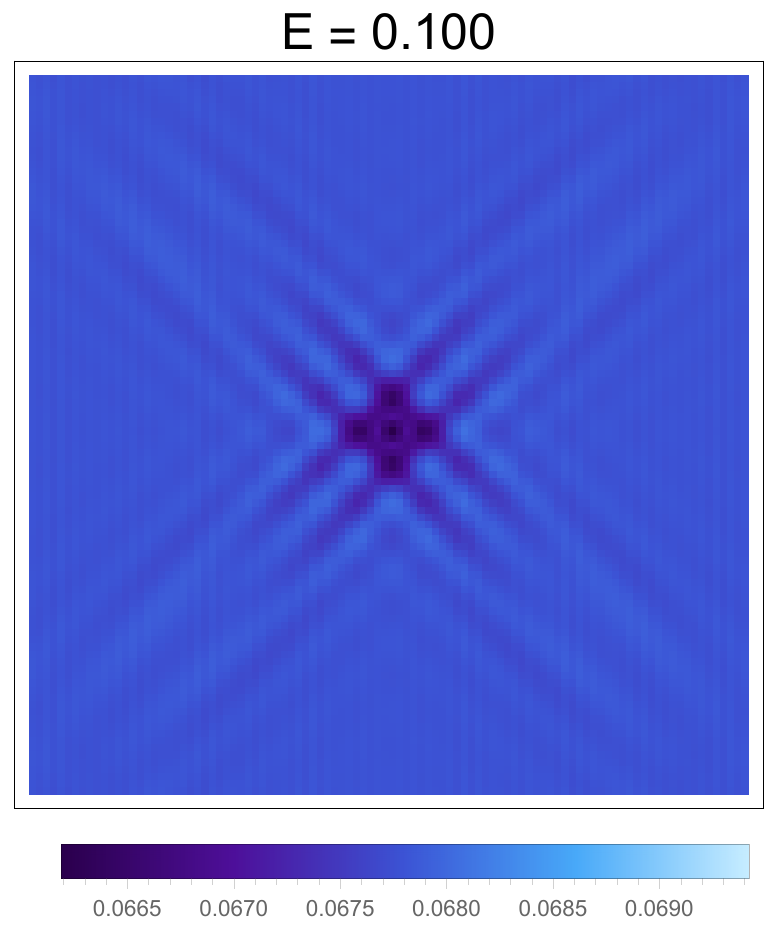}\hfill
	\includegraphics[width=.25\textwidth]{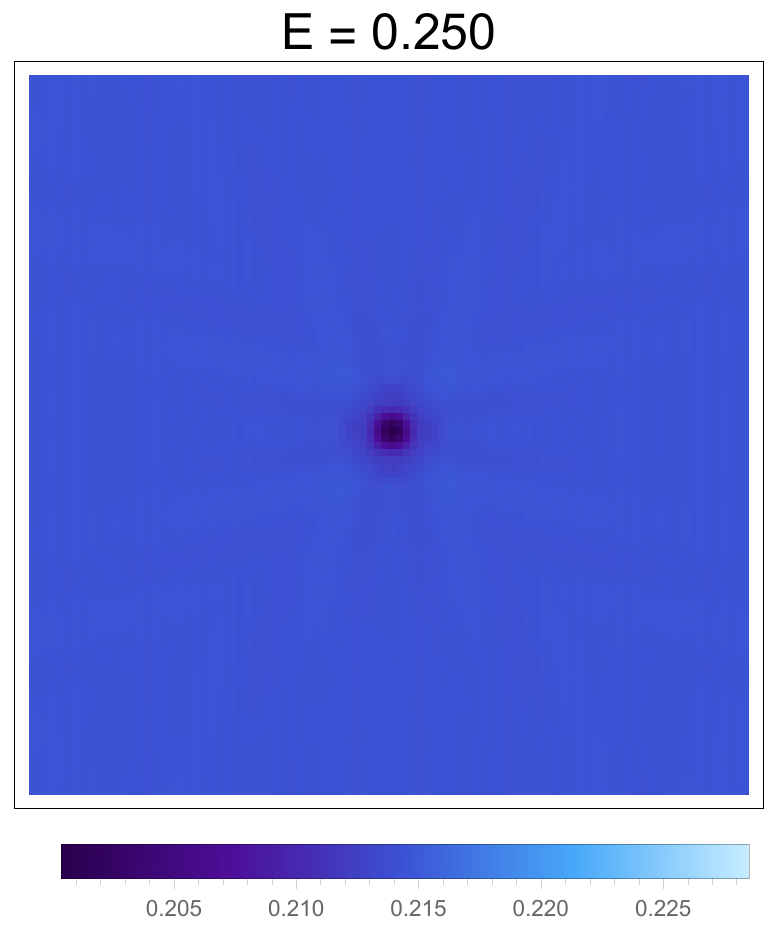}\hfill
	\caption{Real-space LDOS maps for a $d$-wave superconductor with a single smooth-potential scatterer ($V_{sm} = 0.5$, $L = 4$, $z = 2$) located at the center of the field of view and off the CuO$_2$ plane. The field of view is $100 \times 100$, and the energies shown are $E = \pm0.100$ and $E = \pm0.250$.}
	\label{fig:ssreal}
\end{figure*}

\begin{figure*}
	\centering
	\includegraphics[width=.2\textwidth]{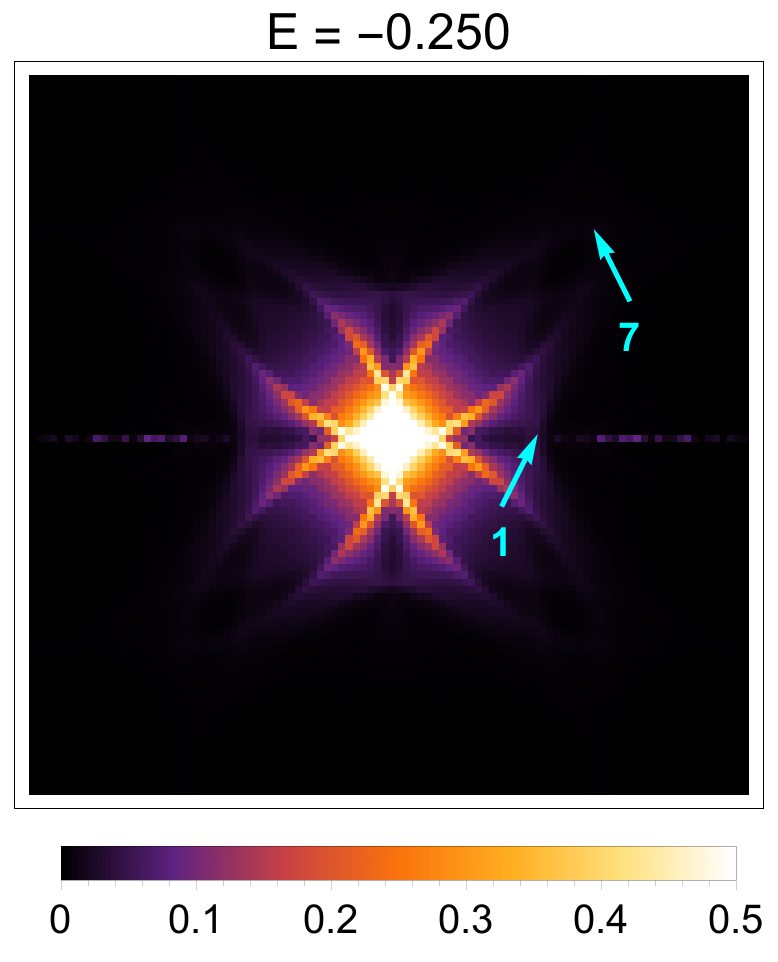}\hfill
	\includegraphics[width=.2\textwidth]{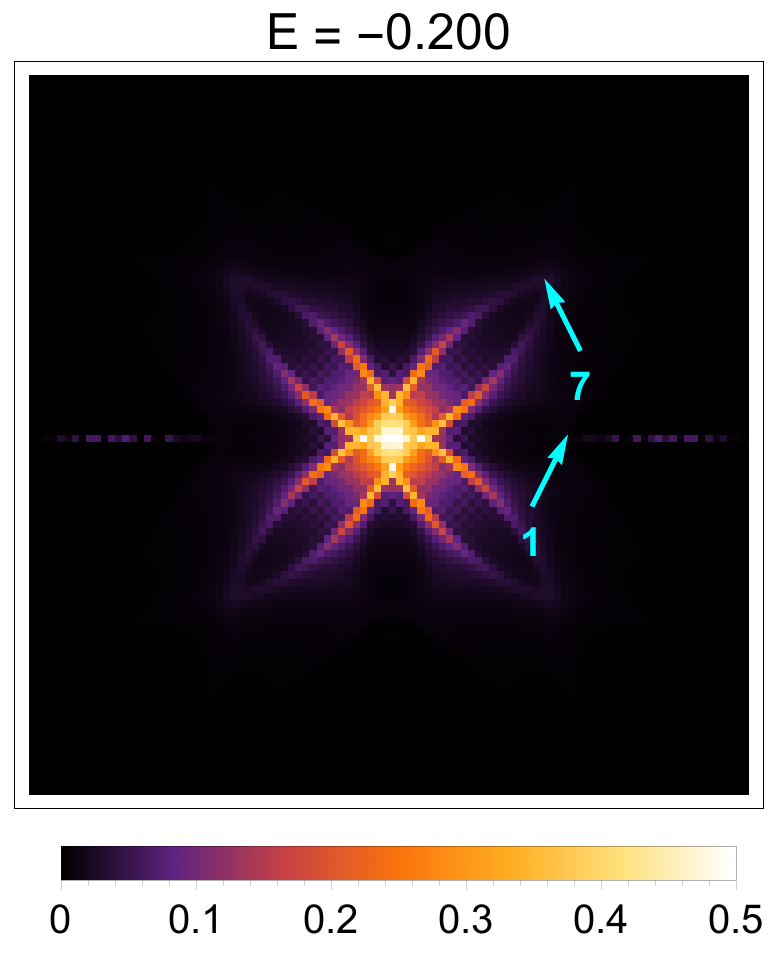}\hfill
	\includegraphics[width=.2\textwidth]{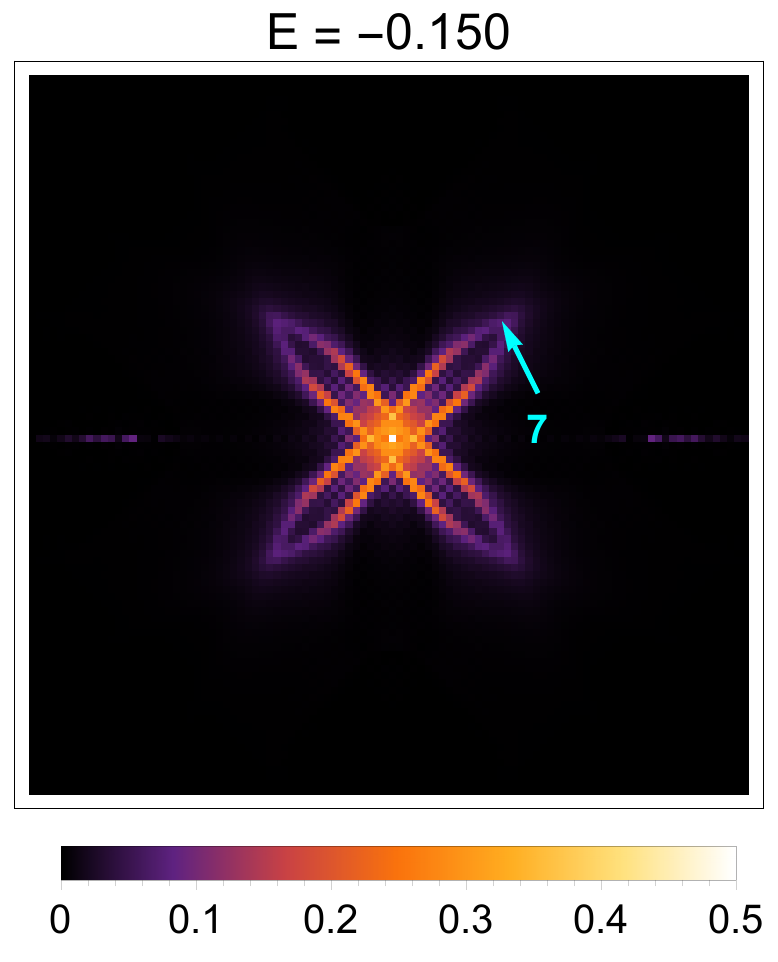}\hfill
	\includegraphics[width=.2\textwidth]{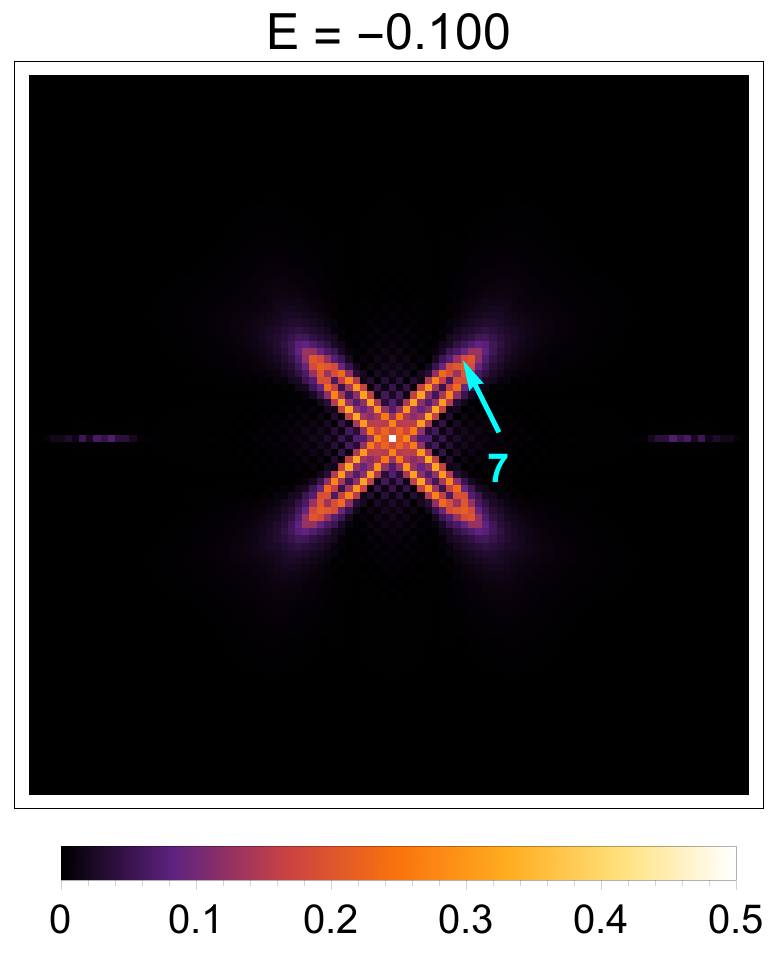}\hfill
	\includegraphics[width=.2\textwidth]{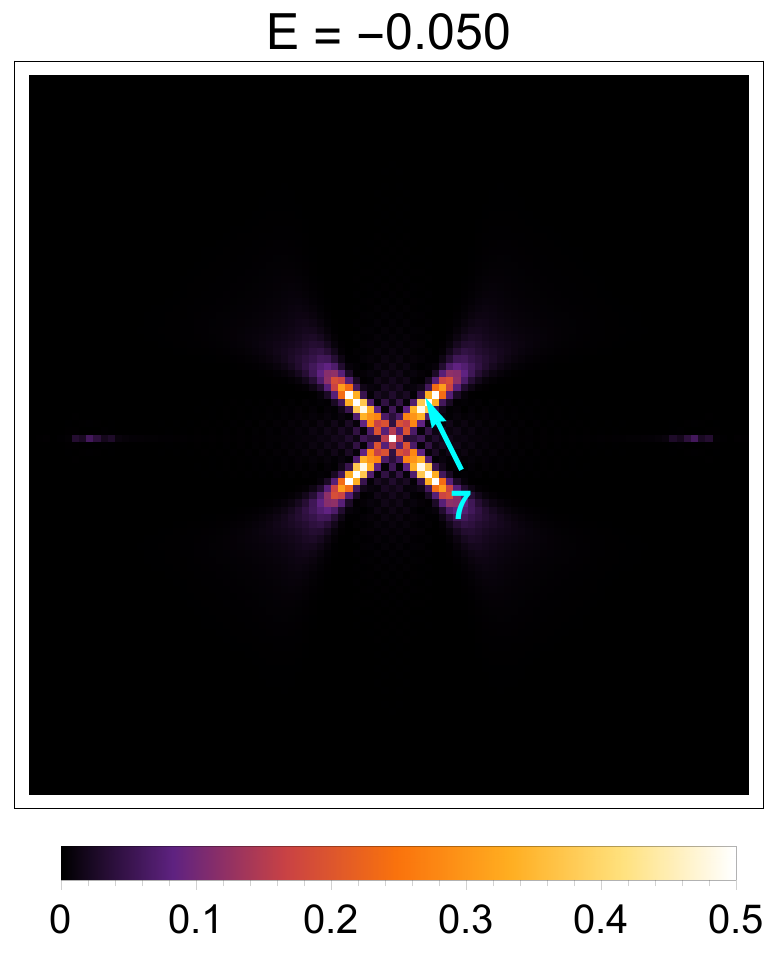}\\
	\includegraphics[width=.2\textwidth]{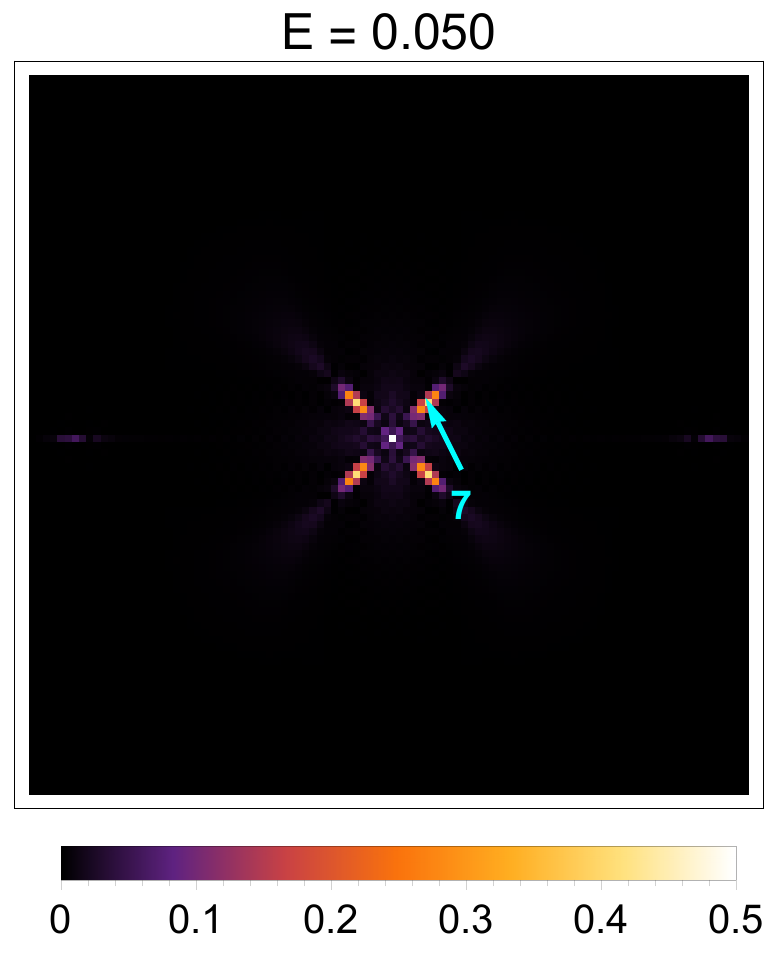}\hfill
	\includegraphics[width=.2\textwidth]{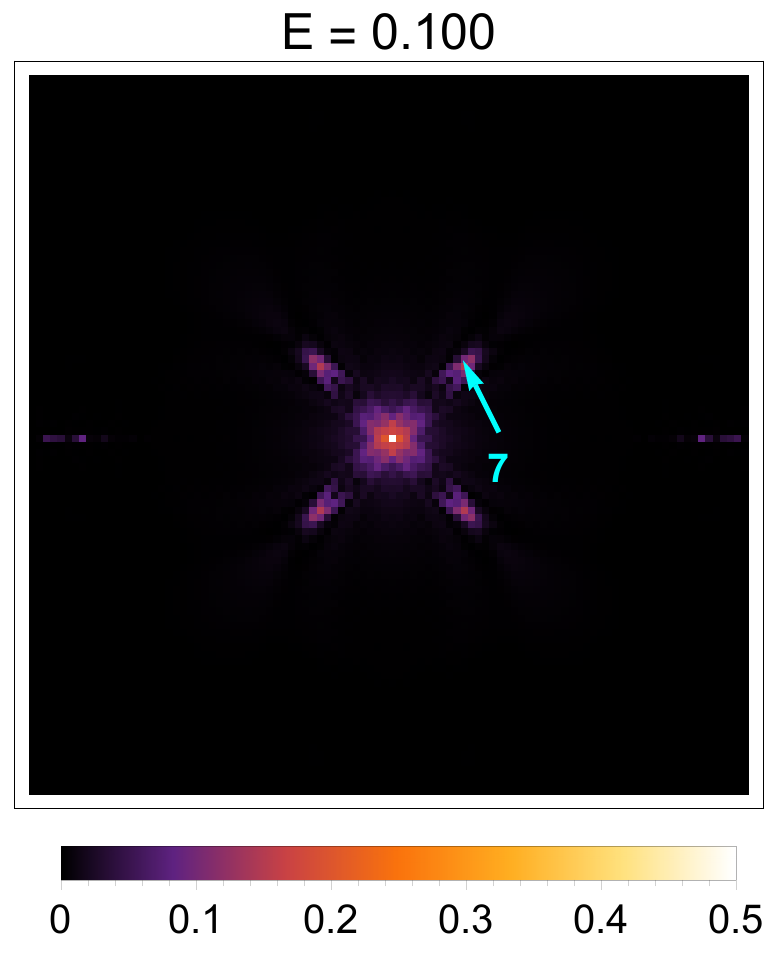}\hfill
	\includegraphics[width=.2\textwidth]{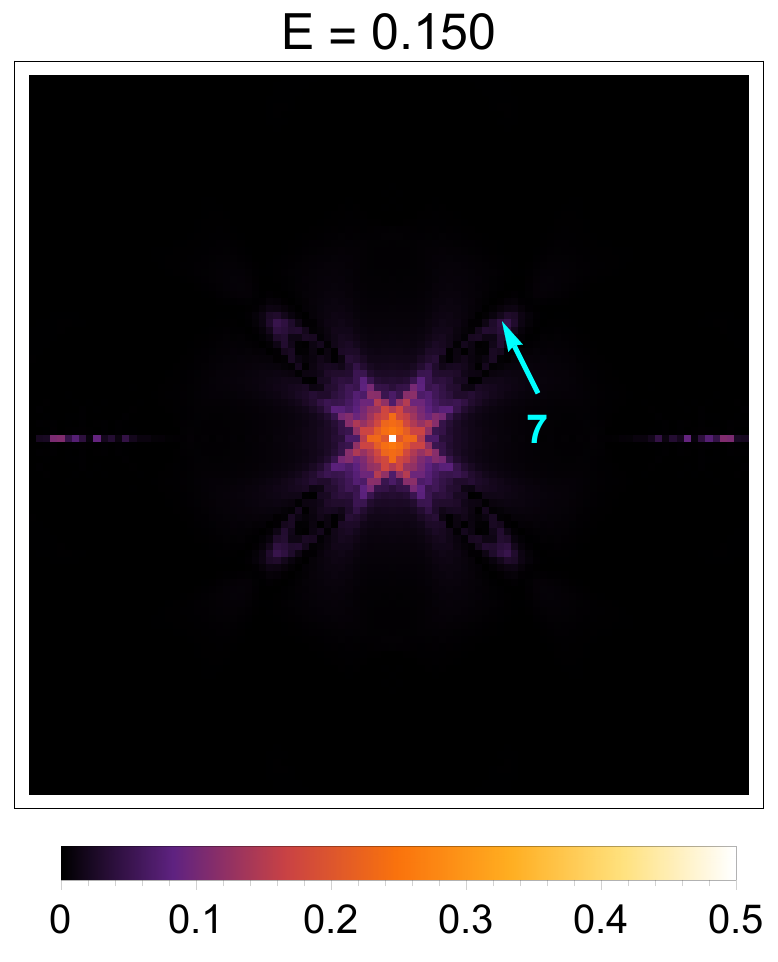}\hfill
	\includegraphics[width=.2\textwidth]{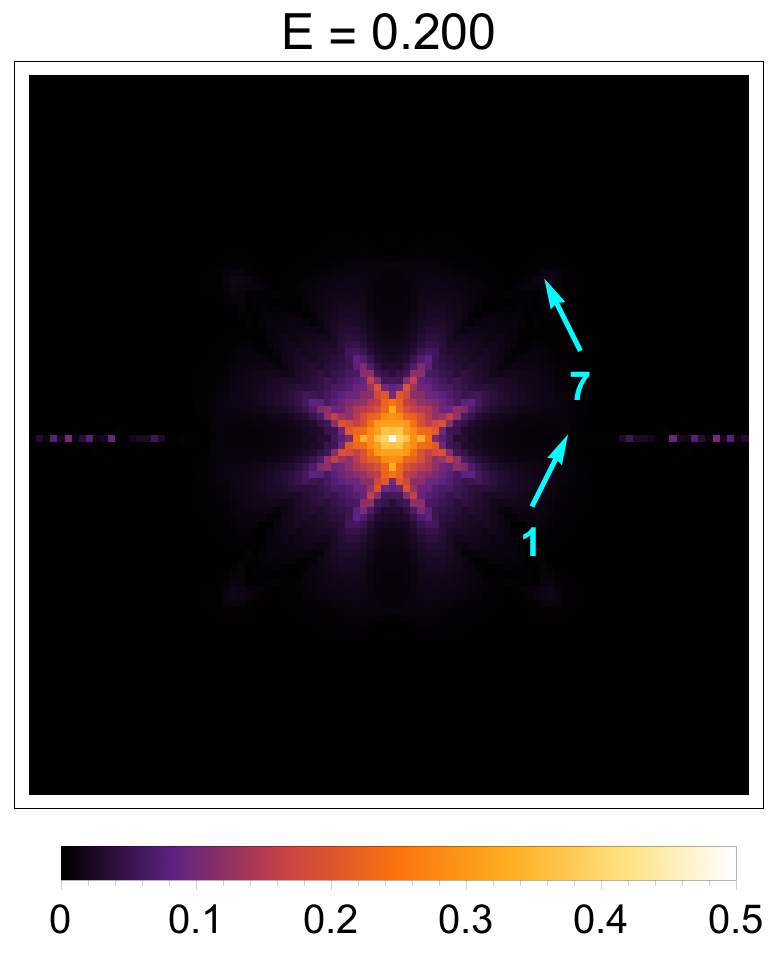}\hfill
	\includegraphics[width=.2\textwidth]{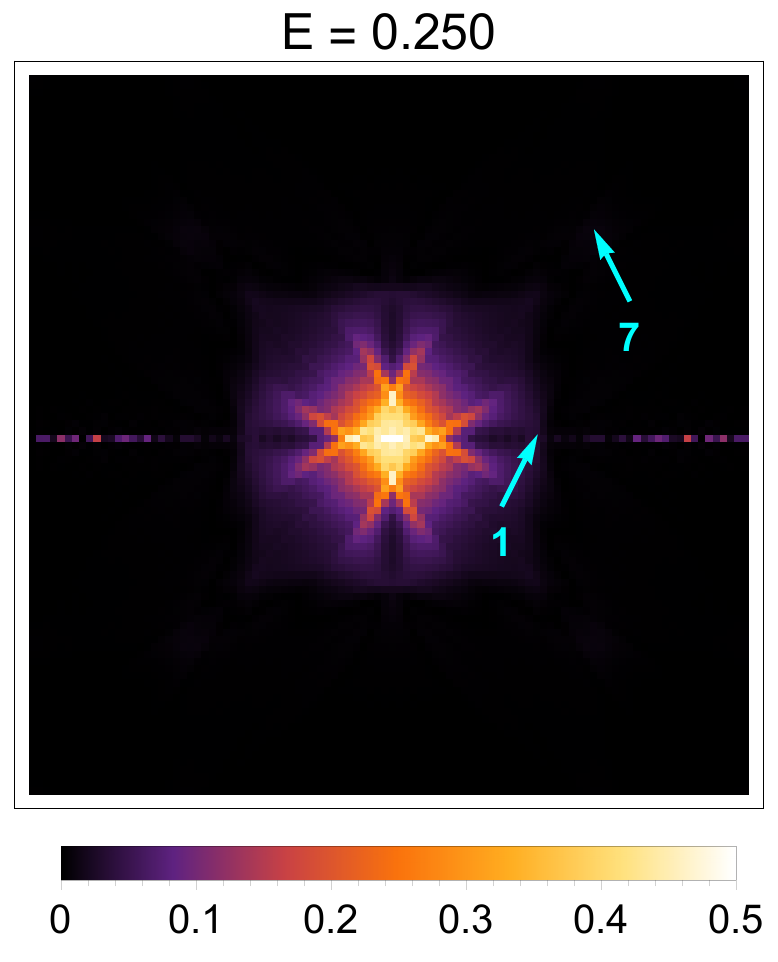}\hfill
	\caption{Fourier-transformed maps for a system with a single smooth-potential scatterer ($V_{sm} = 0.5$, $L = 4$, $z = 2$). Shown are energies ranging from  $E = \pm0.050$ to  $E = \pm0.250$, along with arrows showing where the octet wavevectors are expected to be found. Only $\mathbf{q}_7$ is visible at low energies, while both $\mathbf{q}_1$ and $\mathbf{q}_7$ can be seen at higher energies. The color scale is the same for all energies.}
	\label{fig:ssfourier}
\end{figure*}

When one takes into account the chemistry of intrinsic disorder in the cuprates, it is difficult to justify point-like disorder as a possible source of the effects we study. Since the early history of the high-$T_c$ field, the prevailing understanding is that the doping mechanism is more closely related to modulation doping. The cuprate planes are widely assumed to be chemically very clean. The dopants are located in the ionic/insulating buffer layers some distance away from the metallic planes. The dopants are charged impurities which act as sources for poorly screened Coulomb potentials, which in turn affect the physics on the CuO$_2$ planes. The overall result is a smooth disorder potential which is characterized by small scattering wavevectors.\cite{pan2001microscopic} In a similar way, these dopants can also affect the tilting patterns inside the cuprate planes, and the elastic strain will result in a smooth form of disorder as well.\cite{eisaki2004effect} Smooth disorder potentials have been invoked in explaining the apparent discrepancy between the magnitude of the transport and single-particle lifetimes; the former, which depends heavily on large-momentum scattering, is much smaller than the latter, and hence any scattering that occurs is argued to be forward (\emph{i.e.}, small-momentum) scattering due to impurities located off the CuO$_2$ planes.\cite{abrahams2000angle, varma2001effective, abrahams2003hall} 

Previous theoretical treatments of smooth disorder have been motivated by bulk measurements\cite{huckestein2000quasi,altland2002theories}, but there has been a good amount of work motivated by STS studies as well\cite{zhu2004power,nunner2005dopant,nunner2006fourier}. In particular, Nunner \emph{et al.} provide a comprehensive treatment of the Fourier spectra of a single isolated weak smooth scatterer.\cite{nunner2006fourier} However, in general, work on this form of disorder has not been as extensive as that on point-like disorder, especially in the limit where a very large number of smooth scatterers are present. Following our treatment of point-like scatterers, we will first revisit the case of a single smooth scatterer, first studied by Nunner \emph{et al}., to provide a picture of which scattering processes dominate. We will then discuss the consquences on the LDOS and the power spectrum when one has a large number of these impurities in the sample. We will also look at the sensitivity of the power spectrum to changes in the screening length of Coulomb potentials, especially as such details are not microscopically known.

Smooth potential scatterers in $d$-wave superconductors are not quite as easy to model as point-like scatterers, due to the fact that one cannot apply the $T$-matrix formalism to this form of disorder to obtain the LDOS. The typical method involves extracting the LDOS directly by diagonalizing the Bogoliubov-de Gennes Hamiltonian. This has the restriction that only small systems can usually be accessed. However, smooth scatterers are easily treated by the numerical method that we use, with the advantage that we can scale up the system size to better visualize the LDOS. The flexibility of our method allows us to realistically model a smooth disorder potential \emph{a la} modulation doping. We model smooth disorder using a screened Coulomb potential arising from a source located \emph{outside} the copper-oxygen plane:
\begin{equation}
V(\mathbf{r}) = V_{sm}\frac{e^{\frac{-\sqrt{(\mathbf{r - r}_i)^2 + z^2 }}{L}}}{\sqrt{(\mathbf{r - r}_i)^2 + z^2 }}
\label{smoothpot}
\end{equation}
Here $\mathbf{r}_i$ is the location of the impurity projected onto the CuO$_2$ plane, $z$ is the distance along the $z$-axis from the CuO$_2$ plane to the impurity, and $L$ is the screening length. We will take the potential to be weak, with $V_{sm} = 0.5$, and  as a typical case we set $z = 2$ and $L = 4$ in units of lattice constants, so the length scales are small relative to the system size. 

\subsection{Single Smooth Scatterer}
At the single-impurity level, there are already rather drastic differences between the maps for the smooth scatterer and those for the point-like one. Fig.~\ref{fig:ssreal}  shows real-space LDOS maps for a smooth scatterer for various energies. Note the scale that we used to make the image clearer---the modulations are \emph{much} smaller than in the point-like impurity case. The LDOS is not suppressed above the impurity site, but is reduced from the clean-limit value only by a small amount. There is a pattern of crisscrossing diagonal streaks with four-fold rotational symmetry centered about the impurity site. When one uses the same scale as we used in the point-like case to visualize this, these patterns are quite hard to see.

When one takes the Fourier transform of these LDOS maps, the differences from the point-like case are even more pronounced, as one can see in Fig.~\ref{fig:ssfourier}. Unlike in the case of a point-like scatterer, the Fourier-transformed maps show that only \emph{small-momenta} scattering processes contribute to the LDOS modulations. Large-momenta processes are almost completely suppressed. A closer examination reveals that only \emph{intranodal} scattering processes occur in the presence of smooth potentials at low energies. That is, scattering occurs only between states lying on the same ``banana.'' This can be seen by looking at the surviving peaks. For a broad range of energies, only $\mathbf{q}_7$---the peak corresponding to diagonal tip-to-tip scattering along the same ``banana''---survives. 

With increasing energy, even $\mathbf{q}_7$ becomes suppressed. A faint peak corresponding to $\mathbf{q}_1$ begins to appear in the power spectrum, but it is much less visible than $\mathbf{q}_7$ was at lower energies. The spectrum shows mostly streaks corresponding to small-momenta intranodal processes, as well as peaks in the horizonal and vertical directions where these streaks overlap. The mostly incoherent momenta seen in the power spectrum and the absence of any prominent peaks explain why the real-space picture is largely featureless. There are no longer any  processes corresponding to $\mathbf{q}_7$ that will give rise to periodic modulations along the diagonal directions. As in the previous real-space picture, there is no suppression of the LDOS above the impurity; instead there is only a small reduction of the LDOS.

The takeaway from the single-impurity case is that impurity-induced modulations in the LDOS do occur for smooth scatterers, as they do for point-like scatterers. The crucial difference is that large-momentum scattering is absent, thanks to the smoothness of the potential---even when $V(\mathbf{r})$ is reasonably short-ranged, with a screening length on the order of a few lattice constants.

\subsection{Multiple Smooth Scatterers}

\begin{figure*}[ht]
	\centering
	\includegraphics[width=.25\textwidth]{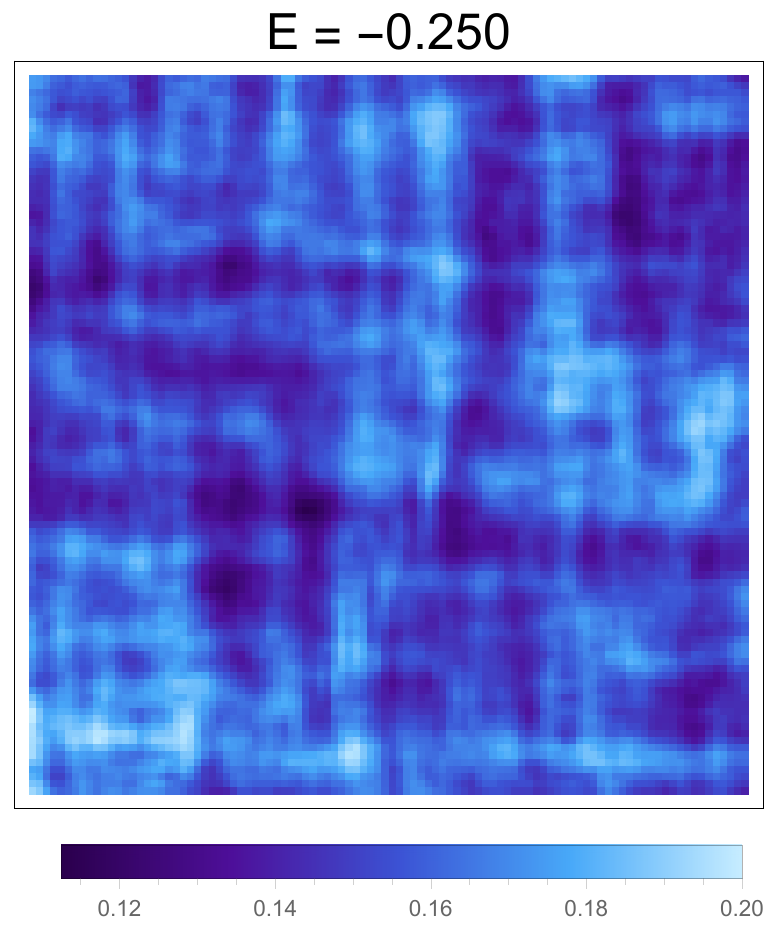}\hfill
	\includegraphics[width=.25\textwidth]{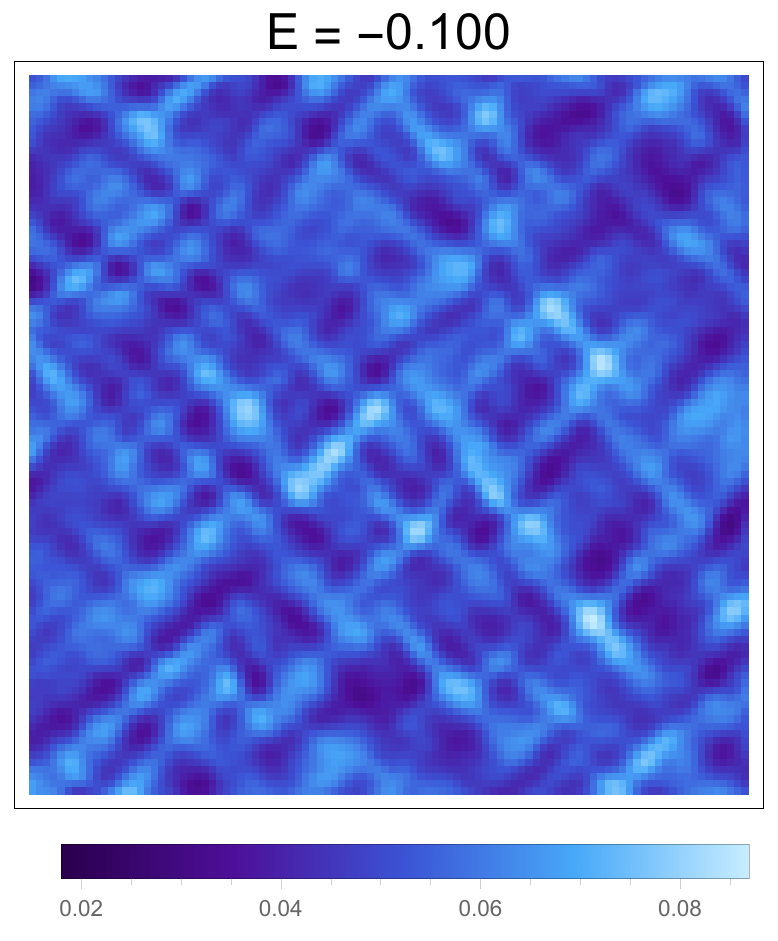}\hfill
	\includegraphics[width=.25\textwidth]{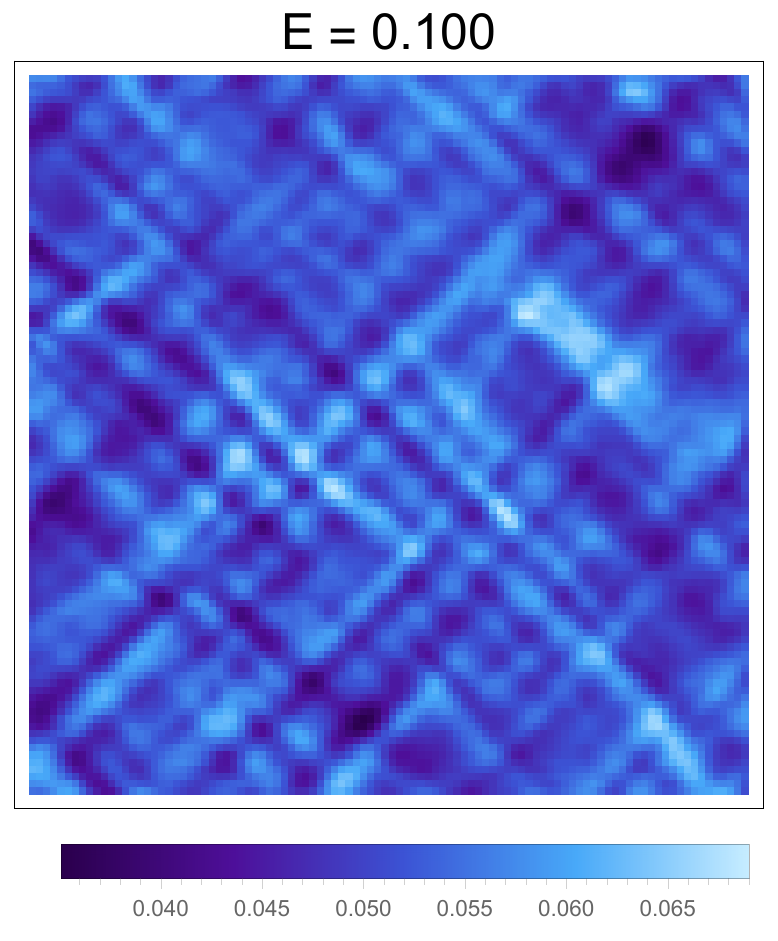}\hfill
	\includegraphics[width=.25\textwidth]{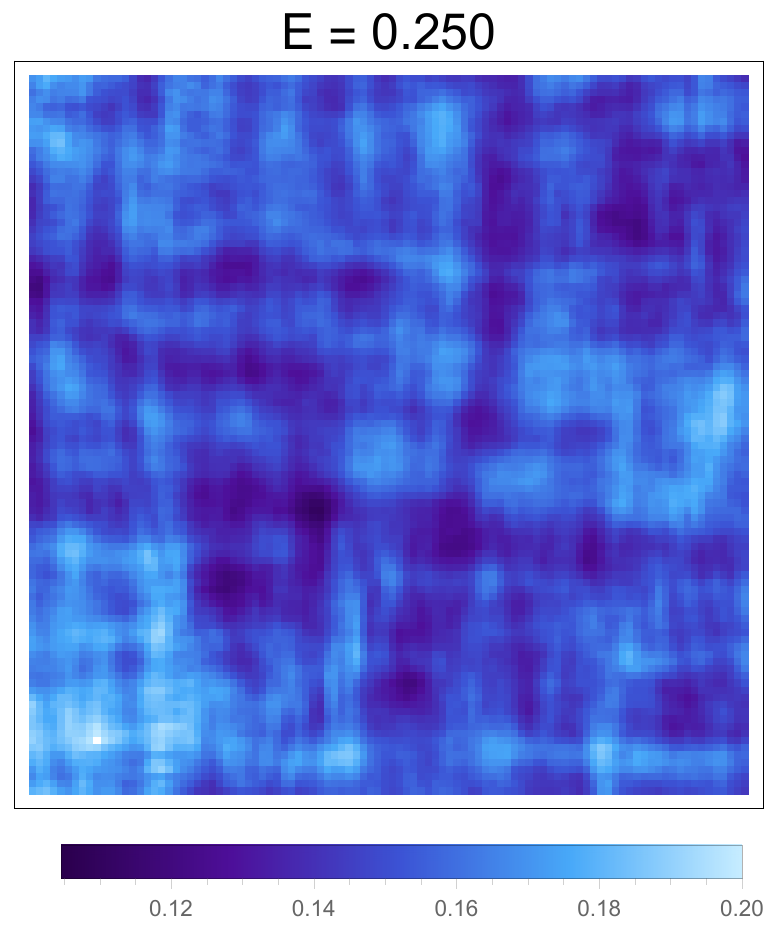}\hfill
	\caption{Real-space LDOS maps for a $d$-wave superconductor with a 20\% concentration of smooth scatterers ($V_{sm} = 0.5$, $L = 4$, $z = 2$) distributed randomly over the buffer planes adjacent to the CuO$_2$ plane. The field of view is $100 \times 100$, and the energies shown are $E = \pm0.100$ and $E = \pm0.250$.}
	\label{fig:msreal}
\end{figure*}

\begin{figure*}
	\centering
	\includegraphics[width=.2\textwidth]{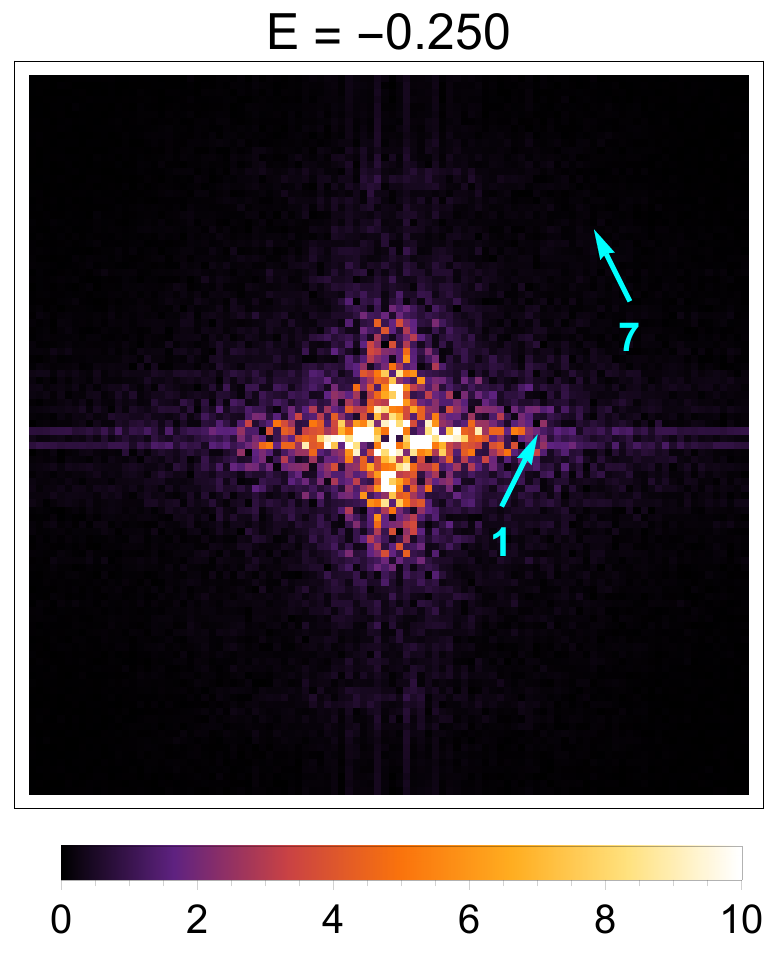}\hfill
	\includegraphics[width=.2\textwidth]{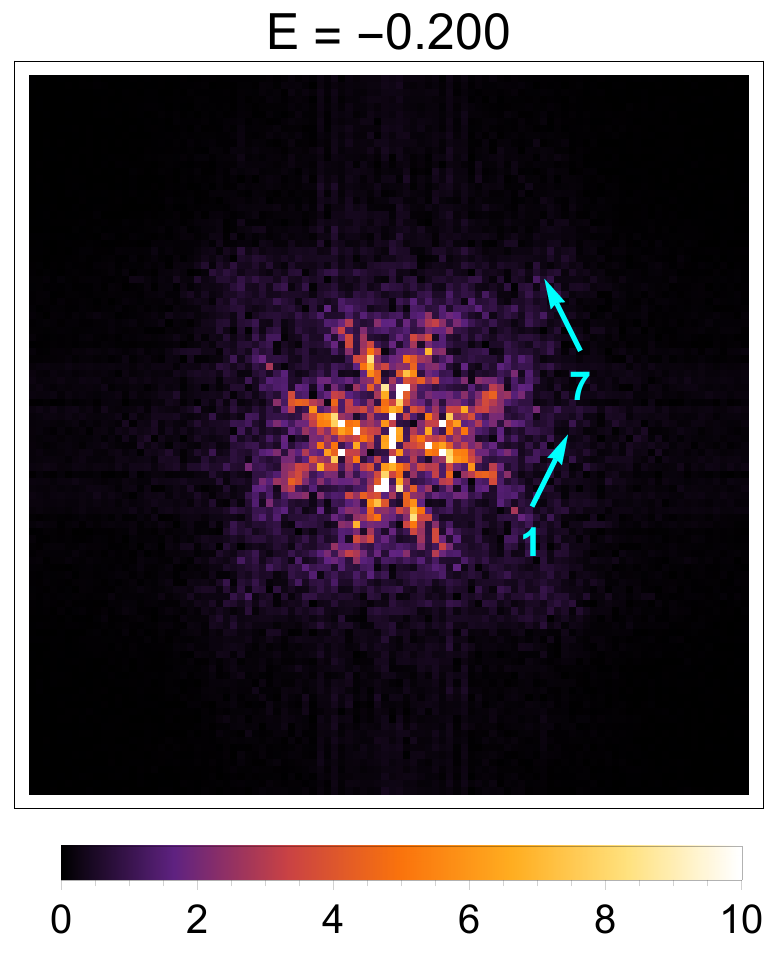}\hfill
	\includegraphics[width=.2\textwidth]{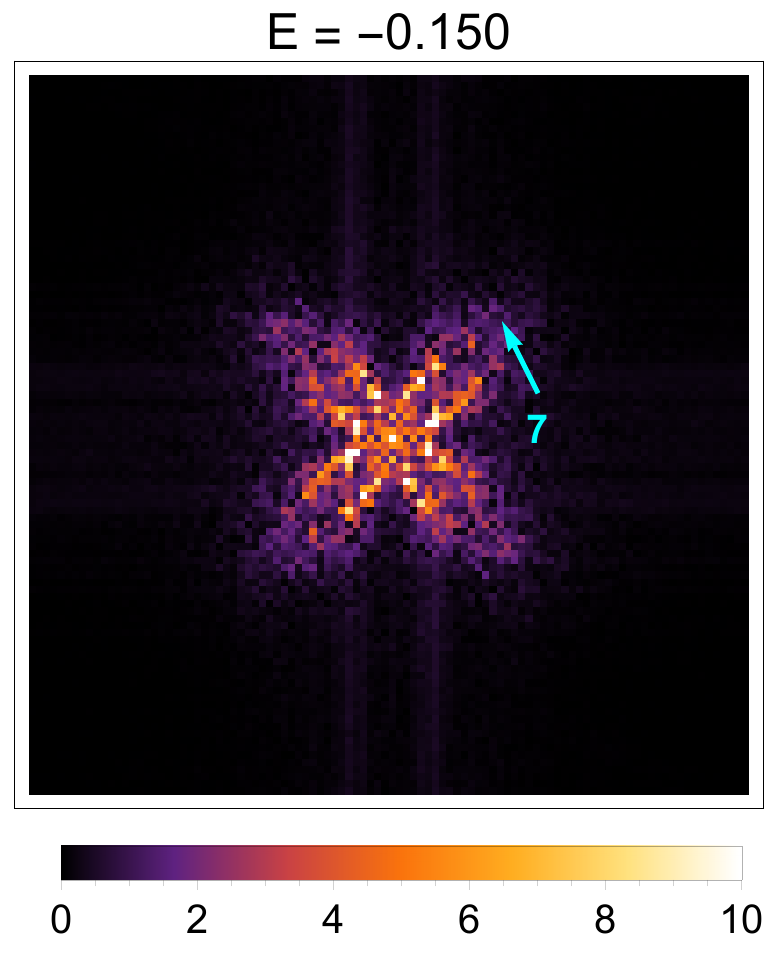}\hfill
	\includegraphics[width=.2\textwidth]{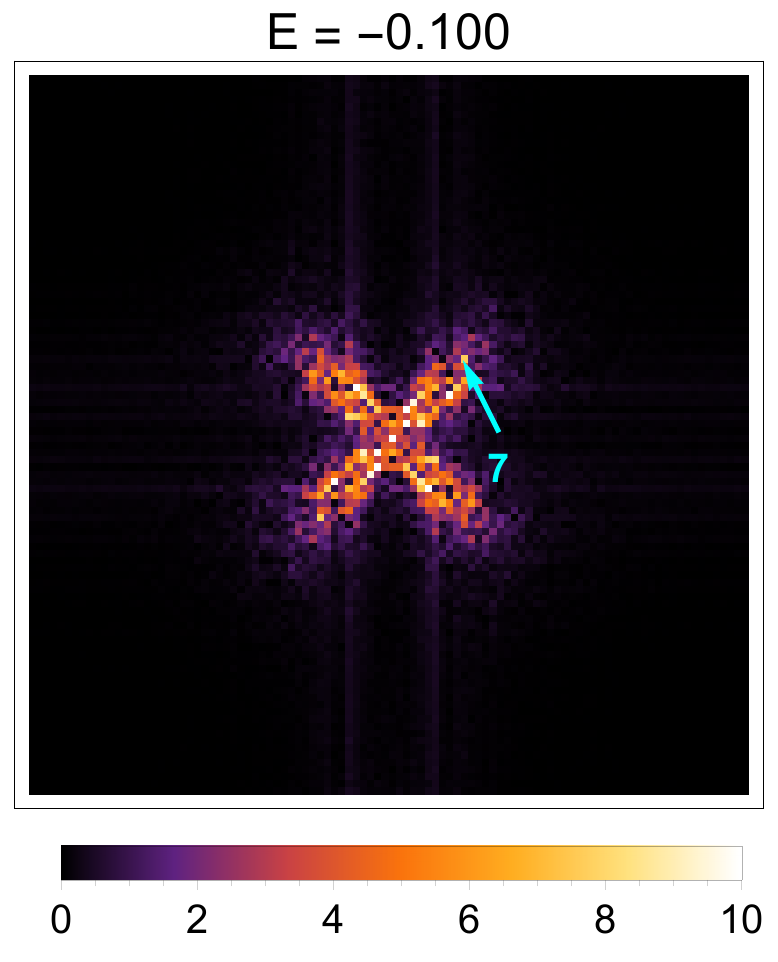}\hfill
	\includegraphics[width=.2\textwidth]{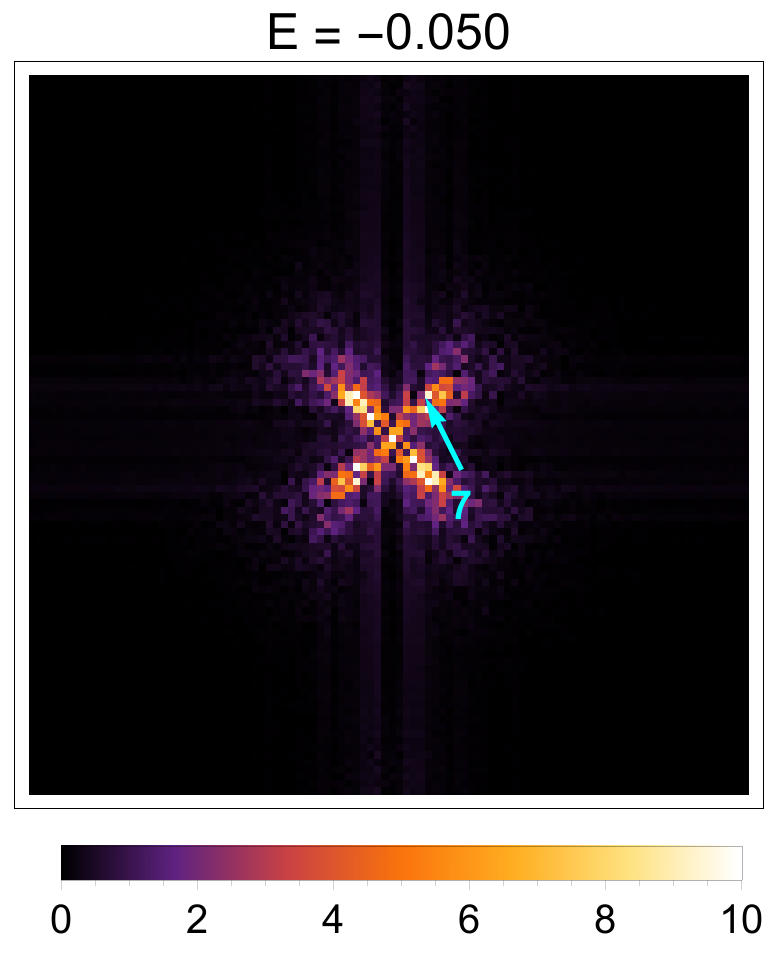}\\
	\includegraphics[width=.2\textwidth]{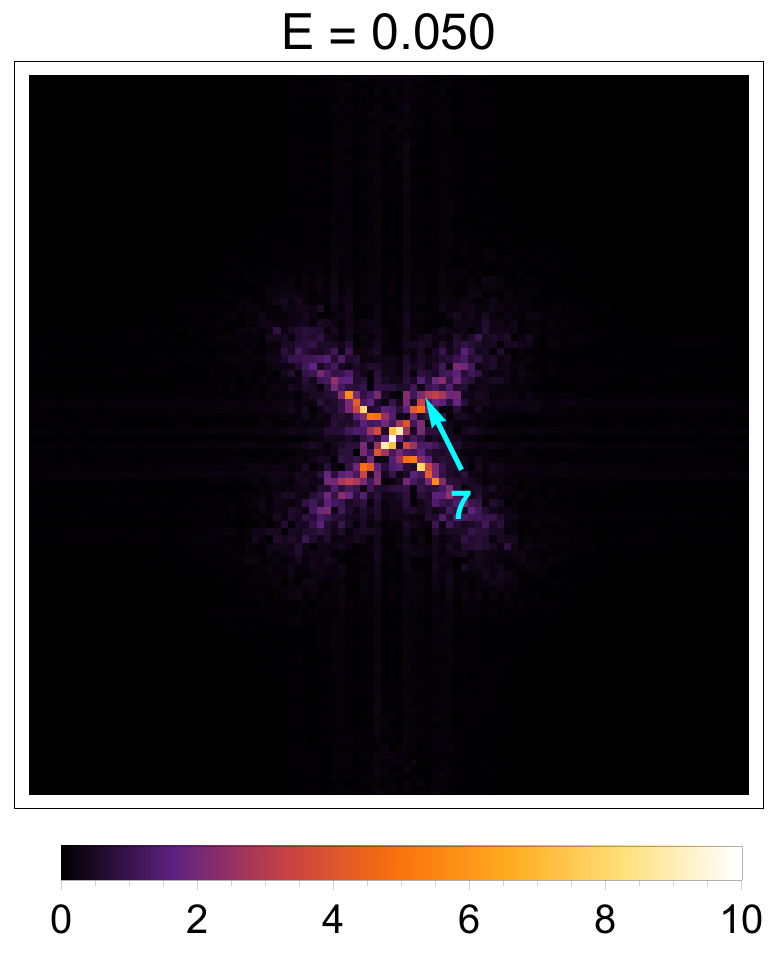}\hfill
	\includegraphics[width=.2\textwidth]{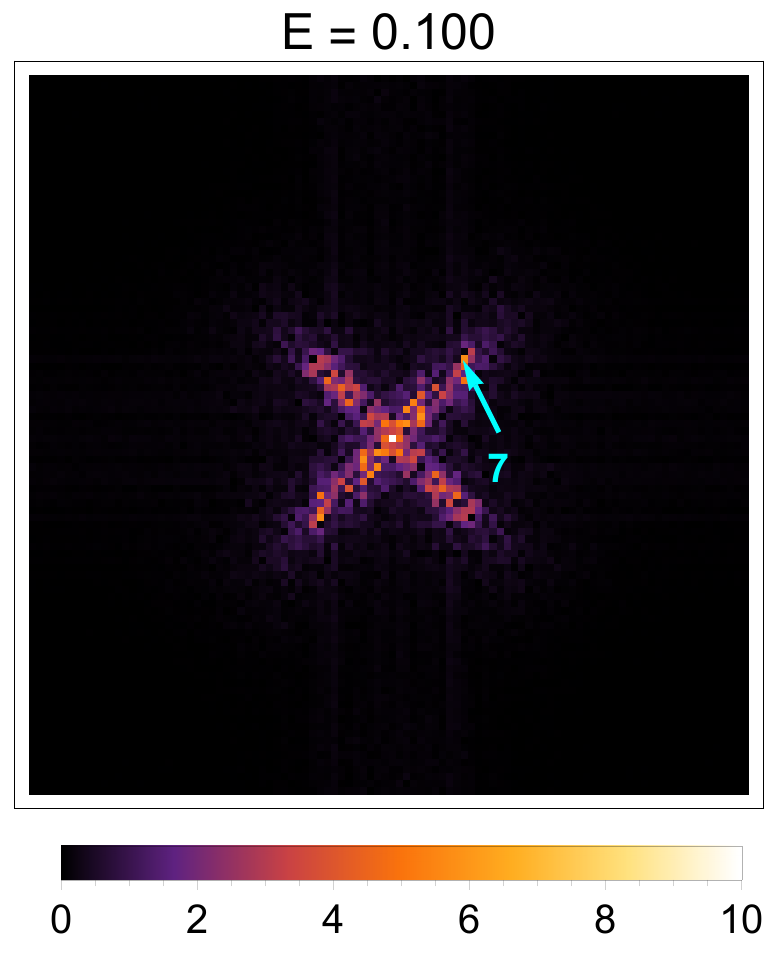}\hfill
	\includegraphics[width=.2\textwidth]{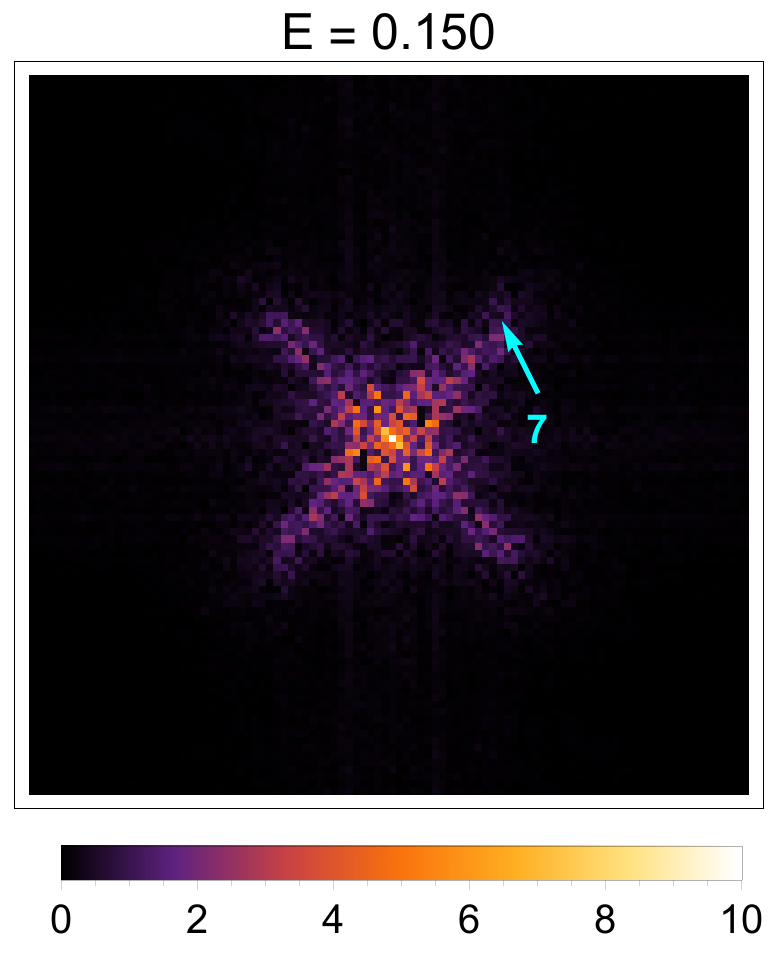}\hfill
	\includegraphics[width=.2\textwidth]{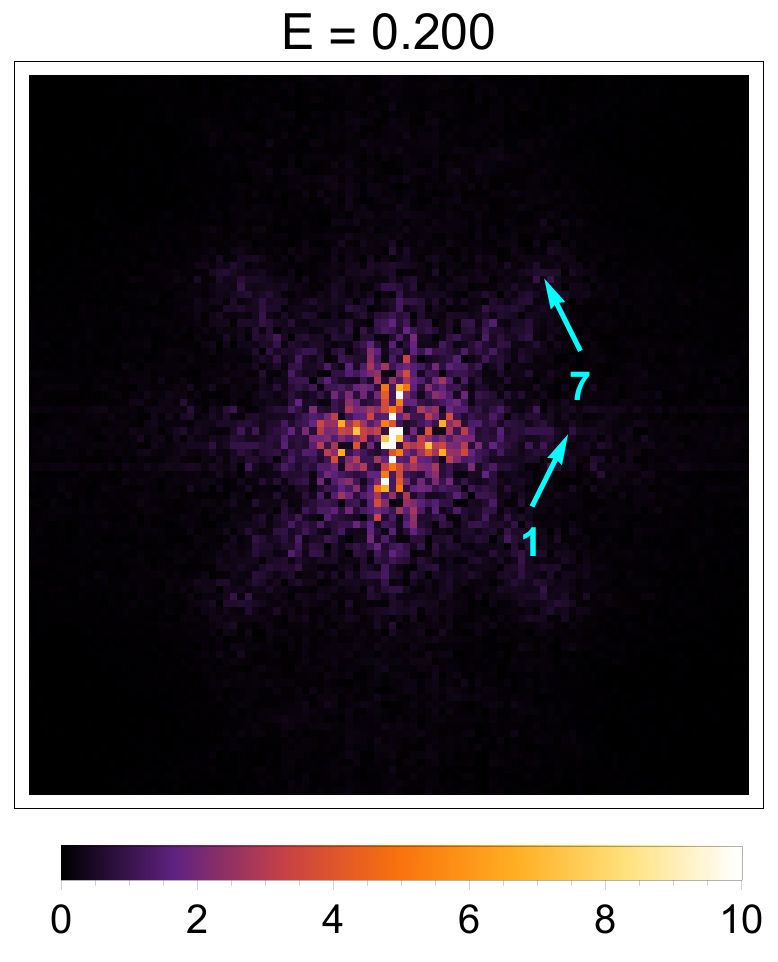}\hfill
	\includegraphics[width=.2\textwidth]{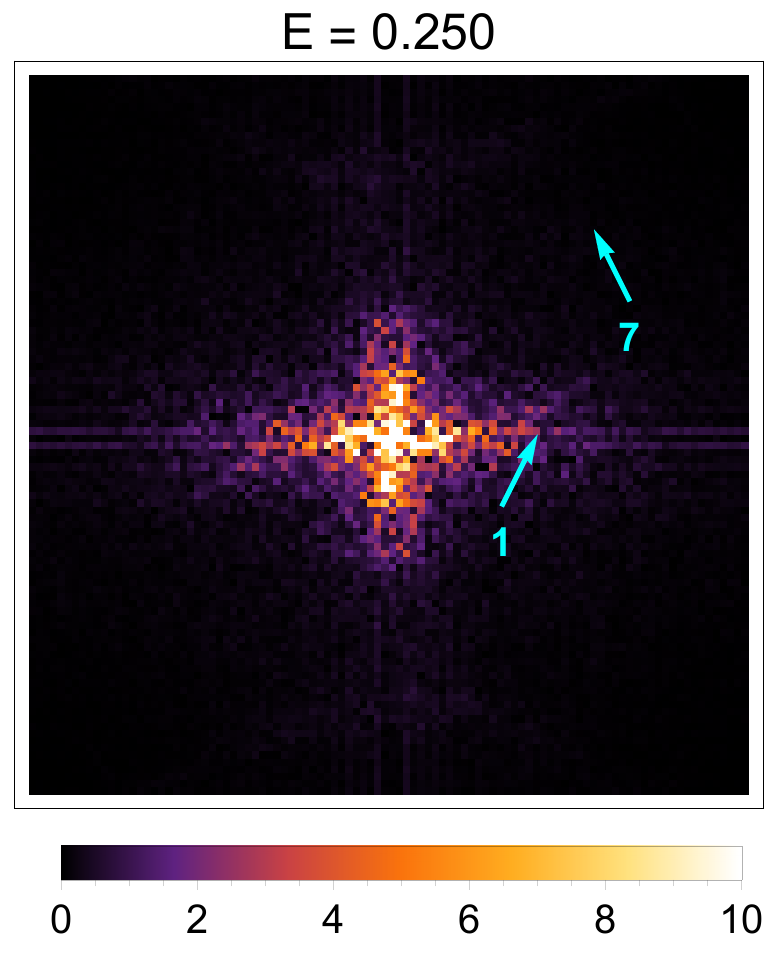}\hfill
	\caption{Fourier-transformed maps for a system with a 20\% concentration of smooth scatterers ($V_{sm} = 0.5$, $L = 4$, $z = 2$). Shown are energies ranging from  $E = \pm0.050$ to  $E = \pm0.250$, along with arrows showing where the octet wavevectors are expected to be found.  Only $\mathbf{q}_7$ is visible at low energies, while both $\mathbf{q}_1$ and $\mathbf{q}_7$ can be seen at higher energies. The color scale is the same for all energies.}
	\label{fig:msfourier}
\end{figure*}

Now that we have intuition about the single smooth scatterer, we can discuss the extension to the case with a very large number of such impurities. We will take the number of smooth scatterers to be 20\% of the total number of lattice sites and randomly place them across the sample. Real-space and Fourier-transformed plots of one realization of disorder are plotted in Figs.~\ref{fig:msreal} and~\ref{fig:msfourier}. 

At low energies the real-space map display stripe-like patterns, featuring modulations in the diagonal directions, which display striking similiarities to STS measurements of BSCCO. One could form the impression that they look even more akin to the stripy QPI patterns in the experimental data than what we found for point-like disorder. Moreover, there is now no discernable sign of the impurity cores. Insofar as these cores were present (albeit difficult to discern) for the single smooth-impurity case, they are now washed away by multi-impurity interference effects at these high concentrations. The absence of clear-cut indications of the precise locations of the off-plane impurities is consistent with the experimental results obtained by McElroy \emph{et al.}, who find that while the positions of the impurities are indeed correlated with the LDOS in that the areas with LDOS suppression at low energies ($|E| < 60$ meV) are likely to be found near the impurities, this correlation is not by any means perfect.\cite{mcelroy2005atomic} The suppression of the LDOS does not imply that an interstitial impurity is above that site; indeed, experiment shows that many regions where the LDOS is suppressed also occur away from impurity sites.

This similarity to real-space experimental images is deceiving, however. Like in the case of the single smooth scatterer, the power spectrum of the many-impurity map here shows suppression of large-momentum internodal scattering processes. The main feature of the power spectrum is a band of wavevectors in the diagonal directions forming a cross in the center of the first Brillouin zone. These arise from intranodal scattering processes between states on one ``banana." These diagonal streaks have a length that is set by $\mathbf{q}_7$. At low energies, no peaks in the spectrum arise from internodal scattering.

As in the single-scatterer case, when energies increase, the diagonal wavevectors become less pronounced in the power spectrum, while wavevectors in the horizontal and vertical directions become more visible. It can be seen that instead of a diagonal cross, one now has a regular cross, with a broad range of wavevectors in the horizontal and vertical directions now being the dominant characteristic of the power spectrum.  These horizontal and vertical streaks feature a length scale roughly set by $\mathbf{q}_1$. This goes hand-in-hand with the fact that the real-space map now features vertical and horizontal stripe-like patterns, instead of diagonal stripes at lower energies. When QPI is the mechanism for the appearance of these stripes, it is expected that the orientation of these patterns will change depending on the energy. This is again different from the case of static stripe order, where stripe patterns remain fixed even when the energy is varied.\cite{kivelson2003detect}

\begin{figure}[ht]
	\centering
	\includegraphics[width=.5\textwidth]{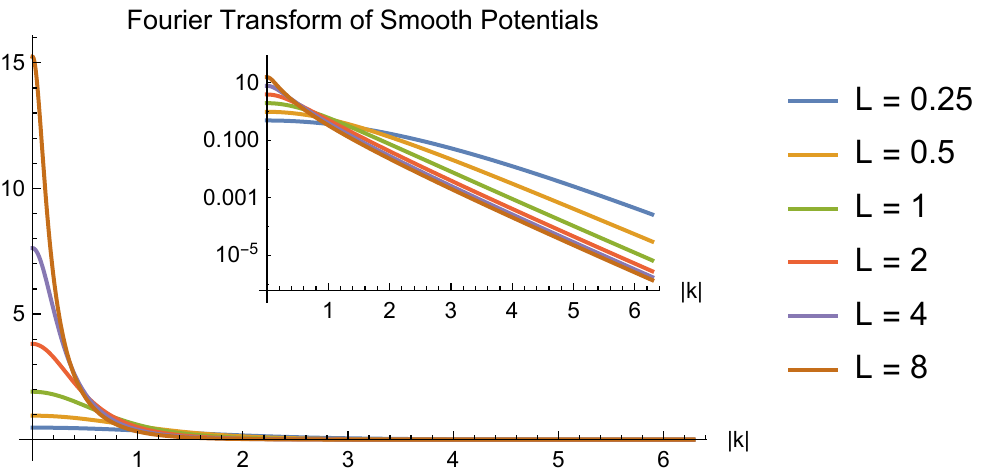}\hfill
	\caption{Plot of the Fourier transforms of smooth potentials (given by Eq.~\ref{smoothpot}) with various screening lengths $L$, shown for momenta in the range $0 \leq |\mathbf{k}| \leq 2\pi$. Inset: semi-log plot of the same quantities. Note that $V_{sm}$ is adjusted so that $V(\mathbf{r = 0})$  is the same for all $L$.}
	\label{fig:hankel}
\end{figure}

\subsection{Quantifying the Range of the Potential}

In hindsight it is clear why the signals for QPI are different for smooth and point-like disorder. The observed power spectra are sensitive to the length scales associated with the disorder potential, since the distribution of the weight in the Fourier maps is set by the characteristic wavevectors of the scattering potential. This is seen in Fig.~\ref{fig:hankel}, which shows the Fourier transform of the scattering potential $V(\mathbf{r})$ for various screening lengths $L$. In plotting these we varied $V_{sm}$ in Eq.~\ref{smoothpot} so that $V(\mathbf{r = 0})$ is the same for different $L$. It can be seen that for all the values of $L$ that we consider, the Fourier transform of $V(\mathbf{r})$ features a very steep dropoff with increasing momentum. The dropoff is most prominent for bigger values of $L$, but is also seen for small ones as well. The Fourier amplitudes at large momenta are larger for small $L$, but they still decrease markedly as $|\mathbf{k}|$ is increased. As this implies that the matrix elements of the scattering potential for large momenta are small, such large-momenta scattering processes will be far less prominent. This explains why, in the power spectra for smooth impurities, $\mathbf{q_7}$ is the only octet-momentum peak visible for low and intermediate energies---as seen in Fig.~\ref{fig:dispersion},  $\mathbf{q_7}$ is the smallest peak for a wide energy range, and its magnitude falls within the range where the Fourier transform of the smooth potential is finite. It is interesting to note that as one moves toward higher energies, $\mathbf{q_1}$ becomes small enough for its magnitude to fall within the aforementioned range of allowed scattering momenta, and its signals are indeed faintly visible in the power spectrum. It is however nowhere near as visible at higher energies as $\mathbf{q_7}$ is at lower energies, a fact that can be attributed to coherence factors that suppress scattering processes between states where the gap has the same sign. All this is to be contrasted with point-like disorder, whose Fourier transform is a constant which depends only on the impurity strength and for which kinematical considerations are the main determinant of the allowed scattering processes.

By measuring carefully not only the dispersions but also the spectral weights of the peaks in the power spectrum, it should be possible in principle to get a quantitative estimate of the typical range of the disorder potential. To the best of our knowledge, this has not been attempted yet. Here we will attempt to quantify in a simple manner the dependence of the power spectrum on the screening length of the Coloumb potential. We introduce a number $w$ that will quantify how much spectral weight is associated with large-momentum scattering processes:
\begin{equation}
w(L, E) = \frac{\sum_{\mathbf{q \in \text{A}}} |\rho(\mathbf{q}, L, E)| - |\rho(\mathbf{q = 0}, L, E)|}{\sum_{\mathbf{q \in \text{BZ}}} |\rho(\mathbf{q}, L, E)| -  |\rho(\mathbf{q = 0}, L, E)|}.
\label{eq:wdef}
\end{equation}
$\rho(\mathbf{q}, L, E)$ is the Fourier map associated with a single smooth scatterer with screening length $L$ in the center of the field of view, taken at energy $E$. As before we will also vary $V_{sm}$ in Eq.~\ref{smoothpot} so that $V(\mathbf{r = 0})$ is independent of $L$. We set $z=2$. $A$ in this instance is defined as the subset of the Brillouin zone centered about the $\Gamma$ point where $-a\pi \leq q_x \leq a\pi$ and  $-a\pi \leq q_y \leq a\pi$, and $a < 1$. We will set $a = 0.4$ in our numerical calculations. 

The point of introducing $w$ is that it is simply the ratio of the integrated power spectrum within $A$ (without the $\mathbf{q = 0}$ contribution) to the integrated power spectrum within the first Brillouin zone (again without the $\mathbf{q = 0}$ part). If most of the weight in the power spectrum is associated with small-momentum scattering processes, $w$ should be close to $1$, whereas if more spectral weight is associated with large-momentum processes, such as in the case of a point-like scatterer, $w$ should be small.

\begin{figure}[ht]
	\centering
	\includegraphics[width=.5\textwidth]{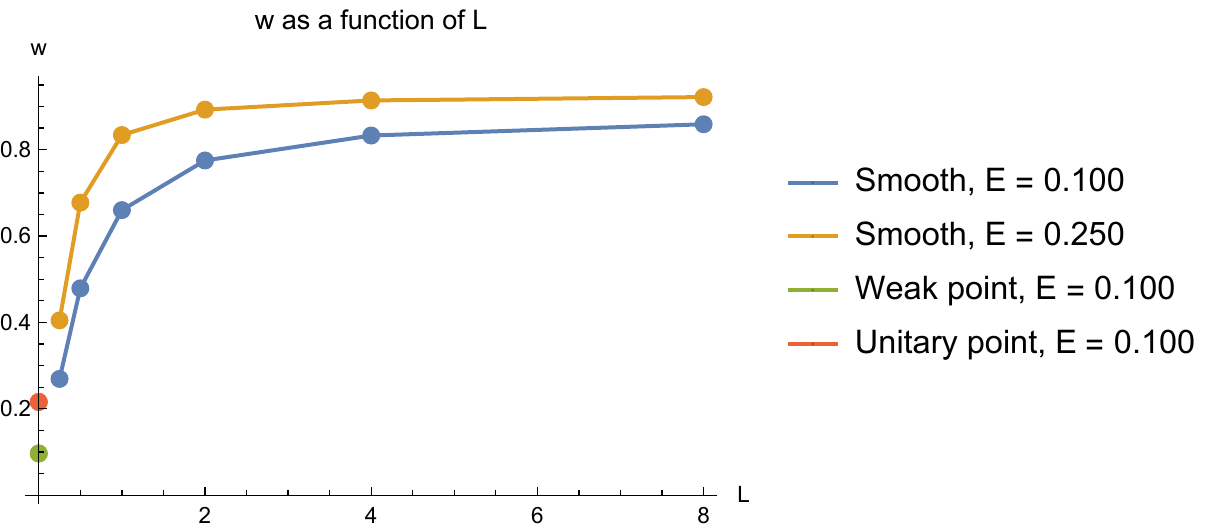}\hfill
	\caption{Plot of $w$ (defined in Eq.~\ref{eq:wdef}) versus the screening length $L$ for $E = 0.100$ and $E = 0.250$. Here we consider a single smooth impurity located in the center of the sample. As discussed in the text, the value of $V_{sm}$ is chosen so that $V(\mathbf{r = 0})$ is the same for all values of $L$. The two data points at $L = 0$ correspond to the values of $w$ obtained for a weak point-like scatterer ($V = 0.5$) and a unitary point-like scatterer ($V = 10$) at energy $E = 0.100$. The smallness of these values indicates that large-momentum processes are a prominent part of the power spectra for these point-like scatterers.}
	\label{fig:wl}
\end{figure}

Fig.~\ref{fig:wl} shows plots of $w$ versus $L$ for energies $E = 0.100$ and $E = 0.250$. It can be seen that when $L$ is large (\emph{i.e.}, $L > 2$), $w$ is large and saturates to a fixed value with increasing $L$. This means that in this regime, the vast majority of the spectral weight is associated with small-momentum processes. On the other hand, when $L$ is small, $w$ becomes small as well, implying that the power spectrum hosts more contributions from large-momentum processes which show up outside $A$ in the power spectrum. We can see that it is only with very small values of $L$ that we start to see behavior resembling that of the point-like scatterer, in which both small- and large-momentum processes figure prominently in the power spectrum.

Although a detailed study of weight distributions in experimentally-obtained Fourier maps has not yet been undertaken, it appears that the strength of large-momentum scattering, at least as evidenced from STS experiments, is actually quite large. These experimental results suggest that disorder is close to the point-scatterer limit. Given what is commonly believed about the nature of intrinsic disorder in the cuprates, this is surprising, if not entirely unreasonable. The reason behind the prominence of the large-momentum peaks in QPI spectra can be considered alongside the problem of the sharpness of the octet-model peaks as two of the primary mysteries of the results of QPI.

\section{Spatially Random On-Site Energies}
\begin{figure*}[ht]
	\centering
	\includegraphics[width=.25\textwidth]{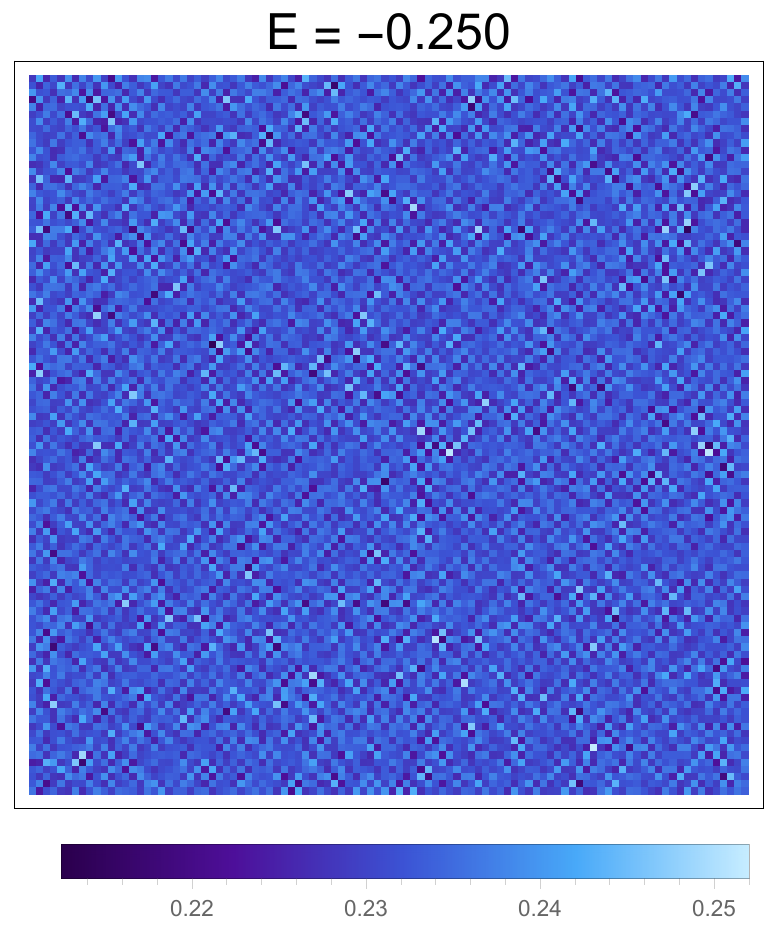}\hfill
	\includegraphics[width=.25\textwidth]{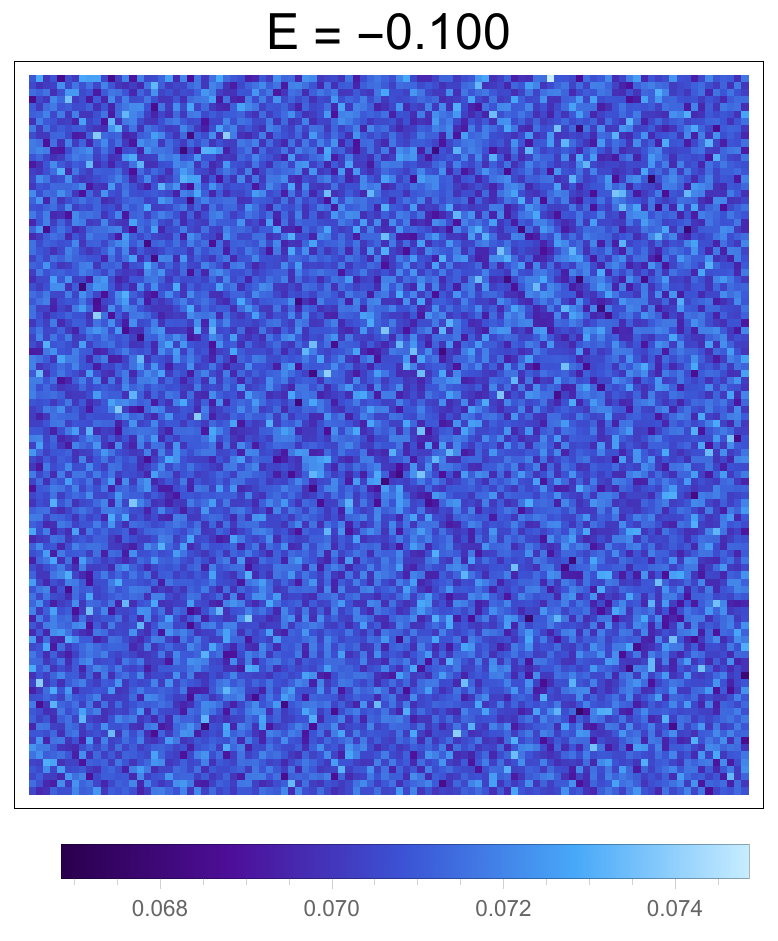}\hfill
	\includegraphics[width=.25\textwidth]{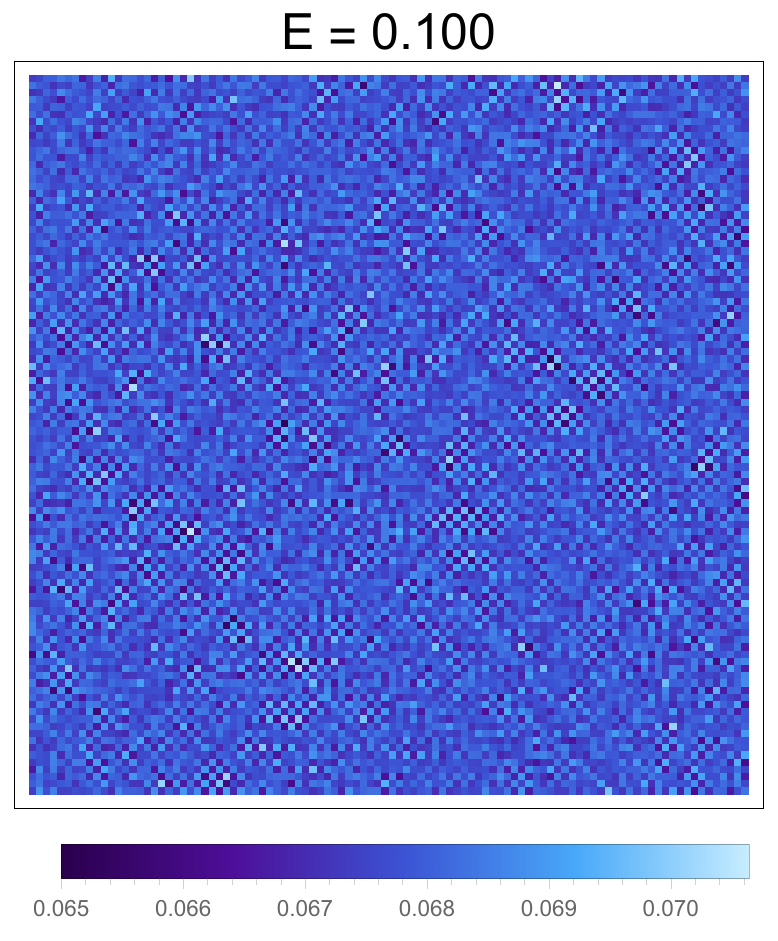}\hfill
	\includegraphics[width=.25\textwidth]{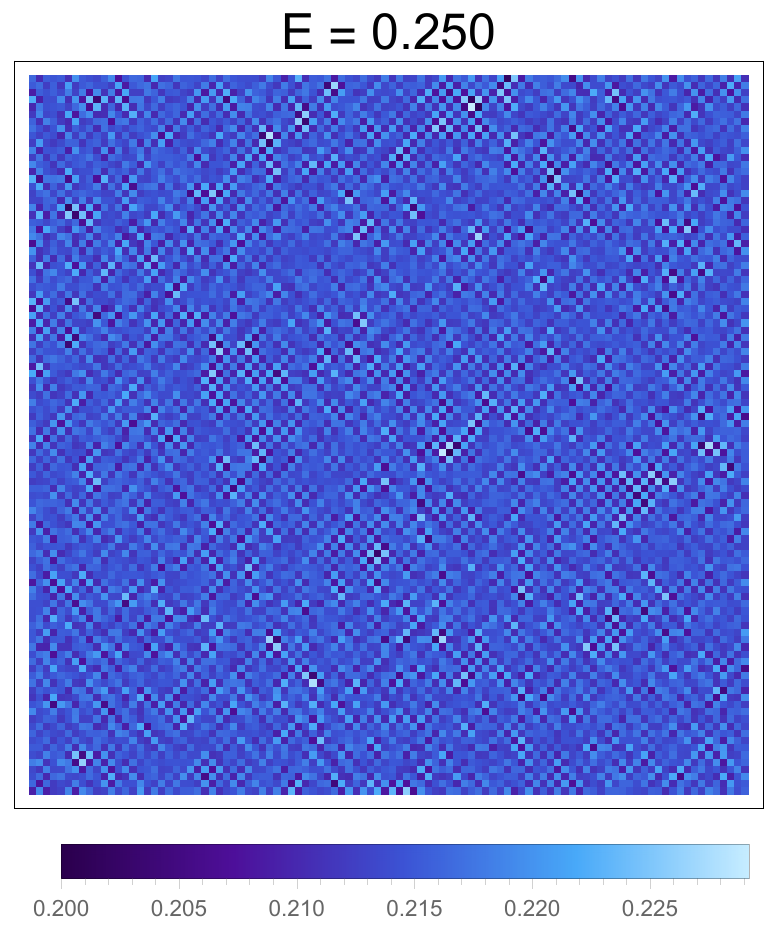}\hfill
	\caption{Real-space LDOS maps for a $d$-wave superconductor with random on-site energies, normally distributed with $M_V = 0.01$. The field of view is $100 \times 100$, and the energies shown are $E = \pm0.100$ and $E = \pm0.250$.}
	\label{fig:rpreal}
\end{figure*}

\begin{figure*}
	\centering
	\includegraphics[width=.2\textwidth]{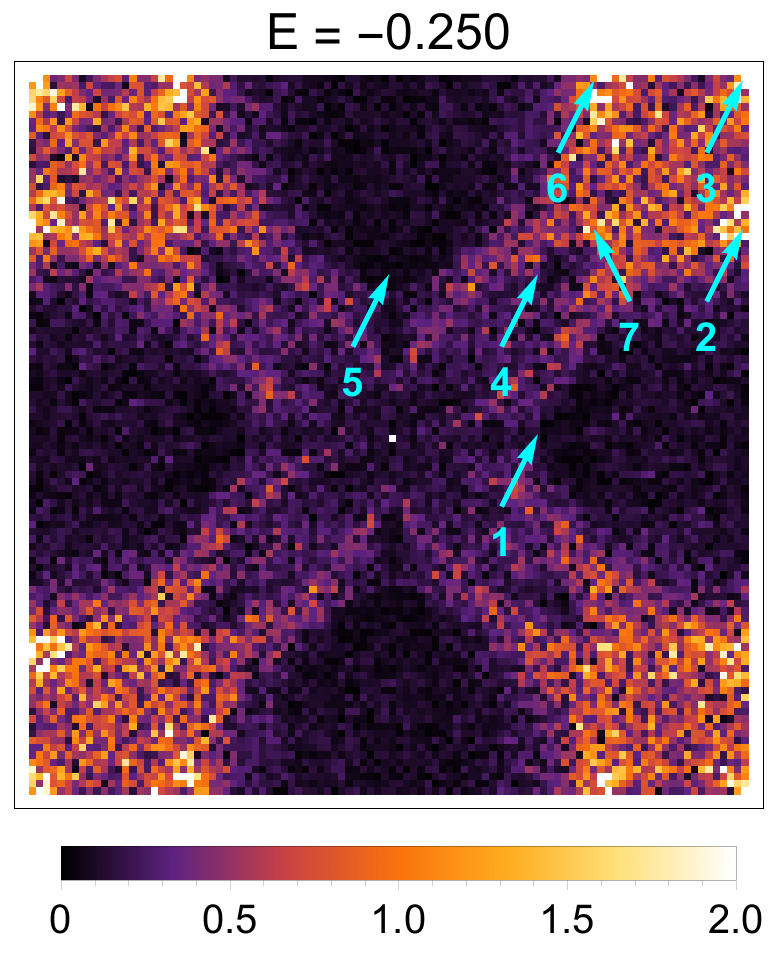}\hfill
	\includegraphics[width=.2\textwidth]{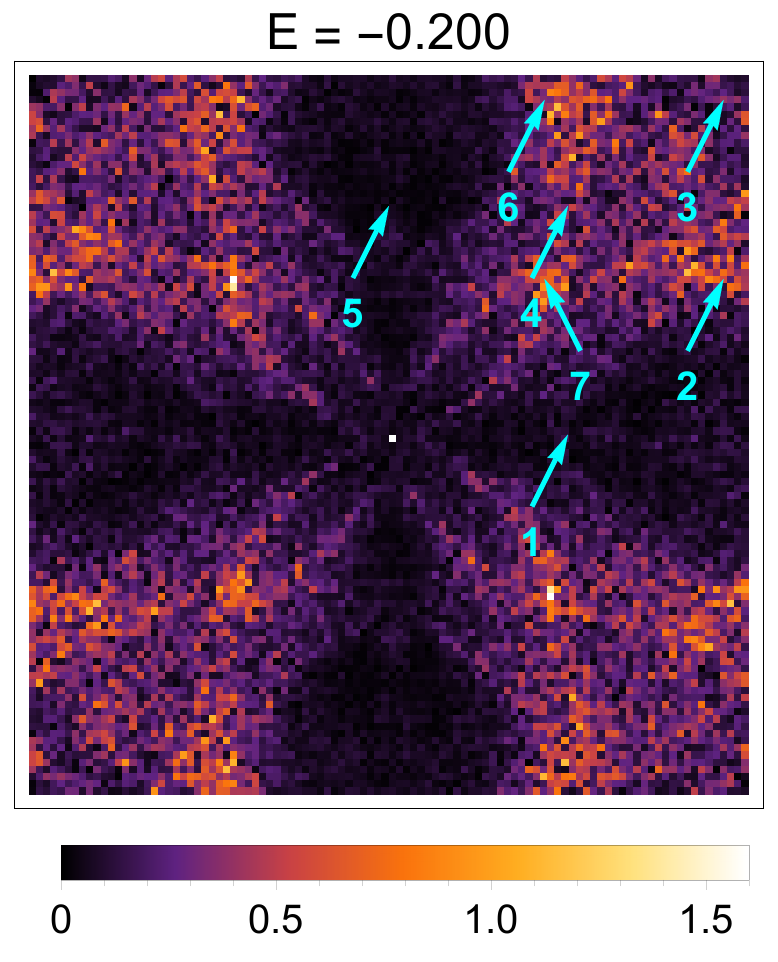}\hfill
	\includegraphics[width=.2\textwidth]{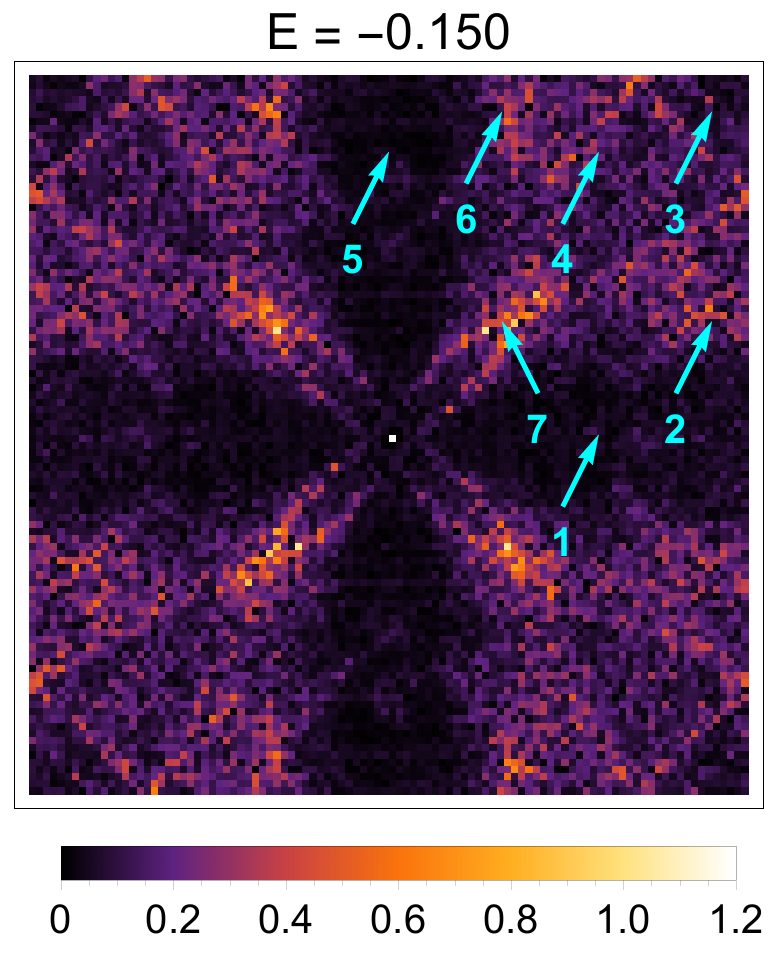}\hfill
	\includegraphics[width=.2\textwidth]{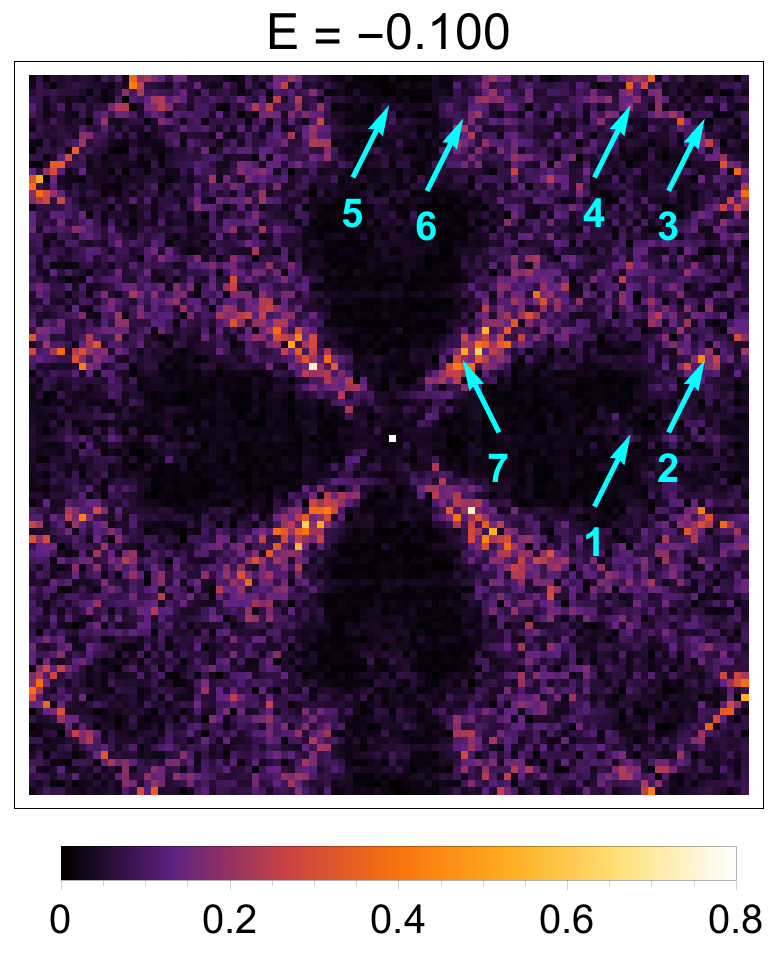}\hfill
	\includegraphics[width=.2\textwidth]{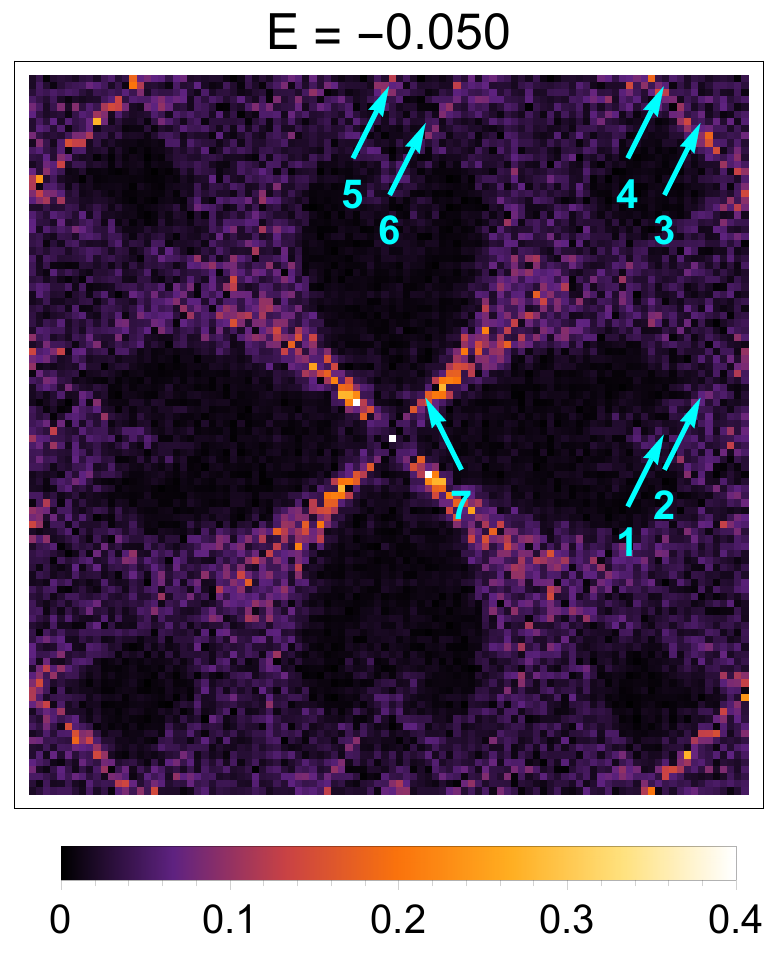}\\
	\includegraphics[width=.2\textwidth]{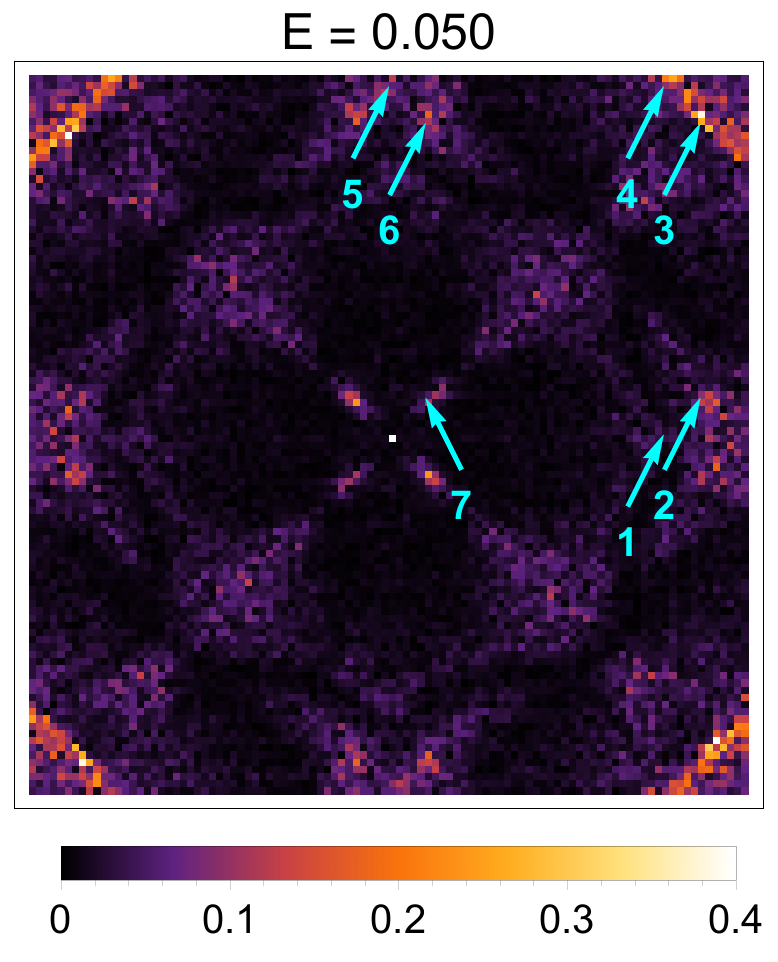}\hfill
	\includegraphics[width=.2\textwidth]{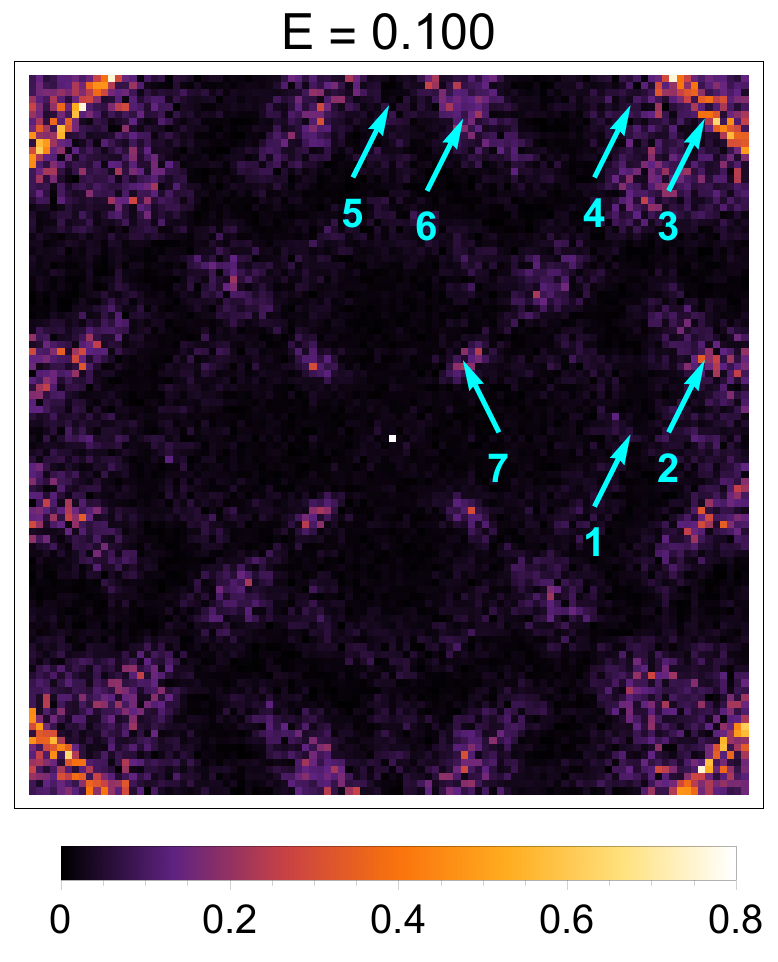}\hfill
	\includegraphics[width=.2\textwidth]{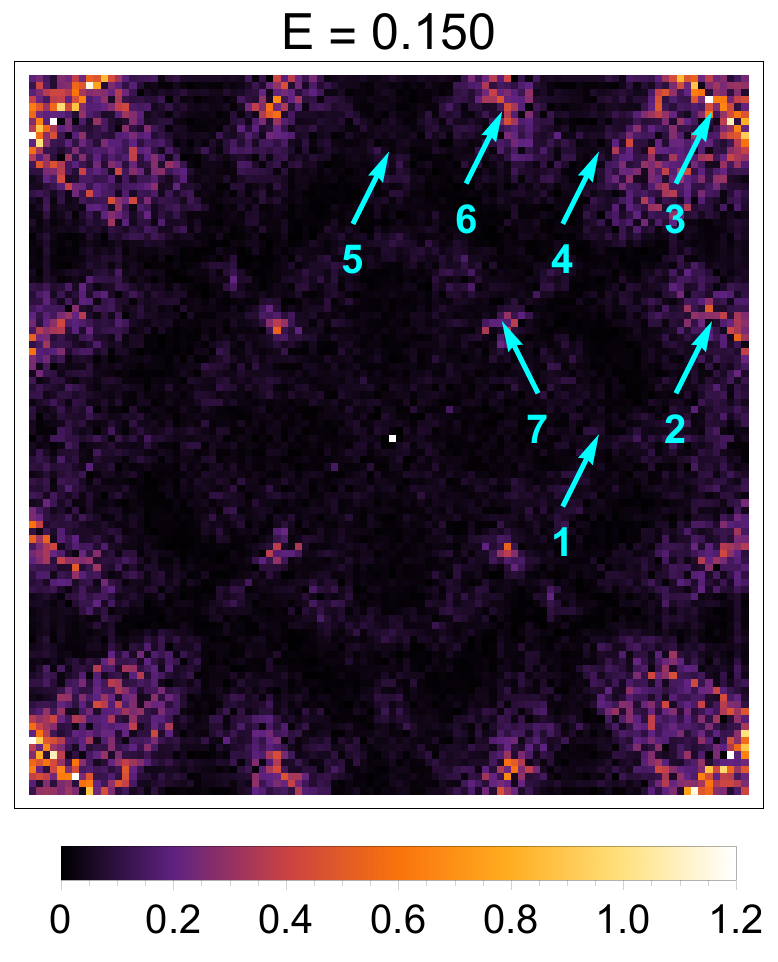}\hfill
	\includegraphics[width=.2\textwidth]{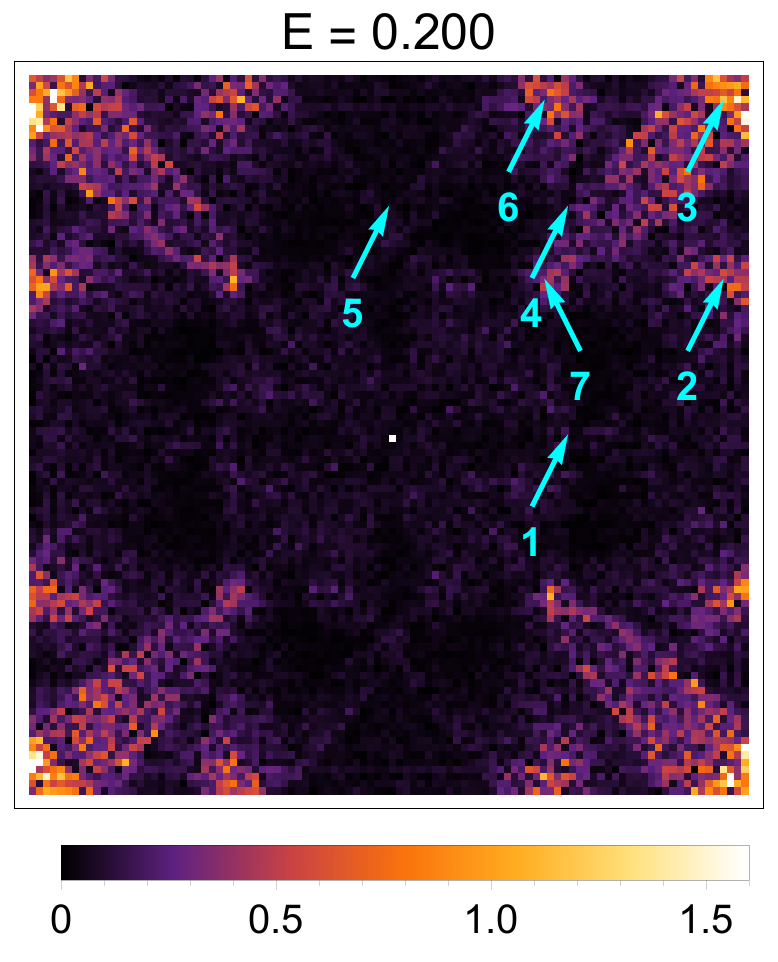}\hfill
	\includegraphics[width=.2\textwidth]{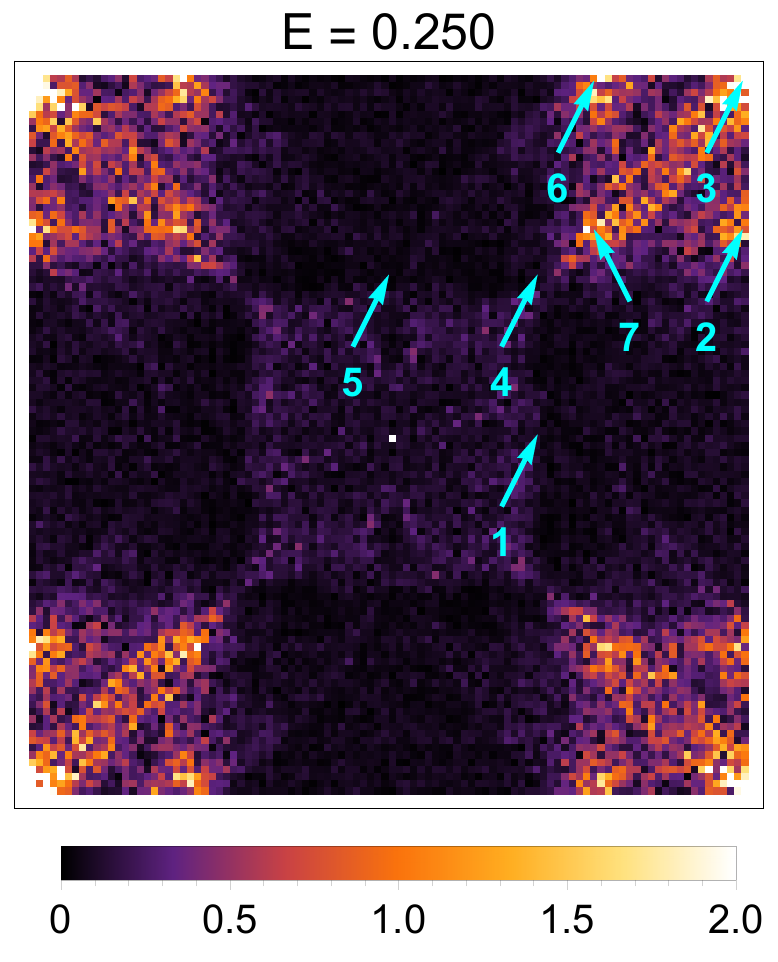}\hfill
	\caption{Fourier-transformed maps for a system with random on-site energies, normally distributed with $M_V = 0.01$. Shown are energies ranging from  $E = \pm0.050$ to  $E = \pm0.250$, along with arrows showing where the octet wavevectors are expected to be found. The color scaling varies linearly with energy.}
	\label{fig:rpfourier}
\end{figure*}
To complete our survey of the effects of various kinds of disorder on STS results, we now turn to yet one more well-known form of disorder: a random and uncorrelated distribution of on-site energies throughout the sample. This is the form of disorder that underlies Anderson localization in metals. Because we do not have isolated impurities in this case, with the on-site energies varying from one site to another and numerous multiple-scattering processes occurring as a result, the $T$-matrix method cannot be easily applied to this problem to obtain the LDOS. In contrast, the numerical method we use here allows us to obtain LDOS maps directly and efficiently.  

To be more specific, on each site we have a \emph{random} perturbation $V_R$ in addition to the spatially uniform mean-field chemical potential $\mu$. In other words the on-site potential at site $\mathbf{r}$ is given by the sum $\mu + V_R(\mathbf{r})$. For simplicity we will take $V_R(\mathbf{r})$ to be drawn from a Gaussian distribution with the following properties:
\begin{eqnarray}
&\langle V_R(\mathbf{r})  \rangle = 0,\\
&\langle V_R(\mathbf{r}) V_R(\mathbf{r'})  \rangle = M^2_{V}\delta_{\mathbf{rr'}}.
\end{eqnarray}
Here the angular brackets denote averaging over disorder realizations. The width of the distribution is parametrized by the standard deviation $M_V$; we will use this to characterize the strength of the disorder potential. 

Why do we pick this form of disorder? From our previous discussion of point-like and smooth potential disorder, it is clear that in order to reproduce both the real-space and Fourier-transformed results from STS measurements, one must have both real-space maps that simulateneously have LDOS modulations and feature no obvious signs of impurity cores; and power spectra that show peaks arising from internodal and intranodal scattering. Here the random-potential model could sidestep the difficulties faced by our previous hypothesized scenarios. First, if we pick our distribution to be sufficiently narrow (and hence weak), there is the possibility that we could have real-space modulations without having visible impurity cores that arise from isolated potential perturbations, as was the case in the point-like case we discussed earlier. Second, this form of disorder, similar to the point-like scatterer, is short-ranged. This would then not have the suppression of internodal scattering that is a feature of smooth potential disorder with finite correlation length. As a result it could potentially feature both small- and large-wavevector peaks in the power spectrum. A similar form of random on-site disorder was considered by Atkinson \emph{et al}.\cite{atkinson2000details}

To check whether these expectations are ultimately borne out, we numerically obtain real-space and Fourier-transformed maps for one realization of random on-site disorder. We make disorder weak by setting the width of the distribution to be narrow. The results are plotted in Figs.~\ref{fig:rpreal} and~\ref{fig:rpfourier}.

The real-space maps feature as before modulations whose structure can be discerned, but not to a similar extent as the point-like- or smooth-scatterer cases. In this particular scenario one cannot tell whether an impurity is present or not---the signatures we have come to expect from the isolated point-like impurity are not present here at all. Instead what we have are modulations, primarily in the diagonal directions, with a crisscrossing pattern slightly similar to that found in the smooth-disorder case. Unlike in the smooth scatterer case, however, the stripe-like patterns are far more subdued. The maps obtained here look very similar to those taken from STS experiments.

This is shown even more so by the Fourier-transformed maps. We see that both small- and large-momentum scattering processes contribute to the observed QPI, as evidenced by peaks at small and large diagonal wavevectors. Interestingly, the power spectrum is very similar to that of the multiple-weak-impurity case. In particular, $\mathbf{q}_2$, $\mathbf{q}_3$, $\mathbf{q}_6$, and $\mathbf{q}_7$ are strongly present, whereas signals of the three remaining $\mathbf{q}$-vectors are quite weak. This can be attributed to the fact that, in the superconducting state, coherence factors enter into the scattering amplitude.\cite{wang2003quasiparticle,pereg2003theory} For scattering off of a weak potential, it turns out that the matrix element between two states with momenta $\mathbf{k}_1$ and $\mathbf{k}_2 $ contains a factor
\begin{equation}
u_{\mathbf{k}_1}u_{\mathbf{k}_2} - v_{\mathbf{k}_1}v_{\mathbf{k}_2},
\end{equation} 
where 
\begin{eqnarray}
&u_{\mathbf{k}} = \text{sgn}(\Delta_k)\sqrt{\frac{1}{2}(1 + \frac{\epsilon_k}{E_{k}})},\\
&v_{\mathbf{k}} = \sqrt{1 - u_k^2}.
\end{eqnarray}
This implies that if $\Delta_{k_1}$ and $\Delta_{k_2}$ have the same sign, the $\mathbf{k}_1 \rightarrow \mathbf{k}_2$ process will be suppressed. Conversely, whenever $\Delta_{k_1}$ and $\Delta_{k_2}$ have the opposite sign, that process will not be suppressed. This explains why the $\mathbf{q}_2$, $\mathbf{q}_3$, $\mathbf{q}_6$, and $\mathbf{q}_7$ wavevectors---which connect states at which the values of the order parameter have opposite sign---are not suppressed, while the remaining ones are.

The possibility that one can have the Fourier-space signatures of QPI while having some qualitative similarities between the theoretical and experimental real-space images suggests that this form of disorder---weak, narrowly distributed random potential disorder---can be responsible for the physics observed in the STS measurements.  Having said this, the peaks in the power spectrum resulting from this form of disorder exhibit the same form of fuzziness as in the weak-impurity scenario. Also, the relative suppression of certain octet-model peaks suggests that even with this form of disorder, the same questions that affect the weak-impurity case affect the random site-energy model as well.

\section{Spatially Random Superconducting Gap}
\begin{figure*}[ht]
	\centering
	\includegraphics[width=.25\textwidth]{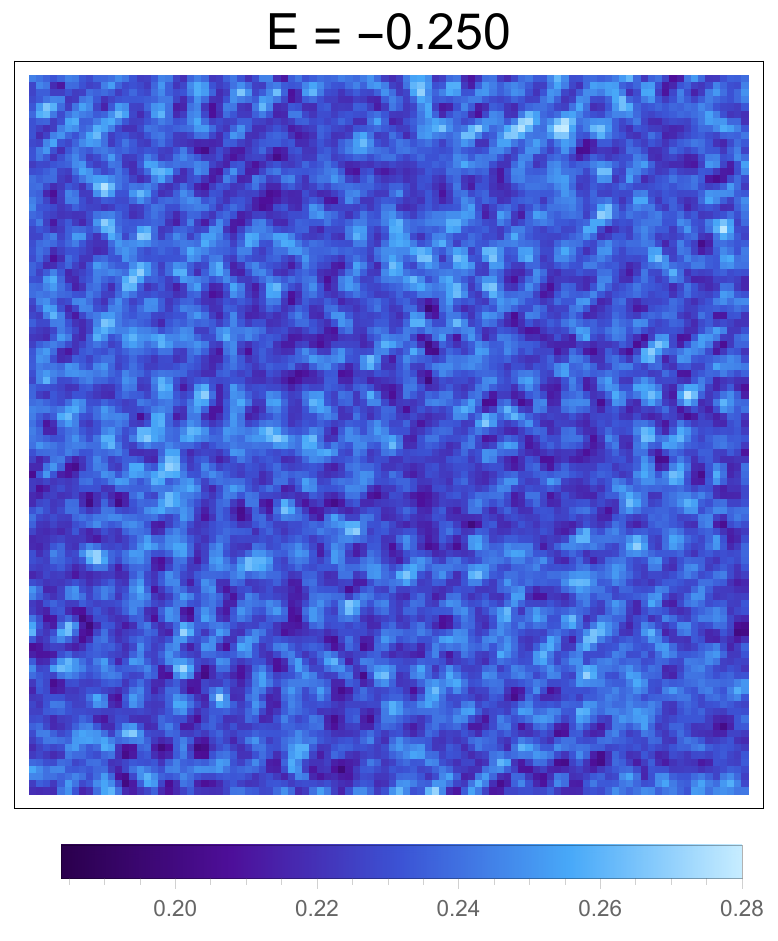}\hfill
	\includegraphics[width=.25\textwidth]{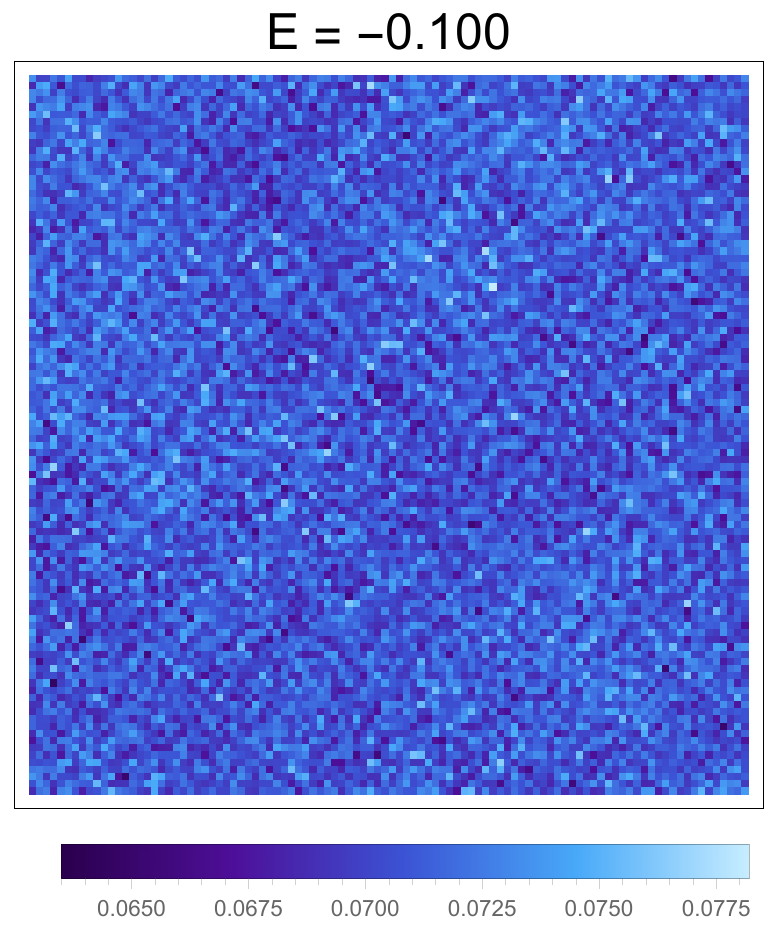}\hfill
	\includegraphics[width=.25\textwidth]{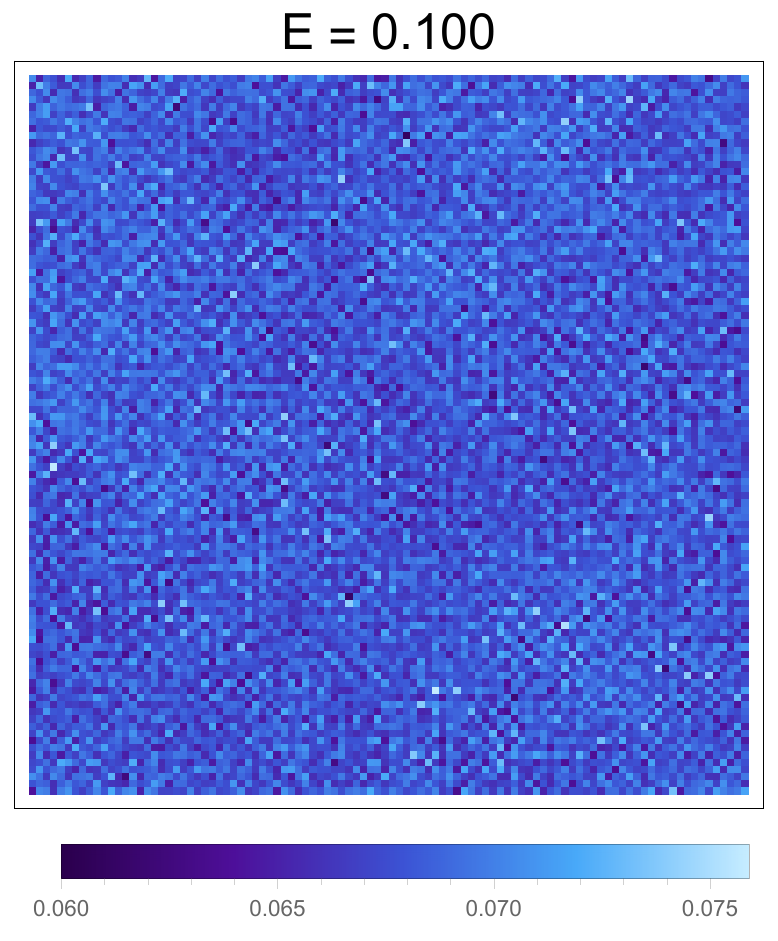}\hfill
	\includegraphics[width=.25\textwidth]{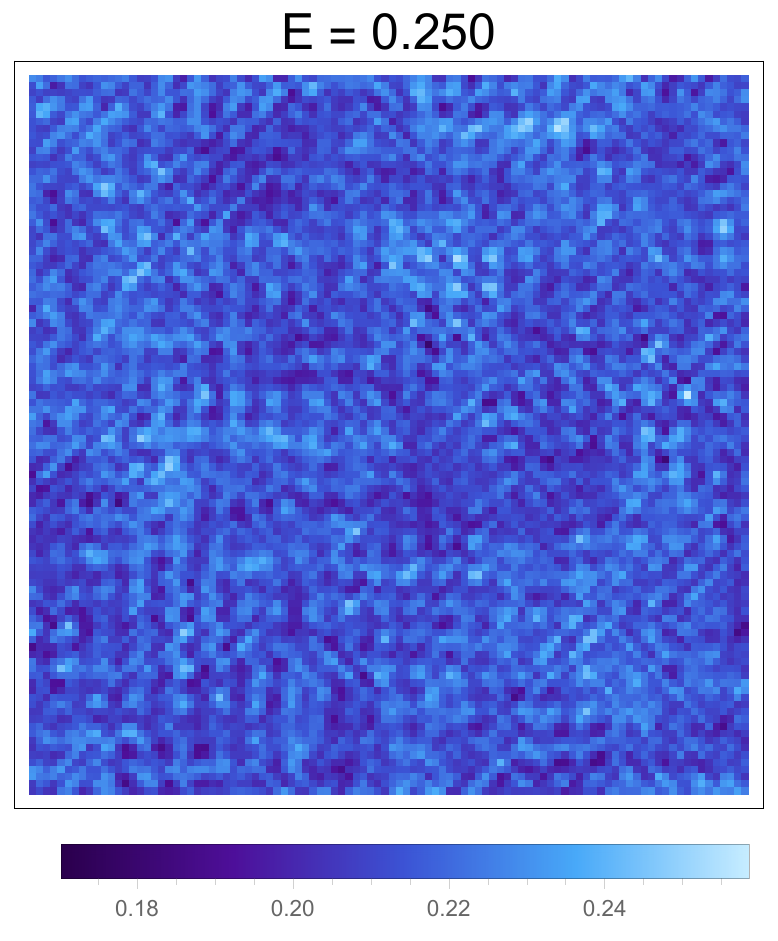}\hfill
	\caption{Real-space LDOS maps for a $d$-wave superconductor with random pairing amplitudes, normally distributed with $M_{\Delta} = 0.01$. The field of view is $100 \times 100$, and the energies shown are $E = \pm0.100$ and $E = \pm0.250$.}
	\label{fig:rgreal}
\end{figure*}

\begin{figure*}
	\centering
	\includegraphics[width=.2\textwidth]{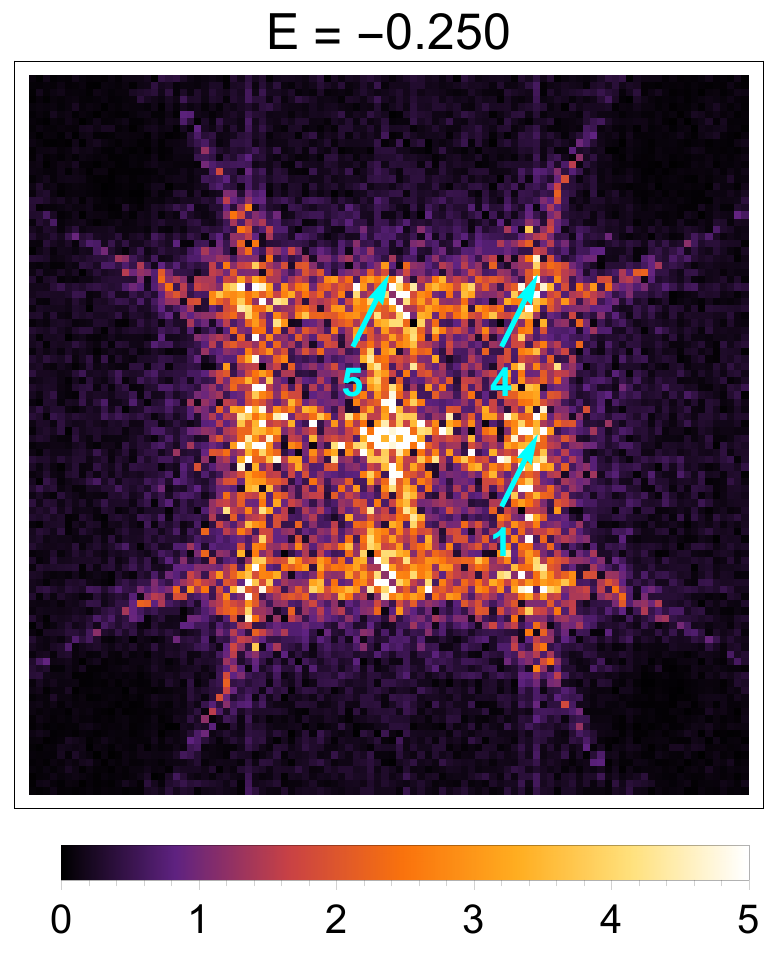}\hfill
	\includegraphics[width=.2\textwidth]{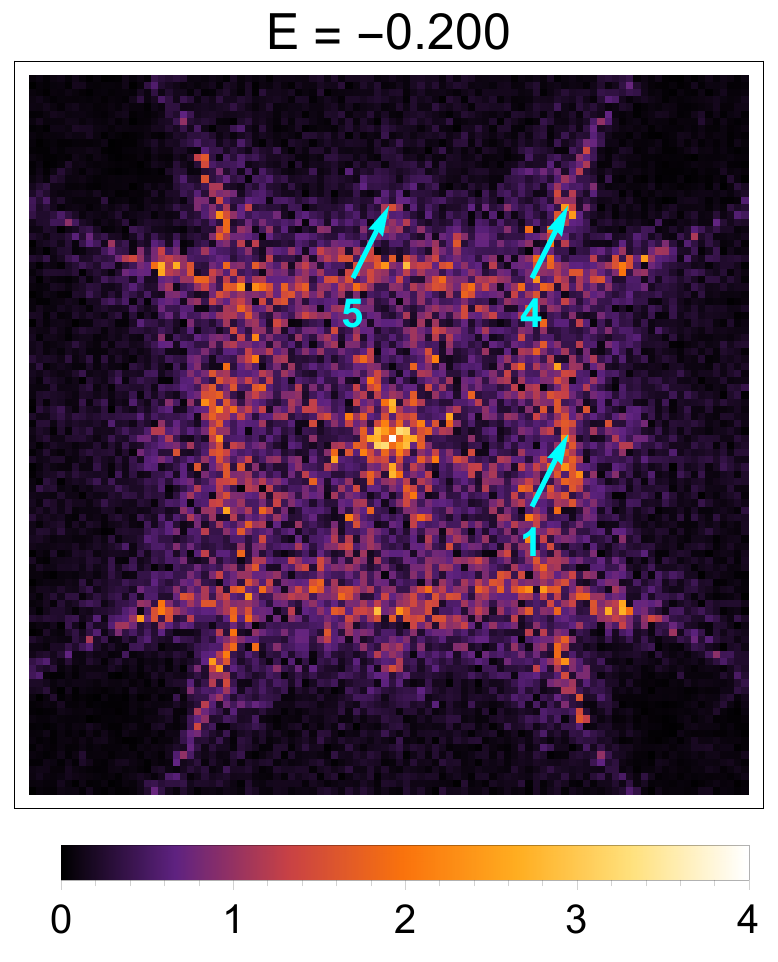}\hfill
	\includegraphics[width=.2\textwidth]{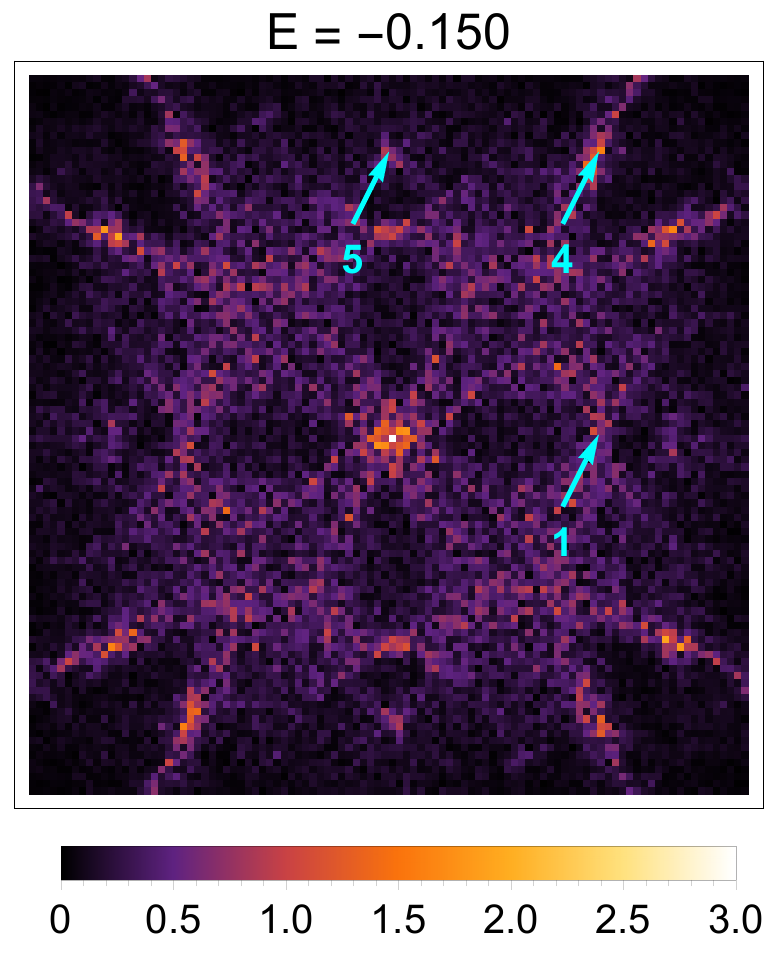}\hfill
	\includegraphics[width=.2\textwidth]{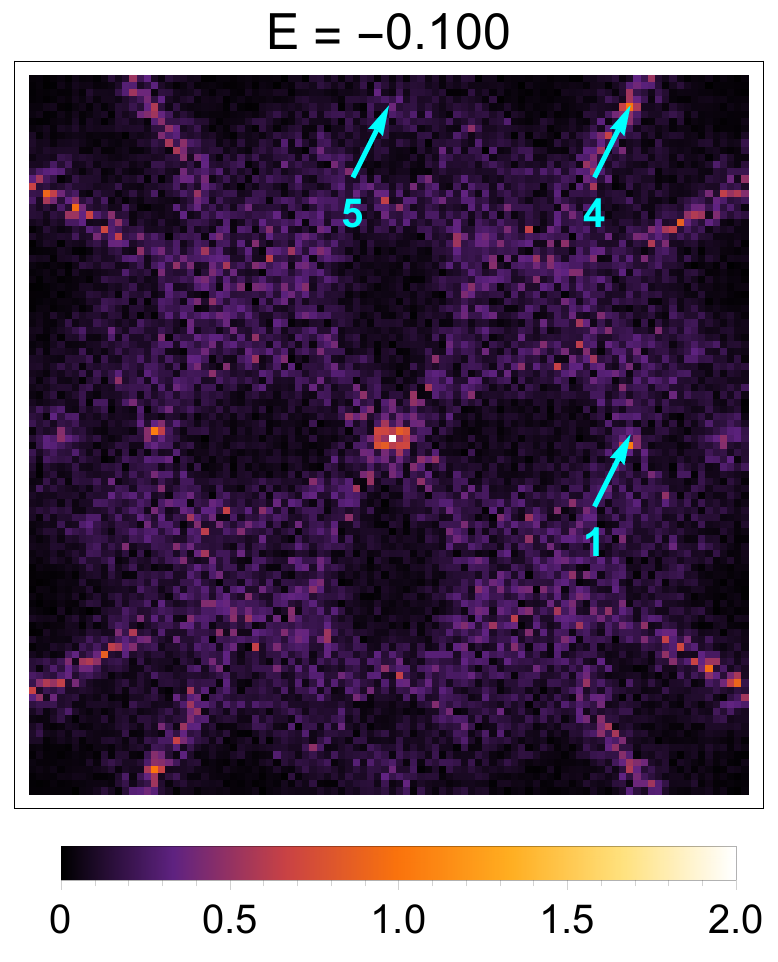}\hfill
	\includegraphics[width=.2\textwidth]{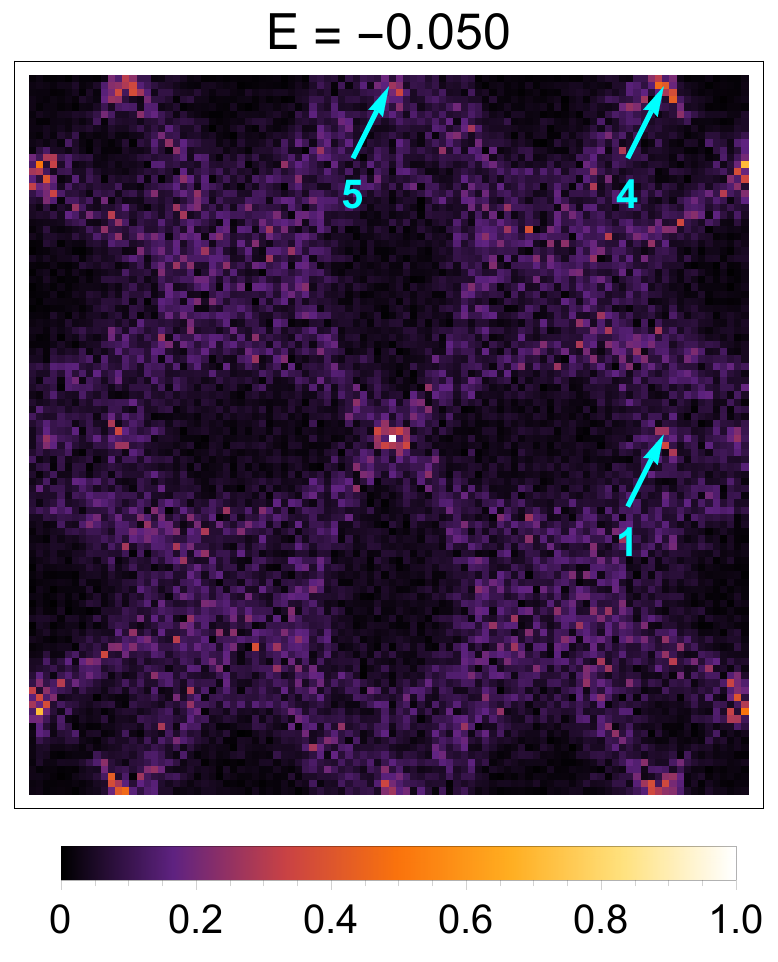}\\
	\includegraphics[width=.2\textwidth]{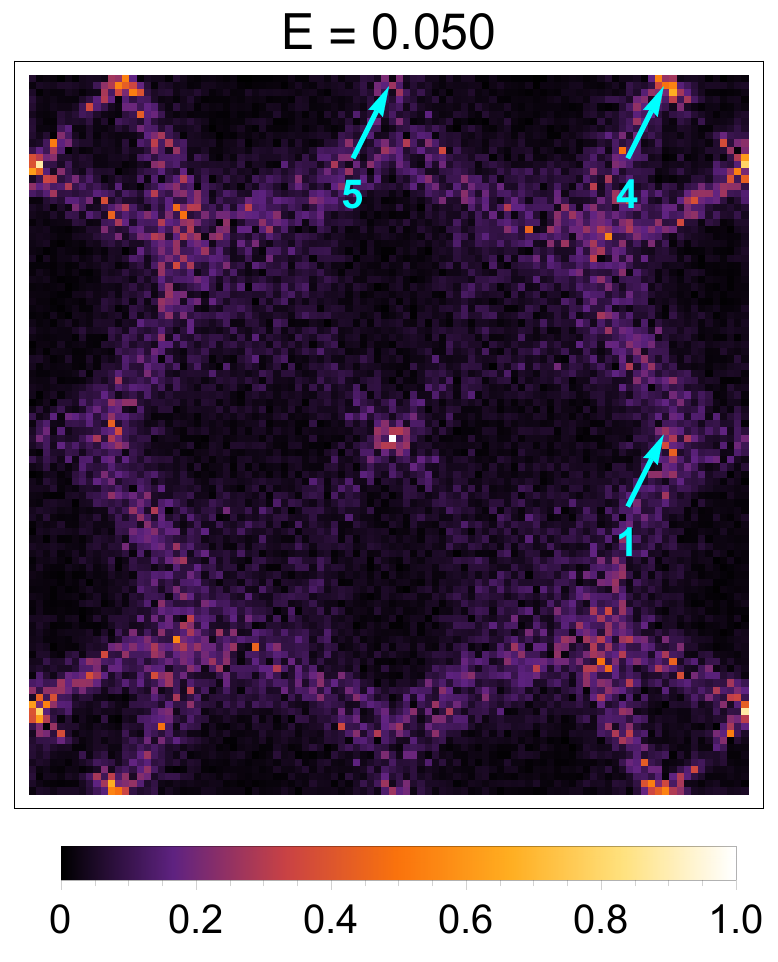}\hfill
	\includegraphics[width=.2\textwidth]{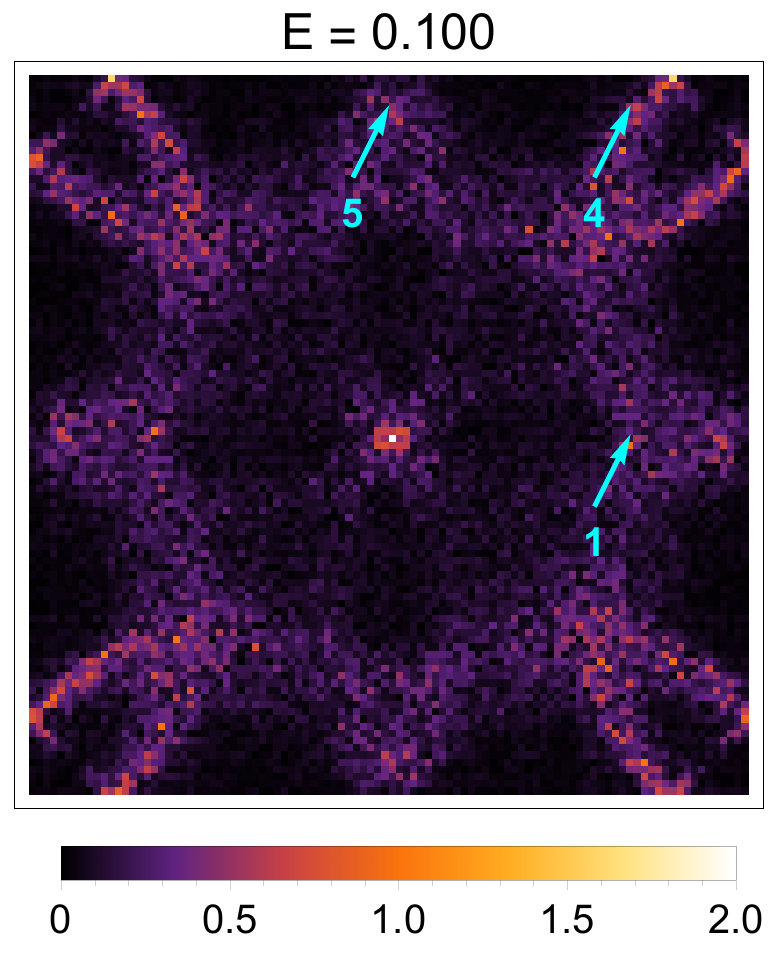}\hfill
	\includegraphics[width=.2\textwidth]{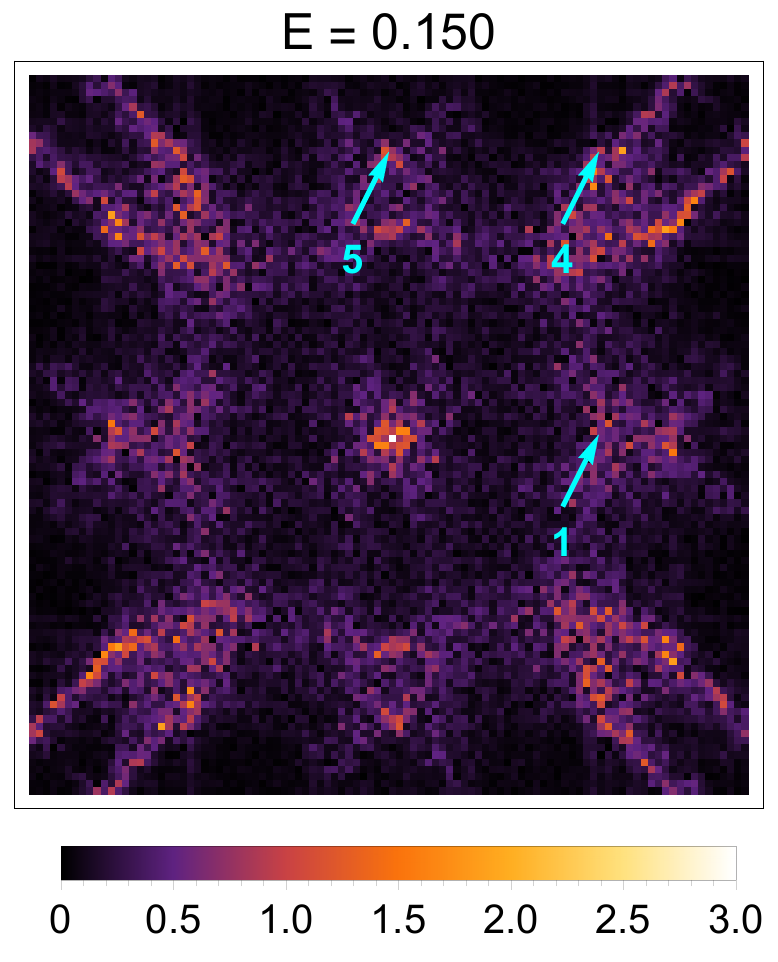}\hfill
	\includegraphics[width=.2\textwidth]{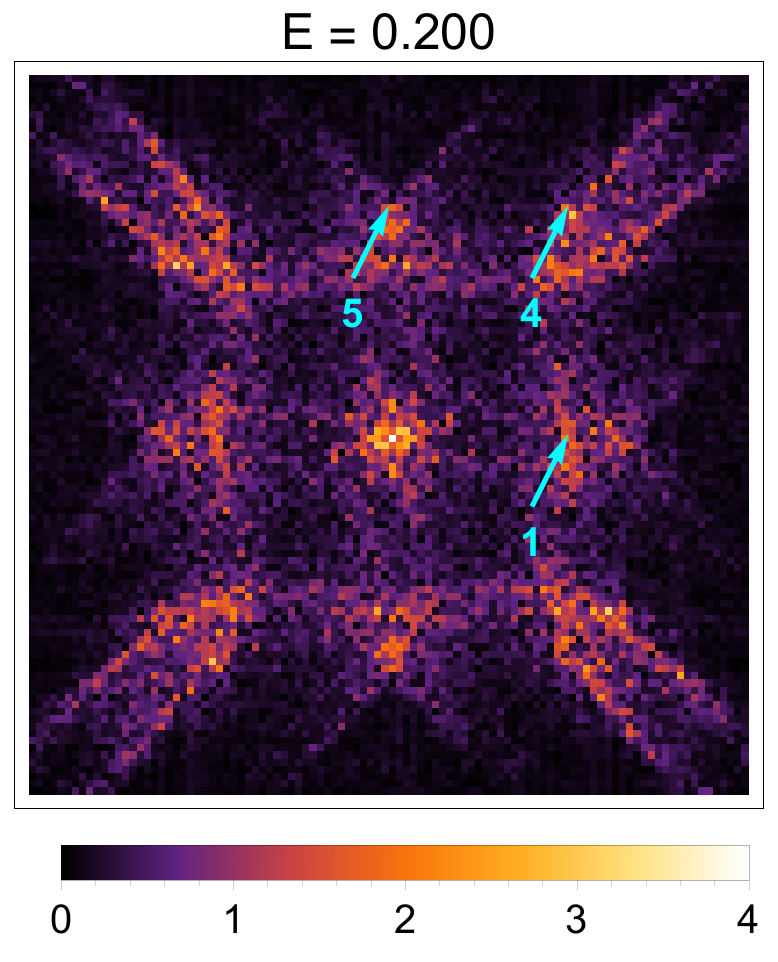}\hfill
	\includegraphics[width=.2\textwidth]{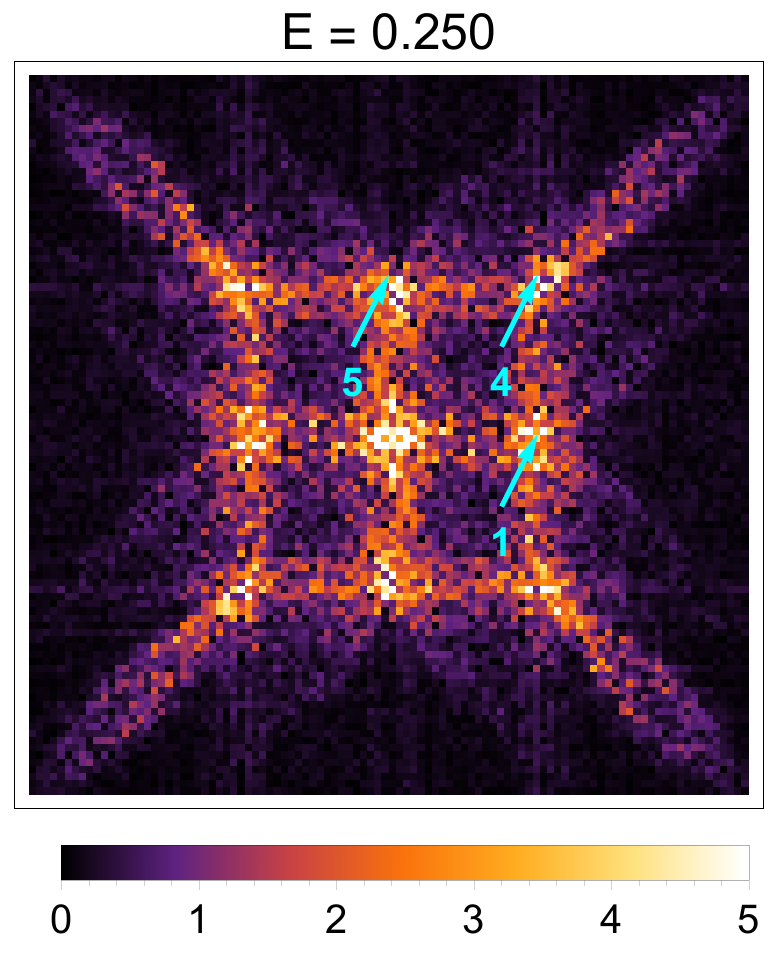}\hfill
	\caption{Fourier-transformed maps for a system with random pairing amplitudes, normally distributed with $M_{\Delta} = 0.01$. Shown are energies ranging from  $E = \pm0.050$ to  $E = \pm0.250$, along with arrows showing where the octet wavevectors are expected to be found. Note that only three peaks---$\mathbf{q}_1$, $\mathbf{q}_4$, and $\mathbf{q}_5$---are visible. The color scaling varies linearly with energy.}
	\label{fig:rgfourier}
\end{figure*}
STS measurements have demonstrated that the superconducting order parameter is in fact inhomogeneous.\cite{lang2002imaging,fang2004periodic,mcelroy2005atomic} It is then worthwhile to ask whether \emph{gap disorder} could also be responsible for QPI. In this section we will consider the case of a $d$-wave superconductor with disorder only in the gap; we keep all other parameters (hoppings and chemical potential) at their mean-field values. This ensures that we can identify the defining characteristics of QPI from pure gap disorder. 

We will assume the simplest model for disorder in the gap that preserves the purely $d$-wave nature of the superconductor. Here only the nearest-neighbor pairing terms are disordered. The pairing amplitude between nearest-neighbor sites $\mathbf{r}$ and $\mathbf{r'}$ is of the form $\Delta_{0, \mathbf{rr'}} + \Delta_{R, \mathbf{rr'}}$, where $\Delta_{0, \mathbf{rr'}}$ is the mean-field pairing amplitude and $\Delta_{R, \mathbf{rr'}}$ is a \emph{random} variable taken from some distribution. Like the random-potential case earlier, we assume that $\Delta_{R, \mathbf{rr'}}$ is normally distributed, with zero mean and a standard deviation $M_{\Delta}$, and, importantly, we will assume that the value of $\Delta_{R, \mathbf{rr'}}$ at one link $(\mathbf{r}, \mathbf{r'})$ is independent of $\Delta_{R, \mathbf{ss'}}$ at any other link $(\mathbf{s}, \mathbf{s'})$. More precisely, 
\begin{eqnarray}
&\langle \Delta_{R, \mathbf{rr'}}  \rangle = 0,\\
&\langle \Delta_{R, \mathbf{rr'}} \Delta_{R, \mathbf{ss'}} \rangle = M^2_{\Delta}(\delta_{\mathbf{rs}}\delta_{\mathbf{r's'}} + \delta_{\mathbf{rs'}}\delta_{\mathbf{r's}}),
\end{eqnarray}
for any two nearest-neighbor links $(\mathbf{r}, \mathbf{r'})$ and $(\mathbf{s}, \mathbf{s'})$. This form of gap disorder is short-ranged and as such should give rise to large-momentum scattering. It should be noted that gap maps from STS measurements do show that the gap variations in space obey a bell-curve-like distribution,\cite{fang2004periodic} which justifies to some extent this choice of distribution.

Plots for this form of gap disorder are shown in Figs.~\ref{fig:rgreal} and~\ref{fig:rgfourier}. We take $M_{\Delta}$ to have a value comparable to that of $M_V$ discussed earlier, so both perturbations are of similar size. The real-space maps exhibit LDOS modulations that are sharper and more noticeable than in the random-potential case. There are stripe-like patterns with streaks in the vertical, horizontal, and diagonal directions. The patterns get far more pronounced with increasing energy. The maps for this case show a marked resemblance to that arising from smooth scatterers, but show considerably more structure in that modulations for more directions are present here than in the smooth-scatterer scenario. There is no signature akin to the single point-like impurity of a localized center of the LDOS modulations. In this sense the results from pure gap disorder match closely real-space maps from experiment.

Like the random-potential case, wavevector peaks corresponding to large-momenta scattering processes are present. It is worth noting that, unlike the unitary point-like scatterer and random-potential cases, only three peaks appear in the power spectrum: $\mathbf{q}_1$, $\mathbf{q}_4$, and $\mathbf{q}_5$. This is because of the fact that scattering due to \emph{weak} gap disorder involves only processes connecting states at which the order parameter has the same sign.\cite{nunner2006fourier,pereg2008magnetic,vishik2009momentum} Out of the seven wavevectors, only the three aforementioned ones correspond to such scattering processes. Curiously, these three momenta are precisely the same ones that are suppressed in the random-potential case. At larger energies, these peaks become more prominent, paralleling the progression in the real-space picture, where the modulations become more and more apparent with increasing energy.

In attributing QPI partially to gap disorder, however, we stress some caution. Our model of disorder involves an order parameter that varies over a length scale of one lattice constant. However, experimentally obtained gap maps show that this is generally not the case. These gap maps feature domains. The average value of the gap from one domain to another can change drastically, but the gap varies slowly within a domain.\cite{lang2002imaging} While the steep change in the gap as one moves from one domain to another is captured well by our simple model, the near-constant nature of the gap within one domain is not. Thus the precise interplay between the smoothness of the gap within a domain and the sharp shifts in the gap from one domain to another cannot be seen from our simple model. Smooth gap disorder would have a similar effect as smooth potential disorder in suppressing large-momentum scattering, and thus a realistic model would very likely feature power spectra dominated by small-momentum process. That said, the problem of modeling the gap inhomogeneities accurately, incorporating both the inter-domain sharpness and intra-domain smoothness of the gap, is an interesting problem for future work.

\section{Discussion and Conclusion}

By utilizing the powerful real-space Green's function method introduced earlier, it is possible to study, in a systematic manner and with large fields of view, the consequences of various forms of distributed disorder on the physics of quasiparticle scattering interference. We made use of the standard method of modeling the low-energy electronic excitations deep in the $d$-wave superconducting state. In addition we also looked at the effects of proposed nontrivial tunneling processes that are potentially of relevance to STS experiments in the cuprates. 

Much of the established intuition regarding the physics of QPI is based on results for a single impurity. When one considers a random distribution of such impurities, however, one deals with the problem of wave interference in a finite-sized random medium. The intuitive expectation is that speckle patterns are formed in the Fourier maps, and this is precisely what we find. 

In the case of a low concentration of impurities with short-ranged potentials, the main difference from the single-impurity case is that the already-diffuse Fourier maps associated with a single impurity turn into speckle patterns that follow closely the weight distribution of the former. This underlies the intuition that the observed QPI can mostly be understood solely on the basis on single-impurity theory. Even in the case of distributed random disorder, in which there is no clear, well-defined sense of an \emph{isolated} impurity, this correspondence with the single-impurity results can be observed clearly. This can be best seen in the case of Gaussian on-site potential disorder, whose power spectra resemble those of the weak single- and multiple-impurity cases. 

The real-space patterns exhibit the characteristic energy-dependent stripe-like interference patterns which are also seen in the raw experimental data. However, upon examining our results more quantitatively, we detect problems that suggest that the present way of interpreting STS experiments may have serious deficiencies. The problem is that the experimental QPI peaks are characterized by a sharpness in momentum space that cannot be reproduced with standard methods of modeling STS experiments, including ours. In the experimental maps, the seven sharp peaks can be discerned and in fact be tracked over a large range of energies. These seven peaks are found to disperse in accordance with the predictions of the octet model. The best-case theoretical scenario is the case of a low concentration of weak point-like impurities, but even here matching our numerically obtained Fourier maps to those obtained from experiment becomes a stretch. The case of a single weak impurity in fact already demonstrates this problem: in addition to peaks, one sees continuous streaks arising from scattering between points on CCEs, whose spectral weight is only enhanced at momenta at the special ``tip-to-tip'' processes. In other words, the peaks in the cases we consider are not observed to be as prominent as in experiment---they happen to be the points which possess the largest spectral weight along the streaks corresponding to scattering between CCEs. When one goes beyond this single-scatterer paradigm and considers other, more general forms of distributed disorder, this sharpness is further reduced.

This is an exceptional circumstance. One usually expects that idealized models like ours will produce outcomes that are \emph{sharper} than experimental data. The incorporation of the most general forms of disorder, which we implement in this work, should have the effect of adding fuzziness in the Fourier-space picture. In our models we ignore complicating factors such as frequency-dependent self-energies that could alter the picture for larger energies. Given the relative lack of complications present in our models, this inability to reproduce the sharpness of experimental data is puzzling.

The outcomes of our simulations for many weak point-like scatterers are perhaps the closest approach to experiment. However, taking these as an explanation is problematic since no impurity cores are seen in experiment. The case of many unitary point-like scatterers is even more rife with problems because in this case impurity cores are much more visible and the strong scattering processes preclude the formation of prominent peaks in the power spectra, showing instead very fuzzy streaks corresponding to inter-CCE scattering. As a next case, there are very good reasons to believe that the intrinsic disorder in cuprates is of a smooth kind. The CuO$_2$ planes themselves are quite clean, lacking disorder from doping, while the chemical sources of disorder are located in the insulating buffer layers located some distance away form the superconducting perovskite planes. Our simulations of smooth disorder show that the large-momentum peaks are suppressed, owing to the fact that in the Fourier decomposition of the screened Coulomb potential the large-wavevector components have very small amplitudes. Our results seem to suggest that in order to reproduce the overall weight distribution seen in the experimental Fourier maps, one needs local, point-like potentials. This is quite puzzling given what is now known about the chemical composition of the cuprates. 

Disorder in the form of randomly distributed on-site energies is another scenario that gives rise to real- and Fourier-space maps that are very similar to those found in experiment. These are found to result in modulations in the LDOS without the presence of visible impurities, and power spectra for this form of disorder show peaks that originate from large-momentum scattering processes. The caveat with this form of disorder however is that, like the many-weak-impurity scenario, not all of the peaks are visible in the Fourier maps. We finally note that as compelling an explanation as this is for the patterns seen in experimental data, it is difficult to argue from microscopic considerations why this form of disorder should exist---unlike point-like and smooth disorder, whose possible origin in the cuprates can at least be justified on the level of chemistry.

We also examined in some detail the influence of gap disorder in the LDOS maps, using a simple model of gap inhomogeneities. It is well-established that in the cuprates, especially the underdoped ones, the gap magnitude is quite inhomogeneous, varying by a large amount in space. Our calculations show that this form of disorder scatters the Bogoliubov quasiparticles efficiently, generating a distinctive power spectrum and visible real-space patterns. These results suggest that gap disorder could potentially generate QPI as well. However, it is also known from experiment that this form of disorder is characterized by a short-distance cutoff scale on the order of the coherence length $\sim 3$ nm.\cite{lang2002imaging, fang2004periodic, howald2001inherent} Gap disorder is therefore a smooth form of disorder, and its effect should thus be similar to that of smooth potential disorder.  

What is the origin of this trouble? One possibility is that the physics underlying QPI in the cuprates is completely different from the standard explanation, which is centered on the quantum-mechanical scattering of Bogoliubov quasiparticles against quenched disorder. One could contemplate exotic possibilities involving the formation of \emph{real} bound states at the special momenta of the octet model---the most obvious way to obtain sharp quantization in momentum space. However, we think that this is far-fetched. Direct, independent evidence for the presence of coherent Bogoliubov quasiparticles with a $d$-wave dispersion exists from angle-resolved photoemission spectroscopy. Moreover, the octet model is qualitatively highly successful in relating the dispersions from QPI to measured dispersions from ARPES. A concrete possibility that builds on the scattering picture of QPI is that nematic quantum-critical fluctuations strongly enhance the amplitudes of the peaks.\cite{kim2008theory, kim2010interference} We suspect that the culprit is the tunneling process itself. On a quantitative level this is sensitive to the details of the microscopic electronic structure. Recent first-principles work demonstrates this vividly.\cite{kreisel2015interpretation} Kreisel \emph{et al.} find nontrivial effects arising from microscopic details, such as the enhancement of large-momentum peaks in the power spectrum. Similarly, it may well be necessary to study disorder in a much more microscopic manner in order to capture the way it affects the microscopic intra-unit cell electronic structure.\cite{vishik2009momentum}  We envisage that it may become possible to extract from such precise modeling of the microscopic tunneling process effective, coarse-grained models which can then be studied in the most general disordered case using the methods we have used in this work. The overarching message is that there is in all likelihood more to the beautiful STS images than meets the eye.

\begin{acknowledgments}
	
We thank S. A. Kivelson and R.-J. Slager for useful discussions. This work was supported by the Netherlands Organisation for Scientific Research (NWO/OCW) as part of the Frontiers of Nanoscience (NanoFront) program.
 
\end{acknowledgments}

\appendix

\section{Single Unitary Point-Like Scatterer}

\begin{figure*}[ht]
	\centering
	\includegraphics[width=.25\textwidth]{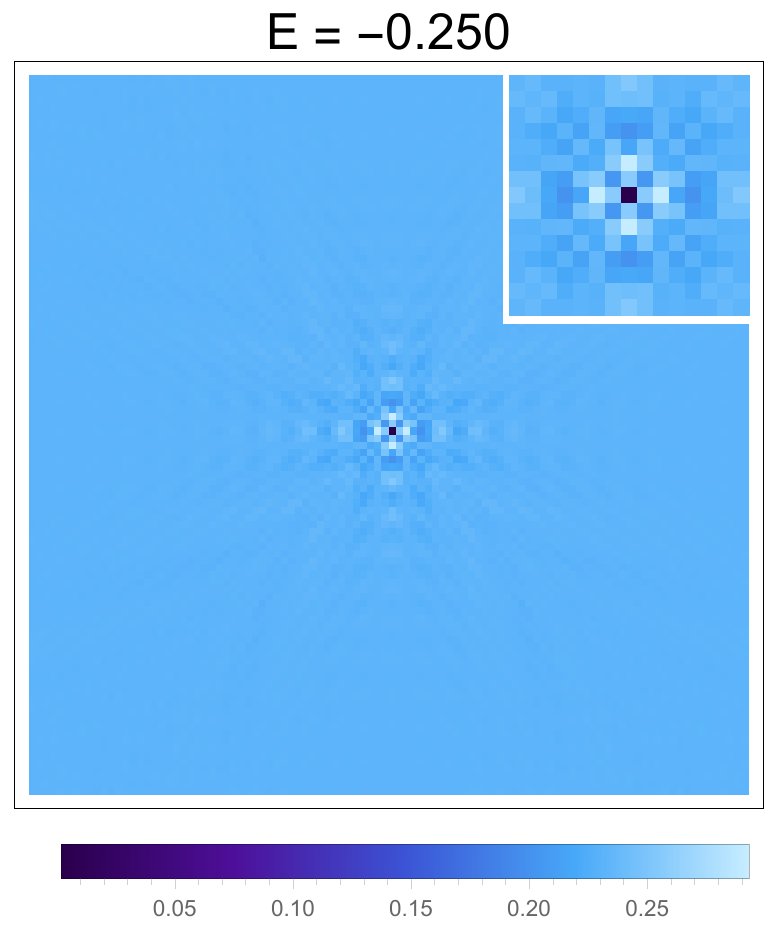}\hfill
	\includegraphics[width=.25\textwidth]{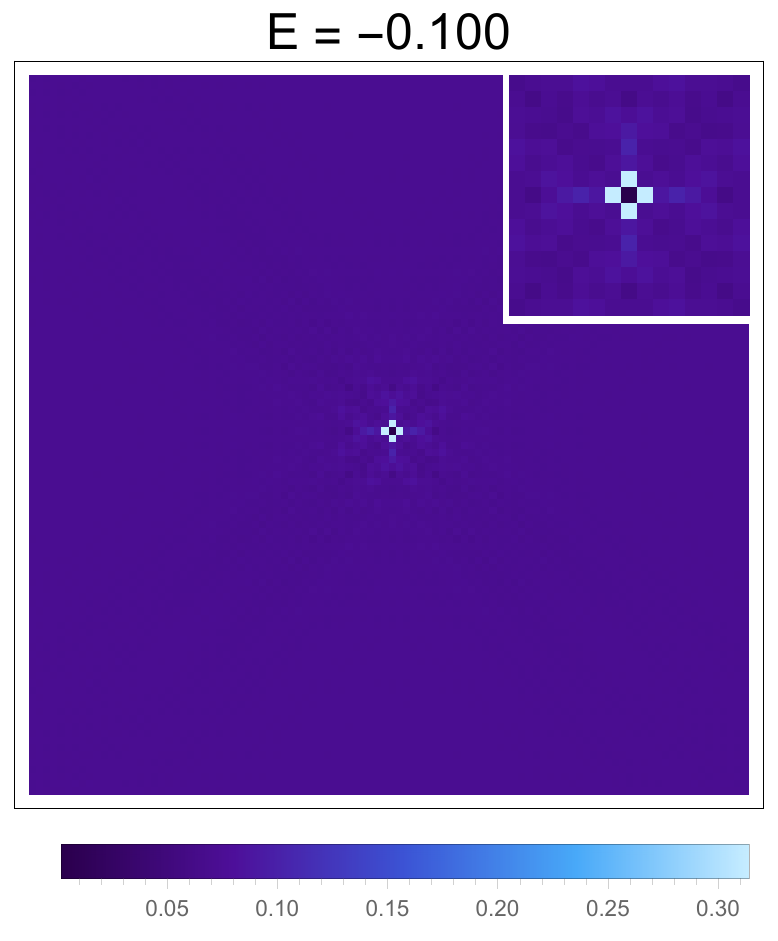}\hfill
	\includegraphics[width=.25\textwidth]{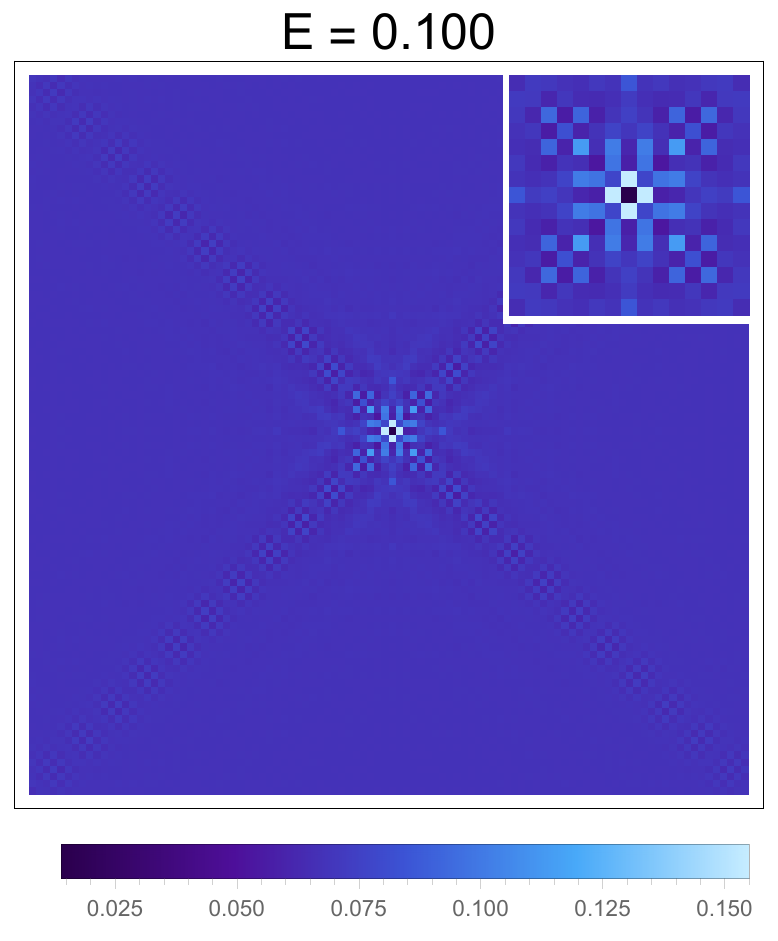}\hfill
	\includegraphics[width=.25\textwidth]{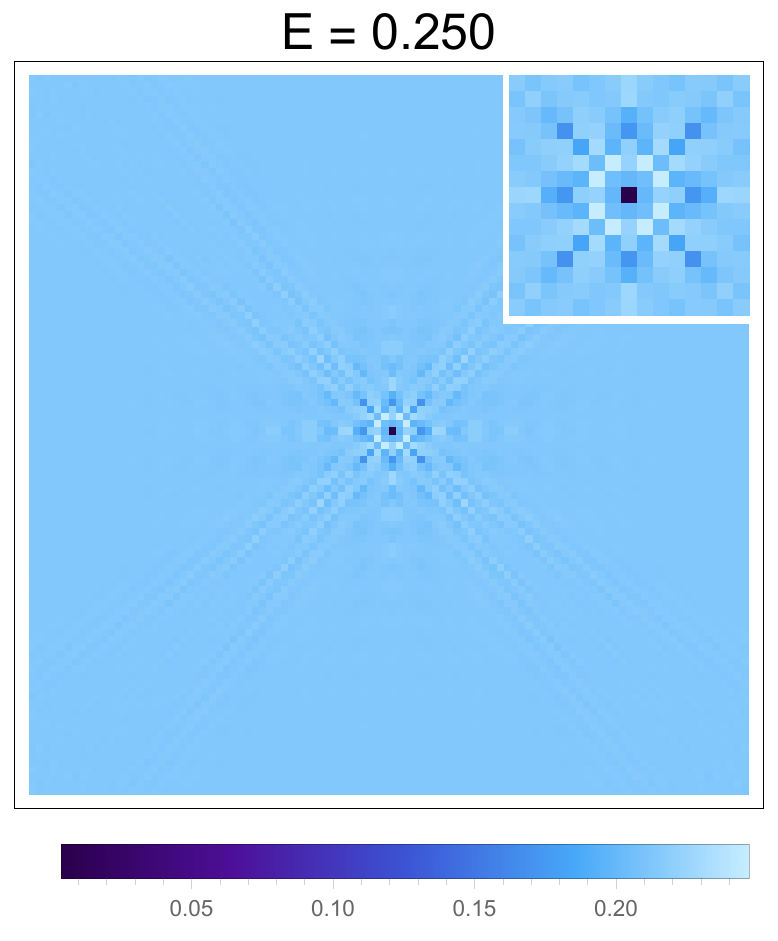}\hfill
	\caption{Real-space LDOS maps for the single unitary point-like scatterer case. Here an isolated point-like impurity ($V = 10$) is placed in the middle of the sample. The field of view is $100 \times 100$. Shown are maps corresponding to energies $E = \pm0.100$ and $E = \pm0.250$. Inset: a close-up view of the impurity.}
	\label{fig:spreal}
\end{figure*}

\begin{figure*}
	\centering
	\includegraphics[width=.2\textwidth]{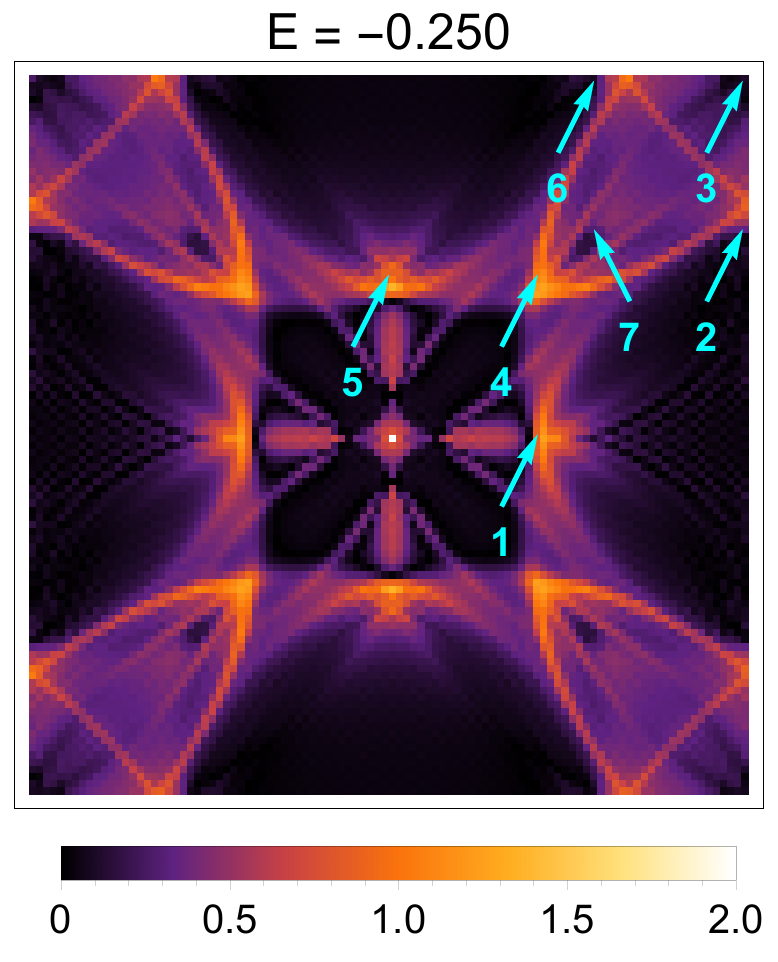}\hfill
	\includegraphics[width=.2\textwidth]{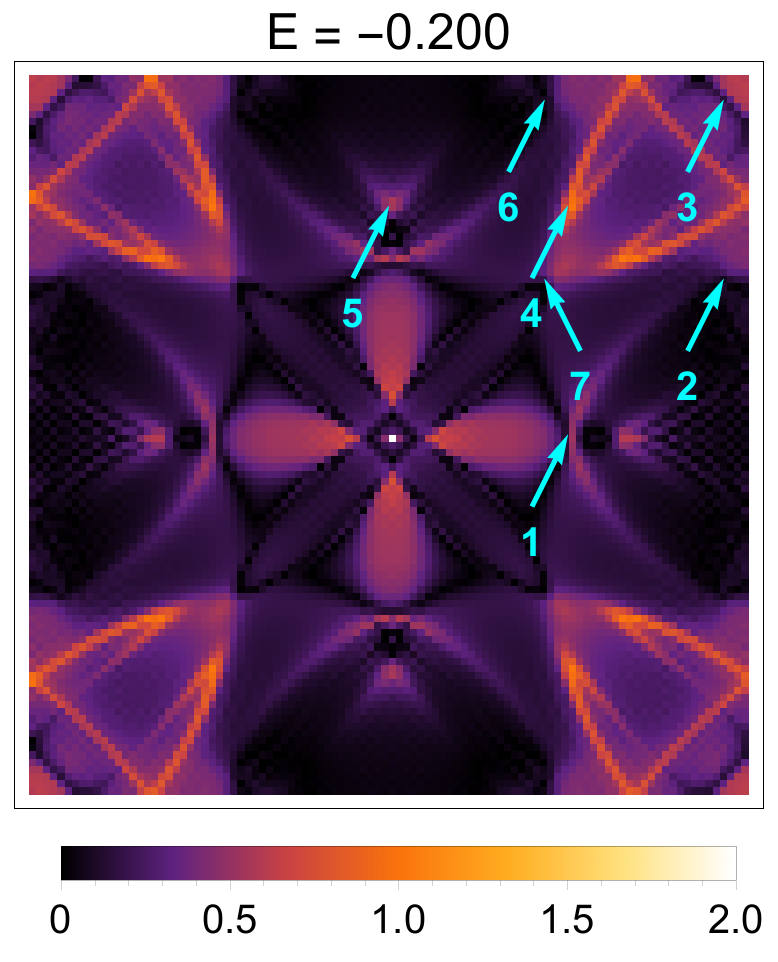}\hfill
	\includegraphics[width=.2\textwidth]{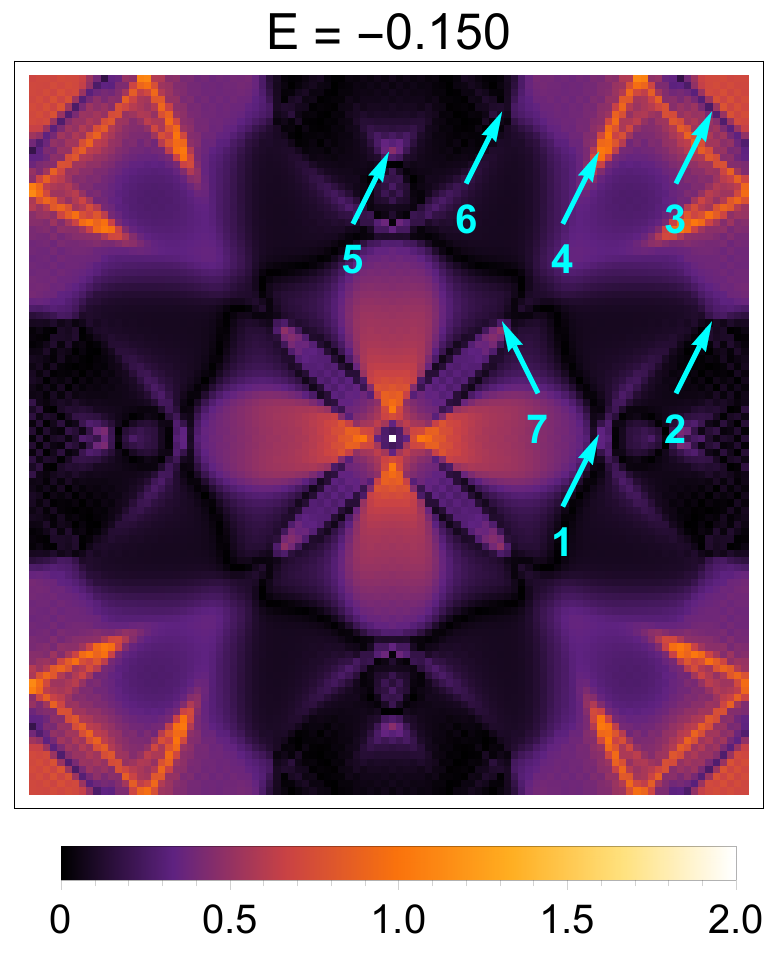}\hfill
	\includegraphics[width=.2\textwidth]{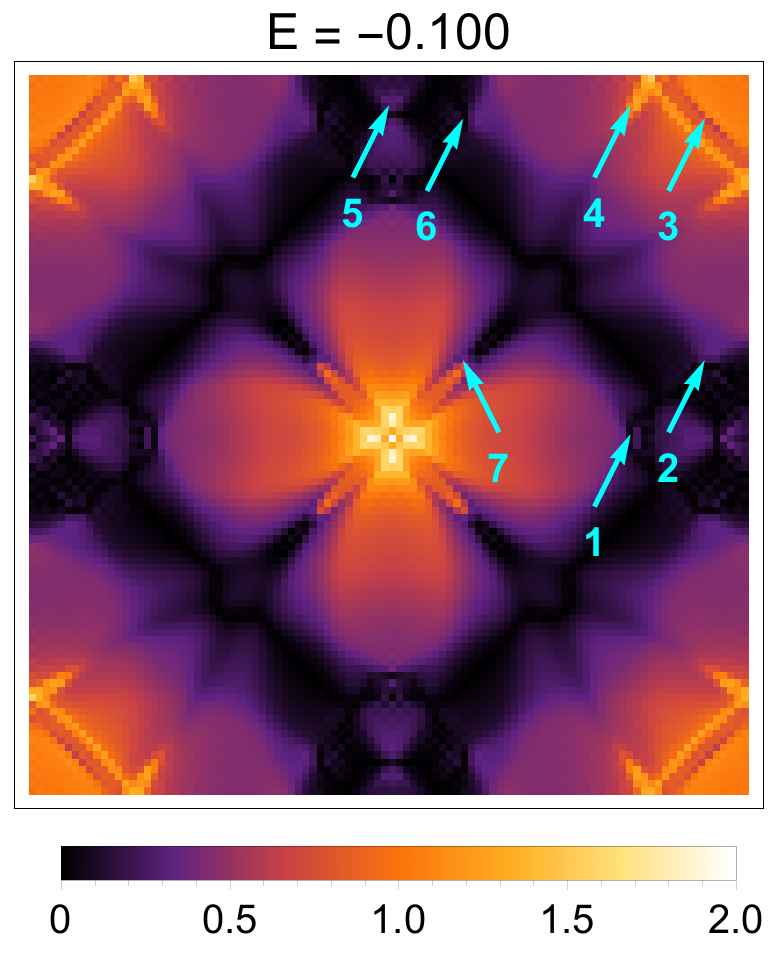}\hfill
	\includegraphics[width=.2\textwidth]{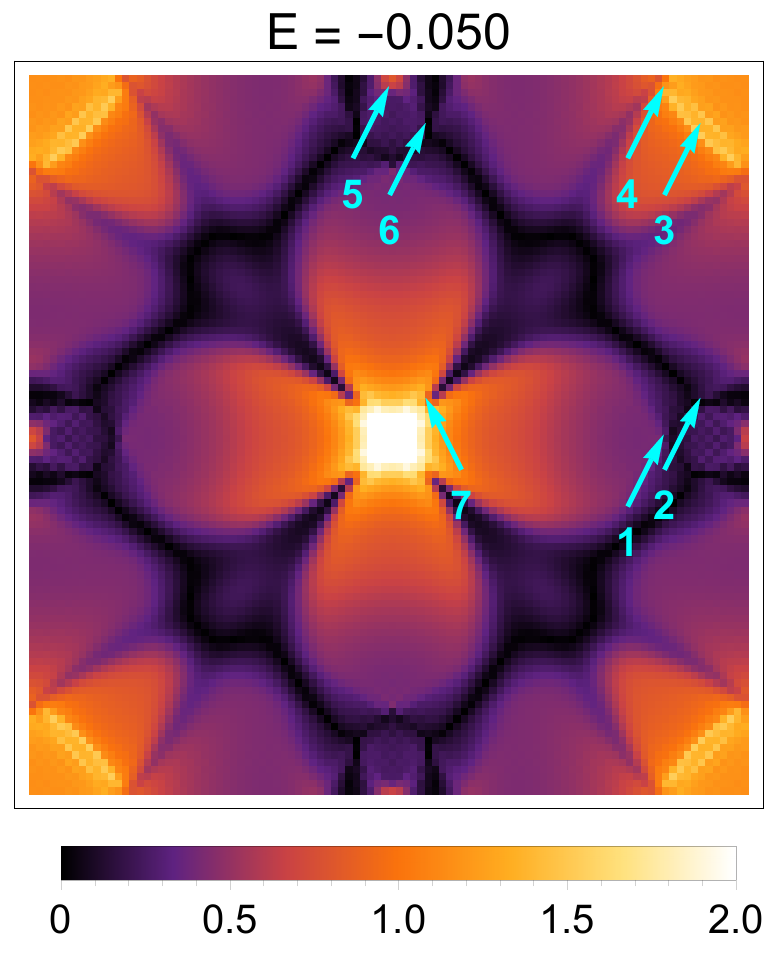}\\
	\includegraphics[width=.2\textwidth]{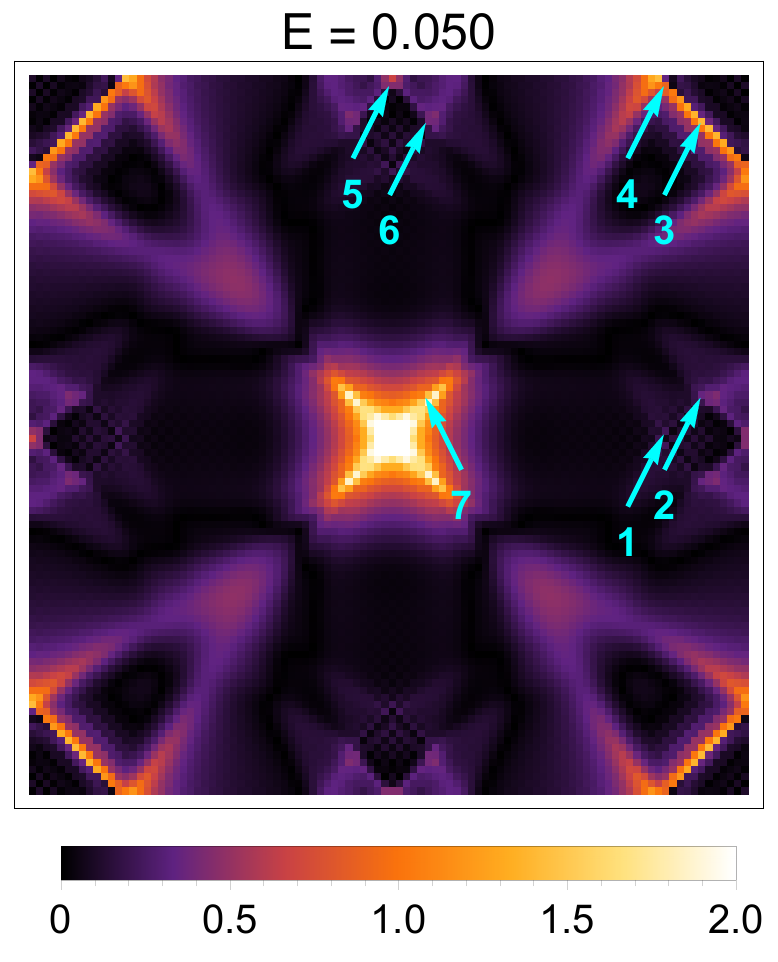}\hfill
	\includegraphics[width=.2\textwidth]{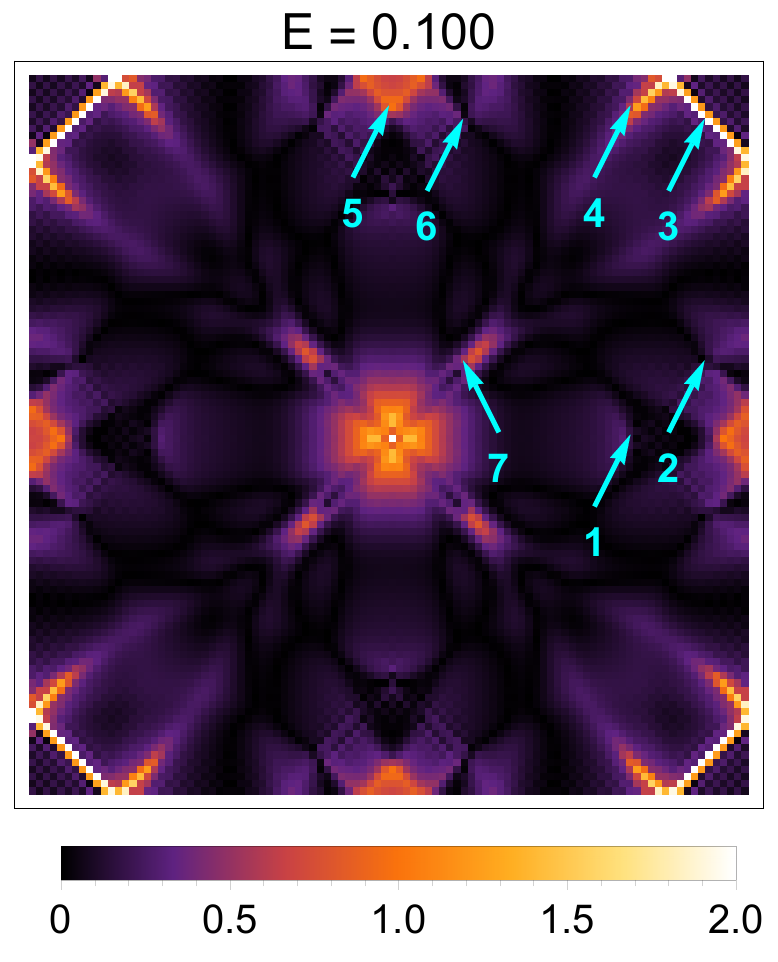}\hfill
	\includegraphics[width=.2\textwidth]{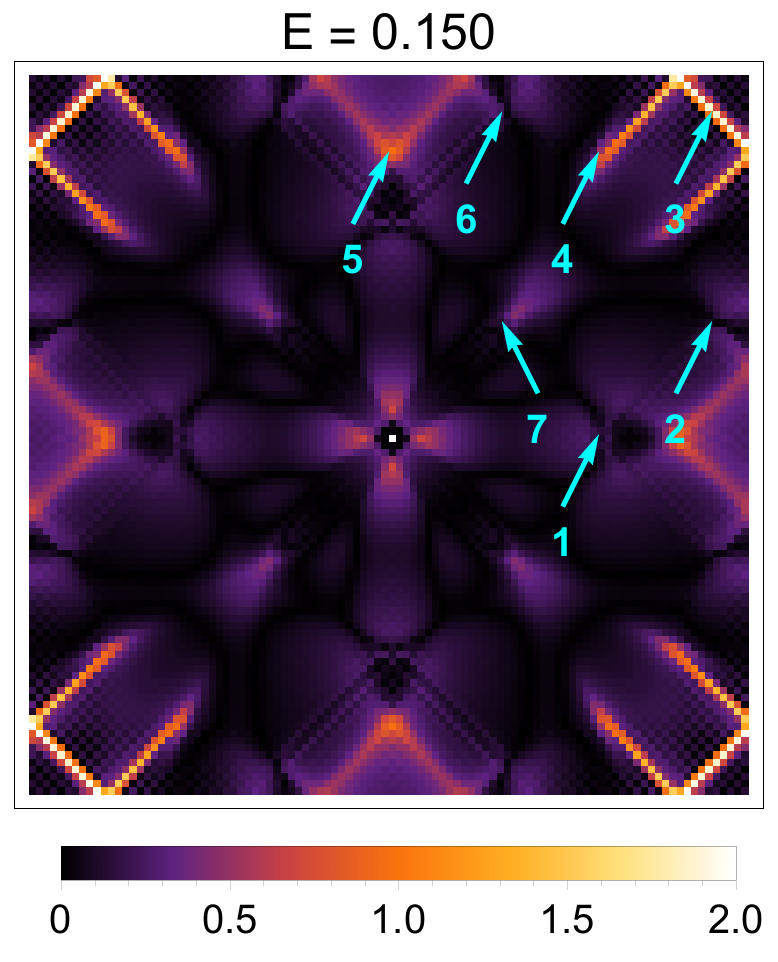}\hfill
	\includegraphics[width=.2\textwidth]{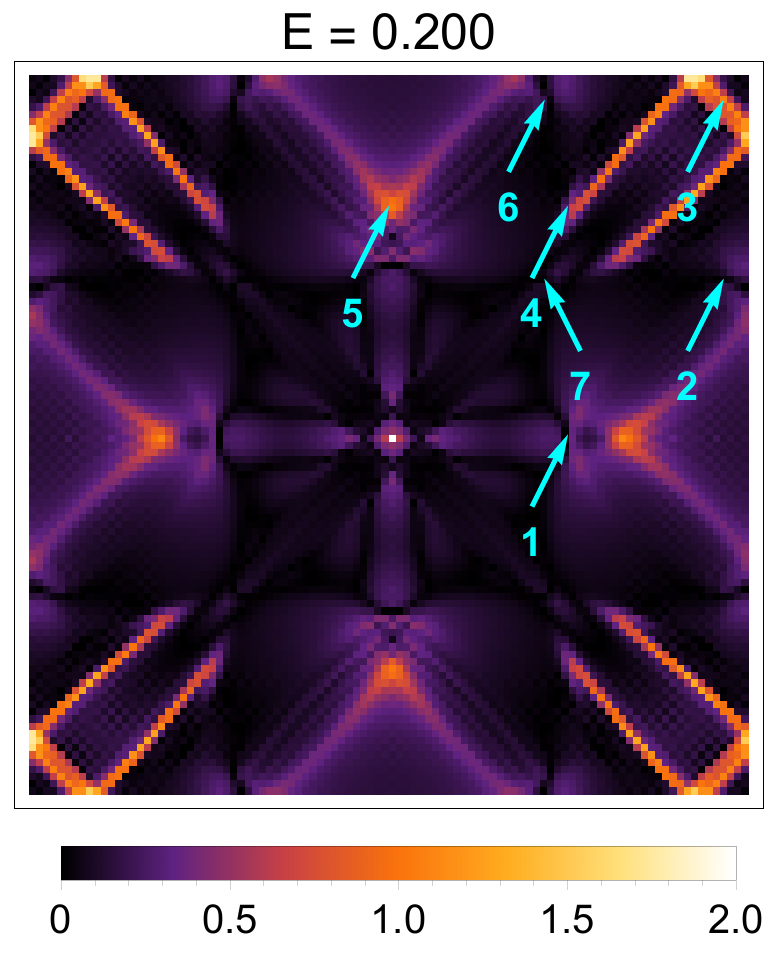}\hfill
	\includegraphics[width=.2\textwidth]{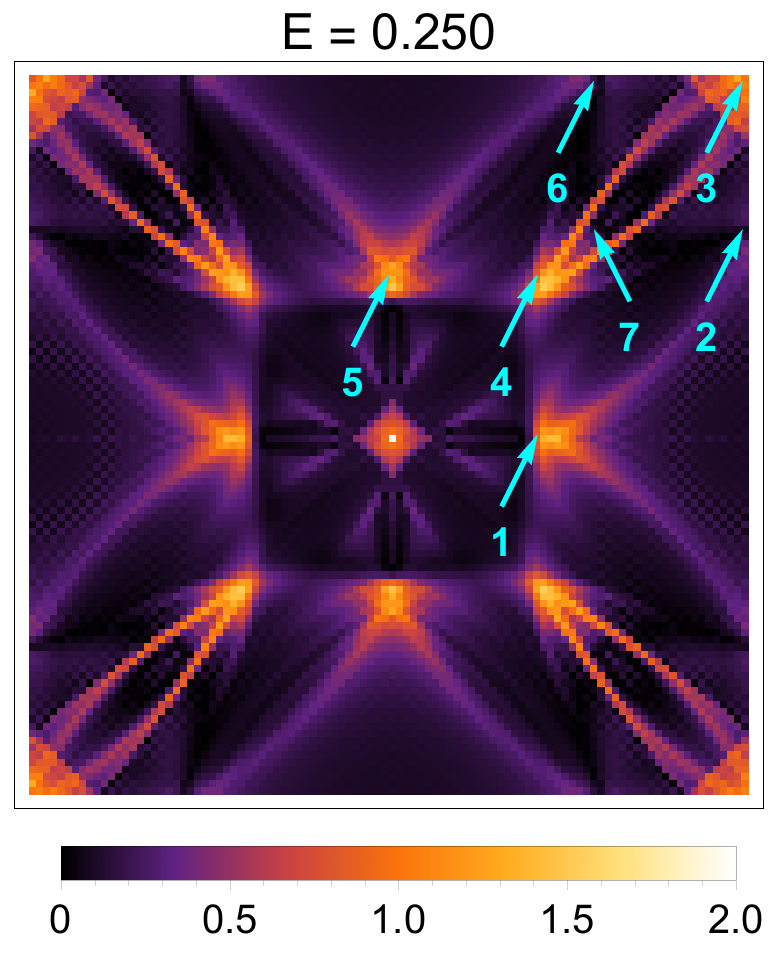}\hfill
	\caption{Fourier-transformed maps for the single unitary point-like scatterer case, with $V = 10$. Power spectra for both positive and negative bias voltages are shown for energies ranging from $E = \pm0.050 $ to $E = \pm0.250$. Arrows indicate where the peaks corresponding to the characteristic momenta of the octet model show up in the upper-right quadrant. The color scale is the same for all energies. }
	\label{fig:spfourier}
\end{figure*}

In this section we will briefly discuss the case of a single \emph{unitary} point-like scatterer. For reasons discussed in depth in the main text, this is not physically relevant for the experimental data we wish to revisit. That said, these are not unphysical---zinc impurities in BSCCO are an example of non-magnetic unitary scatterers, for instance. While QPI in clean cuprates is most likely caused by far weaker impurities, the properties of unitary point-like scatterers are sufficiently different from those of weak ones that it is worth spending a few words delineating some of these differences.

In Figs.~\ref{fig:spreal} and~\ref{fig:spfourier} we plot real-space and Fourier-transformed maps for a single strong impurity embedded in the middle of the sample. We take $V = 10$, ensuring that the impurity is a unitary scatterer. The real-space maps are qualitatively similar to those of the weak-impurity case. In both cases the impurity cores can easily be discerned. The main difference between the unitary- and weak-scatterer cases is that the LDOS at the unitary-impurity site is almost completely suppressed. Recall that in the weak-impurity case, the LDOS at the impurity site is finite.

The second noteworthy feature of the unitary scatterer is apparent in the Fourier-transformed maps. Because the potential is so strong, scattering between any two points lying on CCEs is allowed, resulting in very prominent streaks in the power spectrum. Many of the peaks from the octet model can be seen, similar to the case of the weak scatterer. However, the peaks that are most prominent here differ from those seen in the weak-scatterer case. Observe that when energies become high, $\mathbf{q_2}$, $\mathbf{q_6}$, and $\mathbf{q_7}$ become less visible. Streaks near the corners of the first Brillouin zone corresponding to internodal scattering remains very prominent, but a peak at $\mathbf{q_3}$ is not as visible as it is in the weak-impurity case. In contrast, $\mathbf{q_1}$, $\mathbf{q_4}$, and $\mathbf{q_5}$ become far more visible and in fact become the most dominant wavevectors in the power spectrum. It is interesting to note that $\mathbf{q_4}$ and $\mathbf{q_5}$ are barely visible in the weak-impurity case; this can be attributed to the presence of coherence factors that suppress the amplitudes of these scattering processes.\cite{wang2003quasiparticle, pereg2003theory}

\nocite{*}


\providecommand{\noopsort}[1]{}\providecommand{\singleletter}[1]{#1}%
\begin{thebibliography}{72}%
	\makeatletter
	\providecommand \@ifxundefined [1]{%
		\@ifx{#1\undefined}
	}%
	\providecommand \@ifnum [1]{%
		\ifnum #1\expandafter \@firstoftwo
		\else \expandafter \@secondoftwo
		\fi
	}%
	\providecommand \@ifx [1]{%
		\ifx #1\expandafter \@firstoftwo
		\else \expandafter \@secondoftwo
		\fi
	}%
	\providecommand \natexlab [1]{#1}%
	\providecommand \enquote  [1]{``#1''}%
	\providecommand \bibnamefont  [1]{#1}%
	\providecommand \bibfnamefont [1]{#1}%
	\providecommand \citenamefont [1]{#1}%
	\providecommand \href@noop [0]{\@secondoftwo}%
	\providecommand \href [0]{\begingroup \@sanitize@url \@href}%
	\providecommand \@href[1]{\@@startlink{#1}\@@href}%
	\providecommand \@@href[1]{\endgroup#1\@@endlink}%
	\providecommand \@sanitize@url [0]{\catcode `\\12\catcode `\$12\catcode
		`\&12\catcode `\#12\catcode `\^12\catcode `\_12\catcode `\%12\relax}%
	\providecommand \@@startlink[1]{}%
	\providecommand \@@endlink[0]{}%
	\providecommand \url  [0]{\begingroup\@sanitize@url \@url }%
	\providecommand \@url [1]{\endgroup\@href {#1}{\urlprefix }}%
	\providecommand \urlprefix  [0]{URL }%
	\providecommand \Eprint [0]{\href }%
	\providecommand \doibase [0]{http://dx.doi.org/}%
	\providecommand \selectlanguage [0]{\@gobble}%
	\providecommand \bibinfo  [0]{\@secondoftwo}%
	\providecommand \bibfield  [0]{\@secondoftwo}%
	\providecommand \translation [1]{[#1]}%
	\providecommand \BibitemOpen [0]{}%
	\providecommand \bibitemStop [0]{}%
	\providecommand \bibitemNoStop [0]{.\EOS\space}%
	\providecommand \EOS [0]{\spacefactor3000\relax}%
	\providecommand \BibitemShut  [1]{\csname bibitem#1\endcsname}%
	\let\auto@bib@innerbib\@empty
	\bibitem [{\citenamefont {Schmidt}\ \emph {et~al.}(2011)\citenamefont
		{Schmidt}, \citenamefont {Fujita}, \citenamefont {Kim}, \citenamefont
		{Lawler}, \citenamefont {Eisaki}, \citenamefont {Uchida}, \citenamefont
		{Lee},\ and\ \citenamefont {Davis}}]{schmidt2011electronic}%
	\BibitemOpen
	\bibfield  {author} {\bibinfo {author} {\bibfnamefont {A.~R.}\ \bibnamefont
			{Schmidt}}, \bibinfo {author} {\bibfnamefont {K.}~\bibnamefont {Fujita}},
		\bibinfo {author} {\bibfnamefont {E.-A.}\ \bibnamefont {Kim}}, \bibinfo
		{author} {\bibfnamefont {M.~J.}\ \bibnamefont {Lawler}}, \bibinfo {author}
		{\bibfnamefont {H.}~\bibnamefont {Eisaki}}, \bibinfo {author} {\bibfnamefont
			{S.}~\bibnamefont {Uchida}}, \bibinfo {author} {\bibfnamefont {D.-H.}\
			\bibnamefont {Lee}}, \ and\ \bibinfo {author} {\bibfnamefont {J.~C.}\
			\bibnamefont {Davis}},\ }\href@noop {} {\bibfield  {journal} {\bibinfo
			{journal} {New Journal of Physics}\ }\textbf {\bibinfo {volume} {13}},\
		\bibinfo {pages} {065014} (\bibinfo {year} {2011})}\BibitemShut {NoStop}%
	\bibitem [{\citenamefont {Vershinin}\ \emph {et~al.}(2004)\citenamefont
		{Vershinin}, \citenamefont {Misra}, \citenamefont {Ono}, \citenamefont {Abe},
		\citenamefont {Ando},\ and\ \citenamefont {Yazdani}}]{vershinin2004local}%
	\BibitemOpen
	\bibfield  {author} {\bibinfo {author} {\bibfnamefont {M.}~\bibnamefont
			{Vershinin}}, \bibinfo {author} {\bibfnamefont {S.}~\bibnamefont {Misra}},
		\bibinfo {author} {\bibfnamefont {S.}~\bibnamefont {Ono}}, \bibinfo {author}
		{\bibfnamefont {Y.}~\bibnamefont {Abe}}, \bibinfo {author} {\bibfnamefont
			{Y.}~\bibnamefont {Ando}}, \ and\ \bibinfo {author} {\bibfnamefont
			{A.}~\bibnamefont {Yazdani}},\ }\href@noop {} {\bibfield  {journal} {\bibinfo
			{journal} {Science}\ }\textbf {\bibinfo {volume} {303}},\ \bibinfo {pages}
		{1995} (\bibinfo {year} {2004})}\BibitemShut {NoStop}%
	\bibitem [{\citenamefont {Hanaguri}\ \emph {et~al.}(2004)\citenamefont
		{Hanaguri}, \citenamefont {Lupien}, \citenamefont {Kohsaka}, \citenamefont
		{Lee}, \citenamefont {Azuma}, \citenamefont {Takano}, \citenamefont
		{Takagi},\ and\ \citenamefont {Davis}}]{hanaguri2004checkerboard}%
	\BibitemOpen
	\bibfield  {author} {\bibinfo {author} {\bibfnamefont {T.}~\bibnamefont
			{Hanaguri}}, \bibinfo {author} {\bibfnamefont {C.}~\bibnamefont {Lupien}},
		\bibinfo {author} {\bibfnamefont {Y.}~\bibnamefont {Kohsaka}}, \bibinfo
		{author} {\bibfnamefont {D.-H.}\ \bibnamefont {Lee}}, \bibinfo {author}
		{\bibfnamefont {M.}~\bibnamefont {Azuma}}, \bibinfo {author} {\bibfnamefont
			{M.}~\bibnamefont {Takano}}, \bibinfo {author} {\bibfnamefont
			{H.}~\bibnamefont {Takagi}}, \ and\ \bibinfo {author} {\bibfnamefont {J.~C.}\
			\bibnamefont {Davis}},\ }\href@noop {} {\bibfield  {journal} {\bibinfo
			{journal} {Nature}\ }\textbf {\bibinfo {volume} {430}},\ \bibinfo {pages}
		{1001} (\bibinfo {year} {2004})}\BibitemShut {NoStop}%
	\bibitem [{\citenamefont {Kohsaka}\ \emph {et~al.}(2007)\citenamefont
		{Kohsaka}, \citenamefont {Taylor}, \citenamefont {Fujita}, \citenamefont
		{Schmidt}, \citenamefont {Lupien}, \citenamefont {Hanaguri}, \citenamefont
		{Azuma}, \citenamefont {Takano}, \citenamefont {Eisaki}, \citenamefont
		{Takagi}, \citenamefont {Uchida},\ and\ \citenamefont
		{Davis}}]{kohsaka2007intrinsic}%
	\BibitemOpen
	\bibfield  {author} {\bibinfo {author} {\bibfnamefont {Y.}~\bibnamefont
			{Kohsaka}}, \bibinfo {author} {\bibfnamefont {C.}~\bibnamefont {Taylor}},
		\bibinfo {author} {\bibfnamefont {K.}~\bibnamefont {Fujita}}, \bibinfo
		{author} {\bibfnamefont {A.}~\bibnamefont {Schmidt}}, \bibinfo {author}
		{\bibfnamefont {C.}~\bibnamefont {Lupien}}, \bibinfo {author} {\bibfnamefont
			{T.}~\bibnamefont {Hanaguri}}, \bibinfo {author} {\bibfnamefont
			{M.}~\bibnamefont {Azuma}}, \bibinfo {author} {\bibfnamefont
			{M.}~\bibnamefont {Takano}}, \bibinfo {author} {\bibfnamefont
			{H.}~\bibnamefont {Eisaki}}, \bibinfo {author} {\bibfnamefont
			{H.}~\bibnamefont {Takagi}}, \bibinfo {author} {\bibfnamefont
			{S.}~\bibnamefont {Uchida}}, \ and\ \bibinfo {author} {\bibfnamefont {J.~C.}\
			\bibnamefont {Davis}},\ }\href@noop {} {\bibfield  {journal} {\bibinfo
			{journal} {Science}\ }\textbf {\bibinfo {volume} {315}},\ \bibinfo {pages}
		{1380} (\bibinfo {year} {2007})}\BibitemShut {NoStop}%
	\bibitem [{\citenamefont {Lawler}\ \emph {et~al.}(2010)\citenamefont {Lawler},
		\citenamefont {Fujita}, \citenamefont {Lee}, \citenamefont {Schmidt},
		\citenamefont {Kohsaka}, \citenamefont {Kim}, \citenamefont {Eisaki},
		\citenamefont {Uchida}, \citenamefont {Davis}, \citenamefont {Sethna},\ and\
		\citenamefont {Kim}}]{lawler2010intra}%
	\BibitemOpen
	\bibfield  {author} {\bibinfo {author} {\bibfnamefont {M.~J.}\ \bibnamefont
			{Lawler}}, \bibinfo {author} {\bibfnamefont {K.}~\bibnamefont {Fujita}},
		\bibinfo {author} {\bibfnamefont {J.}~\bibnamefont {Lee}}, \bibinfo {author}
		{\bibfnamefont {A.~R.}\ \bibnamefont {Schmidt}}, \bibinfo {author}
		{\bibfnamefont {Y.}~\bibnamefont {Kohsaka}}, \bibinfo {author} {\bibfnamefont
			{C.~K.}\ \bibnamefont {Kim}}, \bibinfo {author} {\bibfnamefont
			{H.}~\bibnamefont {Eisaki}}, \bibinfo {author} {\bibfnamefont
			{S.}~\bibnamefont {Uchida}}, \bibinfo {author} {\bibfnamefont {J.~C.}\
			\bibnamefont {Davis}}, \bibinfo {author} {\bibfnamefont {J.~P.}\ \bibnamefont
			{Sethna}}, \ and\ \bibinfo {author} {\bibfnamefont {E.-A.}\ \bibnamefont
			{Kim}},\ }\href@noop {} {\bibfield  {journal} {\bibinfo  {journal} {Nature}\
		}\textbf {\bibinfo {volume} {466}},\ \bibinfo {pages} {347} (\bibinfo {year}
		{2010})}\BibitemShut {NoStop}%
	\bibitem [{\citenamefont {Lang}\ \emph {et~al.}(2002)\citenamefont {Lang},
		\citenamefont {Madhavan}, \citenamefont {Hoffman}, \citenamefont {Hudson},
		\citenamefont {Eisaki}, \citenamefont {Uchida},\ and\ \citenamefont
		{Davis}}]{lang2002imaging}%
	\BibitemOpen
	\bibfield  {author} {\bibinfo {author} {\bibfnamefont {K.~M.}\ \bibnamefont
			{Lang}}, \bibinfo {author} {\bibfnamefont {V.}~\bibnamefont {Madhavan}},
		\bibinfo {author} {\bibfnamefont {J.~E.}\ \bibnamefont {Hoffman}}, \bibinfo
		{author} {\bibfnamefont {E.~W.}\ \bibnamefont {Hudson}}, \bibinfo {author}
		{\bibfnamefont {H.}~\bibnamefont {Eisaki}}, \bibinfo {author} {\bibfnamefont
			{S.}~\bibnamefont {Uchida}}, \ and\ \bibinfo {author} {\bibfnamefont {J.~C.}\
			\bibnamefont {Davis}},\ }\href@noop {} {\bibfield  {journal} {\bibinfo
			{journal} {Nature}\ }\textbf {\bibinfo {volume} {415}},\ \bibinfo {pages}
		{412} (\bibinfo {year} {2002})}\BibitemShut {NoStop}%
	\bibitem [{\citenamefont {Fang}\ \emph {et~al.}(2004)\citenamefont {Fang},
		\citenamefont {Howald}, \citenamefont {Kaneko}, \citenamefont {Greven},\ and\
		\citenamefont {Kapitulnik}}]{fang2004periodic}%
	\BibitemOpen
	\bibfield  {author} {\bibinfo {author} {\bibfnamefont {A.}~\bibnamefont
			{Fang}}, \bibinfo {author} {\bibfnamefont {C.}~\bibnamefont {Howald}},
		\bibinfo {author} {\bibfnamefont {N.}~\bibnamefont {Kaneko}}, \bibinfo
		{author} {\bibfnamefont {M.}~\bibnamefont {Greven}}, \ and\ \bibinfo {author}
		{\bibfnamefont {A.}~\bibnamefont {Kapitulnik}},\ }\href@noop {} {\bibfield
		{journal} {\bibinfo  {journal} {Phys. Rev. B}\ }\textbf {\bibinfo {volume}
			{70}},\ \bibinfo {pages} {214514} (\bibinfo {year} {2004})}\BibitemShut
	{NoStop}%
	\bibitem [{\citenamefont {McElroy}\ \emph {et~al.}(2005)\citenamefont
		{McElroy}, \citenamefont {Lee}, \citenamefont {Slezak}, \citenamefont {Lee},
		\citenamefont {Eisaki}, \citenamefont {Uchida},\ and\ \citenamefont
		{Davis}}]{mcelroy2005atomic}%
	\BibitemOpen
	\bibfield  {author} {\bibinfo {author} {\bibfnamefont {K.}~\bibnamefont
			{McElroy}}, \bibinfo {author} {\bibfnamefont {J.}~\bibnamefont {Lee}},
		\bibinfo {author} {\bibfnamefont {J.}~\bibnamefont {Slezak}}, \bibinfo
		{author} {\bibfnamefont {D.-H.}\ \bibnamefont {Lee}}, \bibinfo {author}
		{\bibfnamefont {H.}~\bibnamefont {Eisaki}}, \bibinfo {author} {\bibfnamefont
			{S.}~\bibnamefont {Uchida}}, \ and\ \bibinfo {author} {\bibfnamefont {J.~C.}\
			\bibnamefont {Davis}},\ }\href@noop {} {\bibfield  {journal} {\bibinfo
			{journal} {Science}\ }\textbf {\bibinfo {volume} {309}},\ \bibinfo {pages}
		{1048} (\bibinfo {year} {2005})}\BibitemShut {NoStop}%
	\bibitem [{\citenamefont {Hoffman}\ \emph {et~al.}(2002)\citenamefont
		{Hoffman}, \citenamefont {McElroy}, \citenamefont {Lee}, \citenamefont
		{Lang}, \citenamefont {Eisaki}, \citenamefont {Uchida},\ and\ \citenamefont
		{Davis}}]{hoffman2002imaging}%
	\BibitemOpen
	\bibfield  {author} {\bibinfo {author} {\bibfnamefont {J.~E.}\ \bibnamefont
			{Hoffman}}, \bibinfo {author} {\bibfnamefont {K.}~\bibnamefont {McElroy}},
		\bibinfo {author} {\bibfnamefont {D.-H.}\ \bibnamefont {Lee}}, \bibinfo
		{author} {\bibfnamefont {K.~M.}\ \bibnamefont {Lang}}, \bibinfo {author}
		{\bibfnamefont {H.}~\bibnamefont {Eisaki}}, \bibinfo {author} {\bibfnamefont
			{S.}~\bibnamefont {Uchida}}, \ and\ \bibinfo {author} {\bibfnamefont {J.~C.}\
			\bibnamefont {Davis}},\ }\href@noop {} {\bibfield  {journal} {\bibinfo
			{journal} {Science}\ }\textbf {\bibinfo {volume} {297}},\ \bibinfo {pages}
		{1148} (\bibinfo {year} {2002})}\BibitemShut {NoStop}%
	\bibitem [{\citenamefont {McElroy}\ \emph {et~al.}(2003)\citenamefont
		{McElroy}, \citenamefont {Simmonds}, \citenamefont {Hoffman}, \citenamefont
		{Lee}, \citenamefont {Orenstein}, \citenamefont {Eisaki}, \citenamefont
		{Uchida},\ and\ \citenamefont {Davis}}]{mcelroy2003relating}%
	\BibitemOpen
	\bibfield  {author} {\bibinfo {author} {\bibfnamefont {K.}~\bibnamefont
			{McElroy}}, \bibinfo {author} {\bibfnamefont {R.~W.}\ \bibnamefont
			{Simmonds}}, \bibinfo {author} {\bibfnamefont {J.~E.}\ \bibnamefont
			{Hoffman}}, \bibinfo {author} {\bibfnamefont {D.-H.}\ \bibnamefont {Lee}},
		\bibinfo {author} {\bibfnamefont {J.}~\bibnamefont {Orenstein}}, \bibinfo
		{author} {\bibfnamefont {H.}~\bibnamefont {Eisaki}}, \bibinfo {author}
		{\bibfnamefont {S.}~\bibnamefont {Uchida}}, \ and\ \bibinfo {author}
		{\bibfnamefont {J.~C.}\ \bibnamefont {Davis}},\ }\href@noop {} {\bibfield
		{journal} {\bibinfo  {journal} {Nature}\ }\textbf {\bibinfo {volume} {422}},\
		\bibinfo {pages} {592} (\bibinfo {year} {2003})}\BibitemShut {NoStop}%
	\bibitem [{\citenamefont {Kohsaka}\ \emph {et~al.}(2008)\citenamefont
		{Kohsaka}, \citenamefont {Taylor}, \citenamefont {Wahl}, \citenamefont
		{Schmidt}, \citenamefont {Lee}, \citenamefont {Fujita}, \citenamefont
		{Alldredge}, \citenamefont {McElroy}, \citenamefont {Lee}, \citenamefont
		{Eisaki}, \citenamefont {Uchida}, \citenamefont {Lee},\ and\ \citenamefont
		{Davis}}]{kohsaka2008cooper}%
	\BibitemOpen
	\bibfield  {author} {\bibinfo {author} {\bibfnamefont {Y.}~\bibnamefont
			{Kohsaka}}, \bibinfo {author} {\bibfnamefont {C.}~\bibnamefont {Taylor}},
		\bibinfo {author} {\bibfnamefont {P.}~\bibnamefont {Wahl}}, \bibinfo {author}
		{\bibfnamefont {A.}~\bibnamefont {Schmidt}}, \bibinfo {author} {\bibfnamefont
			{J.}~\bibnamefont {Lee}}, \bibinfo {author} {\bibfnamefont {K.}~\bibnamefont
			{Fujita}}, \bibinfo {author} {\bibfnamefont {J.~W.}\ \bibnamefont
			{Alldredge}}, \bibinfo {author} {\bibfnamefont {K.}~\bibnamefont {McElroy}},
		\bibinfo {author} {\bibfnamefont {J.}~\bibnamefont {Lee}}, \bibinfo {author}
		{\bibfnamefont {H.}~\bibnamefont {Eisaki}}, \bibinfo {author} {\bibfnamefont
			{S.}~\bibnamefont {Uchida}}, \bibinfo {author} {\bibfnamefont {D.-H.}\
			\bibnamefont {Lee}}, \ and\ \bibinfo {author} {\bibfnamefont {J.~C.}\
			\bibnamefont {Davis}},\ }\href@noop {} {\bibfield  {journal} {\bibinfo
			{journal} {Nature}\ }\textbf {\bibinfo {volume} {454}},\ \bibinfo {pages}
		{1072} (\bibinfo {year} {2008})}\BibitemShut {NoStop}%
	\bibitem [{\citenamefont {Howald}\ \emph
		{et~al.}(2003{\natexlab{a}})\citenamefont {Howald}, \citenamefont {Eisaki},
		\citenamefont {Kaneko},\ and\ \citenamefont
		{Kapitulnik}}]{howald2003coexistence}%
	\BibitemOpen
	\bibfield  {author} {\bibinfo {author} {\bibfnamefont {C.}~\bibnamefont
			{Howald}}, \bibinfo {author} {\bibfnamefont {H.}~\bibnamefont {Eisaki}},
		\bibinfo {author} {\bibfnamefont {N.}~\bibnamefont {Kaneko}}, \ and\ \bibinfo
		{author} {\bibfnamefont {A.}~\bibnamefont {Kapitulnik}},\ }\href@noop {}
	{\bibfield  {journal} {\bibinfo  {journal} {Proceedings of the National
				Academy of Sciences}\ }\textbf {\bibinfo {volume} {100}},\ \bibinfo {pages}
		{9705} (\bibinfo {year} {2003}{\natexlab{a}})}\BibitemShut {NoStop}%
	\bibitem [{\citenamefont {Howald}\ \emph
		{et~al.}(2003{\natexlab{b}})\citenamefont {Howald}, \citenamefont {Eisaki},
		\citenamefont {Kaneko}, \citenamefont {Greven},\ and\ \citenamefont
		{Kapitulnik}}]{howald2003periodic}%
	\BibitemOpen
	\bibfield  {author} {\bibinfo {author} {\bibfnamefont {C.}~\bibnamefont
			{Howald}}, \bibinfo {author} {\bibfnamefont {H.}~\bibnamefont {Eisaki}},
		\bibinfo {author} {\bibfnamefont {N.}~\bibnamefont {Kaneko}}, \bibinfo
		{author} {\bibfnamefont {M.}~\bibnamefont {Greven}}, \ and\ \bibinfo {author}
		{\bibfnamefont {A.}~\bibnamefont {Kapitulnik}},\ }\href@noop {} {\bibfield
		{journal} {\bibinfo  {journal} {Phys. Rev. B}\ }\textbf {\bibinfo {volume}
			{67}},\ \bibinfo {pages} {014533} (\bibinfo {year}
		{2003}{\natexlab{b}})}\BibitemShut {NoStop}%
	\bibitem [{\citenamefont {Wang}\ and\ \citenamefont
		{Lee}(2003)}]{wang2003quasiparticle}%
	\BibitemOpen
	\bibfield  {author} {\bibinfo {author} {\bibfnamefont {Q.-H.}\ \bibnamefont
			{Wang}}\ and\ \bibinfo {author} {\bibfnamefont {D.-H.}\ \bibnamefont {Lee}},\
	}\href@noop {} {\bibfield  {journal} {\bibinfo  {journal} {Phys. Rev. B}\
	}\textbf {\bibinfo {volume} {67}},\ \bibinfo {pages} {020511} (\bibinfo
	{year} {2003})}\BibitemShut {NoStop}%
\bibitem [{\citenamefont {Capriotti}\ \emph {et~al.}(2003)\citenamefont
	{Capriotti}, \citenamefont {Scalapino},\ and\ \citenamefont
	{Sedgewick}}]{capriotti2003wave}%
\BibitemOpen
\bibfield  {author} {\bibinfo {author} {\bibfnamefont {L.}~\bibnamefont
		{Capriotti}}, \bibinfo {author} {\bibfnamefont {D.~J.}\ \bibnamefont
		{Scalapino}}, \ and\ \bibinfo {author} {\bibfnamefont {R.~D.}\ \bibnamefont
		{Sedgewick}},\ }\href@noop {} {\bibfield  {journal} {\bibinfo  {journal}
		{Phys. Rev. B}\ }\textbf {\bibinfo {volume} {68}},\ \bibinfo {pages} {014508}
	(\bibinfo {year} {2003})}\BibitemShut {NoStop}%
\bibitem [{\citenamefont {Zaanen}(2003)}]{zaanen2003superconductivity}%
\BibitemOpen
\bibfield  {author} {\bibinfo {author} {\bibfnamefont {J.}~\bibnamefont
		{Zaanen}},\ }\href@noop {} {\bibfield  {journal} {\bibinfo  {journal}
		{Nature}\ }\textbf {\bibinfo {volume} {422}},\ \bibinfo {pages} {569}
	(\bibinfo {year} {2003})}\BibitemShut {NoStop}%
\bibitem [{\citenamefont {Hanaguri}\ \emph {et~al.}(2007)\citenamefont
	{Hanaguri}, \citenamefont {Kohsaka}, \citenamefont {Davis}, \citenamefont
	{Lupien}, \citenamefont {Yamada}, \citenamefont {Azuma}, \citenamefont
	{Takano}, \citenamefont {Ohishi}, \citenamefont {Ono},\ and\ \citenamefont
	{Takagi}}]{hanaguri2007quasiparticle}%
\BibitemOpen
\bibfield  {author} {\bibinfo {author} {\bibfnamefont {T.}~\bibnamefont
		{Hanaguri}}, \bibinfo {author} {\bibfnamefont {Y.}~\bibnamefont {Kohsaka}},
	\bibinfo {author} {\bibfnamefont {J.~C.}\ \bibnamefont {Davis}}, \bibinfo
	{author} {\bibfnamefont {C.}~\bibnamefont {Lupien}}, \bibinfo {author}
	{\bibfnamefont {I.}~\bibnamefont {Yamada}}, \bibinfo {author} {\bibfnamefont
		{M.}~\bibnamefont {Azuma}}, \bibinfo {author} {\bibfnamefont
		{M.}~\bibnamefont {Takano}}, \bibinfo {author} {\bibfnamefont
		{K.}~\bibnamefont {Ohishi}}, \bibinfo {author} {\bibfnamefont
		{M.}~\bibnamefont {Ono}}, \ and\ \bibinfo {author} {\bibfnamefont
		{H.}~\bibnamefont {Takagi}},\ }\href@noop {} {\bibfield  {journal} {\bibinfo
		{journal} {Nature Physics}\ }\textbf {\bibinfo {volume} {3}},\ \bibinfo
	{pages} {865} (\bibinfo {year} {2007})}\BibitemShut {NoStop}%
\bibitem [{\citenamefont {Allan}\ \emph {et~al.}(2012)\citenamefont {Allan},
	\citenamefont {Rost}, \citenamefont {Mackenzie}, \citenamefont {Xie},
	\citenamefont {Davis}, \citenamefont {Kihou}, \citenamefont {Lee},
	\citenamefont {Iyo}, \citenamefont {Eisaki},\ and\ \citenamefont
	{Chuang}}]{allan2012anisotropic}%
\BibitemOpen
\bibfield  {author} {\bibinfo {author} {\bibfnamefont {M.~P.}\ \bibnamefont
		{Allan}}, \bibinfo {author} {\bibfnamefont {A.~W.}\ \bibnamefont {Rost}},
	\bibinfo {author} {\bibfnamefont {A.~P.}\ \bibnamefont {Mackenzie}}, \bibinfo
	{author} {\bibfnamefont {Y.}~\bibnamefont {Xie}}, \bibinfo {author}
	{\bibfnamefont {J.~C.}\ \bibnamefont {Davis}}, \bibinfo {author}
	{\bibfnamefont {K.}~\bibnamefont {Kihou}}, \bibinfo {author} {\bibfnamefont
		{C.~H.}\ \bibnamefont {Lee}}, \bibinfo {author} {\bibfnamefont
		{A.}~\bibnamefont {Iyo}}, \bibinfo {author} {\bibfnamefont {H.}~\bibnamefont
		{Eisaki}}, \ and\ \bibinfo {author} {\bibfnamefont {T.-M.}\ \bibnamefont
		{Chuang}},\ }\href@noop {} {\bibfield  {journal} {\bibinfo  {journal}
		{Science}\ }\textbf {\bibinfo {volume} {336}},\ \bibinfo {pages} {563}
	(\bibinfo {year} {2012})}\BibitemShut {NoStop}%
\bibitem [{\citenamefont {Allan}\ \emph
	{et~al.}(2013{\natexlab{a}})\citenamefont {Allan}, \citenamefont {Chuang},
	\citenamefont {Massee}, \citenamefont {Xie}, \citenamefont {Ni},
	\citenamefont {Bud’ko}, \citenamefont {Boebinger}, \citenamefont {Wang},
	\citenamefont {Dessau}, \citenamefont {Canfield}, \citenamefont {Golden},\
	and\ \citenamefont {Davis}}]{allan2013anisotropic}%
\BibitemOpen
\bibfield  {author} {\bibinfo {author} {\bibfnamefont {M.~P.}\ \bibnamefont
		{Allan}}, \bibinfo {author} {\bibfnamefont {T.}~\bibnamefont {Chuang}},
	\bibinfo {author} {\bibfnamefont {F.}~\bibnamefont {Massee}}, \bibinfo
	{author} {\bibfnamefont {Y.}~\bibnamefont {Xie}}, \bibinfo {author}
	{\bibfnamefont {N.}~\bibnamefont {Ni}}, \bibinfo {author} {\bibfnamefont
		{S.~L.}\ \bibnamefont {Bud’ko}}, \bibinfo {author} {\bibfnamefont {G.~S.}\
		\bibnamefont {Boebinger}}, \bibinfo {author} {\bibfnamefont {Q.}~\bibnamefont
		{Wang}}, \bibinfo {author} {\bibfnamefont {D.~S.}\ \bibnamefont {Dessau}},
	\bibinfo {author} {\bibfnamefont {P.~C.}\ \bibnamefont {Canfield}}, \bibinfo
	{author} {\bibfnamefont {M.~S.}\ \bibnamefont {Golden}}, \ and\ \bibinfo
	{author} {\bibfnamefont {J.~C.}\ \bibnamefont {Davis}},\ }\href@noop {}
{\bibfield  {journal} {\bibinfo  {journal} {Nature Physics}\ }\textbf
	{\bibinfo {volume} {9}},\ \bibinfo {pages} {220} (\bibinfo {year}
	{2013}{\natexlab{a}})}\BibitemShut {NoStop}%
\bibitem [{\citenamefont {Allan}\ \emph {et~al.}(2015)\citenamefont {Allan},
	\citenamefont {Lee}, \citenamefont {Rost}, \citenamefont {Fischer},
	\citenamefont {Massee}, \citenamefont {Kihou}, \citenamefont {Lee},
	\citenamefont {Iyo}, \citenamefont {Eisaki}, \citenamefont {Chuang},
	\citenamefont {Davis},\ and\ \citenamefont {Kim}}]{allan2015identifying}%
\BibitemOpen
\bibfield  {author} {\bibinfo {author} {\bibfnamefont {M.~P.}\ \bibnamefont
		{Allan}}, \bibinfo {author} {\bibfnamefont {K.}~\bibnamefont {Lee}}, \bibinfo
	{author} {\bibfnamefont {A.~W.}\ \bibnamefont {Rost}}, \bibinfo {author}
	{\bibfnamefont {M.~H.}\ \bibnamefont {Fischer}}, \bibinfo {author}
	{\bibfnamefont {F.}~\bibnamefont {Massee}}, \bibinfo {author} {\bibfnamefont
		{K.}~\bibnamefont {Kihou}}, \bibinfo {author} {\bibfnamefont {C.-H.}\
		\bibnamefont {Lee}}, \bibinfo {author} {\bibfnamefont {A.}~\bibnamefont
		{Iyo}}, \bibinfo {author} {\bibfnamefont {H.}~\bibnamefont {Eisaki}},
	\bibinfo {author} {\bibfnamefont {T.-M.}\ \bibnamefont {Chuang}}, \bibinfo
	{author} {\bibfnamefont {J.~C.}\ \bibnamefont {Davis}}, \ and\ \bibinfo
	{author} {\bibfnamefont {E.~A.}\ \bibnamefont {Kim}},\ }\href@noop {}
{\bibfield  {journal} {\bibinfo  {journal} {Nature Physics}\ }\textbf
	{\bibinfo {volume} {11}},\ \bibinfo {pages} {177} (\bibinfo {year}
	{2015})}\BibitemShut {NoStop}%
\bibitem [{\citenamefont {Lee}\ \emph {et~al.}(2009)\citenamefont {Lee},
	\citenamefont {Allan}, \citenamefont {Wang}, \citenamefont {Farrell},
	\citenamefont {Grigera}, \citenamefont {Baumberger}, \citenamefont {Davis},\
	and\ \citenamefont {Mackenzie}}]{lee2009heavy}%
\BibitemOpen
\bibfield  {author} {\bibinfo {author} {\bibfnamefont {J.}~\bibnamefont
		{Lee}}, \bibinfo {author} {\bibfnamefont {M.~P.}\ \bibnamefont {Allan}},
	\bibinfo {author} {\bibfnamefont {M.~A.}\ \bibnamefont {Wang}}, \bibinfo
	{author} {\bibfnamefont {J.}~\bibnamefont {Farrell}}, \bibinfo {author}
	{\bibfnamefont {S.~A.}\ \bibnamefont {Grigera}}, \bibinfo {author}
	{\bibfnamefont {F.}~\bibnamefont {Baumberger}}, \bibinfo {author}
	{\bibfnamefont {J.~C.}\ \bibnamefont {Davis}}, \ and\ \bibinfo {author}
	{\bibfnamefont {A.~P.}\ \bibnamefont {Mackenzie}},\ }\href@noop {} {\bibfield
	{journal} {\bibinfo  {journal} {Nature Physics}\ }\textbf {\bibinfo {volume}
		{5}},\ \bibinfo {pages} {800} (\bibinfo {year} {2009})}\BibitemShut {NoStop}%
\bibitem [{\citenamefont {Schmidt}\ \emph {et~al.}(2010)\citenamefont
	{Schmidt}, \citenamefont {Hamidian}, \citenamefont {Wahl}, \citenamefont
	{Meier}, \citenamefont {Balatsky}, \citenamefont {Garrett}, \citenamefont
	{Williams}, \citenamefont {Luke},\ and\ \citenamefont
	{Davis}}]{schmidt2010imaging}%
\BibitemOpen
\bibfield  {author} {\bibinfo {author} {\bibfnamefont {A.~R.}\ \bibnamefont
		{Schmidt}}, \bibinfo {author} {\bibfnamefont {M.~H.}\ \bibnamefont
		{Hamidian}}, \bibinfo {author} {\bibfnamefont {P.}~\bibnamefont {Wahl}},
	\bibinfo {author} {\bibfnamefont {F.}~\bibnamefont {Meier}}, \bibinfo
	{author} {\bibfnamefont {A.~V.}\ \bibnamefont {Balatsky}}, \bibinfo {author}
	{\bibfnamefont {J.~D.}\ \bibnamefont {Garrett}}, \bibinfo {author}
	{\bibfnamefont {T.~J.}\ \bibnamefont {Williams}}, \bibinfo {author}
	{\bibfnamefont {G.~M.}\ \bibnamefont {Luke}}, \ and\ \bibinfo {author}
	{\bibfnamefont {J.~C.}\ \bibnamefont {Davis}},\ }\href@noop {} {\bibfield
	{journal} {\bibinfo  {journal} {Nature}\ }\textbf {\bibinfo {volume} {465}},\
	\bibinfo {pages} {570} (\bibinfo {year} {2010})}\BibitemShut {NoStop}%
\bibitem [{\citenamefont {Allan}\ \emph
	{et~al.}(2013{\natexlab{b}})\citenamefont {Allan}, \citenamefont {Massee},
	\citenamefont {Morr}, \citenamefont {Van~Dyke}, \citenamefont {Rost},
	\citenamefont {Mackenzie}, \citenamefont {Petrovic},\ and\ \citenamefont
	{Davis}}]{allan2013imaging}%
\BibitemOpen
\bibfield  {author} {\bibinfo {author} {\bibfnamefont {M.~P.}\ \bibnamefont
		{Allan}}, \bibinfo {author} {\bibfnamefont {F.}~\bibnamefont {Massee}},
	\bibinfo {author} {\bibfnamefont {D.~K.}\ \bibnamefont {Morr}}, \bibinfo
	{author} {\bibfnamefont {J.}~\bibnamefont {Van~Dyke}}, \bibinfo {author}
	{\bibfnamefont {A.~W.}\ \bibnamefont {Rost}}, \bibinfo {author}
	{\bibfnamefont {A.~P.}\ \bibnamefont {Mackenzie}}, \bibinfo {author}
	{\bibfnamefont {C.}~\bibnamefont {Petrovic}}, \ and\ \bibinfo {author}
	{\bibfnamefont {J.~C.}\ \bibnamefont {Davis}},\ }\href@noop {} {\bibfield
	{journal} {\bibinfo  {journal} {Nature Physics}\ }\textbf {\bibinfo {volume}
		{9}},\ \bibinfo {pages} {468} (\bibinfo {year}
	{2013}{\natexlab{b}})}\BibitemShut {NoStop}%
\bibitem [{\citenamefont {Van~Dyke}\ \emph {et~al.}(2014)\citenamefont
	{Van~Dyke}, \citenamefont {Massee}, \citenamefont {Allan}, \citenamefont
	{Davis}, \citenamefont {Petrovic},\ and\ \citenamefont
	{Morr}}]{van2014direct}%
\BibitemOpen
\bibfield  {author} {\bibinfo {author} {\bibfnamefont {J.~S.}\ \bibnamefont
		{Van~Dyke}}, \bibinfo {author} {\bibfnamefont {F.}~\bibnamefont {Massee}},
	\bibinfo {author} {\bibfnamefont {M.~P.}\ \bibnamefont {Allan}}, \bibinfo
	{author} {\bibfnamefont {J.~C.~S.}\ \bibnamefont {Davis}}, \bibinfo {author}
	{\bibfnamefont {C.}~\bibnamefont {Petrovic}}, \ and\ \bibinfo {author}
	{\bibfnamefont {D.~K.}\ \bibnamefont {Morr}},\ }\href@noop {} {\bibfield
	{journal} {\bibinfo  {journal} {Proceedings of the National Academy of
			Sciences}\ }\textbf {\bibinfo {volume} {111}},\ \bibinfo {pages} {11663}
	(\bibinfo {year} {2014})}\BibitemShut {NoStop}%
\bibitem [{\citenamefont {Pereg-Barnea}\ and\ \citenamefont
	{Franz}(2003)}]{pereg2003theory}%
\BibitemOpen
\bibfield  {author} {\bibinfo {author} {\bibfnamefont {T.}~\bibnamefont
		{Pereg-Barnea}}\ and\ \bibinfo {author} {\bibfnamefont {M.}~\bibnamefont
		{Franz}},\ }\href@noop {} {\bibfield  {journal} {\bibinfo  {journal} {Phys.
			Rev. B}\ }\textbf {\bibinfo {volume} {68}},\ \bibinfo {pages} {180506}
	(\bibinfo {year} {2003})}\BibitemShut {NoStop}%
\bibitem [{\citenamefont {Pereg-Barnea}\ and\ \citenamefont
	{Franz}(2005)}]{pereg2005quasiparticle}%
\BibitemOpen
\bibfield  {author} {\bibinfo {author} {\bibfnamefont {T.}~\bibnamefont
		{Pereg-Barnea}}\ and\ \bibinfo {author} {\bibfnamefont {M.}~\bibnamefont
		{Franz}},\ }\href@noop {} {\bibfield  {journal} {\bibinfo  {journal}
		{International Journal of Modern Physics B}\ }\textbf {\bibinfo {volume}
		{19}},\ \bibinfo {pages} {731} (\bibinfo {year} {2005})}\BibitemShut
{NoStop}%
\bibitem [{\citenamefont {Misra}\ \emph {et~al.}(2004)\citenamefont {Misra},
	\citenamefont {Vershinin}, \citenamefont {Phillips},\ and\ \citenamefont
	{Yazdani}}]{misra2004failure}%
\BibitemOpen
\bibfield  {author} {\bibinfo {author} {\bibfnamefont {S.}~\bibnamefont
		{Misra}}, \bibinfo {author} {\bibfnamefont {M.}~\bibnamefont {Vershinin}},
	\bibinfo {author} {\bibfnamefont {P.}~\bibnamefont {Phillips}}, \ and\
	\bibinfo {author} {\bibfnamefont {A.}~\bibnamefont {Yazdani}},\ }\href@noop
{} {\bibfield  {journal} {\bibinfo  {journal} {Phys. Rev. B}\ }\textbf
	{\bibinfo {volume} {70}},\ \bibinfo {pages} {220503} (\bibinfo {year}
	{2004})}\BibitemShut {NoStop}%
\bibitem [{\citenamefont {Bena}\ \emph {et~al.}(2004)\citenamefont {Bena},
	\citenamefont {Chakravarty}, \citenamefont {Hu},\ and\ \citenamefont
	{Nayak}}]{bena2004quasiparticle}%
\BibitemOpen
\bibfield  {author} {\bibinfo {author} {\bibfnamefont {C.}~\bibnamefont
		{Bena}}, \bibinfo {author} {\bibfnamefont {S.}~\bibnamefont {Chakravarty}},
	\bibinfo {author} {\bibfnamefont {J.}~\bibnamefont {Hu}}, \ and\ \bibinfo
	{author} {\bibfnamefont {C.}~\bibnamefont {Nayak}},\ }\href@noop {}
{\bibfield  {journal} {\bibinfo  {journal} {Phys. Rev. B}\ }\textbf {\bibinfo
		{volume} {69}},\ \bibinfo {pages} {134517} (\bibinfo {year}
	{2004})}\BibitemShut {NoStop}%
\bibitem [{\citenamefont {Bena}\ and\ \citenamefont
	{Kivelson}(2005)}]{bena2005quasiparticle}%
\BibitemOpen
\bibfield  {author} {\bibinfo {author} {\bibfnamefont {C.}~\bibnamefont
		{Bena}}\ and\ \bibinfo {author} {\bibfnamefont {S.~A.}\ \bibnamefont
		{Kivelson}},\ }\href@noop {} {\bibfield  {journal} {\bibinfo  {journal}
		{Phys. Rev. B}\ }\textbf {\bibinfo {volume} {72}},\ \bibinfo {pages} {125432}
	(\bibinfo {year} {2005})}\BibitemShut {NoStop}%
\bibitem [{\citenamefont {Fu}(2009)}]{fu2009hexagonal}%
\BibitemOpen
\bibfield  {author} {\bibinfo {author} {\bibfnamefont {L.}~\bibnamefont
		{Fu}},\ }\href@noop {} {\bibfield  {journal} {\bibinfo  {journal} {Phys. Rev.
			Lett.}\ }\textbf {\bibinfo {volume} {103}},\ \bibinfo {pages} {266801}
	(\bibinfo {year} {2009})}\BibitemShut {NoStop}%
\bibitem [{\citenamefont {Roushan}\ \emph {et~al.}(2009)\citenamefont
	{Roushan}, \citenamefont {Seo}, \citenamefont {Parker}, \citenamefont {Hor},
	\citenamefont {Hsieh}, \citenamefont {Qian}, \citenamefont {Richardella},
	\citenamefont {Hasan}, \citenamefont {Cava},\ and\ \citenamefont
	{Yazdani}}]{roushan2009topological}%
\BibitemOpen
\bibfield  {author} {\bibinfo {author} {\bibfnamefont {P.}~\bibnamefont
		{Roushan}}, \bibinfo {author} {\bibfnamefont {J.}~\bibnamefont {Seo}},
	\bibinfo {author} {\bibfnamefont {C.~V.}\ \bibnamefont {Parker}}, \bibinfo
	{author} {\bibfnamefont {Y.~S.}\ \bibnamefont {Hor}}, \bibinfo {author}
	{\bibfnamefont {D.}~\bibnamefont {Hsieh}}, \bibinfo {author} {\bibfnamefont
		{D.}~\bibnamefont {Qian}}, \bibinfo {author} {\bibfnamefont {A.}~\bibnamefont
		{Richardella}}, \bibinfo {author} {\bibfnamefont {M.~Z.}\ \bibnamefont
		{Hasan}}, \bibinfo {author} {\bibfnamefont {R.~J.}\ \bibnamefont {Cava}}, \
	and\ \bibinfo {author} {\bibfnamefont {A.}~\bibnamefont {Yazdani}},\
}\href@noop {} {\bibfield  {journal} {\bibinfo  {journal} {Nature}\ }\textbf
{\bibinfo {volume} {460}},\ \bibinfo {pages} {1106} (\bibinfo {year}
{2009})}\BibitemShut {NoStop}%
\bibitem [{\citenamefont {Guo}\ and\ \citenamefont
	{Franz}(2010)}]{guo2010theory}%
\BibitemOpen
\bibfield  {author} {\bibinfo {author} {\bibfnamefont {H.-M.}\ \bibnamefont
		{Guo}}\ and\ \bibinfo {author} {\bibfnamefont {M.}~\bibnamefont {Franz}},\
}\href@noop {} {\bibfield  {journal} {\bibinfo  {journal} {Phys. Rev. B}\
}\textbf {\bibinfo {volume} {81}},\ \bibinfo {pages} {041102} (\bibinfo
{year} {2010})}\BibitemShut {NoStop}%
\bibitem [{\citenamefont {Crommie}\ \emph {et~al.}(1993)\citenamefont
	{Crommie}, \citenamefont {Lutz},\ and\ \citenamefont
	{Eigler}}]{crommie1993imaging}%
\BibitemOpen
\bibfield  {author} {\bibinfo {author} {\bibfnamefont {M.~F.}\ \bibnamefont
		{Crommie}}, \bibinfo {author} {\bibfnamefont {C.~P.}\ \bibnamefont {Lutz}}, \
	and\ \bibinfo {author} {\bibfnamefont {D.~M.}\ \bibnamefont {Eigler}},\
}\href@noop {} {\bibfield  {journal} {\bibinfo  {journal} {Nature}\ }\textbf
{\bibinfo {volume} {363}},\ \bibinfo {pages} {524} (\bibinfo {year}
{1993})}\BibitemShut {NoStop}%
\bibitem [{\citenamefont {Sprunger}\ \emph {et~al.}(1997)\citenamefont
	{Sprunger}, \citenamefont {Petersen}, \citenamefont {Plummer}, \citenamefont
	{L{\ae}gsgaard},\ and\ \citenamefont {Besenbacher}}]{sprunger1997giant}%
\BibitemOpen
\bibfield  {author} {\bibinfo {author} {\bibfnamefont {P.~T.}\ \bibnamefont
		{Sprunger}}, \bibinfo {author} {\bibfnamefont {L.}~\bibnamefont {Petersen}},
	\bibinfo {author} {\bibfnamefont {E.~W.}\ \bibnamefont {Plummer}}, \bibinfo
	{author} {\bibfnamefont {E.}~\bibnamefont {L{\ae}gsgaard}}, \ and\ \bibinfo
	{author} {\bibfnamefont {F.}~\bibnamefont {Besenbacher}},\ }\href@noop {}
{\bibfield  {journal} {\bibinfo  {journal} {Science}\ }\textbf {\bibinfo
		{volume} {275}},\ \bibinfo {pages} {1764} (\bibinfo {year}
	{1997})}\BibitemShut {NoStop}%
\bibitem [{\citenamefont {Hofmann}\ \emph {et~al.}(1997)\citenamefont
	{Hofmann}, \citenamefont {Briner}, \citenamefont {Doering}, \citenamefont
	{Rust}, \citenamefont {Plummer},\ and\ \citenamefont
	{Bradshaw}}]{hofmann1997anisotropic}%
\BibitemOpen
\bibfield  {author} {\bibinfo {author} {\bibfnamefont {P.}~\bibnamefont
		{Hofmann}}, \bibinfo {author} {\bibfnamefont {B.~G.}\ \bibnamefont {Briner}},
	\bibinfo {author} {\bibfnamefont {M.}~\bibnamefont {Doering}}, \bibinfo
	{author} {\bibfnamefont {H.-P.}\ \bibnamefont {Rust}}, \bibinfo {author}
	{\bibfnamefont {E.~W.}\ \bibnamefont {Plummer}}, \ and\ \bibinfo {author}
	{\bibfnamefont {A.~M.}\ \bibnamefont {Bradshaw}},\ }\href@noop {} {\bibfield
	{journal} {\bibinfo  {journal} {Phys. Rev. Lett.}\ }\textbf {\bibinfo
		{volume} {79}},\ \bibinfo {pages} {265} (\bibinfo {year} {1997})}\BibitemShut
{NoStop}%
\bibitem [{\citenamefont {Petersen}\ \emph {et~al.}(1998)\citenamefont
	{Petersen}, \citenamefont {Sprunger}, \citenamefont {Hofmann}, \citenamefont
	{L{\ae}gsgaard}, \citenamefont {Briner}, \citenamefont {Doering},
	\citenamefont {Rust}, \citenamefont {Bradshaw}, \citenamefont {Besenbacher},\
	and\ \citenamefont {Plummer}}]{petersen1998direct}%
\BibitemOpen
\bibfield  {author} {\bibinfo {author} {\bibfnamefont {L.}~\bibnamefont
		{Petersen}}, \bibinfo {author} {\bibfnamefont {P.~T.}\ \bibnamefont
		{Sprunger}}, \bibinfo {author} {\bibfnamefont {P.}~\bibnamefont {Hofmann}},
	\bibinfo {author} {\bibfnamefont {E.}~\bibnamefont {L{\ae}gsgaard}}, \bibinfo
	{author} {\bibfnamefont {B.~G.}\ \bibnamefont {Briner}}, \bibinfo {author}
	{\bibfnamefont {M.}~\bibnamefont {Doering}}, \bibinfo {author} {\bibfnamefont
		{H.-P.}\ \bibnamefont {Rust}}, \bibinfo {author} {\bibfnamefont {A.~M.}\
		\bibnamefont {Bradshaw}}, \bibinfo {author} {\bibfnamefont {F.}~\bibnamefont
		{Besenbacher}}, \ and\ \bibinfo {author} {\bibfnamefont {E.~W.}\ \bibnamefont
		{Plummer}},\ }\href@noop {} {\bibfield  {journal} {\bibinfo  {journal} {Phys.
			Rev. B}\ }\textbf {\bibinfo {volume} {57}},\ \bibinfo {pages} {R6858}
	(\bibinfo {year} {1998})}\BibitemShut {NoStop}%
\bibitem [{\citenamefont {Zhu}\ \emph {et~al.}(2004)\citenamefont {Zhu},
	\citenamefont {Atkinson},\ and\ \citenamefont {Hirschfeld}}]{zhu2004power}%
\BibitemOpen
\bibfield  {author} {\bibinfo {author} {\bibfnamefont {L.}~\bibnamefont
		{Zhu}}, \bibinfo {author} {\bibfnamefont {W.~A.}\ \bibnamefont {Atkinson}}, \
	and\ \bibinfo {author} {\bibfnamefont {P.~J.}\ \bibnamefont {Hirschfeld}},\
}\href@noop {} {\bibfield  {journal} {\bibinfo  {journal} {Phys. Rev. B}\
}\textbf {\bibinfo {volume} {69}},\ \bibinfo {pages} {060503} (\bibinfo
{year} {2004})}\BibitemShut {NoStop}%
\bibitem [{\citenamefont {Nunner}\ \emph {et~al.}(2006)\citenamefont {Nunner},
	\citenamefont {Chen}, \citenamefont {Andersen}, \citenamefont {Melikyan},\
	and\ \citenamefont {Hirschfeld}}]{nunner2006fourier}%
\BibitemOpen
\bibfield  {author} {\bibinfo {author} {\bibfnamefont {T.~S.}\ \bibnamefont
		{Nunner}}, \bibinfo {author} {\bibfnamefont {W.}~\bibnamefont {Chen}},
	\bibinfo {author} {\bibfnamefont {B.~M.}\ \bibnamefont {Andersen}}, \bibinfo
	{author} {\bibfnamefont {A.}~\bibnamefont {Melikyan}}, \ and\ \bibinfo
	{author} {\bibfnamefont {P.~J.}\ \bibnamefont {Hirschfeld}},\ }\href@noop {}
{\bibfield  {journal} {\bibinfo  {journal} {Phys. Rev. B}\ }\textbf {\bibinfo
		{volume} {73}},\ \bibinfo {pages} {104511} (\bibinfo {year}
	{2006})}\BibitemShut {NoStop}%
\bibitem [{\citenamefont {Vishik}\ \emph {et~al.}(2009)\citenamefont {Vishik},
	\citenamefont {Nowadnick}, \citenamefont {Lee}, \citenamefont {Shen},
	\citenamefont {Moritz}, \citenamefont {Devereaux}, \citenamefont {Tanaka},
	\citenamefont {Sasagawa},\ and\ \citenamefont {Fujii}}]{vishik2009momentum}%
\BibitemOpen
\bibfield  {author} {\bibinfo {author} {\bibfnamefont {I.~M.}\ \bibnamefont
		{Vishik}}, \bibinfo {author} {\bibfnamefont {E.~A.}\ \bibnamefont
		{Nowadnick}}, \bibinfo {author} {\bibfnamefont {W.~S.}\ \bibnamefont {Lee}},
	\bibinfo {author} {\bibfnamefont {Z.~X.}\ \bibnamefont {Shen}}, \bibinfo
	{author} {\bibfnamefont {B.}~\bibnamefont {Moritz}}, \bibinfo {author}
	{\bibfnamefont {T.~P.}\ \bibnamefont {Devereaux}}, \bibinfo {author}
	{\bibfnamefont {K.}~\bibnamefont {Tanaka}}, \bibinfo {author} {\bibfnamefont
		{T.}~\bibnamefont {Sasagawa}}, \ and\ \bibinfo {author} {\bibfnamefont
		{T.}~\bibnamefont {Fujii}},\ }\href@noop {} {\bibfield  {journal} {\bibinfo
		{journal} {Nature Physics}\ }\textbf {\bibinfo {volume} {5}},\ \bibinfo
	{pages} {718} (\bibinfo {year} {2009})}\BibitemShut {NoStop}%
\bibitem [{\citenamefont {Kreisel}\ \emph {et~al.}(2015)\citenamefont
	{Kreisel}, \citenamefont {Choubey}, \citenamefont {Berlijn}, \citenamefont
	{Ku}, \citenamefont {Andersen},\ and\ \citenamefont
	{Hirschfeld}}]{kreisel2015interpretation}%
\BibitemOpen
\bibfield  {author} {\bibinfo {author} {\bibfnamefont {A.}~\bibnamefont
		{Kreisel}}, \bibinfo {author} {\bibfnamefont {P.}~\bibnamefont {Choubey}},
	\bibinfo {author} {\bibfnamefont {T.}~\bibnamefont {Berlijn}}, \bibinfo
	{author} {\bibfnamefont {W.}~\bibnamefont {Ku}}, \bibinfo {author}
	{\bibfnamefont {B.~M.}\ \bibnamefont {Andersen}}, \ and\ \bibinfo {author}
	{\bibfnamefont {P.~J.}\ \bibnamefont {Hirschfeld}},\ }\href@noop {}
{\bibfield  {journal} {\bibinfo  {journal} {Phys. Rev. Lett.}\ }\textbf
	{\bibinfo {volume} {114}},\ \bibinfo {pages} {217002} (\bibinfo {year}
	{2015})}\BibitemShut {NoStop}%
\bibitem [{\citenamefont {Martin}\ \emph {et~al.}(2002)\citenamefont {Martin},
	\citenamefont {Balatsky},\ and\ \citenamefont {Zaanen}}]{martin2002impurity}%
\BibitemOpen
\bibfield  {author} {\bibinfo {author} {\bibfnamefont {I.}~\bibnamefont
		{Martin}}, \bibinfo {author} {\bibfnamefont {A.~V.}\ \bibnamefont
		{Balatsky}}, \ and\ \bibinfo {author} {\bibfnamefont {J.}~\bibnamefont
		{Zaanen}},\ }\href@noop {} {\bibfield  {journal} {\bibinfo  {journal} {Phys.
			Rev. Lett.}\ }\textbf {\bibinfo {volume} {88}},\ \bibinfo {pages} {097003}
	(\bibinfo {year} {2002})}\BibitemShut {NoStop}%
\bibitem [{\citenamefont {Pan}\ \emph {et~al.}(2000)\citenamefont {Pan},
	\citenamefont {Hudson}, \citenamefont {Lang}, \citenamefont {Eisaki},
	\citenamefont {Uchida},\ and\ \citenamefont {Davis}}]{pan2000imaging}%
\BibitemOpen
\bibfield  {author} {\bibinfo {author} {\bibfnamefont {S.~H.}\ \bibnamefont
		{Pan}}, \bibinfo {author} {\bibfnamefont {E.~W.}\ \bibnamefont {Hudson}},
	\bibinfo {author} {\bibfnamefont {K.~M.}\ \bibnamefont {Lang}}, \bibinfo
	{author} {\bibfnamefont {H.}~\bibnamefont {Eisaki}}, \bibinfo {author}
	{\bibfnamefont {S.}~\bibnamefont {Uchida}}, \ and\ \bibinfo {author}
	{\bibfnamefont {J.~C.}\ \bibnamefont {Davis}},\ }\href@noop {} {\bibfield
	{journal} {\bibinfo  {journal} {Nature}\ }\textbf {\bibinfo {volume} {403}},\
	\bibinfo {pages} {746} (\bibinfo {year} {2000})}\BibitemShut {NoStop}%
\bibitem [{\citenamefont {Balatsky}\ \emph {et~al.}(2006)\citenamefont
	{Balatsky}, \citenamefont {Vekhter},\ and\ \citenamefont
	{Zhu}}]{balatsky2006impurity}%
\BibitemOpen
\bibfield  {author} {\bibinfo {author} {\bibfnamefont {A.~V.}\ \bibnamefont
		{Balatsky}}, \bibinfo {author} {\bibfnamefont {I.}~\bibnamefont {Vekhter}}, \
	and\ \bibinfo {author} {\bibfnamefont {J.-X.}\ \bibnamefont {Zhu}},\
}\href@noop {} {\bibfield  {journal} {\bibinfo  {journal} {Rev. Mod. Phys.}\
}\textbf {\bibinfo {volume} {78}},\ \bibinfo {pages} {373} (\bibinfo {year}
{2006})}\BibitemShut {NoStop}%
\bibitem [{\citenamefont {Hudson}\ \emph {et~al.}(2001)\citenamefont {Hudson},
	\citenamefont {Lang}, \citenamefont {Madhavan}, \citenamefont {Pan},
	\citenamefont {Eisaki}, \citenamefont {Uchida},\ and\ \citenamefont
	{Davis}}]{hudson2001interplay}%
\BibitemOpen
\bibfield  {author} {\bibinfo {author} {\bibfnamefont {E.~W.}\ \bibnamefont
		{Hudson}}, \bibinfo {author} {\bibfnamefont {K.~M.}\ \bibnamefont {Lang}},
	\bibinfo {author} {\bibfnamefont {V.}~\bibnamefont {Madhavan}}, \bibinfo
	{author} {\bibfnamefont {S.~H.}\ \bibnamefont {Pan}}, \bibinfo {author}
	{\bibfnamefont {H.}~\bibnamefont {Eisaki}}, \bibinfo {author} {\bibfnamefont
		{S.}~\bibnamefont {Uchida}}, \ and\ \bibinfo {author} {\bibfnamefont {J.~C.}\
		\bibnamefont {Davis}},\ }\href@noop {} {\bibfield  {journal} {\bibinfo
		{journal} {Nature}\ }\textbf {\bibinfo {volume} {411}},\ \bibinfo {pages}
	{920} (\bibinfo {year} {2001})}\BibitemShut {NoStop}%
\bibitem [{\citenamefont {Pan}\ \emph {et~al.}(2001)\citenamefont {Pan},
	\citenamefont {O'Neal}, \citenamefont {Badzey}, \citenamefont {Chamon},
	\citenamefont {Ding}, \citenamefont {Engelbrecht}, \citenamefont {Wang},
	\citenamefont {Eisaki}, \citenamefont {Uchida}, \citenamefont {Gupta},
	\citenamefont {Ng}, \citenamefont {Hudson}, \citenamefont {Lang},\ and\
	\citenamefont {Davis}}]{pan2001microscopic}%
\BibitemOpen
\bibfield  {author} {\bibinfo {author} {\bibfnamefont {S.~H.}\ \bibnamefont
		{Pan}}, \bibinfo {author} {\bibfnamefont {J.~P.}\ \bibnamefont {O'Neal}},
	\bibinfo {author} {\bibfnamefont {R.~L.}\ \bibnamefont {Badzey}}, \bibinfo
	{author} {\bibfnamefont {C.}~\bibnamefont {Chamon}}, \bibinfo {author}
	{\bibfnamefont {H.}~\bibnamefont {Ding}}, \bibinfo {author} {\bibfnamefont
		{J.~R.}\ \bibnamefont {Engelbrecht}}, \bibinfo {author} {\bibfnamefont
		{Z.}~\bibnamefont {Wang}}, \bibinfo {author} {\bibfnamefont {H.}~\bibnamefont
		{Eisaki}}, \bibinfo {author} {\bibfnamefont {S.}~\bibnamefont {Uchida}},
	\bibinfo {author} {\bibfnamefont {A.~K.}\ \bibnamefont {Gupta}}, \bibinfo
	{author} {\bibfnamefont {K.-W.}\ \bibnamefont {Ng}}, \bibinfo {author}
	{\bibfnamefont {E.~W.}\ \bibnamefont {Hudson}}, \bibinfo {author}
	{\bibfnamefont {K.~M.}\ \bibnamefont {Lang}}, \ and\ \bibinfo {author}
	{\bibfnamefont {J.~C.}\ \bibnamefont {Davis}},\ }\href@noop {} {\bibfield
	{journal} {\bibinfo  {journal} {Nature}\ }\textbf {\bibinfo {volume} {413}},\
	\bibinfo {pages} {282} (\bibinfo {year} {2001})}\BibitemShut {NoStop}%
\bibitem [{\citenamefont {Eisaki}\ \emph {et~al.}(2004)\citenamefont {Eisaki},
	\citenamefont {Kaneko}, \citenamefont {Feng}, \citenamefont {Damascelli},
	\citenamefont {Mang}, \citenamefont {Shen}, \citenamefont {Shen},\ and\
	\citenamefont {Greven}}]{eisaki2004effect}%
\BibitemOpen
\bibfield  {author} {\bibinfo {author} {\bibfnamefont {H.}~\bibnamefont
		{Eisaki}}, \bibinfo {author} {\bibfnamefont {N.}~\bibnamefont {Kaneko}},
	\bibinfo {author} {\bibfnamefont {D.~L.}\ \bibnamefont {Feng}}, \bibinfo
	{author} {\bibfnamefont {A.}~\bibnamefont {Damascelli}}, \bibinfo {author}
	{\bibfnamefont {P.~K.}\ \bibnamefont {Mang}}, \bibinfo {author}
	{\bibfnamefont {K.~M.}\ \bibnamefont {Shen}}, \bibinfo {author}
	{\bibfnamefont {Z.-X.}\ \bibnamefont {Shen}}, \ and\ \bibinfo {author}
	{\bibfnamefont {M.}~\bibnamefont {Greven}},\ }\href@noop {} {\bibfield
	{journal} {\bibinfo  {journal} {Phys. Rev. B}\ }\textbf {\bibinfo {volume}
		{69}},\ \bibinfo {pages} {064512} (\bibinfo {year} {2004})}\BibitemShut
{NoStop}%
\bibitem [{\citenamefont {Atkinson}\ \emph {et~al.}(2000)\citenamefont
	{Atkinson}, \citenamefont {Hirschfeld}, \citenamefont {MacDonald},\ and\
	\citenamefont {Ziegler}}]{atkinson2000details}%
\BibitemOpen
\bibfield  {author} {\bibinfo {author} {\bibfnamefont {W.~A.}\ \bibnamefont
		{Atkinson}}, \bibinfo {author} {\bibfnamefont {P.~J.}\ \bibnamefont
		{Hirschfeld}}, \bibinfo {author} {\bibfnamefont {A.~H.}\ \bibnamefont
		{MacDonald}}, \ and\ \bibinfo {author} {\bibfnamefont {K.}~\bibnamefont
		{Ziegler}},\ }\href@noop {} {\bibfield  {journal} {\bibinfo  {journal} {Phys.
			Rev. Lett.}\ }\textbf {\bibinfo {volume} {85}},\ \bibinfo {pages} {3926}
	(\bibinfo {year} {2000})}\BibitemShut {NoStop}%
\bibitem [{\citenamefont {Dalla~Torre}\ \emph {et~al.}(2016)\citenamefont
	{Dalla~Torre}, \citenamefont {Benjamin}, \citenamefont {He}, \citenamefont
	{Dentelski},\ and\ \citenamefont {Demler}}]{dalla2016friedel}%
\BibitemOpen
\bibfield  {author} {\bibinfo {author} {\bibfnamefont {E.~G.}\ \bibnamefont
		{Dalla~Torre}}, \bibinfo {author} {\bibfnamefont {D.}~\bibnamefont
		{Benjamin}}, \bibinfo {author} {\bibfnamefont {Y.}~\bibnamefont {He}},
	\bibinfo {author} {\bibfnamefont {D.}~\bibnamefont {Dentelski}}, \ and\
	\bibinfo {author} {\bibfnamefont {E.}~\bibnamefont {Demler}},\ }\href@noop {}
{\bibfield  {journal} {\bibinfo  {journal} {Phys. Rev. B}\ }\textbf {\bibinfo
		{volume} {93}},\ \bibinfo {pages} {205117} (\bibinfo {year}
	{2016})}\BibitemShut {NoStop}%
\bibitem [{\citenamefont {Sulangi}\ and\ \citenamefont
	{Zaanen}()}]{sulangi2017upcoming}%
\BibitemOpen
\bibfield  {author} {\bibinfo {author} {\bibfnamefont {M.}~\bibnamefont
		{Sulangi}}\ and\ \bibinfo {author} {\bibfnamefont {J.}~\bibnamefont
		{Zaanen}},\ }\href@noop {} {}\bibinfo {note} {(unpublished)}\BibitemShut
{NoStop}%
\bibitem [{\citenamefont {Norman}\ \emph {et~al.}(1995)\citenamefont {Norman},
	\citenamefont {Randeria}, \citenamefont {Ding},\ and\ \citenamefont
	{Campuzano}}]{norman1995phenomenological}%
\BibitemOpen
\bibfield  {author} {\bibinfo {author} {\bibfnamefont {M.~R.}\ \bibnamefont
		{Norman}}, \bibinfo {author} {\bibfnamefont {M.}~\bibnamefont {Randeria}},
	\bibinfo {author} {\bibfnamefont {H.}~\bibnamefont {Ding}}, \ and\ \bibinfo
	{author} {\bibfnamefont {J.~C.}\ \bibnamefont {Campuzano}},\ }\href@noop {}
{\bibfield  {journal} {\bibinfo  {journal} {Phys. Rev. B}\ }\textbf {\bibinfo
		{volume} {52}},\ \bibinfo {pages} {615} (\bibinfo {year} {1995})}\BibitemShut
{NoStop}%
\bibitem [{\citenamefont {Godfrin}(1991)}]{godfrin1991method}%
\BibitemOpen
\bibfield  {author} {\bibinfo {author} {\bibfnamefont {E.~M.}\ \bibnamefont
		{Godfrin}},\ }\href@noop {} {\bibfield  {journal} {\bibinfo  {journal}
		{Journal of Physics: Condensed Matter}\ }\textbf {\bibinfo {volume} {3}},\
	\bibinfo {pages} {7843} (\bibinfo {year} {1991})}\BibitemShut {NoStop}%
\bibitem [{\citenamefont {Reuter}\ and\ \citenamefont
	{Hill}(2012)}]{reuter2012efficient}%
\BibitemOpen
\bibfield  {author} {\bibinfo {author} {\bibfnamefont {M.~G.}\ \bibnamefont
		{Reuter}}\ and\ \bibinfo {author} {\bibfnamefont {J.~C.}\ \bibnamefont
		{Hill}},\ }\href@noop {} {\bibfield  {journal} {\bibinfo  {journal}
		{Computational Science \& Discovery}\ }\textbf {\bibinfo {volume} {5}},\
	\bibinfo {pages} {014009} (\bibinfo {year} {2012})}\BibitemShut {NoStop}%
\bibitem [{\citenamefont {Hod}\ \emph {et~al.}(2006)\citenamefont {Hod},
	\citenamefont {Peralta},\ and\ \citenamefont {Scuseria}}]{hod2006first}%
\BibitemOpen
\bibfield  {author} {\bibinfo {author} {\bibfnamefont {O.}~\bibnamefont
		{Hod}}, \bibinfo {author} {\bibfnamefont {J.~E.}\ \bibnamefont {Peralta}}, \
	and\ \bibinfo {author} {\bibfnamefont {G.~E.}\ \bibnamefont {Scuseria}},\
}\href@noop {} {\bibfield  {journal} {\bibinfo  {journal} {The Journal of
		Chemical Physics}\ }\textbf {\bibinfo {volume} {125}},\ \bibinfo {pages}
{114704} (\bibinfo {year} {2006})}\BibitemShut {NoStop}%
\bibitem [{\citenamefont {Drouvelis}\ \emph {et~al.}(2006)\citenamefont
	{Drouvelis}, \citenamefont {Schmelcher},\ and\ \citenamefont
	{Bastian}}]{drouvelis2006parallel}%
\BibitemOpen
\bibfield  {author} {\bibinfo {author} {\bibfnamefont {P.~S.}\ \bibnamefont
		{Drouvelis}}, \bibinfo {author} {\bibfnamefont {P.}~\bibnamefont
		{Schmelcher}}, \ and\ \bibinfo {author} {\bibfnamefont {P.}~\bibnamefont
		{Bastian}},\ }\href@noop {} {\bibfield  {journal} {\bibinfo  {journal}
		{Journal of Computational Physics}\ }\textbf {\bibinfo {volume} {215}},\
	\bibinfo {pages} {741} (\bibinfo {year} {2006})}\BibitemShut {NoStop}%
\bibitem [{\citenamefont {Petersen}\ \emph {et~al.}(2008)\citenamefont
	{Petersen}, \citenamefont {S{\o}rensen}, \citenamefont {Hansen},
	\citenamefont {Skelboe},\ and\ \citenamefont {Stokbro}}]{petersen2008block}%
\BibitemOpen
\bibfield  {author} {\bibinfo {author} {\bibfnamefont {D.~E.}\ \bibnamefont
		{Petersen}}, \bibinfo {author} {\bibfnamefont {H.~H.~B.}\ \bibnamefont
		{S{\o}rensen}}, \bibinfo {author} {\bibfnamefont {P.~C.}\ \bibnamefont
		{Hansen}}, \bibinfo {author} {\bibfnamefont {S.}~\bibnamefont {Skelboe}}, \
	and\ \bibinfo {author} {\bibfnamefont {K.}~\bibnamefont {Stokbro}},\
}\href@noop {} {\bibfield  {journal} {\bibinfo  {journal} {Journal of
		Computational Physics}\ }\textbf {\bibinfo {volume} {227}},\ \bibinfo {pages}
{3174} (\bibinfo {year} {2008})}\BibitemShut {NoStop}%
\bibitem [{\citenamefont {Wimmer}\ and\ \citenamefont
	{Richter}(2009)}]{wimmer2009optimal}%
\BibitemOpen
\bibfield  {author} {\bibinfo {author} {\bibfnamefont {M.}~\bibnamefont
		{Wimmer}}\ and\ \bibinfo {author} {\bibfnamefont {K.}~\bibnamefont
		{Richter}},\ }\href@noop {} {\bibfield  {journal} {\bibinfo  {journal}
		{Journal of Computational Physics}\ }\textbf {\bibinfo {volume} {228}},\
	\bibinfo {pages} {8548} (\bibinfo {year} {2009})}\BibitemShut {NoStop}%
\bibitem [{\citenamefont {Li}\ and\ \citenamefont
	{Darve}(2012)}]{li2012extension}%
\BibitemOpen
\bibfield  {author} {\bibinfo {author} {\bibfnamefont {S.}~\bibnamefont
		{Li}}\ and\ \bibinfo {author} {\bibfnamefont {E.}~\bibnamefont {Darve}},\
}\href@noop {} {\bibfield  {journal} {\bibinfo  {journal} {Journal of
		Computational Physics}\ }\textbf {\bibinfo {volume} {231}},\ \bibinfo {pages}
{1121} (\bibinfo {year} {2012})}\BibitemShut {NoStop}%
\bibitem [{\citenamefont {Li}\ \emph {et~al.}(2013)\citenamefont {Li},
	\citenamefont {Wu},\ and\ \citenamefont {Darve}}]{li2013fast}%
\BibitemOpen
\bibfield  {author} {\bibinfo {author} {\bibfnamefont {S.}~\bibnamefont
		{Li}}, \bibinfo {author} {\bibfnamefont {W.}~\bibnamefont {Wu}}, \ and\
	\bibinfo {author} {\bibfnamefont {E.}~\bibnamefont {Darve}},\ }\href@noop {}
{\bibfield  {journal} {\bibinfo  {journal} {Journal of Computational
			Physics}\ }\textbf {\bibinfo {volume} {242}},\ \bibinfo {pages} {915}
	(\bibinfo {year} {2013})}\BibitemShut {NoStop}%
\bibitem [{\citenamefont {Kuzmin}\ \emph {et~al.}(2013)\citenamefont {Kuzmin},
	\citenamefont {Luisier},\ and\ \citenamefont {Schenk}}]{kuzmin2013fast}%
\BibitemOpen
\bibfield  {author} {\bibinfo {author} {\bibfnamefont {A.}~\bibnamefont
		{Kuzmin}}, \bibinfo {author} {\bibfnamefont {M.}~\bibnamefont {Luisier}}, \
	and\ \bibinfo {author} {\bibfnamefont {O.}~\bibnamefont {Schenk}},\ }in\
\href@noop {} {\emph {\bibinfo {booktitle} {European Conference on Parallel
			Processing}}}\ (\bibinfo {organization} {Springer},\ \bibinfo {year} {2013})\
pp.\ \bibinfo {pages} {533--544}\BibitemShut {NoStop}%
\bibitem [{\citenamefont {Atkinson}\ \emph {et~al.}(2003)\citenamefont
	{Atkinson}, \citenamefont {Hirschfeld},\ and\ \citenamefont
	{Zhu}}]{atkinson2003quantum}%
\BibitemOpen
\bibfield  {author} {\bibinfo {author} {\bibfnamefont {W.~A.}\ \bibnamefont
		{Atkinson}}, \bibinfo {author} {\bibfnamefont {P.~J.}\ \bibnamefont
		{Hirschfeld}}, \ and\ \bibinfo {author} {\bibfnamefont {L.}~\bibnamefont
		{Zhu}},\ }\href@noop {} {\bibfield  {journal} {\bibinfo  {journal} {Phys.
			Rev. B}\ }\textbf {\bibinfo {volume} {68}},\ \bibinfo {pages} {054501}
	(\bibinfo {year} {2003})}\BibitemShut {NoStop}%
\bibitem [{\citenamefont {Zhu}\ \emph {et~al.}(2003)\citenamefont {Zhu},
	\citenamefont {Atkinson},\ and\ \citenamefont {Hirschfeld}}]{zhu2003two}%
\BibitemOpen
\bibfield  {author} {\bibinfo {author} {\bibfnamefont {L.}~\bibnamefont
		{Zhu}}, \bibinfo {author} {\bibfnamefont {W.~A.}\ \bibnamefont {Atkinson}}, \
	and\ \bibinfo {author} {\bibfnamefont {P.~J.}\ \bibnamefont {Hirschfeld}},\
}\href@noop {} {\bibfield  {journal} {\bibinfo  {journal} {Phys. Rev. B}\
}\textbf {\bibinfo {volume} {67}},\ \bibinfo {pages} {094508} (\bibinfo
{year} {2003})}\BibitemShut {NoStop}%
\bibitem [{\citenamefont {Abrahams}\ and\ \citenamefont
	{Varma}(2000)}]{abrahams2000angle}%
\BibitemOpen
\bibfield  {author} {\bibinfo {author} {\bibfnamefont {E.}~\bibnamefont
		{Abrahams}}\ and\ \bibinfo {author} {\bibfnamefont {C.~M.}\ \bibnamefont
		{Varma}},\ }\href@noop {} {\bibfield  {journal} {\bibinfo  {journal}
		{Proceedings of the National Academy of Sciences}\ }\textbf {\bibinfo
		{volume} {97}},\ \bibinfo {pages} {5714} (\bibinfo {year}
	{2000})}\BibitemShut {NoStop}%
\bibitem [{\citenamefont {Varma}\ and\ \citenamefont
	{Abrahams}(2001)}]{varma2001effective}%
\BibitemOpen
\bibfield  {author} {\bibinfo {author} {\bibfnamefont {C.~M.}\ \bibnamefont
		{Varma}}\ and\ \bibinfo {author} {\bibfnamefont {E.}~\bibnamefont
		{Abrahams}},\ }\href@noop {} {\bibfield  {journal} {\bibinfo  {journal}
		{Phys. Rev. Lett.}\ }\textbf {\bibinfo {volume} {86}},\ \bibinfo {pages}
	{4652} (\bibinfo {year} {2001})}\BibitemShut {NoStop}%
\bibitem [{\citenamefont {Abrahams}\ and\ \citenamefont
	{Varma}(2003)}]{abrahams2003hall}%
\BibitemOpen
\bibfield  {author} {\bibinfo {author} {\bibfnamefont {E.}~\bibnamefont
		{Abrahams}}\ and\ \bibinfo {author} {\bibfnamefont {C.~M.}\ \bibnamefont
		{Varma}},\ }\href@noop {} {\bibfield  {journal} {\bibinfo  {journal} {Phys.
			Rev. B}\ }\textbf {\bibinfo {volume} {68}},\ \bibinfo {pages} {094502}
	(\bibinfo {year} {2003})}\BibitemShut {NoStop}%
\bibitem [{\citenamefont {Huckestein}\ and\ \citenamefont
	{Altland}(2000)}]{huckestein2000quasi}%
\BibitemOpen
\bibfield  {author} {\bibinfo {author} {\bibfnamefont {B.}~\bibnamefont
		{Huckestein}}\ and\ \bibinfo {author} {\bibfnamefont {A.}~\bibnamefont
		{Altland}},\ }\href@noop {} {} (\bibinfo {year} {2000}),\ \Eprint
{http://arxiv.org/abs/cond-mat/0007413} {cond-mat/0007413} \BibitemShut
{NoStop}%
\bibitem [{\citenamefont {Altland}\ \emph {et~al.}(2002)\citenamefont
	{Altland}, \citenamefont {Simons},\ and\ \citenamefont
	{Zirnbauer}}]{altland2002theories}%
\BibitemOpen
\bibfield  {author} {\bibinfo {author} {\bibfnamefont {A.}~\bibnamefont
		{Altland}}, \bibinfo {author} {\bibfnamefont {B.~D.}\ \bibnamefont {Simons}},
	\ and\ \bibinfo {author} {\bibfnamefont {M.~R.}\ \bibnamefont {Zirnbauer}},\
}\href@noop {} {\bibfield  {journal} {\bibinfo  {journal} {Physics Reports}\
}\textbf {\bibinfo {volume} {359}},\ \bibinfo {pages} {283} (\bibinfo {year}
{2002})}\BibitemShut {NoStop}%
\bibitem [{\citenamefont {Nunner}\ \emph {et~al.}(2005)\citenamefont {Nunner},
	\citenamefont {Andersen}, \citenamefont {Melikyan},\ and\ \citenamefont
	{Hirschfeld}}]{nunner2005dopant}%
\BibitemOpen
\bibfield  {author} {\bibinfo {author} {\bibfnamefont {T.~S.}\ \bibnamefont
		{Nunner}}, \bibinfo {author} {\bibfnamefont {B.~M.}\ \bibnamefont
		{Andersen}}, \bibinfo {author} {\bibfnamefont {A.}~\bibnamefont {Melikyan}},
	\ and\ \bibinfo {author} {\bibfnamefont {P.~J.}\ \bibnamefont {Hirschfeld}},\
}\href@noop {} {\bibfield  {journal} {\bibinfo  {journal} {Phys. Rev. Lett.}\
}\textbf {\bibinfo {volume} {95}},\ \bibinfo {pages} {177003} (\bibinfo
{year} {2005})}\BibitemShut {NoStop}%
\bibitem [{\citenamefont {Kivelson}\ \emph {et~al.}(2003)\citenamefont
	{Kivelson}, \citenamefont {Bindloss}, \citenamefont {Fradkin}, \citenamefont
	{Oganesyan}, \citenamefont {Tranquada}, \citenamefont {Kapitulnik},\ and\
	\citenamefont {Howald}}]{kivelson2003detect}%
\BibitemOpen
\bibfield  {author} {\bibinfo {author} {\bibfnamefont {S.~A.}\ \bibnamefont
		{Kivelson}}, \bibinfo {author} {\bibfnamefont {I.~P.}\ \bibnamefont
		{Bindloss}}, \bibinfo {author} {\bibfnamefont {E.}~\bibnamefont {Fradkin}},
	\bibinfo {author} {\bibfnamefont {V.}~\bibnamefont {Oganesyan}}, \bibinfo
	{author} {\bibfnamefont {J.~M.}\ \bibnamefont {Tranquada}}, \bibinfo {author}
	{\bibfnamefont {A.}~\bibnamefont {Kapitulnik}}, \ and\ \bibinfo {author}
	{\bibfnamefont {C.}~\bibnamefont {Howald}},\ }\href@noop {} {\bibfield
	{journal} {\bibinfo  {journal} {Rev. Mod. Phys.}\ }\textbf {\bibinfo {volume}
		{75}},\ \bibinfo {pages} {1201} (\bibinfo {year} {2003})}\BibitemShut
{NoStop}%
\bibitem [{\citenamefont {Pereg-Barnea}\ and\ \citenamefont
	{Franz}(2008)}]{pereg2008magnetic}%
\BibitemOpen
\bibfield  {author} {\bibinfo {author} {\bibfnamefont {T.}~\bibnamefont
		{Pereg-Barnea}}\ and\ \bibinfo {author} {\bibfnamefont {M.}~\bibnamefont
		{Franz}},\ }\href@noop {} {\bibfield  {journal} {\bibinfo  {journal} {Phys.
			Rev. B}\ }\textbf {\bibinfo {volume} {78}},\ \bibinfo {pages} {020509}
	(\bibinfo {year} {2008})}\BibitemShut {NoStop}%
\bibitem [{\citenamefont {Howald}\ \emph {et~al.}(2001)\citenamefont {Howald},
	\citenamefont {Fournier},\ and\ \citenamefont
	{Kapitulnik}}]{howald2001inherent}%
\BibitemOpen
\bibfield  {author} {\bibinfo {author} {\bibfnamefont {C.}~\bibnamefont
		{Howald}}, \bibinfo {author} {\bibfnamefont {P.}~\bibnamefont {Fournier}}, \
	and\ \bibinfo {author} {\bibfnamefont {A.}~\bibnamefont {Kapitulnik}},\
}\href@noop {} {\bibfield  {journal} {\bibinfo  {journal} {Phys. Rev. B}\
}\textbf {\bibinfo {volume} {64}},\ \bibinfo {pages} {100504} (\bibinfo
{year} {2001})}\BibitemShut {NoStop}%
\bibitem [{\citenamefont {Kim}\ \emph {et~al.}(2008)\citenamefont {Kim},
	\citenamefont {Lawler}, \citenamefont {Oreto}, \citenamefont {Sachdev},
	\citenamefont {Fradkin},\ and\ \citenamefont {Kivelson}}]{kim2008theory}%
\BibitemOpen
\bibfield  {author} {\bibinfo {author} {\bibfnamefont {E.-A.}\ \bibnamefont
		{Kim}}, \bibinfo {author} {\bibfnamefont {M.~J.}\ \bibnamefont {Lawler}},
	\bibinfo {author} {\bibfnamefont {P.}~\bibnamefont {Oreto}}, \bibinfo
	{author} {\bibfnamefont {S.}~\bibnamefont {Sachdev}}, \bibinfo {author}
	{\bibfnamefont {E.}~\bibnamefont {Fradkin}}, \ and\ \bibinfo {author}
	{\bibfnamefont {S.~A.}\ \bibnamefont {Kivelson}},\ }\href@noop {} {\bibfield
	{journal} {\bibinfo  {journal} {Phys. Rev. B}\ }\textbf {\bibinfo {volume}
		{77}},\ \bibinfo {pages} {184514} (\bibinfo {year} {2008})}\BibitemShut
{NoStop}%
\bibitem [{\citenamefont {Kim}\ and\ \citenamefont
	{Lawler}(2010)}]{kim2010interference}%
\BibitemOpen
\bibfield  {author} {\bibinfo {author} {\bibfnamefont {E.-A.}\ \bibnamefont
		{Kim}}\ and\ \bibinfo {author} {\bibfnamefont {M.~J.}\ \bibnamefont
		{Lawler}},\ }\href@noop {} {\bibfield  {journal} {\bibinfo  {journal} {Phys.
			Rev. B}\ }\textbf {\bibinfo {volume} {81}},\ \bibinfo {pages} {132501}
	(\bibinfo {year} {2010})}\BibitemShut {NoStop}%
\end{thebibliography}%

\providecommand{\noopsort}[1]{}\providecommand{\singleletter}[1]{#1}%

\end{document}